\documentclass{article}

\usepackage[english]{babel}
\usepackage[T1]{fontenc}
\usepackage[utf8]{inputenc}
\usepackage{booktabs}

\addto{\captionsenglish}{}

\usepackage{amsmath}
\usepackage{lscape}
\usepackage[colorinlistoftodos]{todonotes}
\usepackage{authblk}
\usepackage{siunitx}
\usepackage[margin=10pt,font=small,labelfont=bf,labelsep=endash]{caption} 

\usepackage{graphicx}
\usepackage[font=footnotesize,margin=0.5cm]{subfig} 
\usepackage{array}
\newcolumntype{H}{>{\setbox0=\hbox\bgroup}c<{\egroup}@{}} 
\usepackage{ragged2e}
\newcolumntype{R}[1]{>{\RaggedLeft\arraybackslash}p{#1}} 
\usepackage[flushleft]{threeparttable} 

\usepackage{arydshln}
\usepackage[letterpaper,margin=1in,marginparwidth=0.75in]{geometry}
\usepackage{float} 

\usepackage[backend=biber,style=nature,sorting=none,maxbibnames=99]{biblatex}
\newcommand{\approximately}{\raise.17ex\hbox{$\scriptstyle\sim$}}

\newcommand\lonestar{\texttt{TACC LONESTAR4}}
\newcommand\wrangler{\texttt{TACC WRANGLER}}
\newcommand\jetstream{\texttt{IU JETSTREAM}}
\newcommand\ranger{\texttt{TACC RANGER}}
\newcommand\stampede{\texttt{TACC STAMPEDE}}
\newcommand\stampedetwo{\texttt{TACC STAMPEDE2}}
\newcommand\comet{\texttt{SDSC COMET}}
\newcommand\cometgpu{\texttt{SDSC COMET GPU}}
\newcommand\gordon{\texttt{SDSC GORDON}}
\newcommand\kraken{\texttt{NICS KRAKEN}}
\newcommand{\blacklight}{\texttt{PSC BLACKLIGHT}}
\newcommand{\bridges}{\texttt{PSC BRIDGES}}

\newcommand{\bridgeslarge}{\texttt{PSC BRIDGES LARGE}}
\newcommand{\supermic}{\texttt{CCT LSU SUPERMIC}}
\newcommand{\xstream}{\texttt{STANFORD XSTREAM}}
\newcommand{\osg}{\texttt{OSG}}
\newcommand{\darter}{\texttt{NICS DARTER}}

\title{A Workload Analysis of NSF's Innovative HPC Resources Using XDMoD}

\author{Nikolay A. Simakov}
\author{Joseph P. White}
\author{Robert L. DeLeon}
\author{Steven M. Gallo}
\author{Matthew D. Jones} 
\author{Jeffrey T. Palmer} 
\author{Benjamin Plessinger}
\author{Thomas R. Furlani}
\affil{Center for Computational Research,
University at Buffalo, Buffalo NY USA 14203
}        

\begin{document}
\maketitle

\begin{abstract}
Workload characterization is an integral part of performance analysis of high performance computing (HPC) systems. An understanding of workload properties sheds light on resource utilization and can be used to inform performance optimization both at the software and system configuration levels.  It can provide information on how computational science usage modalities are changing that could potentially aid holistic capacity planning for the wider HPC ecosystem. Here, we report on the results of a detailed workload analysis of the portfolio of supercomputers comprising the NSF Innovative HPC program in order to characterize its past and current workload and look for trends to understand the nature of how the broad portfolio of computational science research is being supported and how it is changing over time.  The workload analysis also sought to illustrate a wide variety of usage patterns and performance requirements for jobs running on these systems.  File system performance, memory utilization and the types of parallelism employed by users (MPI, threads, etc) were also studied for all systems for which job level performance data was available.  

Unless stated otherwise, the analysis covered the date range 2011-07-01 to 2017-09-30, the start date of which coincides with that of the XSEDE program. In addition to job accounting data for almost all of the production systems during this time period, job level performance data was available for \ranger{}, \lonestar{}, \stampede{}, \stampedetwo{}, \supermic{}, \darter{}, \comet{}, and \gordon{}, as described in Appendix \ref{appendix:ResourceCharacteristics}.  Highlights of the analysis are as follows.

\vspace{\baselineskip}
\noindent
\textbf{Trends in Utilization}
\begin{itemize} 

\item
Overall, the allocation utilization of NSF Innovative HPC resources is high, with only 10\% of allocations going unused by researchers.  

\item Since 2011, utilization over most NSF directorates in terms of XD SUs has increased by two orders of magnitude.
\item
The  Mathematical and Physical Sciences (MPS) and Biological Sciences (BIO) Directorates account for about 70\% of XD SUs consumed and this percentage has remained  constant over time.  

\item Molecular Biosciences, Physics, and Materials Research account for half of all XD SUs consumed by Parent Science.  

\item  Behavioral and Neural Sciences and Integrative Biology and Neuroscience account for over 50\% of all jobs run, with the bulk of those on Open Science Grid.  

\item
27\% of XSEDE projects are responsible for 73\% of the utilization.

\end{itemize}

\vspace{\baselineskip}
\noindent
\textbf{Job Characteristics}

\begin{itemize}

\item
Average job size decreased from a high of about 9000 cores in 2011 to about 1000 cores in 2017 (prior to the introduction of \stampedetwo{}).  Several factors contributed to this decrease including, the retirement of \kraken{}, the availability of Blue Waters for capability class computing, improved core performance, and resource policies limiting the maximum core count.

\item 
A very recent trend, with the addition of the \stampedetwo{} Xeon Phi resource, shows increasing percentages for the largest job sizes (8k and larger in core count).

\item 
The percentage of single node jobs (excluding OSG jobs) increased significantly from 5\% of all consumed XD SUs in 2011 to 21\% in 2017.  While the percentage of single node jobs fluctuates from year to year, it is typically greater than 50\% of all jobs run.  In 2017, serial jobs comprised at least 14\% of XD SUs consumed by all single node jobs (which corresponds to 3\% of all XD SUs consumed).

\item
HPC job throughput may be improved through node sharing.

\item
Most parallel jobs efficiently use all or almost all of the allocated CPU cores.  

\item
Many of the most heavily utilized HPC community applications have modest per core memory requirements of less than 2 GB per core.   However, many custom codes developed by researchers require substantially more than 2 GB per core to run.  

\item
The majority of parallel HPC jobs use multiple single-threaded processes rather than multi-threaded processes.

\item
Initial early data from \stampedetwo{} suggests that a significant proportion of jobs are running 32 or fewer processes/threads per compute node and may not yet be making  optimal use of its multicore architecture.

\item
Lustre file system usage as measured by reads, writes and file opens rates is independent of job size.  Total daily aggregate Lustre reads and writes are approximately equal for most resources.

\end{itemize}

\vspace{\baselineskip}
\noindent
\textbf{Applications}
\begin{itemize}
\item
Of identified, non-proprietary applications, the top five in terms of consumed XD SUs are \texttt{NAMD}, \texttt{GROMACS},  
\texttt{CACTUS}, \texttt{LAMMPS}, and \texttt{WRF}.  
\item
Most of the identified applications with the greatest utilization (XD SUs) use less than 1.5 GB per core, with the majority falling in the range of 0.4--1.1 GB per core.   
\item
The top twenty applications in terms of memory use have a per core memory use that lies in the range of 1.3 GB--3.1 GB.   
\item
The average memory usage by application is relatively constant throughout the resource's lifetime.  
\item
Applications show a wide range of interconnect usage, with the more intensive averaging more than 1GB/s of bidirectional bandwidth, bursting considerably higher.
Data indicates, at least within the sampling intervals, that applications are not bandwidth limited.
\end{itemize}

\vspace{\baselineskip}
\noindent
\textbf{Science Gateways}
\begin{itemize}
\item
Gateway usage has increased 5-fold from 2011--2017.  
\item
The Biological (82\%) and Mathematical and Physical Sciences (16\%) directorates account for almost 98\% of Science Gateway XD SUs consumed, up from 70\% in 2011.  
\item
While Science Gateways consume only about 3\% of the total XD SUs, the number of active Gateway users is growing more rapidly than the active HPC users, surpassing the number of active HPC users in 2015. 
\item
Due to the targeted nature of Science Gateways, the gateway job mix in terms of applications run and fields of science is more narrow than that of HPC users. 
\item
The average size of non-Gateway jobs is roughly twice that of Gateway jobs.  However, Gateway jobs tend to run longer than non-gateway jobs.
\end{itemize}

\vspace{\baselineskip}
\noindent
\textbf{Job Submission Patterns \& Over-subscription}

\begin{itemize}
\item
A 10\% to 20\% increase in the size (number of nodes) of current HPC resources would allow 95\% of all back logged jobs to run immediately.  This analysis assumes that user allocations remain at their current level.  
\item
Queue limits in terms of the total number of jobs a user is allowed to have in their queue were found to have no impact on job throughput since the majority of users do not reach the maximum queue limit.
\end{itemize}

\end{abstract}

\newpage

\renewcommand{\baselinestretch}{0.9}\normalsize 
\tableofcontents
\renewcommand{\baselinestretch}{1.0}\normalsize
\newpage
\section{Introduction}
The NSF Innovative HPC program is an integrated (currently through the XSEDE program~\cite{towns2014xsede} and previously TeraGrid and the PACI programs~\cite{reed2003grids,catlett2008teragrid}) collection of state-of-the-art digital resources that enable researchers, engineers, and scholars to conduct computational and data intensive research in a diverse range of disciplines. The resources provided by the program are intended to be technically diverse, reflecting changing and growing use of computation in both the research and education process.  With over 7000 users from more than 800 institutions directly using these systems, this program fulfills an important need in the U.S.\ cyberinfrastructure ecosystem by providing researchers with access to computational and data intensive resources that are beyond the capabilities of most campus based systems.  Evidence of this is given by the following plots which serve to underscore the important and unique role that these resources play in supporting computationally and data intensive research in the U.S.

\begin{figure}[H]
\centering
\subfloat[\label{fig:NSF_CCR}]{%
\includegraphics[width=0.5\textwidth]{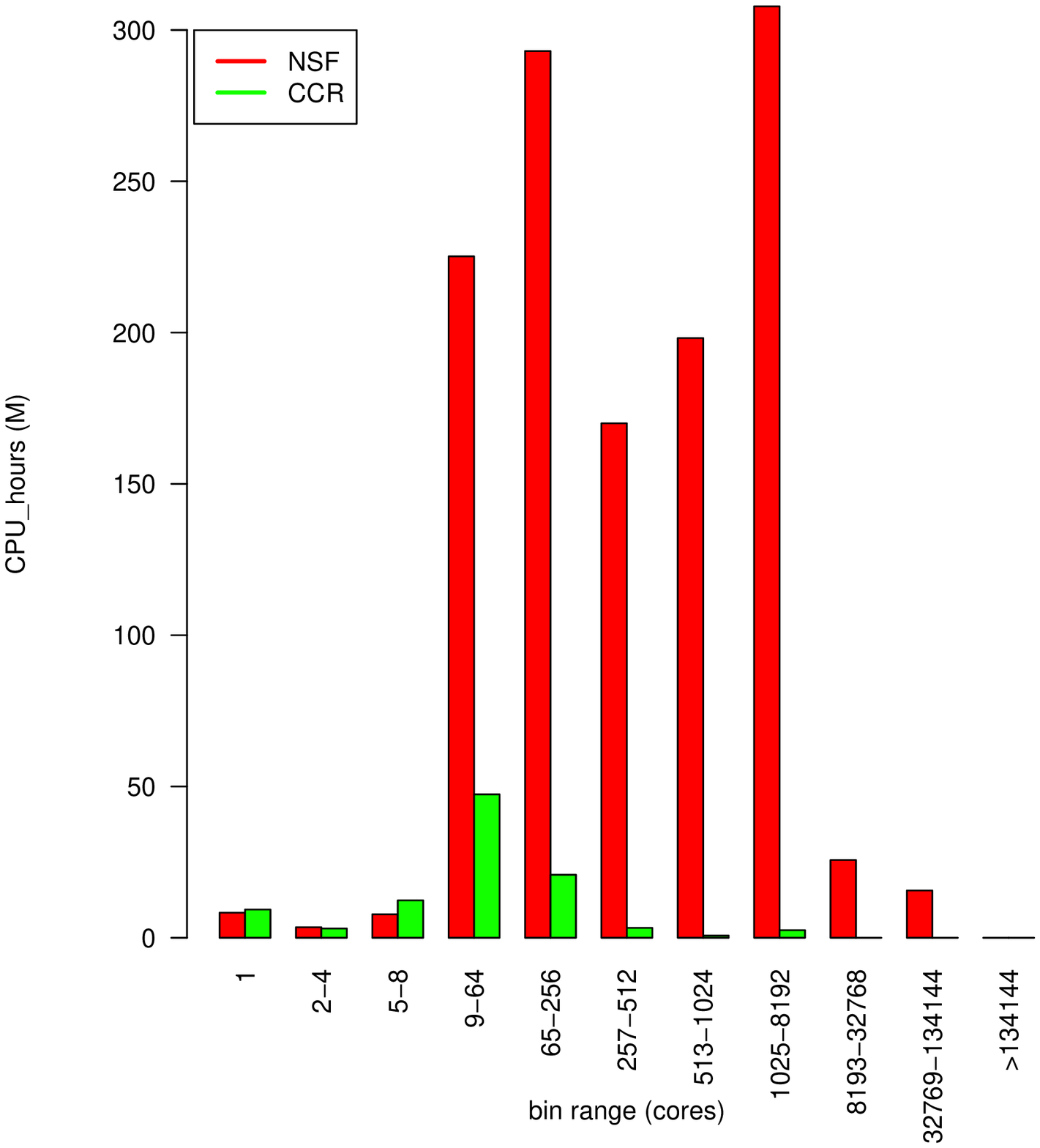}
}
\subfloat[\label{fig:NSF_CCR_BW}]{%
\includegraphics[width=0.5\textwidth]{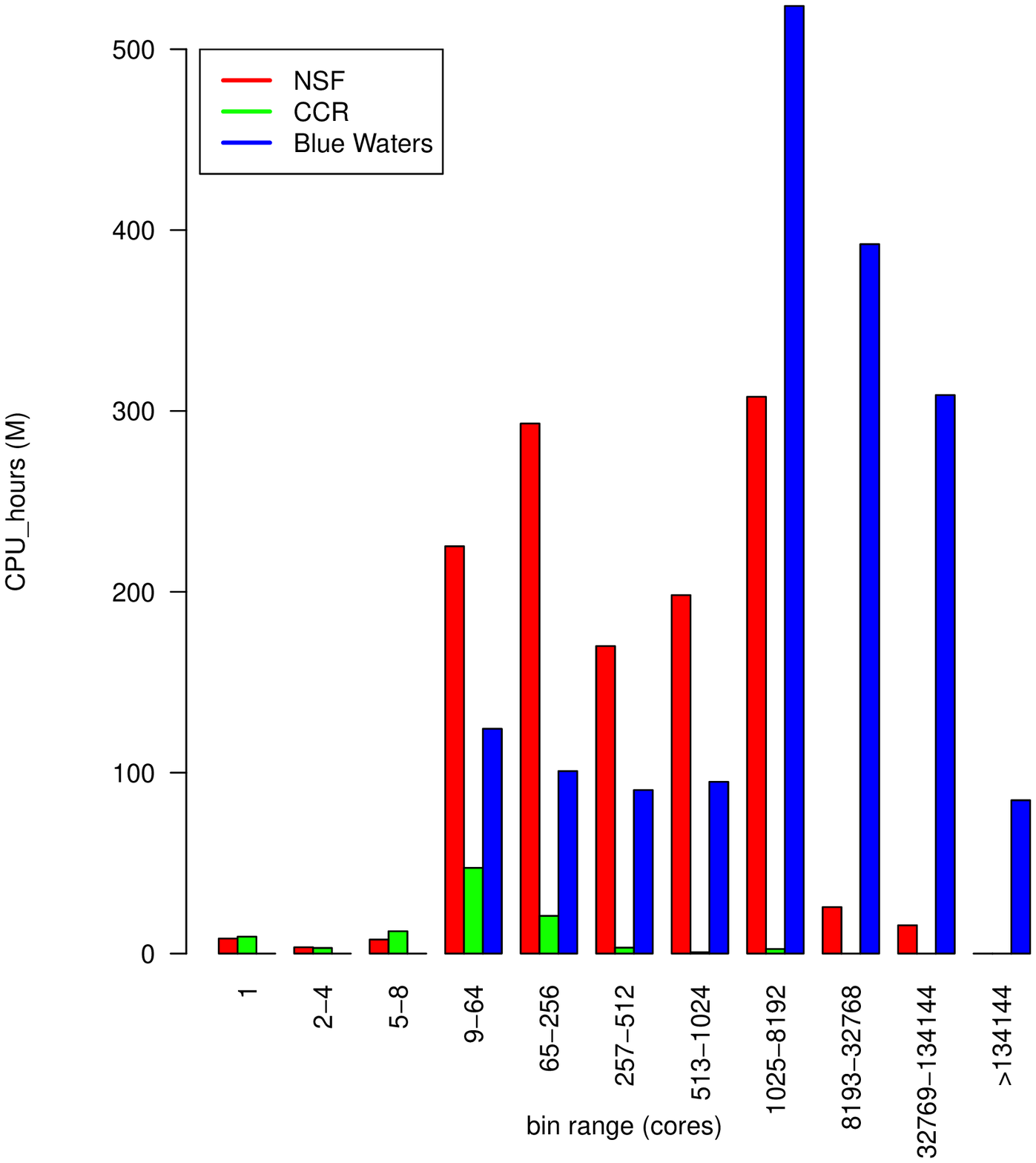}
}
\caption{Comparison of CPU hours versus job size for a typical academic HPC center (CCR) versus the NSF Innovative HPC resources (a) and Blue Waters (b) in 2016.}
\end{figure}

Figure \ref{fig:NSF_CCR} is a comparison of the distribution of CPU hours consumed by job size (number of cores) for the resources provided through the NSF Innovative HPC program versus that for an academic HPC center, namely the Center for Computational Research (CCR) at the University at Buffalo.  Blue Waters, NSF’s capability class HPC system, is also shown in Figure \ref{fig:NSF_CCR_BW} for comparison.  Assuming that CCR’s job mix is representative of that of a typical campus based HPC system, Figure \ref{fig:NSF_CCR} clearly shows the differences in the scale of the jobs run on the two systems, with NSF Innovative HPC resources providing researchers with the ability to consistently run much larger jobs. Figure \ref{fig:NSF_CCR_BW} shows a more complete picture the unique roles that these 3 components of the NSF cyberinfrastructure ecosystem play for scientific research in the U.S.  It should be noted that the NSF Innovative HPC Program also supports a wide range of more specialized services such as visualization, storage, large memory and data intensive facilities not emphasized in the computation-based figure \ref{fig:NSF_CCR_BW}.

\begin{figure}[H]
\centering
\includegraphics[width=0.8\textwidth]{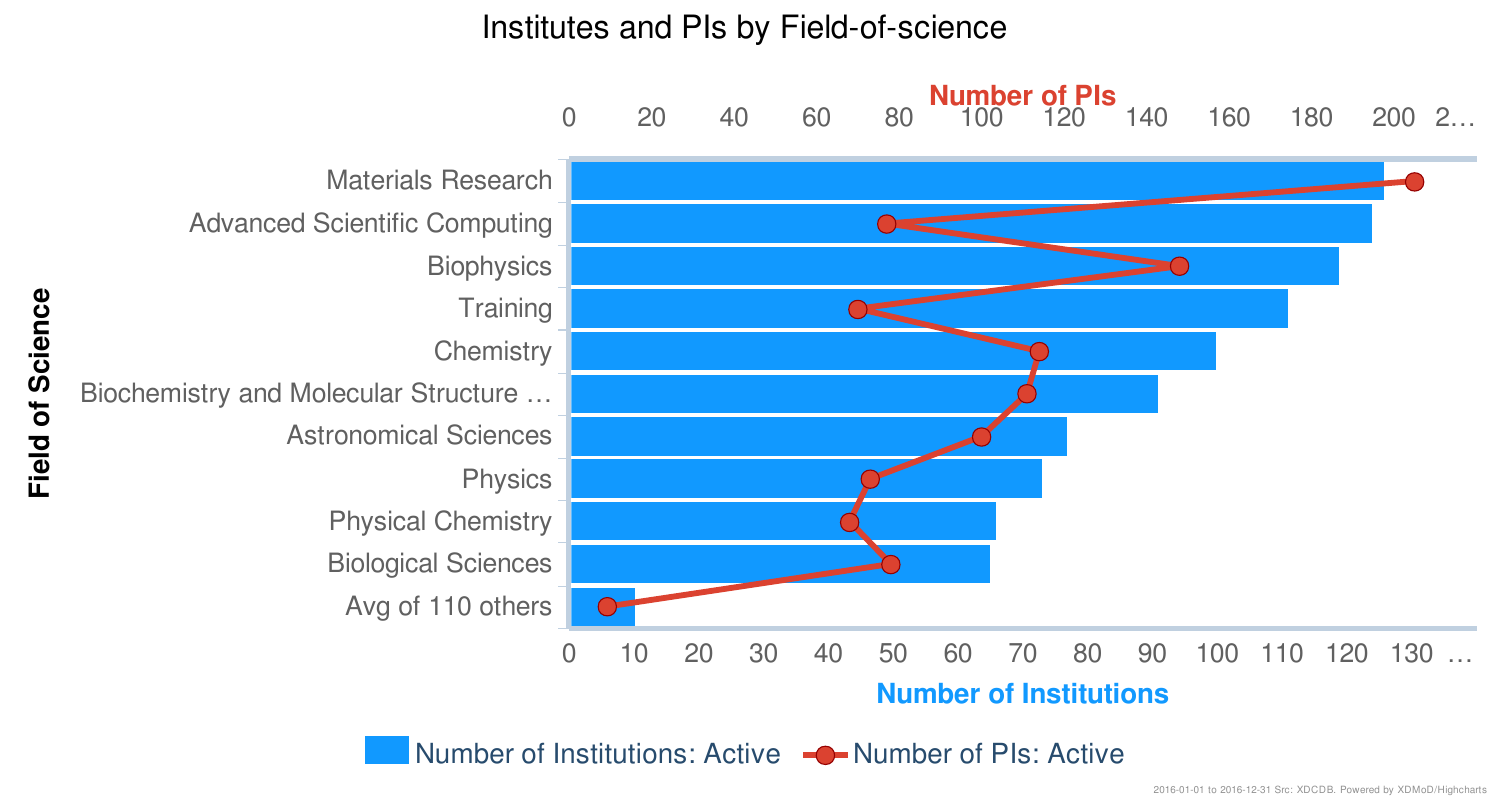}
\caption{\label{fig:xwl_plot_182.pdf}Number of Institutions and PIs using NSF Resources by Field of Science in 2016}
\end{figure}

Figure \ref{fig:xwl_plot_182.pdf} is a plot of the number of institutions and principal investigators using NSF Innovative HPC Program resources broken out by field of science for 2016.  It serves to underscore the extent of the NSF Innovative HPC resources’ user base and the diversity of the areas of science served by this program.

\begin{figure}[H]
\centering
\includegraphics[width=0.8\textwidth]{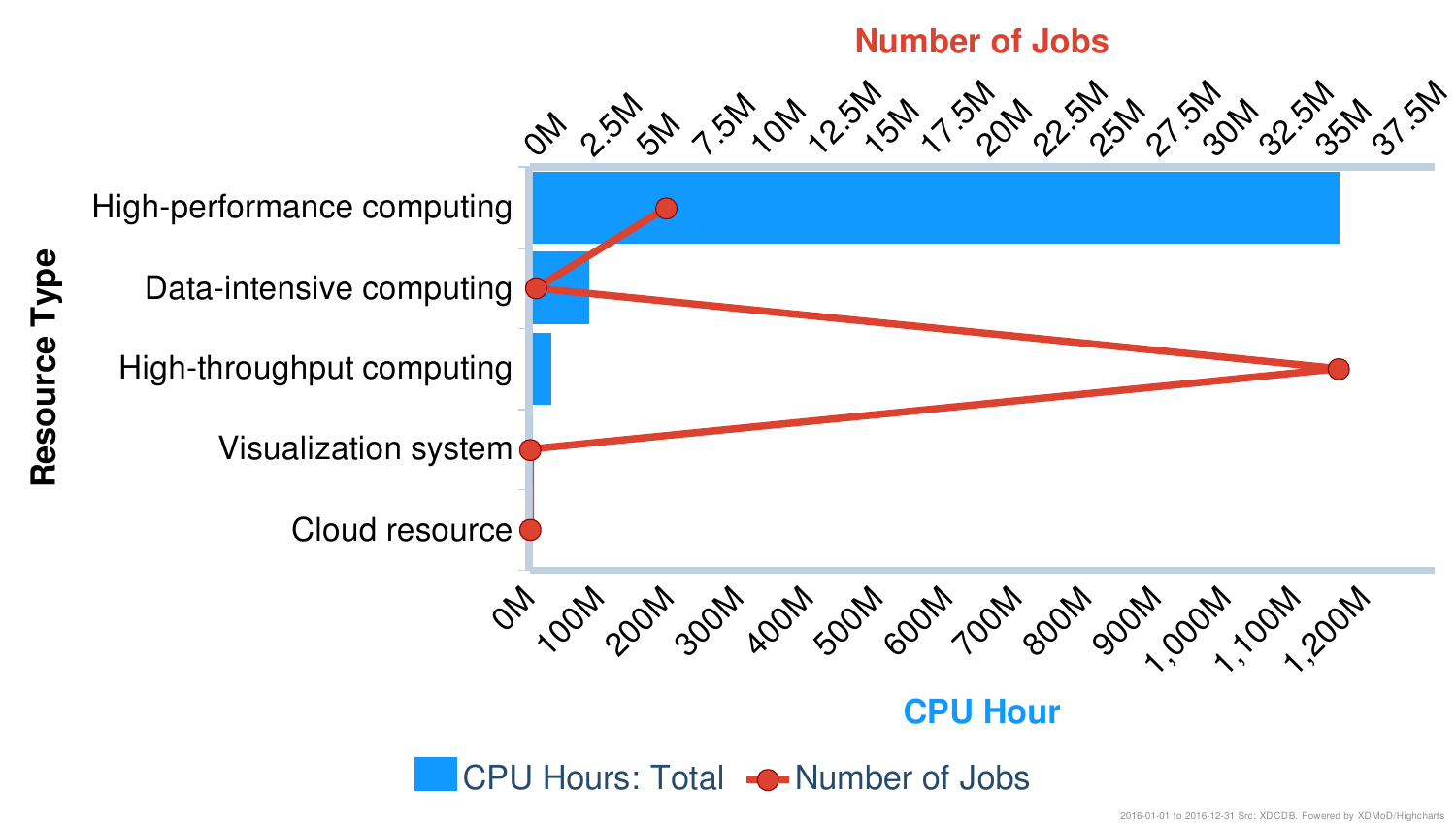}
\caption{\label{fig:ResourceType_2016}CPU hours consumed and number of jobs run on the Types of NSF Innovative HPC Program Resources}
\end{figure}

Figure \ref{fig:ResourceType_2016} is a plot of CPU hours consumed and number of jobs run by resource type and provides an indication of the diversity of digital resources provided by the NSF Innovative HPC program to enable research.   While there is a long history of HPC and HTC computing by U.S.\ researchers, other areas such as cloud and data intensive computing are growing and NSF’s Innovative HPC program is expanding offerings in these areas to help keep pace with the growing demand.  Note that the job type classification scheme employed to create Figure \ref{fig:ResourceType_2016} is based on the resource type classification taken from the XSEDE central database, and not the result of an analysis of individual research awards.   For example, in the past three years all jobs run on \gordon{}, \bridgeslarge{}, and \blacklight{} are classified as Data Intensive.  Likewise, all Open Science Grid jobs are classified as high throughput computing jobs, and all \jetstream{} jobs are classified as cloud jobs.

\begin{figure}
\centering
\includegraphics[width=0.8\textwidth]{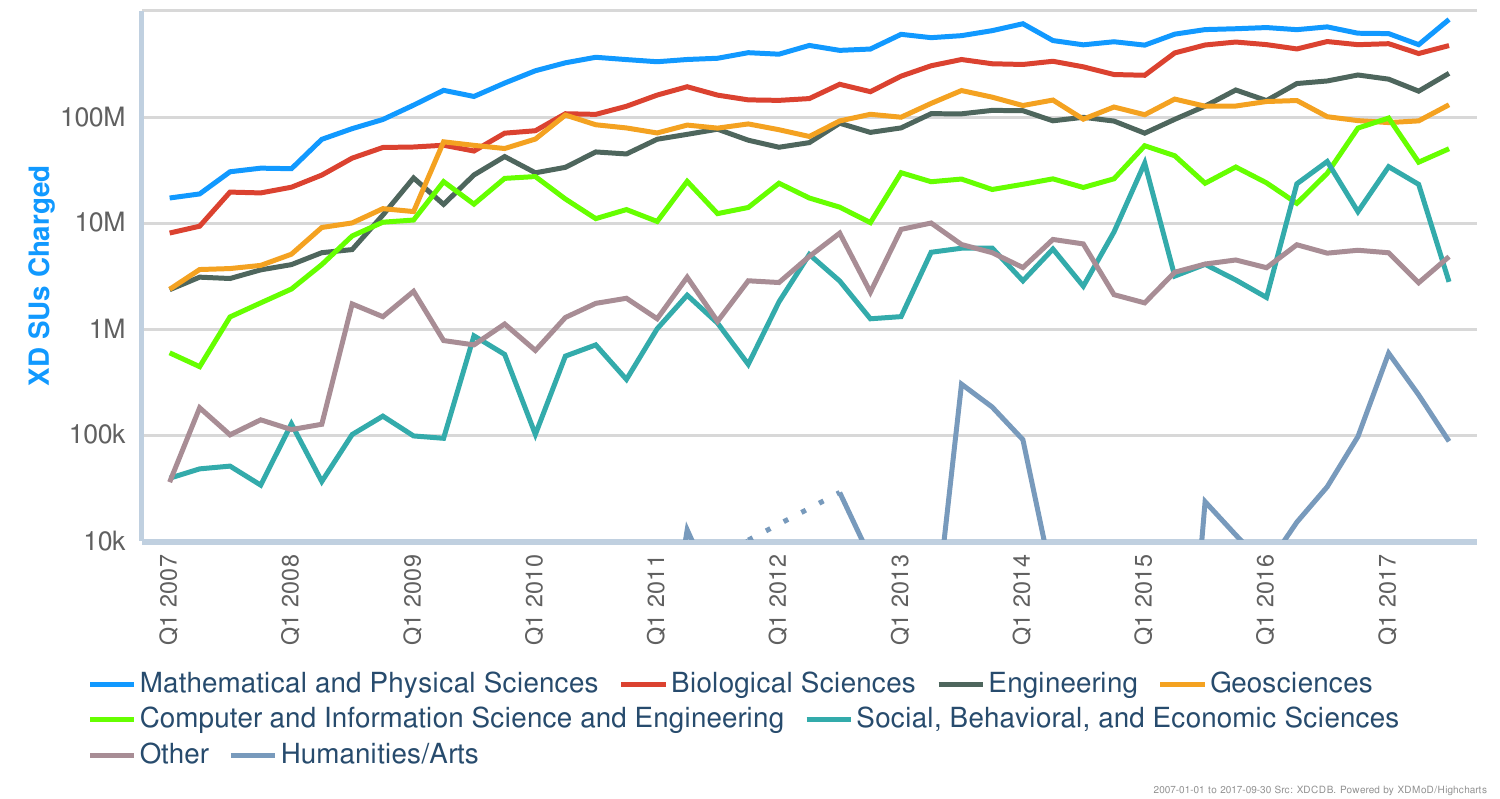}
\caption{\label{fig:SUs_Charged_by_NSF_Directorate}Growth in NSF Directorate Utilization in XD SUs on NSF Innovative HPC Program resources from 2007-2017. The $y$-axis is log scale.}
\end{figure}

Further evidence of the impact and increasing importance that computation and data play in science, including non-traditional areas, can be found in Figure \ref{fig:SUs_Charged_by_NSF_Directorate}, which shows the remarkable growth in XD SUs consumed (approximately 2 orders of magnitude) on the NSF Innovative HPC resources by the NSF directorates over the past 10 years (2007-2017).

Given the pivotal role that the NSF Innovative HPC program plays in the advancement of science and engineering in the United States, it is therefore important to characterize its workload. An understanding of workload properties sheds light on resource utilization and can be used to guide performance optimization both at the software and system configuration levels leading to greater overall throughput for end-users.  Here, we report on the results of a a detailed workload analysis of the portfolio of supercomputers comprising the Innovative HPC program in order to characterize its past and current workload and look for trends to understand the nature of how the broad portfolio of computational science research is being supported and how it is changing over time.

\newpage
\section{Workload Analysis Goals}

This analysis, which was modeled after a similar analysis carried out on Blue Waters~\cite{xdmod-bw2016}, and a 2014 NERSC Workload Analysis~\cite{nersc2014}, builds on prior workload characterizations of TeraGrid and XSEDE carried out by Hart~\cite{Hart2011DeepAndWideMetrics, Hart2011b, Hart2012b}.   The workload analysis targeted the following high-level questions:

\begin{enumerate}
\item What is the proportional mix of disciplines (field of science, parent science, NSF division, NSF directorate) and and how is it changing over the lifetime of the NSF Innovative HPC program, including job sizes/concurrency, and key resource utilization (wall time, memory, GPUs, etc.)?  How does this mix differ among the resources?  Section \ref{sec:trends}.  
\item What if any trends are there in allocations versus awards by resource and discipline?  Section \ref{sec:trends}.
\item What fraction of the portfolio resources are used for data analytics/data intensive computing (hadoop, spark, etc.)?  Is the trend going up or down? How does it vary among the resources? Section \ref{sec:jobs}.
\item How much of the resources usage is consumed by high throughput applications (large numbers of loosely-coupled serial, single and small node count jobs) and gateway applications, and is this changing over time?  Sections \ref{sec:jobs} and \ref{sec:apps}.
\item Are jobs using a larger number of cores over time? Are there differences of job size by discipline or application or Innovative HPC Program resource?  Section \ref{sec:jobs}.
\item Is job memory usage increasing/decreasing over time?  Section \ref{sec:jobs}.
\begin{itemize}
\item Are there specific discipline differences in memory usage?
\item Are there specific memory usage differences in the most used applications?
\item Are there memory usage differences among the resources and does this impact throughput (i.e., result in a bottleneck)?
\end{itemize}
\item Are there important differences among job types, e.g., interactive jobs, gateway jobs, etc.?  Section \ref{sec:gateways}.
\item What are the characteristics of gateway jobs?  Section \ref{sec:gateways}.
\begin{itemize}
\item What are the characteristics and trends for CPU core hours per job, node counts and types, memory and interconnect usage?
\item Are the parallel jobs simply ensembles (many independent jobs)?
\item What is the gateway job distribution by resource?
\item How many new (unique) users are using gateways?
\item Are the usage patterns of gateway users significantly different than traditional HPC users?
\item How does gateway utilization and growth differ by discipline
\end{itemize}
\item Are jobs constrained by resource policy limits such as queue length, user limits or node sharing?  Section \ref{sec:patterns}.
\begin{itemize}
\item How does this vary by resource?
\item Do these limits affect the analysis?
\end{itemize}
\item Are there differences in the job mixes among the resources and if so, how does this impact job throughput?  Section \ref{sec:patterns}.
\begin{itemize}
\item What is the relative proportion of these jobs between systems?
\item Is this the result of allocation decisions, or something else we can determine? 
\end{itemize}
\item How do wait times, throughput and queue length vary among the resources?  Has this changed over time?  Section \ref{sec:patterns}.
\item What is the run-time over-subscription? What is the breakdown by resource and resource type?  Section \ref{sec:patterns}.
\end{enumerate}

\newpage
\section {Workload Analysis Tools}
This analysis leveraged our unique capability in the comprehensive management of HPC systems through XDMoD. 

\subsection {XMS/XDMoD} 
The XD Metrics Service (XMS) for High Performance Computing (HPC) systems supports the comprehensive resource management of XSEDE and the associated computational resources of the NSF Innovative HPC program, and accordingly was employed for the workload analysis presented here.  The analysis was carried out primarily through the XD Metrics on Demand (XDMoD) tool, which was developed under the XMS program. 

The XDMoD tool provides stakeholders with access to  utilization, performance, and quality of service data for high performance computing (HPC) resources~\cite{furlani2013using}. Originally developed to provide independent audit capability for the XSEDE program, XDMoD was later open-sourced and is widely used by university, government, and industrial HPC centers~\cite{Palmer:2015}. XDMoD enables users, managers, and operations staff to monitor, assess and maintain quality of service for their computational resources. To do this, XDMoD harvests data from the various resources and displays the resulting job, usage, and accounting metrics, using the XDMoD portal (\url{https://xdmod.ccr.buffalo.edu}) and its rich array of visual analysis and charting tools.

Metrics provided by XDMoD include: number of jobs, CPU cycles consumed, wait time, and wall time, in addition to many others. These metrics can be broken down in many
different ways such as: field of science, institution, job size, job wall time, NSF directorate, NSF user status, parent science, person, principal investigator, and by resource.  A context-sensitive drill-down capability is available for many charts allowing users to access additional related information simply by clicking inside a plot and then selecting the desired metric to explore.  Another key feature is the ability to make a custom plot of any metric or combination of metrics filtered or aggregated as desired.
 
The XDMoD tool is also designed to preemptively identify under-performing hardware and software by deploying customized, computationally lightweight ``application kernels'' that continuously monitor HPC system performance and reliability from the end-users’ point of view. Accordingly, through XDMoD, system managers have the ability to pro-actively monitor system performance as opposed to having to rely on users to report failures or under-performing hardware and software.  

In addition to the application kernels, which provide quality of service level metrics, XDMoD allows system support personnel and end-users to obtain detailed performance metrics aggregated by node, user, and application for every job run on an HPC cluster. These metrics cover the usage of CPU, memory and cache, and network and input-output devices. Performance metrics are obtained from hardware performance counters and general UNIX/Linux monitoring tools with no need to recompile codes, which is  highly desirable for performance and practical reasons.  In addition to characterizing the cluster’s workload, XDMoD can also be used to identify poorly running applications that can be subsequently tuned to improve performance and overall machine throughput. 

The data sources for XDMoD include job accounting data from the Teragrid/XSEDE Central Database
(XDCDB), allocation information from the XSEDE Resource Allocation System (XRAS)
and job performance data from \texttt{tacc\_stats}~\cite{evans2014comprehensive}. A full
list of data sources with detailed descriptions is provided in Appendix~\ref{appendix:data-sources}.  

In addition to job accounting data for most of the production systems during the period covered by this report, job node level performance data was available for \ranger{}, \lonestar{}, \stampede{}, \stampedetwo{}, \supermic{}, \darter{}, \comet{}, and \gordon{}, as described in Appendix \ref{appendix:ResourceCharacteristics}.  The collection of job and node level performance data has not yet been implemented for the current production systems \bridges{}, \jetstream{}, and \wrangler{}. Accounting data is also not available for \wrangler{}.

\newpage
\section{Trends in Utilization}\label{sec:trends}
\textit{
Goals Addressed in Section
\begin{enumerate}
\item What is the proportional mix of disciplines (field of science, parent science, NSF division, NSF directorate) and how is it changing over the lifetime of the NSF Innovative HPC program, including job sizes/concurrency, and key resource utilization (wall time, memory, GPUs, etc.)?  How does this mix differ among the resources?
\item What if any trends are there in allocations versus awards by resource and discipline?
\end{enumerate}
}

\subsection{Allocations and Usage}
In this section we present historical allocation and usage data for the NSF Innovative HPC Program resources.  Unless stated otherwise, all plots are over the date range 2011-07-01 to 2017-09-30.  The end date was chosen to include the transition from \stampede{} to \stampedetwo{}.  The start date was selected to include 6 years of data (multiple generations of HPC resources) and coincides with the start of the XSEDE program.    

In many of the figures included in this analysis, utilization is measured in terms of XSEDE Service Units (XD SUs) instead of CPU hours.  XD SUs are proportional to CPU hours consumed but contain a scaling factor (for each resource) which attempts to account for changes in CPU processing power over time and between resources, and therefore in theory allows for more meaningful comparisons of utilization between different resources and over different time periods.  See Appendix \ref{appendix:SU Conversions} for greater detail.   Wherever possible, we report \textbf{percent utilization} for XD SUs, which to first order has the advantage of removing the influence of the XD SU scaling factors on the shape of curves when compared to CPU hours, that is plots done in percent utilization using either metric are very similar.

We begin by examining summary statistics for allocations awarded through the end of 2016 in Table~\ref{table:allocation_stats_summary}. For the purpose of this summary we are defining an allocation as the combination of a project and a resource. For example, the project with charge number TG-MCA06N060 that has access to three resources is considered three distinct allocations.  We include only allocations awarded from 2011-07 - 2016-12 so as not to penalize those that were recently awarded and have used only a small portion of the SUs available. Over this period there were 13,226 allocations awarded totaling 7.9B SUs including new allocations, renewals, and supplements to existing allocations. In addition to the total number of allocations made and the mean allocation size for the top 1\%, 5\%, 25\% and bottom 25\% by allocation size, we also examine the number of allocations that did not use any of the SUs that were awarded. The percentage allocation utilization will be defined as the allocation percentage weighted by SU. Note that allocations are awarded in \textbf{SUs local to a particular resource} so the values presented in the these tables are also local SUs and not XSEDE SUs (XD SUs). A local SU is defined as 1 SU = 1 core hour for the resources discussed in this report with the exception of \stampedetwo{} and \wrangler{} where 1 SU = 1 node hour.


\begin{table}[ht]
\centering
\begin{threeparttable}
\caption{Summary statistics for the top 1\%, 5\%, 25\% and bottom 25\% of allocations based on the total number of local SUs allocated for the period 2011-07 - 2016-12.}\label{table:allocation_stats_summary}
\begin{tabular}{l *7r}
\hline
Allocation & \#     & \#     & Allocation        & \multicolumn{4}{c}{Allocation Statistics (Millions of SU)}\\
Size (SU)  &  Alloc & Unused & Utilization\tnote{1} & Total & Mean & Median & Variance \\
\hline
All         & 13,226 & 4,688 & 88.66 & 7,921 & 0.59  & 0.05  & 6,734,474   \\
Top 1\%     & 133    & 0     & 97.04 & 2,816 & 21.10 & 17.10 & 123,533,752 \\
Top 5\%     & 662    & 3     & 96.45 & 5,470 & 8.20  & 5.20  & 69,959,810  \\
Top 25\%    & 3,525  & 594   & 91.16 & 7,441 & 2.10  & 0.64  & 22,151,371  \\
Bottom 25\% & 3,512  & 1,923 & 39.32 & 49    & 0.01  & 0.01  & 130         \\
\hline
\end{tabular}
\begin{tablenotes}
\item[1] Allocation utilization is calculated by weighting the percentage used by the number of local SUs allocated.
\end{tablenotes}
\end{threeparttable}
\end{table}

Table \ref{table:allocation_stats_type} examines the allocations broken down by type. The large TRAC/XRAC/Research allocations have a higher utilization than smaller allocations.  Among the smaller allocations, the Campus Champion allocations have a very low overall utilization of 9\% with 77\% going completely unused.  This is partly a reflection of the open-ended nature of these allocations.

\begin{table}[ht]
\centering
\begin{threeparttable}
\caption{Summary statistics allocation types based on the total number of local SUs allocated for the period 2011-07 to 2016-12.}
\label{table:allocation_stats_type}
\begin{tabular}{l *7r}
\toprule
Allocation & \#     & \#     & Allocation        & \multicolumn{4}{c}{Allocation Statistics (SUs)}\\
Size       &  Alloc & Unused & Utilization\tnote{1} & Total (M) & Mean & Median & Variance (M) \\
\midrule
XRAC               & 2,238 & 208   & 94.00 & 3,981.73 & 1,779,148 & 500,000   & 17,359,208 \\
Research           & 1,706 & 130   & 93.76 & 2,647.90 & 1,552,108 & 541,502   & 8,366,503 \\
TRAC               & 190   & 28    & 87.71 & 386.90   & 2,036,315 & 550,500   & 17,278,763 \\
Startup            & 5,447 & 1,552 & 67.29 & 330.11   & 60,604    & 50,000    & 12,447 \\
Campus Champions   & 3,273 & 2,519 & 9.00  & 267.69   & 81,788    & 50,000    & 11,173 \\
Staff              & 39    & 15    & 53.47 & 69.27    & 1,776,105 & 1,200,000 & 17,426,190 \\
Educational        & 525   & 124   & 59.53 & 31.80    & 60,570    & 40,000    & 7,445 \\
Discretionary      & 5     & 0     & 23.26 & 11.90    & 2,380,000 & 700,000   & 7,827,000 \\
XSEDE2 Staff Alloc & 96    & 59    & 13.78 & 3.76     & 39,142    & 10,000    & 7,593 \\
Software Testbeds  & 5     & 5     & 0.00  & 0.06     & 12,000    & 10,000    & 20 \\
\bottomrule
\end{tabular}
\begin{tablenotes}
\item[1] Allocation utilization is calculated by weighting the percentage used by the number of local SUs allocated.
\end{tablenotes}
\end{threeparttable}
\end{table}

Table \ref{table:allocaton_stats_discipline} shows allocations broken down by discipline, using the NSF directorate when available. The top 4 disciplines comprise 92.5\% of the total SUs awarded (7.3M).  Of note is Social, Behavioral, and Economic Sciences (SBE) with a 248\% allocation utilization, which is largely attributed to a single 3.0M SU allocation (TG-IBN130001 on \osg{}) overcharging their account by more than 2000\%. This is likely due to policies at the SP level allowing users to overcharge the allocation. Also of note is that over 60\% of allocations in all other disciplines go completely unused, although the average size of the unused allocations is small (73,734 SU).

\begin{table}[ht]
\centering
\begin{threeparttable}
\caption{Summary statistics for allocations by discipline for the period 2011-07 to 2016-12 in local SUs.}
\label{table:allocaton_stats_discipline}
\begin{tabular}{l *7r}
\toprule
Discipline & \#    & \#     & Allocation      & \multicolumn{4}{c}{Allocation Statistics (SUs)}\\
           & Alloc & Unused & Utilization\tnote{1} & Total (M) & Mean & Median & Variance (M) \\
\midrule
MPS              & 4,039 & 785   & 92.66  & 3,738 & 925,461 & 60,000 & 10,917,846 \\
BIO              & 2,537 & 623   & 94.12  & 1,926 & 759,035 & 50,000 & 8,349,869 \\
GEO              & 881   & 219   & 82.36  & 874   & 991,824 & 75,000 & 14,630,775 \\
ENG              & 1,201 & 263   & 87.12  & 790   & 657,697 & 60,380 & 6,190,157 \\
CIE              & 2,269 & 1,194 & 55.67  & 311   & 136,929 & 50,000 & 426,908 \\
Training         & 1,708 & 1,286 & 12.91  & 153   & 89,631  & 50,000 & 31,761 \\
Center Sys Staff & 166   & 114   & 49.12  & 76    & 460,467 & 50,000 & 4,550,722 \\
SBE              & 246   & 105   & 248.36 & 43    & 173,986 & 50,000 & 252,561 \\
Sci and Eng Edu  & 87    & 45    & 24.68  & 5     & 62,776  & 44,800 & 4,611 \\
Other            & 43    & 33    & 19.86  & 2     & 56,140  & 50,000 & 2,800 \\
HUA              & 40    & 18    & 19.22  & 2     & 50,226  & 50,000 & 1,966 \\
\bottomrule
\end{tabular}
\begin{tablenotes}
\item[1] Allocation utilization is calculated by weighting the percentage used by the number of local SUs allocated.
\end{tablenotes}
\end{threeparttable}
\end{table}

Allocation statistics by resource are shown in Table \ref{table:allocation_stats_resource} where we can see that all resources providing at least 100M total local SUs (from 2011-07 to 2016-12) have at least an 80\% allocation utilization rate with \stampede{}, \kraken{}, and \ranger{}, having over 90\%.

\begin{table}[ht]
\centering
\begin{threeparttable}
\caption{Summary statistics for allocations by resource for the period 2011-07 to 2016-12 in local SUs.}
\label{table:allocation_stats_resource}
{\small
\begin{tabular}{l *7r}
\toprule
Resource & \#    & \#     & Allocation        & \multicolumn{4}{c}{Allocation Statistics (SUs)}\\
         & Alloc & Unused & Utilization\tnote{1} & Total (M) & Mean & Median & Variance (M) \\
\midrule
TACC-STAMPEDE     & 3,113 & 373 & 94.23  & 2,758.48 & 886,116   & 100,000   & 5,923,008 \\
NICS-KRAKEN       & 770   & 151 & 90.85  & 1,952.77 & 2,536,061 & 200,000   & 38,039,724 \\
SDSC-COMET        & 1,187 & 239 & 85.81  & 684.05   & 576,285   & 50,000    & 3,008,691 \\
TACC-RANGER       & 551   & 54  & 94.82  & 506.79   & 919,767   & 200,000   & 4,138,302 \\
SDSC-GORDON       & 1,268 & 425 & 83.62  & 493.96   & 389,561   & 50,000    & 2,217,706 \\
SDSC-TRESTLES     & 780   & 264 & 85.05  & 229.01   & 293,601   & 50,000    & 946,801 \\
TACC-LONESTAR4    & 714   & 222 & 79.97  & 179.16   & 250,925   & 50,000    & 367,455 \\
PSC-BRIDGES       & 519   & 175 & 80.15  & 167.40   & 322,552   & 50,000    & 1,646,446 \\
NICS-DARTER       & 70    & 15  & 86.75  & 165.02   & 2,357,490 & 1,302,481 & 8,860,043 \\
OSG               & 463   & 403 & 110.79 & 134.02   & 289,453   & 200,000   & 194,997 \\
PSC-BLACKLIGHT    & 1,042 & 339 & 66.64  & 97.34    & 93,418    & 30,000    & 65,403 \\
CCT-LSU-SUPERMIC  & 400   & 263 & 63.83  & 70.63    & 176,578   & 50,000    & 187,884 \\
PURDUE-STEELE     & 258   & 133 & 70.55  & 69.99    & 271,267   & 135,000   & 1,038,079 \\
TG-STAFF          & 33    & 11  & 53.52  & 69.19    & 2,096,518 & 1,200,000 & 20,005,400 \\
PURDUE-CONDOR     & 135   & 111 & 11.32  & 30.89    & 228,847   & 200,000   & 233,999 \\
TACC-MAVERICK2    & 429   & 281 & 55.08  & 28.41    & 66,216    & 3,000     & 100,425 \\
IU-JETSTREAM    & 254   & 147 & 52.84  & 27.49    & 108,235   & 50,000    & 45,870 \\
TACC-LONGHORN     & 244   & 172 & 32.49  & 16.71    & 68,474    & 5,000     & 98,654 \\
NICS-NAUTILUS     & 214   & 132 & 32.89  & 13.33    & 62,298    & 30,000    & 14,746 \\
GATECH-KEENELAND  & 104   & 34  & 79.37  & 11.37    & 109,342   & 30,000    & 33,915 \\
NCSA-FORGE        & 145   & 84  & 20.20  & 7.81     & 53,892    & 30,000    & 6,932 \\
TACC-SPUR         & 98    & 83  & 1.96   & 2.06     & 20,994    & 20,000    & 488 \\
PSC-GREENFIELD    & 134   & 106 & 10.52  & 1.74     & 12,978    & 4,000     & 2,520 \\
NCSA-EMBER        & 24    & 17  & 10.18  & 1.71     & 71,157    & 30,000    & 17,117 \\
Stanford-XSTREAM  & 102   & 70  & 57.85  & 1.46     & 14,289    & 5,000     & 1,669 \\
PSC-BRIDGES-LARGE & 210   & 151 & 25.82  & 1.24     & 5,923     & 1,000     & 1,049 \\
IU-MASON          & 33    & 33  & 0.00   & 0.73     & 22,236    & 15,000    & 654 \\
SDSC-DASH         & 29    & 24  & 1.09   & 0.57     & 19,653    & 30,000    & 148 \\
TACC-WRANGLER     & 90    & 55  & 12.50   & 0.56     & 6,254     & 1,000     & 366 \\
GATECH-KIDS       & 45    & 23  & 51.83  & 0.39     & 8,703     & 10,000    & 22 \\
NCSA-LINCOLN      & 12    & 10  & 0.46   & 0.23     & 19,333    & 15,000    & 216 \\
PURDUE-WISPY      & 14    & 14  & 0.00   & 0.22     & 15,793    & 10,000    & 250 \\
\bottomrule
\end{tabular}
}
\begin{tablenotes}
\item[1] Allocation utilization is calculated by weighting the percentage used by the number of local SUs allocated.
\end{tablenotes}
\end{threeparttable}
\end{table}

\begin{figure}[htb]
\centering
\includegraphics [width=\textwidth] {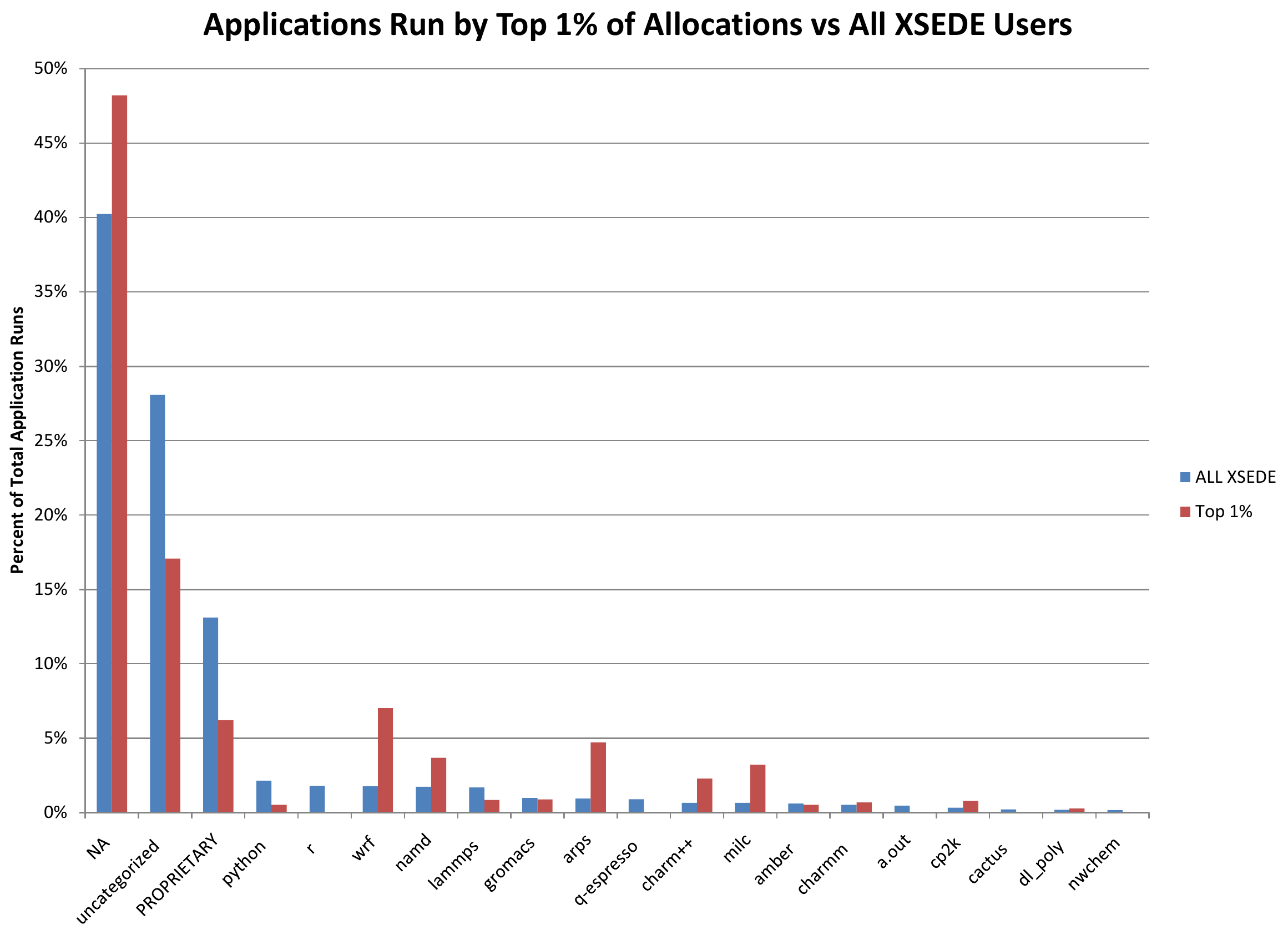}
\caption{\label{fig:applications_run_by_top_1pct}A comparison of the most executed codes for top 1\% of allocations versus all of XSEDE. Note that the top 1\% of allocations utilize community codes such as \texttt{WRF}, \texttt{ARPS}, \texttt{NAMD}, \texttt{CHARM++}, and \texttt{MILC} more heavily than the general XSEDE user base.}
\end{figure}

We can study the characteristics of the users with the largest allocations versus that of the average user to see if there are significant differences in their utilization patterns.  For example, are there differences in the average job size, job length or the applications they use?  To answer these questions, we compared the usage of all XSEDE users to those running under one of the top 1\% of allocations by local SU.  We note that the average job size is 88 cores for general XSEDE users and 136 cores for users in the top 1\%. Job durations do not change significantly between the two groups with general XSEDE users averaging 3.4 hours  per job compared to 3.3 hours for the top 1\%. However, in terms of applications, users in the top 1\% are more likely to run known community codes such as \texttt{WRF}, \texttt{ARPS}, \texttt{NAMD}, \texttt{CHARM++}, and \texttt{MILC} and less likely to run  uncategorized or other custom-developed codes as shown in Figure \ref{fig:applications_run_by_top_1pct} (blue columns are for all of XSEDE and red columns are for the top 1\%). 

\begin{figure}[H]
\centering
\includegraphics{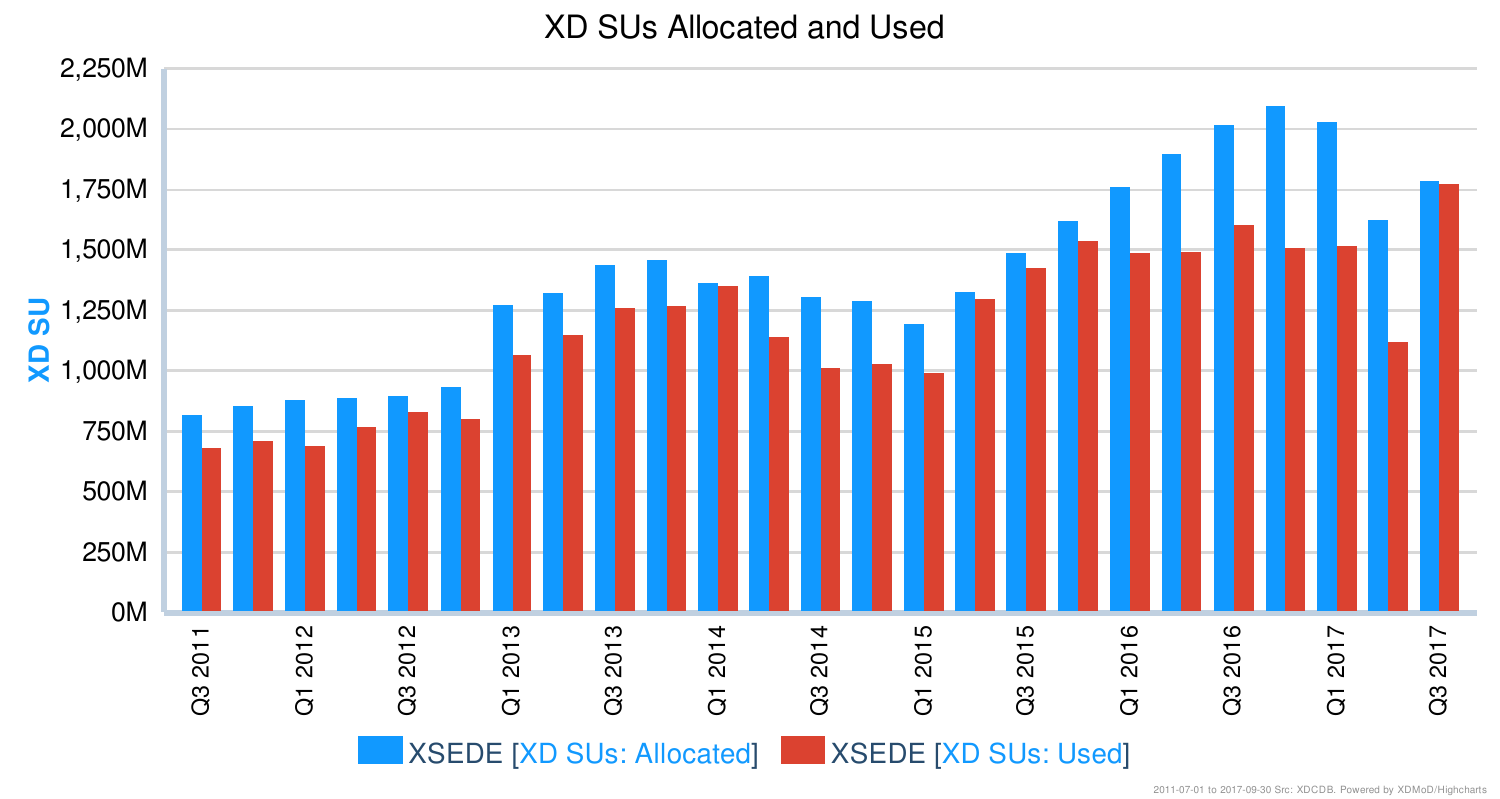}
\caption{\label{fig:XD_SUs_Allocated}XD SUs Allocated and Used for the period 2011-07 to 2017-09.}
\end{figure}

Figure \ref{fig:XD_SUs_Allocated}, which is a historical plot of allocated and used XD SUs, shows an upward trend in allocation and consumed XD SUs reflecting the NSF Innovative HPC program's continued effort to provide sufficient resources to meet the ever increasing computational and data needs of U.S. researchers.   As we demonstrate below, downward trends typically coincide with end-of-life resources going off-line prior to the full integration of new resources.   

\begin{figure}[H]
\centering
\includegraphics{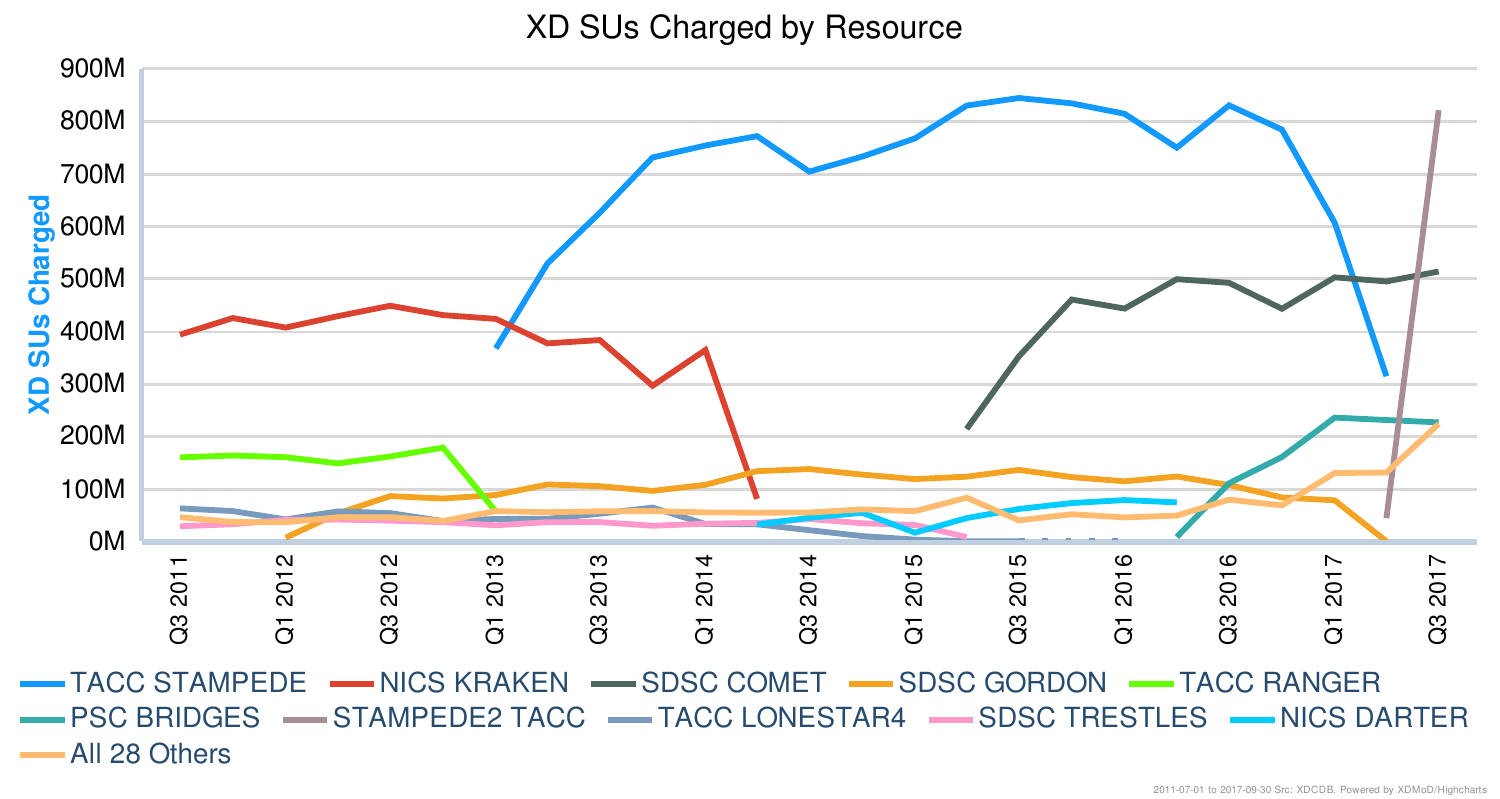}
\caption{\label{fig:SUs_by_Resource}XD SUs Charged by Resource for the period 2011-07 to 2017-09.}
\end{figure}

Figure \ref{fig:SUs_by_Resource}, which shows the time history of XD SUs charged by resource, provides a time-line for the NSF Innovative HPC Program resources.  The transition from \ranger{} to \stampede{} and most recently to \stampedetwo{} is shown, as are the recent additions of \bridges{} and \comet{}.   It is now evident that the 2014 downward trend in allocated and used XD SUs in Figure \ref{fig:XD_SUs_Allocated} is due to \kraken{} going off-line.  Similarly, the increase in XD SUs beginning in 2015 is attributable to \comet{} coming on-line.   

\begin{figure}[H]
\centering
\includegraphics{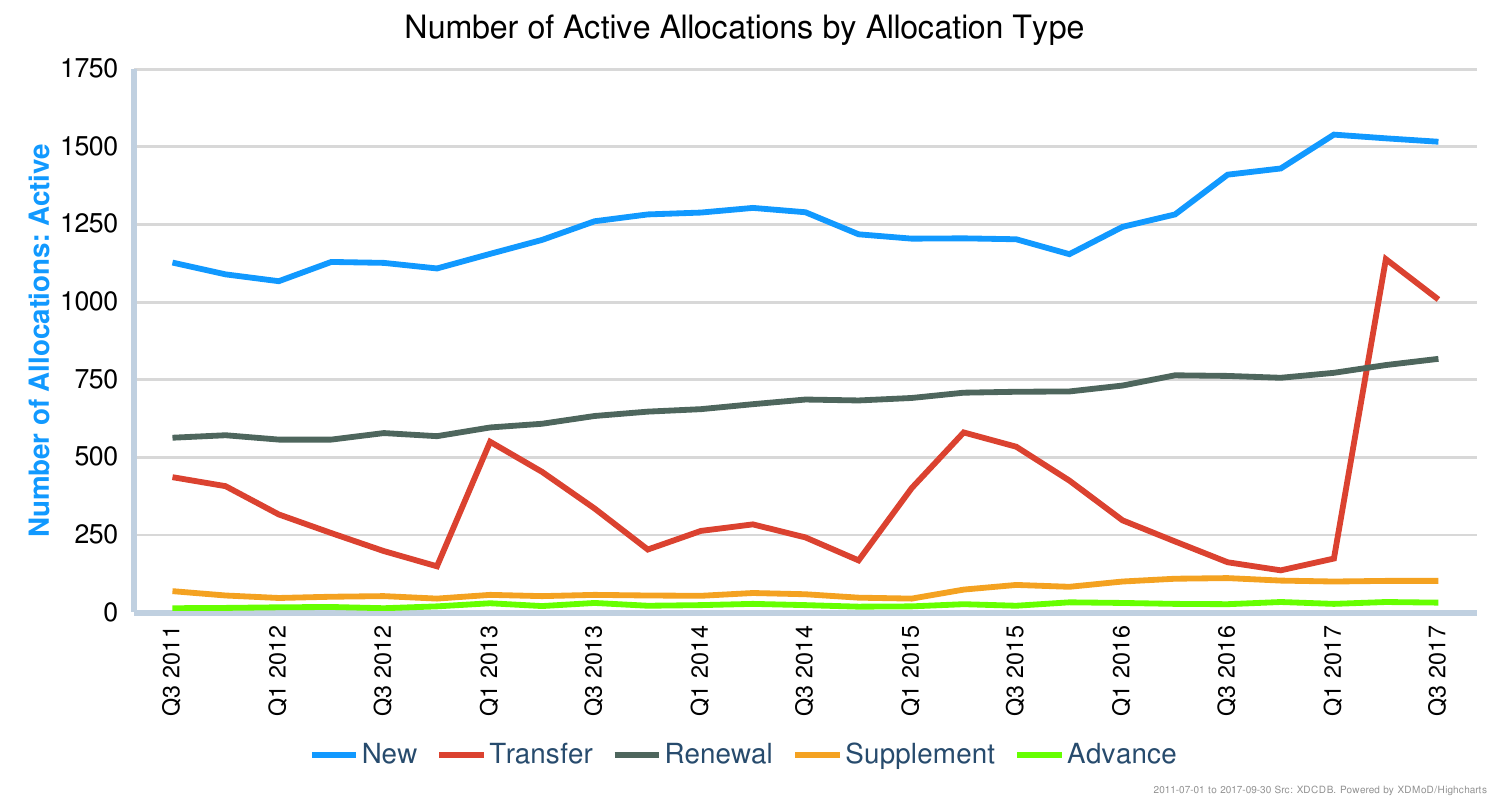}
\caption{\label{fig:Allocations_by_allocationType}Number of allocations granted by Allocation Type for the period 2011-07 to 2017-09.}
\end{figure}

Figure \ref{fig:Allocations_by_allocationType} shows the time history of the types of allocations that are possible on NSF Innovative HPC program resources, namely New, Transfer, Renewal, Supplemental, and Advanced. Not surprisingly, renewal and new accounts show an upward trend.  Transfer accounts, which are migrations from one resource to another resources show a non-linear trend and are a refection  of the integration of new systems and the subsequent transfer of accounts to them.  For example, the large increase in transfers in Q1 2017 are a reflection of the transition to \stampede2{}.

\begin{figure}[H]
\centering
\includegraphics{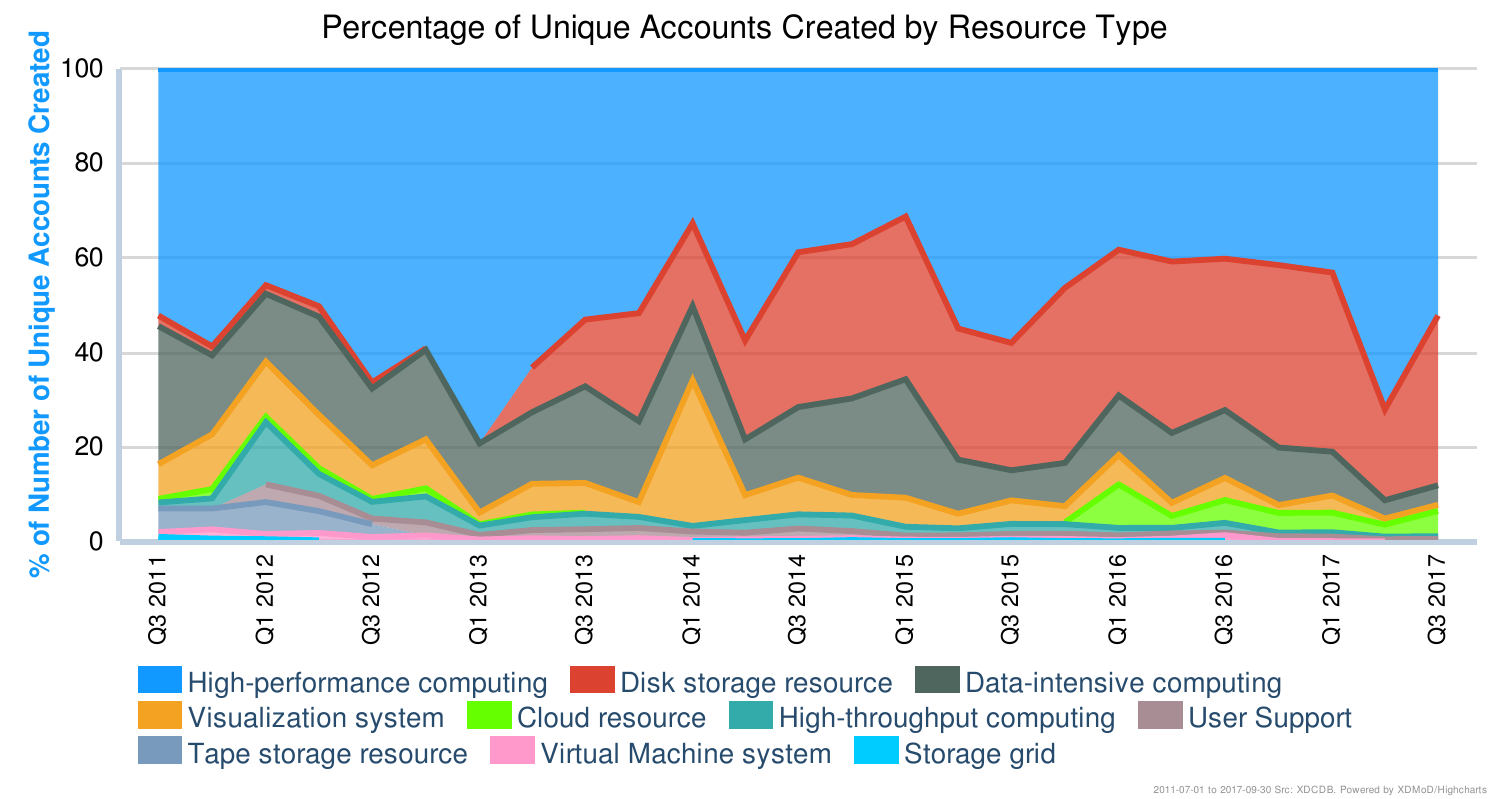}
\caption{\label{fig:PercentageUniqueAccounts}Percentage Unique Accounts Created for the period 2011-07 to 2017-09.}
\end{figure}

Figure \ref{fig:PercentageUniqueAccounts} shows the time history of unique accounts as a percentage of all accounts created.  Noteworthy is the increase in accounts for disk storage resources that begins in 2013 and shows an increasing trend. The addition of cloud resources to the NSF HPC portfolio is evident by the appearance of cloud accounts starting in 2016.  

\begin{figure}[H]
\centering
\includegraphics{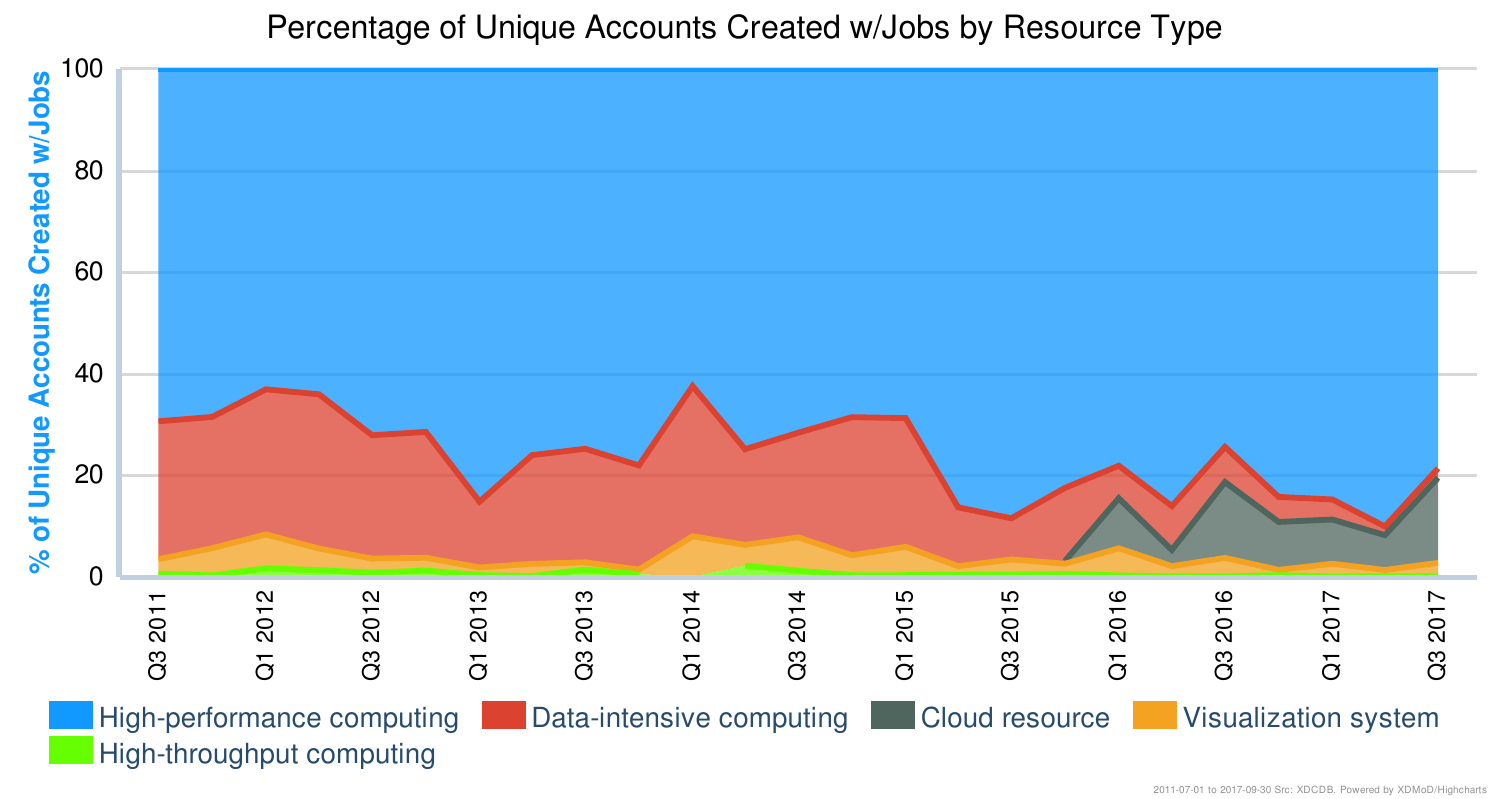}
\caption{\label{fig:PercentageUniqueAccountsWithJobs}Percentage Unique Accounts Created With Jobs for the period 2011-07 to 2017-09.}
\end{figure}

Figure \ref{fig:PercentageUniqueAccountsWithJobs} shows the time history of unique accounts by resource type as a percentage of all accounts.  Resource types include high performance computing (HPC), high throughput computing (HTC), data intensive computing, visualization systems, and cloud computing. Note that as described in the introduction, the job type classification scheme employed to create Figure \ref{fig:PercentageUniqueAccountsWithJobs} is based on the resource type classification for each resource taken from the XSEDE central database. The classification therefore does not reflect an analysis of the research projects to determine the type of computing carried out by the research group but rather is a reflection of the resource requested by the PI or assigned by the allocation committee. During the time period shown, HPC resource accounts for 70 - 90\% of all accounts, with cloud computing resource accounts showing a growing presence.  

\begin{figure}[H]
\centering
\includegraphics{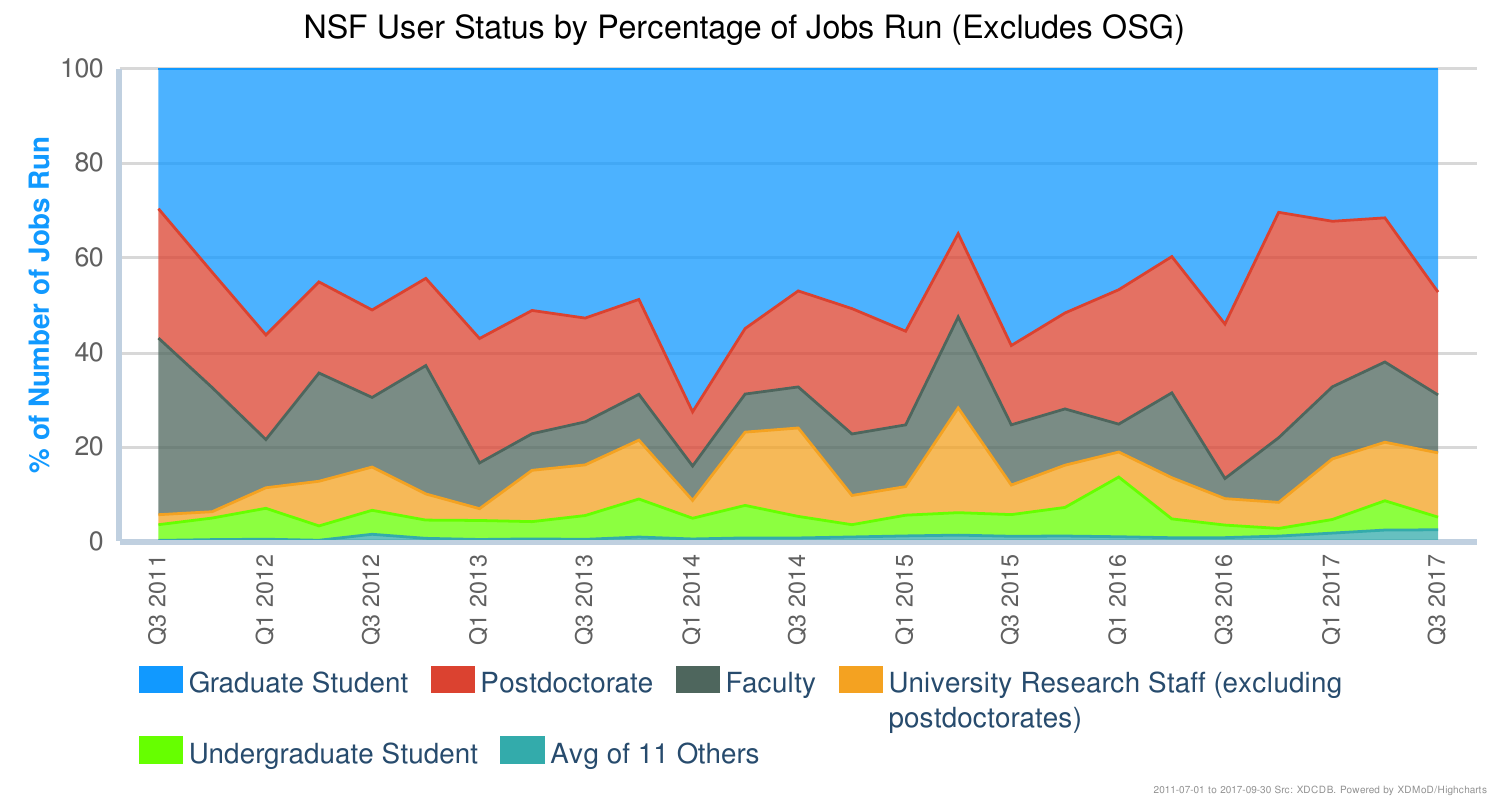}
\caption{\label{fig:NSF_UserStatus_TimeHistory}Percentage of jobs run by NSF User Status for the period 2011-07 to 2017-09.}
\end{figure}

Figure \ref{fig:NSF_UserStatus_TimeHistory} shows the time history of the end-users running jobs on NSF HPC Innovation program resources by the type of end-user (graduate student, post doc, faculty, etc).  This plot excludes Open Science Grid (\osg{}) jobs.  It is interesting to note that in 2011 faculty account for about 40\% of all jobs run but by 2017 as a group they account only for about 10\%.  Over this time period, the utilization of resources by graduate students, post docs, and university research staff increases as a percentage of all users, which speaks well for NSF's emphasis on  fostering the development of the U.S.'s next generation of computational and data scientists.  

\begin{figure}[H]
\centering
\includegraphics{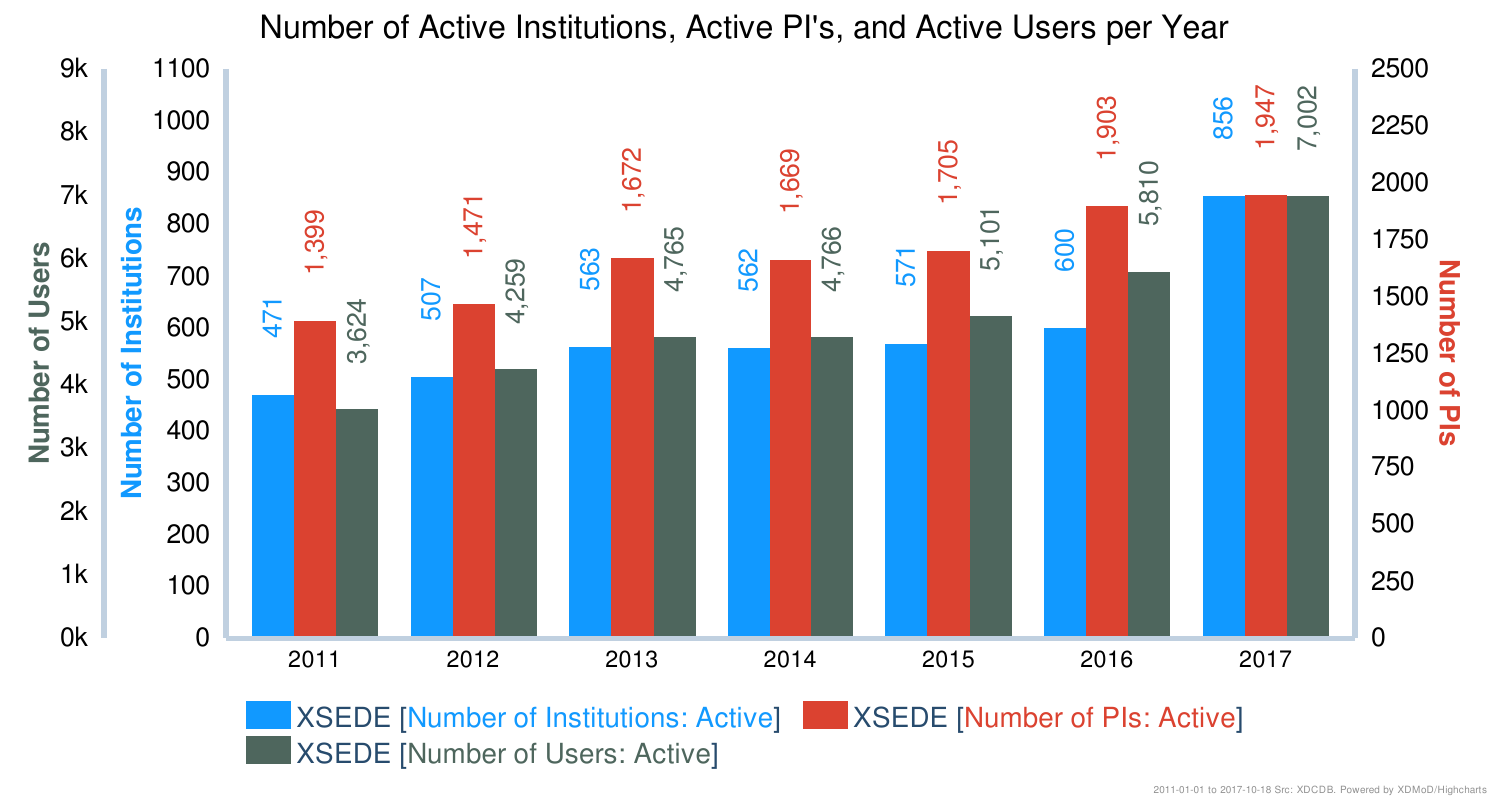}
\caption{\label{fig:TimeHistory_PI_Institutions}Number of Active PI's, Users, and Institutions (excluding Science Gateway users) per year for the period 2011-07 to 2017-09.}
\end{figure}

Figure \ref{fig:TimeHistory_PI_Institutions} shows the time history of active institutions, PI's, and Users running on NSF Innovative HPC Program resources per year.   Over the 6 years shown, the number of PI's utilizing the resources increases by more than 540 (a 40\% increase), the number of users by over 3,300 (a 93\% increase) and the number of institutions by about 380 (an 80\% increase).  Note that this plot does not include gateway utilization.

\subsection{NSF Directorate, Parent Science and Field of Science Trends}
\begin{figure}[H]
\centering
\includegraphics{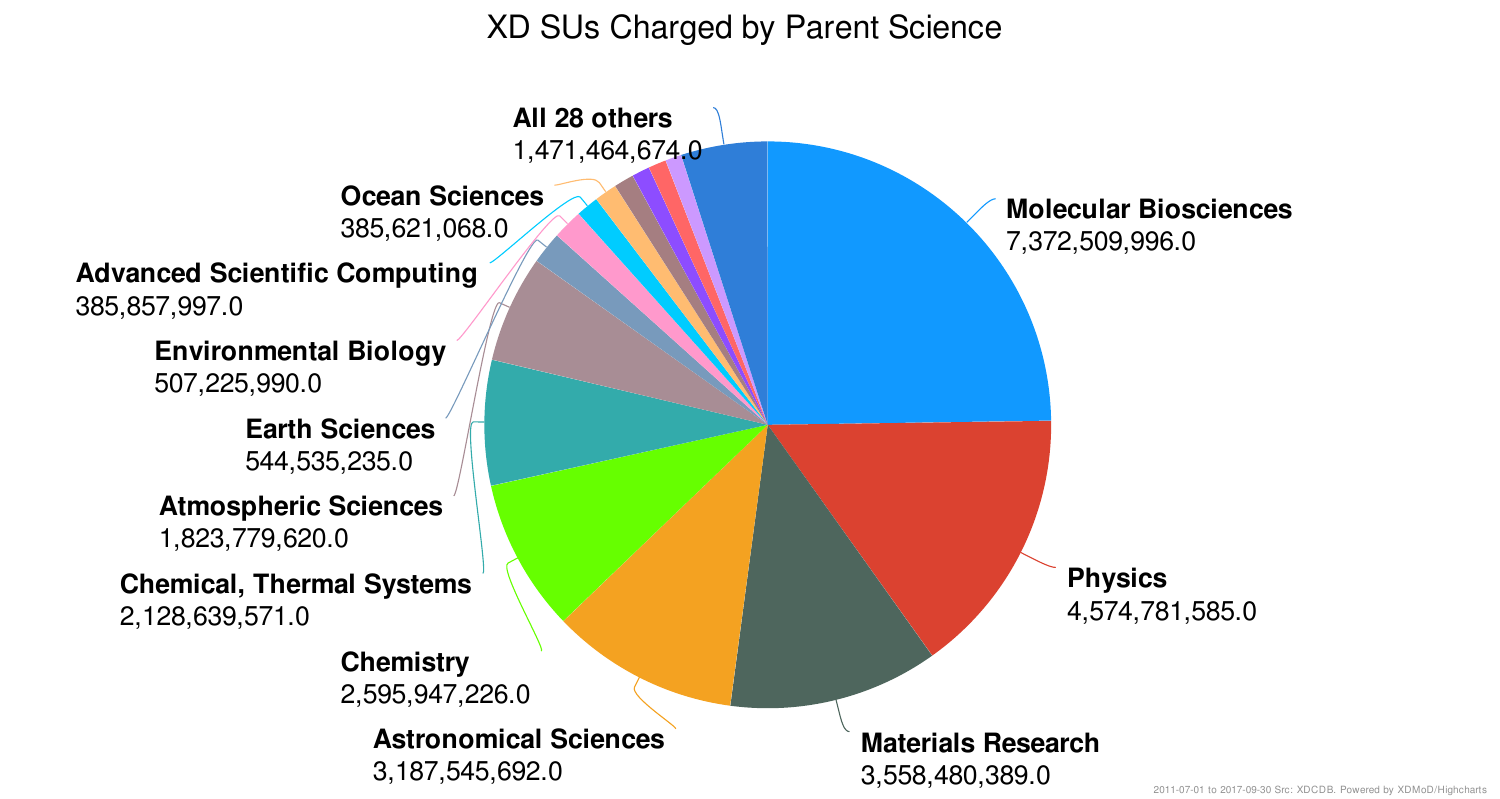}
\caption{\label{fig:XD_SU_Parent_Science_PieChart}Breakdown by Parent Science in terms of XD SUs charged for the period 2011-07 to 2017-09.}
\end{figure}

\begin{figure}[H]
\centering
\includegraphics{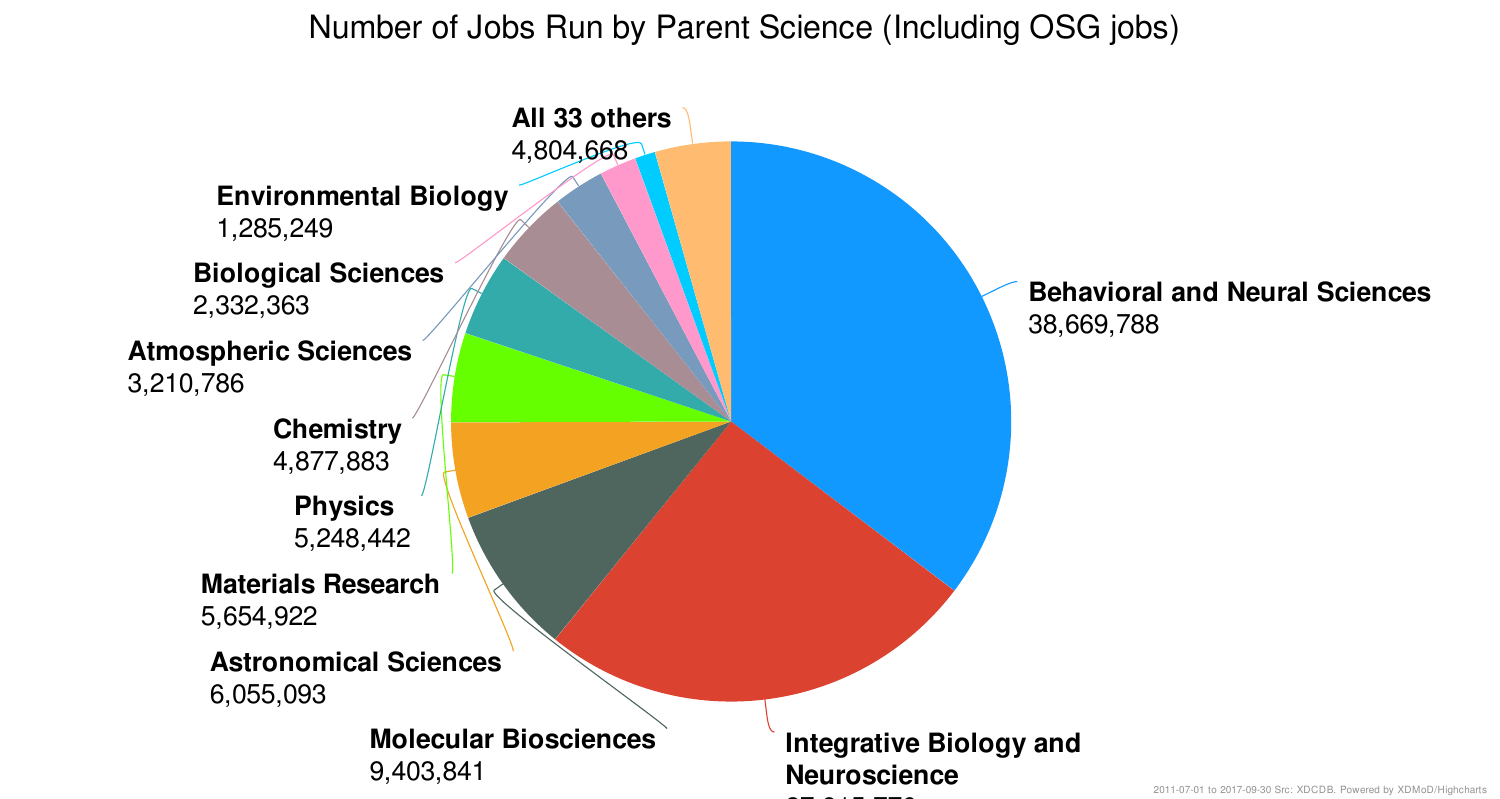}
\caption{\label{fig:Jobs_Parent_Science_PieChart}Breakdown by Parent Science in terms of Number of Jobs for the period 2011-07 to 2017-09. (Including \osg{} Jobs)}
\end{figure}

\begin{figure}[H]
\centering
\includegraphics{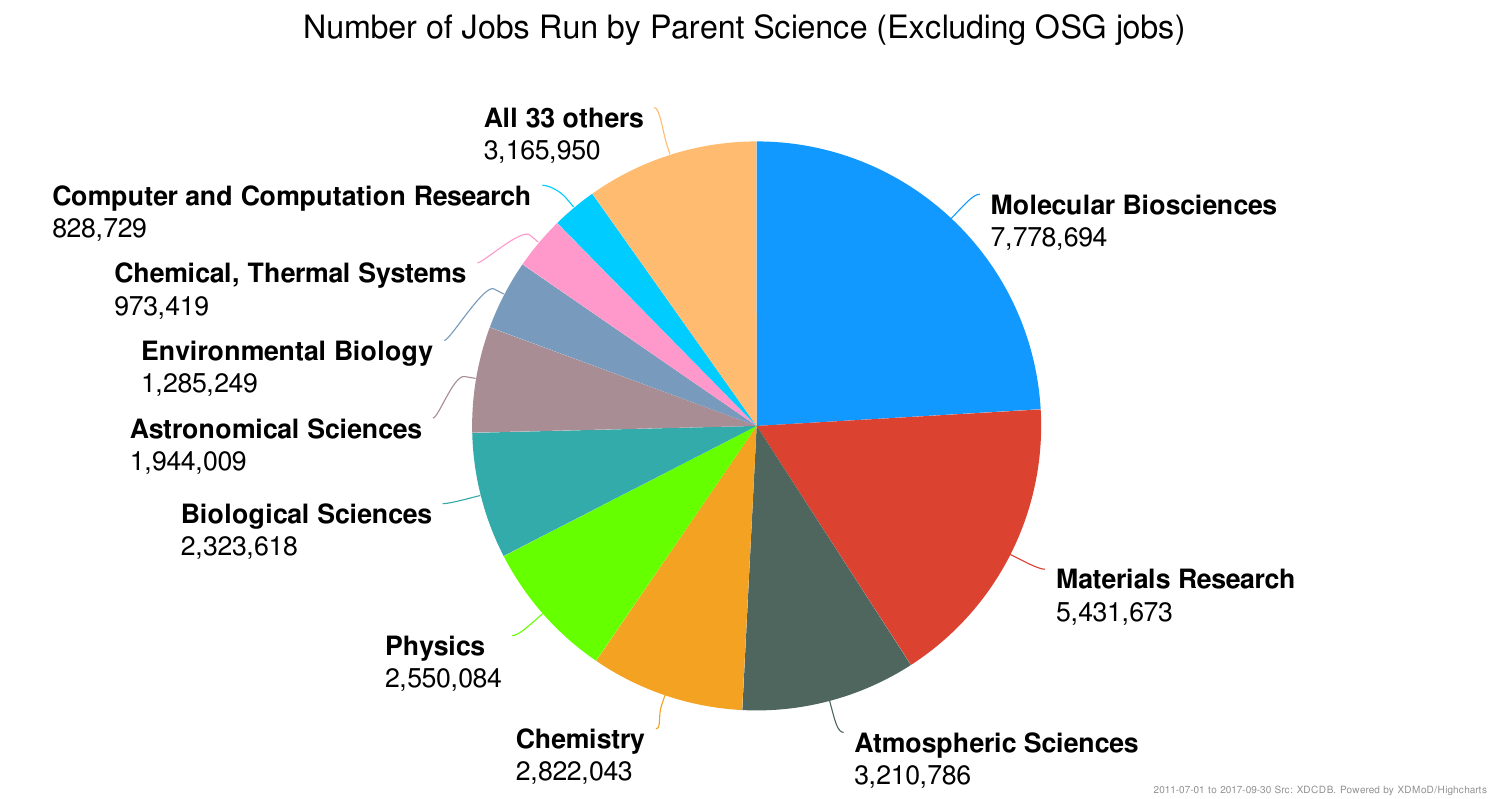}
\caption{\label{fig:Jobs_Parent_Science_No_OSG_PieChart}Breakdown by Parent Science in terms of Number of Jobs for the period 2011-07 to 2017-09.  Excluding \osg{} jobs)}
\end{figure}

\begin{figure}[H]
\centering
\includegraphics{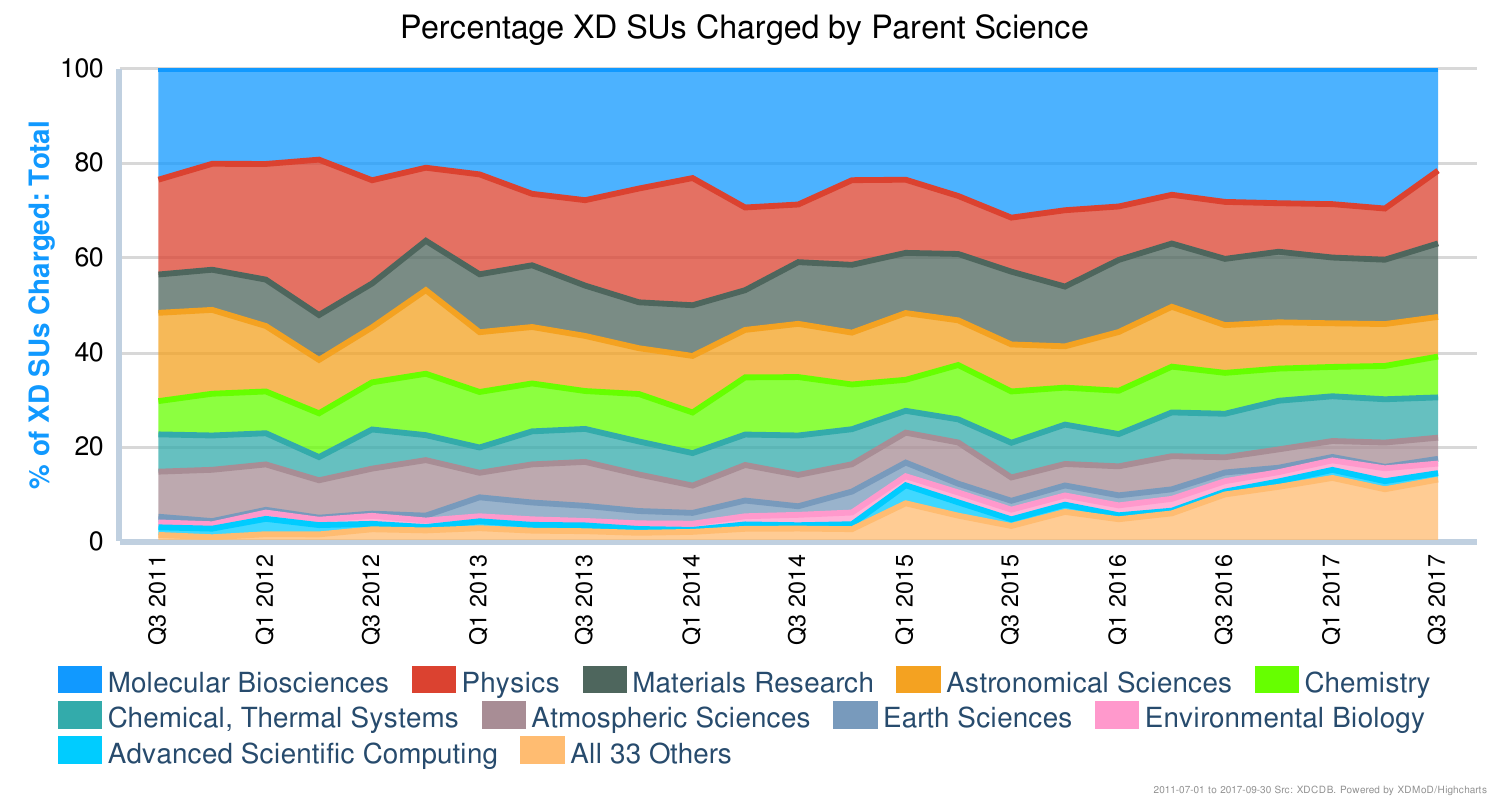}
\caption{\label{fig:Parent_Science_TimeHistory}Time History of Parent Science for the period 2011-07 to 2017-09.}
\end{figure}

\begin{figure}[H]
\centering
\includegraphics{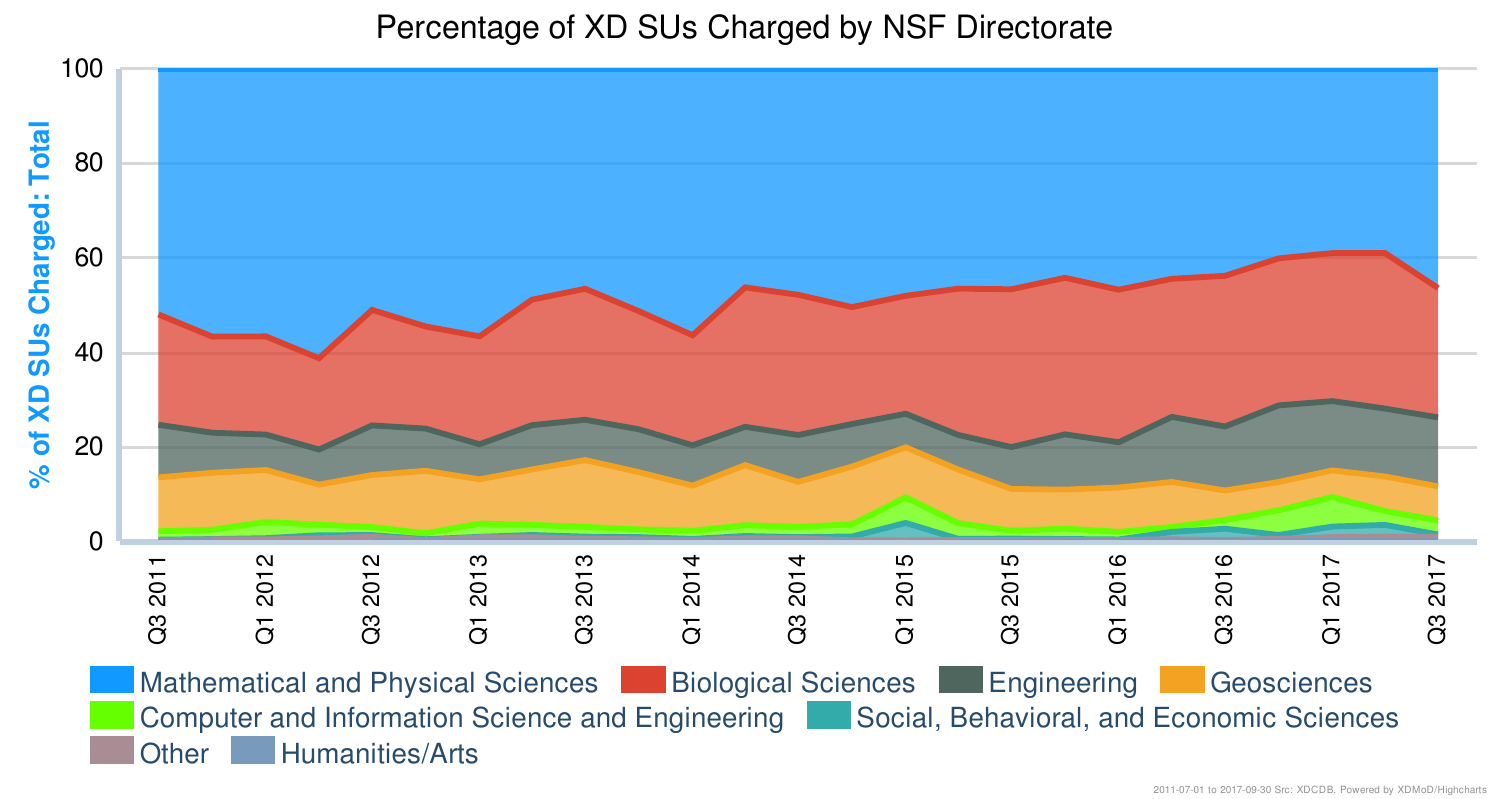}
\caption{\label{fig:NSF_Directorate_Utilization}Percentage Utilization by XD SUs charged by NSF Directorate for the period 2011-07 to 2017-09.}
\end{figure}

\begin{figure}[H]
\centering
\includegraphics{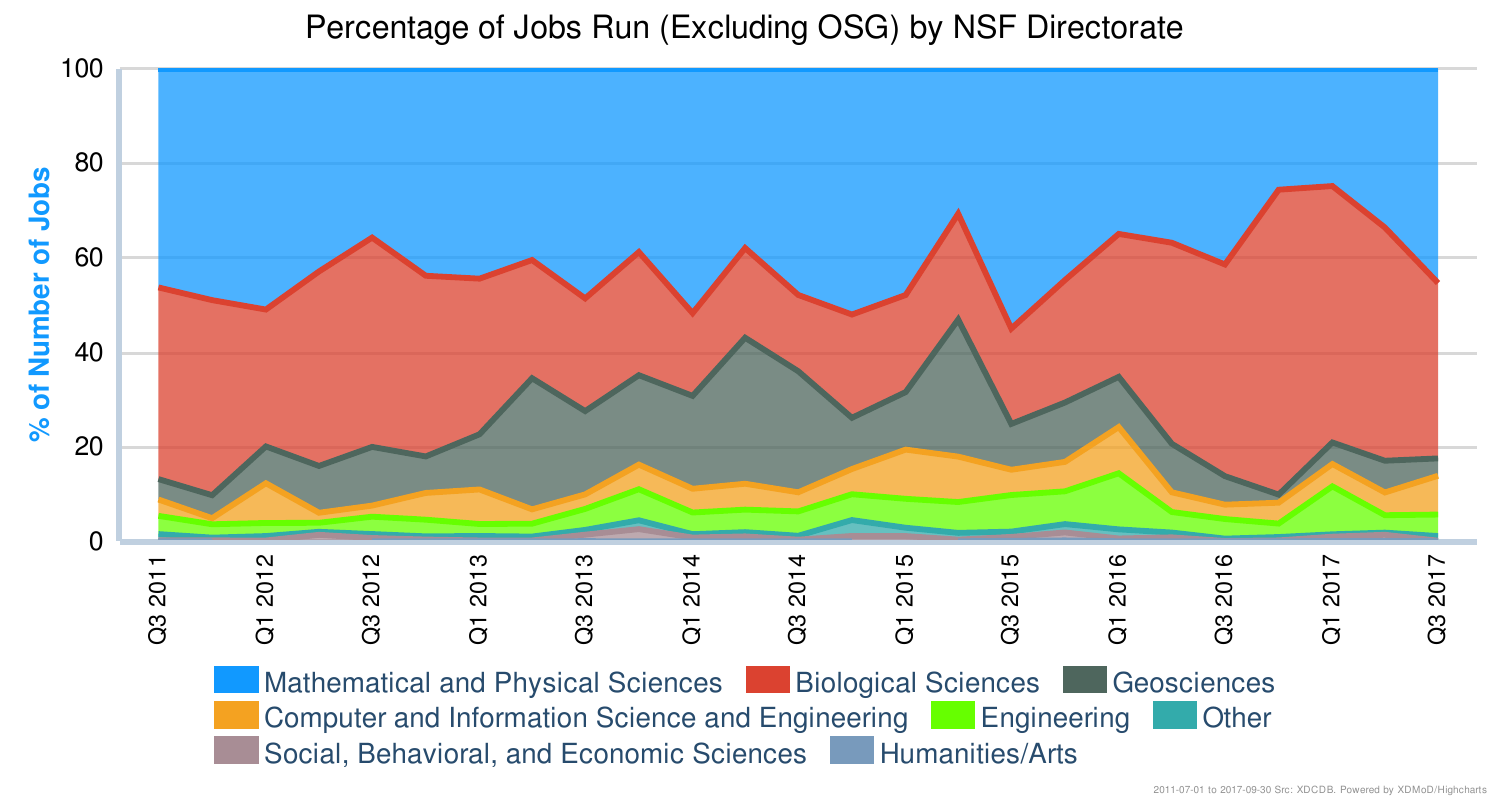}
\caption{\label{fig:NSF_Directorate_UtilizationJobs}Percentage Utilization by number of jobs run by NSF Directorate for the period 2011-07 to 2017-09. (Excludes \osg{} jobs)}
\end{figure}

\begin{figure}[H]
\centering
\includegraphics{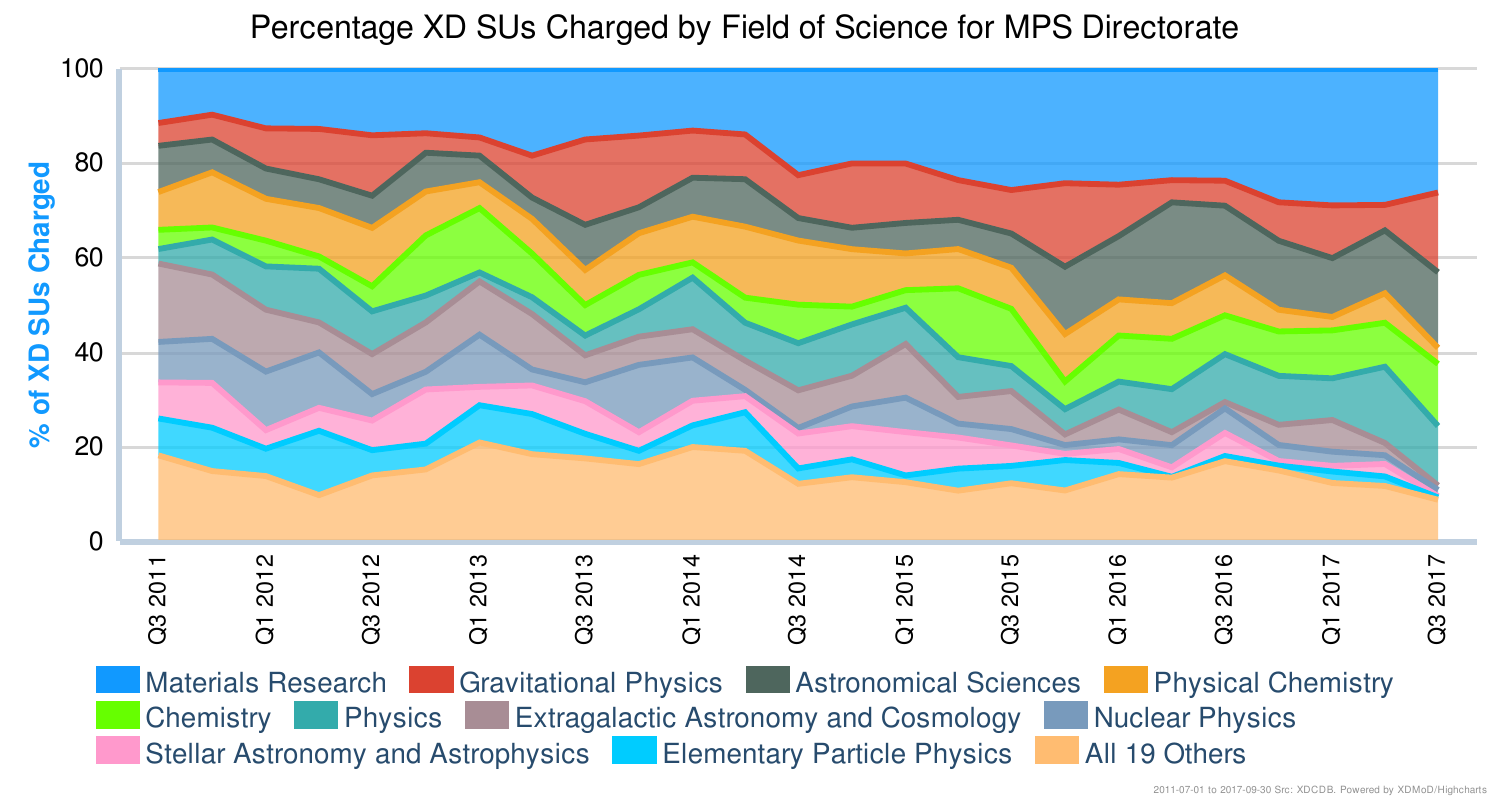}
\caption{\label{fig:XD_SUs_by_FOS_MPS}Percentage XD SUs charged by Field of Science for MPS Directorate for the period 2011-07 to 2017-09.}
\end{figure}

\begin{figure}[H]
\centering
\includegraphics{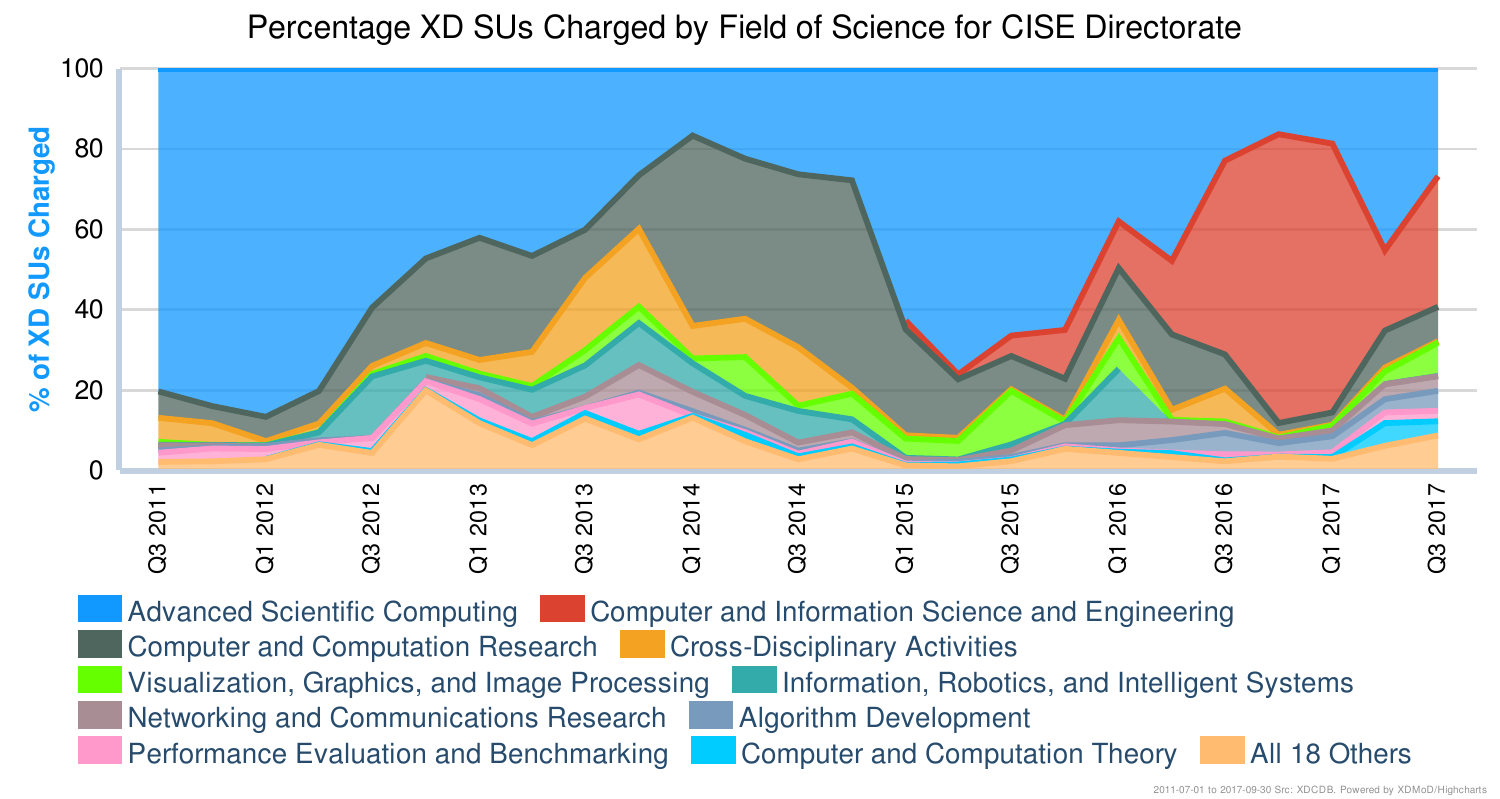}
\caption{\label{fig:XD_SUs_by_FOS_CISE}Percentage XD SUs charged by Field of Science for CISE Directorate for the period 2011-07 to 2017-09.}
\end{figure}

\begin{figure}[H]
\centering
\includegraphics{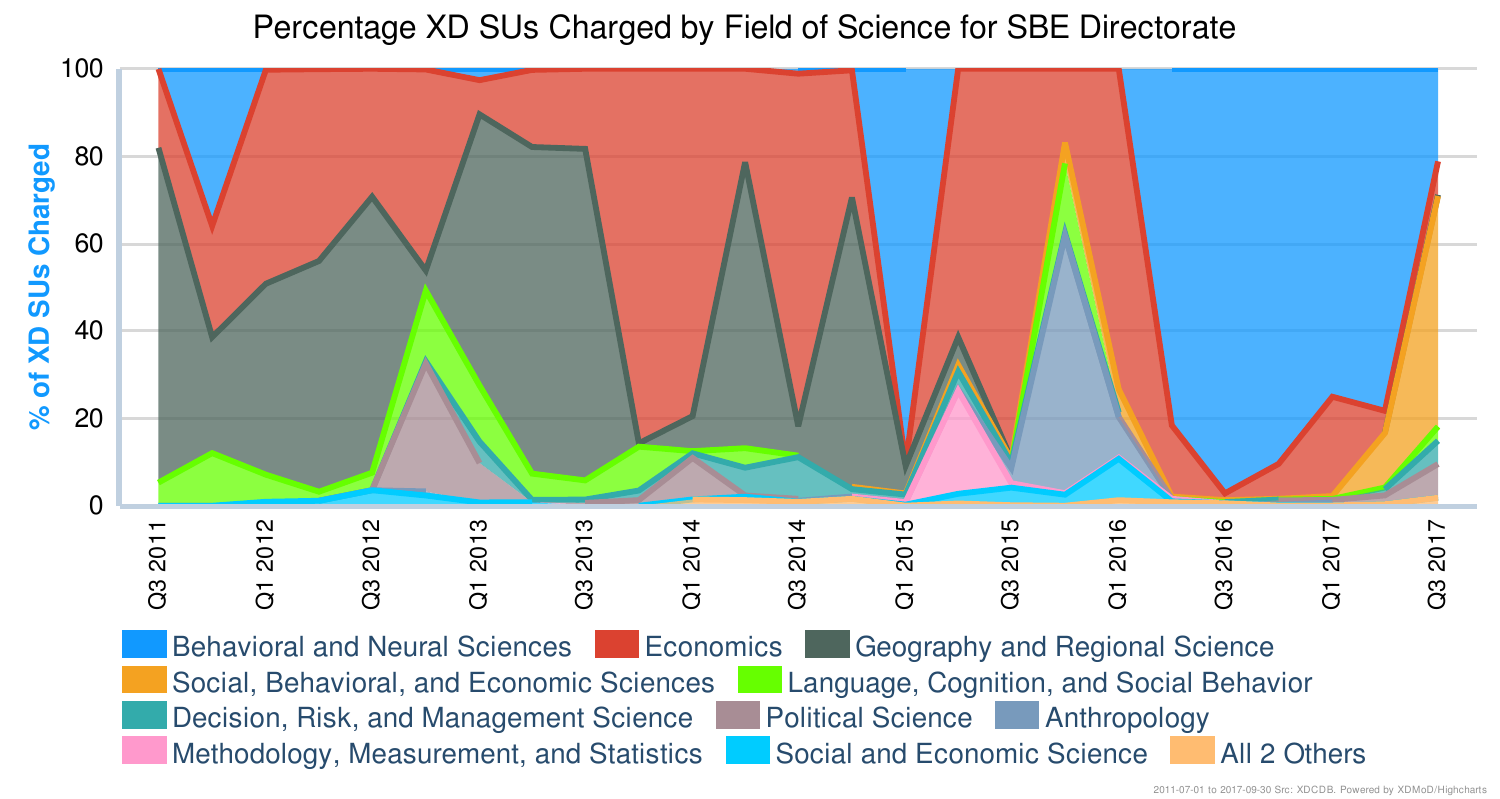}
\caption{\label{fig:XD_SUs_by_FOS_Social}Percentage XD SUs charged by Field of Science for Social, Behavioral, and Economic Sciences Directorate for the period 2011-07 to 2017-09.}
\end{figure}

Figures \ref{fig:XD_SU_Parent_Science_PieChart} - \ref{fig:XD_SUs_by_FOS_Social} focus on trends in utilization by the NSF directorates, parent sciences, and fields of science over time.   Figure \ref{fig:XD_SU_Parent_Science_PieChart} is a pie chart showing the aggregate XD SUs charged by parent science for the time period 2011-07-01 to 2017-09-30.  While Molecular Biosciences, Physics, and Materials Research account for half of all XD SUs consumed, there is substantial utilization by many of the other parent sciences. Note that since \osg{} jobs consume few XD SUs compared to HPC jobs, Figure \ref{fig:XD_SU_Parent_Science_PieChart} is unchanged if \osg{} jobs are excluded from the analysis.   In addition to XD SUs, utilization can equally well be measured by the number of jobs run.  Using the number of jobs as the utilization metric results in a remarkable difference in the observed parent science utilization as shown in Figure \ref{fig:Jobs_Parent_Science_PieChart}.  In this case, the dominant Parent Sciences are Behavioral and Neural Sciences and Integrative Biology and Neuroscience, which when taken together account for over 50\% of all jobs run (75 million jobs).  Figure \ref{fig:Jobs_Parent_Science_PieChart} includes HTC jobs run through Open Science Grid (\osg{}). While \osg{} jobs don't consume a large fraction of CPU hours when compared to the HPC jobs, they clearly play an important role for many of the parent sciences.  Excluding \osg{} jobs from this plot results in Figure \ref{fig:Jobs_Parent_Science_No_OSG_PieChart} for which the trend is more in line with that of Figure \ref{fig:XD_SU_Parent_Science_PieChart}.  However, there are substantial differences, namely utilization by Physics and Astronomical Sciences has diminished and Atmospheric Sciences, Biological Sciences and Chemistry have increased.    

Note that the \osg{} utilization reported here is restricted to community access of \osg{} via the XRAC allocation process (XSEDE), and the majority of community use of \osg{} comes from other means. 

The time history of the percentage of XD SUs charged by parent science for the same period covered in Figure \ref{fig:XD_SU_Parent_Science_PieChart} is shown in Figure \ref{fig:Parent_Science_TimeHistory}. Not surprisingly, the top three parent sciences are the same as in Figure \ref{fig:XD_SU_Parent_Science_PieChart}. However, during this time period, the parent science Molecular Biosciences increases by 10\% (from 20\% to 30\%) at the expense of Physics (20\% to 10\%).  In addition, the thirty-three ``Other'' parent sciences not individually labeled in this figure increases from 2\% to more than 10\%, reflecting the increasing reliance of many of the parent sciences on access to advanced computing resources.

Figure \ref{fig:NSF_Directorate_Utilization} shows the time history of the utilization of NSF Innovative HPC Program resources by the NSF directorates in terms of percent of total utilization. Mathematical and Physical Sciences (MPS) and Biological Sciences account for about 70\% of the utilization as measured by XD SUs consumed.  During this time period MPS utilization decreases by 10\% while there was a corresponding 10\% increase in utilization by researchers in the Biological Sciences directorate.  This trend is also evident in the time history of parent sciences (Figure \ref{fig:Parent_Science_TimeHistory}).  Also noteworthy is the growth in utilization by the Computer and Information Sciences Directorate (CISE). 

As was the case for the analysis of parent science utilization, the time history of utilization by NSF directorate is considerably different if we consider the number of jobs as the utilization metric as opposed to XD SUs, as is shown in Figure \ref{fig:NSF_Directorate_UtilizationJobs}.  Compared to Figure \ref{fig:NSF_Directorate_Utilization}, the order of the third, fourth, and fifth ranked directorates in terms of utilization changes from Eng, Geo, and CISE to Geo, CISE, Eng, respectively. 

We now consider historical trends in utilization as measured by XD SUs charged for various fields of science (FoS).  We begin with a breakdown of the the fields of science for the MPS directorate, which as was shown in Figure \ref{fig:NSF_Directorate_Utilization}, accounts for the greatest number of XD SUs consumed among the directorates.  Utilization by FoS in Figure \ref{fig:XD_SUs_by_FOS_MPS} is fairly constant for many of the FoS's.  However, Materials Research shows a three fold increase in the percentage of XD SUs consumed relative to the other FoS's (10\% to 30\%). Similarly, Astronomical Sciences increases in percentage, albeit not as substantially as Materials Research.
 
As was noted in Figure \ref{fig:NSF_Directorate_Utilization}, utilization in the CISE directorate increases during the time period studied.  Figure \ref{fig:XD_SUs_by_FOS_CISE} shows a breakdown of utilization by FoS for the CISE directorate.  Unlike the FoS breakdown for the MPS directorate, there is great variability in utilization over time among the fields of science making up the CISE directorate.  For example, in terms of percentage of total XD SUs charged in the CISE directorate, Advanced Scientific Computing varies from a high of 80\% to a low of 20\%.   Likewise, Computer and Information Science and Engineering and Computer and Computation Research show fluctuations of 70\% and 50\% respectively.

The FoS breakdown for the Social, Behavioral, and Economic Science directorate shows even greater variability as is evident in Figure \ref{fig:XD_SUs_by_FOS_Social}.  The variability of the fields of science in terms of the percentage of XD SUs consumed within the directorate is remarkable.   Given the relatively low XD SUs consumed by this directorate compared to other directorates (see Figure \ref{fig:NSF_Directorate_Utilization}), this variability is likely a reflection of the time history of individual awards within the fields of science within this directorate. 

\subsection{Deep and Wide Metrics for HPC Resource Capability and Project Usage}\label{subsec:depthwidth}

The TeraGrid project team coined the terms ``deep'' and ``wide'' computing to describe the needs of users and their science problems \cite{Catlett2005TeraGrid}. Deep problems are generally defined to be problems which require capability computing, while ``wide'' refers to the community of researchers whose individually computing needs may be modest but when taken together represent a large capacity of computational work.  In order to better understand the ability of the NSF Innovative HPC Program resources to simultaneously meet these needs, we carried out an analysis similar to that conducted by Hart \cite{Hart2011DeepAndWideMetrics}. Project depth was defined as the maximal number of cores used by any job within a project. The reasoning behind defining project depth this way is that the jobs with highest core count are the most computationally demanding part of the project and are essential for completing project goals.  Smaller core count jobs within the same project often times corresponds to pre- and post-processing, low resolution preliminary calculations and other less computationally intense types of calculations for the project. This way the maximal core count for each project corresponds to the required computational depth. To quantify the width of a resource or group of resources, we follow the definition proposed by Hart, namely the fraction of projects at or below a given depth.  In order to quantify the capability utilization of a resource, we adopt the NERSC definition, namely the fraction of use by projects above a given depth. 

\begin{figure}[ht]
\centering
\includegraphics[width=0.8\textwidth]{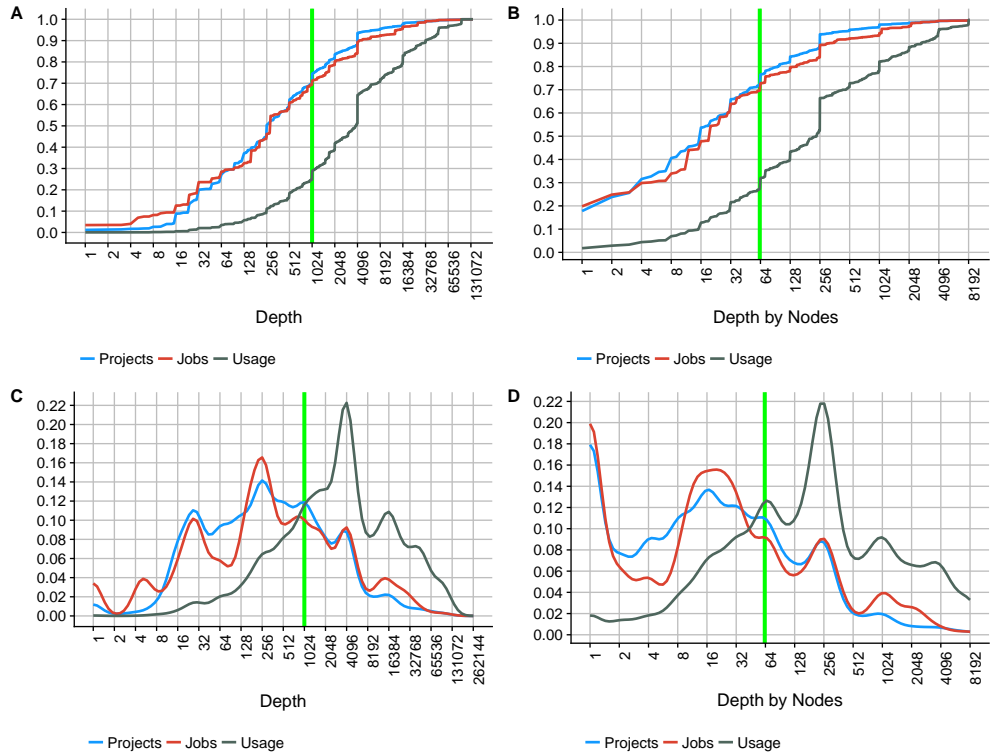} 
\caption{\label{fig:depth_and_width}XSEDE overall projects depth and width. A - cumulative distribution of project by their depth (maximal core count for any job of the project), unweighted labeled as project and shown by blue line, weighted by job count labeled as jobs and shown by red line and weighted by core-hours and shown by dark gray line. Vertical green line shows depth where the joint ratio occurs. B -  same as A but for nodes instead of cores, C - distribution of projects unweighted and weighted by jobs and usage (total CPU hours)  D - same as C but for nodes instead of cores.}
\end{figure}

The distributions of projects are shown in Figure \ref{fig:depth_and_width}-C. In addition to the unweighted distribution, distributions weighted by jobs and by utilization (as measured by CPU hours) were calculated to take into account the project size as measured by job count and total CPU hours.  The integration of these three distributions produces the cumulative plot shown in \ref{fig:depth_and_width}-A.

The plots are over all HPC resources for the duration of the XSEDE program. Here we find it useful to compute (as was done in Reference \cite{Hart2011DeepAndWideMetrics}) the ``joint ratio'' - namely the point on the x-axis in Figure \ref{fig:depth_and_width}-A where the percentage utilization and the percentage of projects total 100\%.  It is the point at which x\% of the projects use y\% of the resources and the remaining y\% of the projects consume the other x\% of the resources.  In terms of cores, the ``joint ratio'' for all NSF Innovative HPC Program resources during the XSEDE program is 73:27 at 1023 cores (green line in \ref{fig:depth_and_width}-A), meaning that 27\% of XSEDE projects account for 73\% of the utilization, while the remaining 73\% of the projects account for only 27\% of the utilization.  In addition to cores, Figure \ref{fig:depth_and_width}-D\&-B includes the analogous plots in terms of nodes (joint ratio at 64 nodes).  

\begin{figure}[H]
\centering
\includegraphics[width=0.8\textwidth]{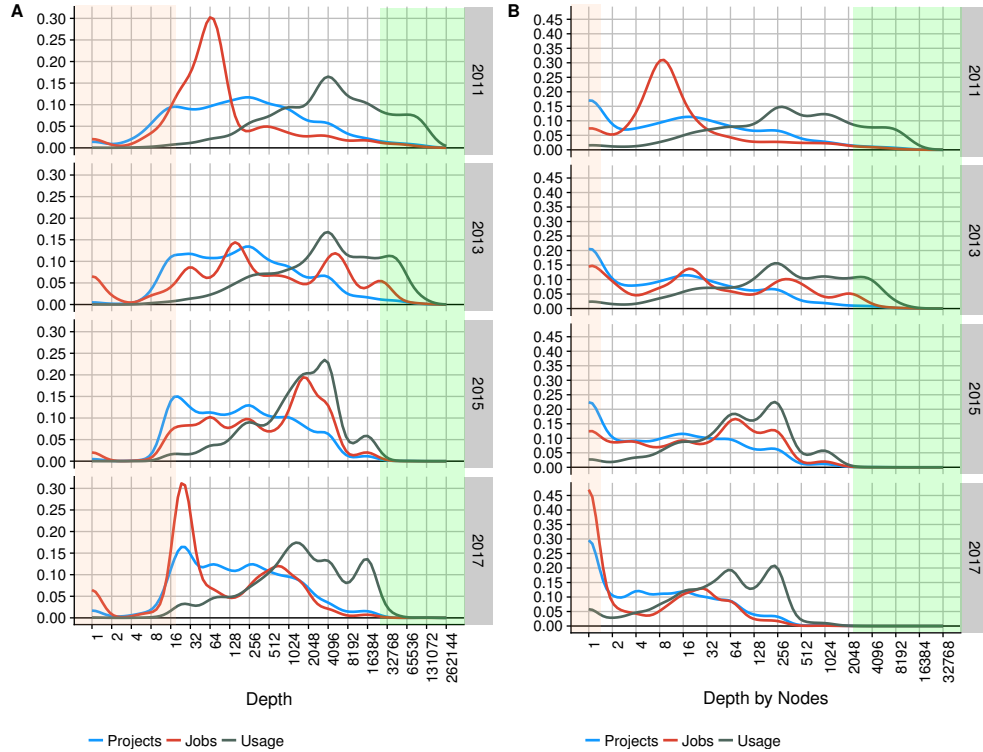} 
\caption{\label{fig:depth_and_width_by_years} Projects depth distribution by years. Project depth calculated as the maximum core count (\textbf{A}) for a project and as the maximum node count (\textbf{B}) for a project. Similar to Figure \ref{fig:depth_and_width}, the blue line corresponds to unweighted distribution of projects over their depth, the red line is weighted by job count distribution of projects, and dark gray line shows weighted by core-hours distribution of projects. 
Light orange vertical shading highlights the increase in single node jobs that occur over time and the green vertical shading highlights the decrease in large core (capability class) jobs over time}
\end{figure}

It is informative to study the changes in the project depth, width, and joint ratio over time as well as differences in these metrics among the HPC resources.  The change in these metrics over time are shown in Table \ref{table:depth_and_width_by_years} and Figure \ref{fig:depth_and_width_by_years}. It is important to note that the metrics computed in this table are obtained by summation over each project only within each calendar year and not over the entire lifetime of the project as was the case for the joint ratio, depth, and width shown in Figure \ref{fig:depth_and_width}.   Accordingly, these metrics should be considered ``per year'' metrics.  

While the joint ratio changes little over time, the ``per year'' project depth decreases by almost a factor of 2. That is, in 2017, 73\% of projects had a maximum core count that was about 50\% that of the maximum core count in 2011. In terms of capability class computing, the most interesting observation is a significant reduction of high core-count jobs after \kraken{} was removed from service in the second quarter of 2014. Although many newer systems offer core-counts comparable to \kraken{}, recent projects rarely utilize more than 30,000 cores.  While GPU or Xeon Phi accelerator use was not accounted for here, our analysis for \stampede{} showed that accelerator use was not significant in large core count jobs. The substantial decrease in the size of capability jobs can be partially explained by (1) improved per core performance of most systems (for example \stampede{} single core performance is 2 times greater than \kraken{} single core performance based on theoretical FLOP/S and 50\% faster as measured by XDMoD's \texttt{NAMD} application kernel) and (2) by the introduction of Blue Waters~\cite{Bode:2013}, a capability specialized resource. The latter is manifested in resource policies that limit the maximum job size allowed for users (see Table \ref{table:user-limits-along-resource-summary}, for example for \stampede{} it is 16384 cores, which is 16\% of \stampede{}). Another factor is the steady increase in projects that only run single node jobs.  These trends are evident in Figure \ref{fig:depth_and_width_by_years}, which shows the project depth distributions at the core level and node level broken out by year.   The light orange area in this figure highlights single node jobs, which show an increase over time, while the light green area highlights the decrease in large core (and node) jobs that occur from 2011 to 2017.   This trend is also supported by Figure \ref{fig:ave-core-counts} which shows the average job size by core count over the same period.

In addition to considering the changes over time in these metrics, we can also study the differences in project depth, width, and joint ratio by individual resource as is shown in Table \ref{table:depth_and_width_by_resources}. Most resources have a joint ratio close to 70:30. However, the depth at the joint ratio varies considerably.  For example \kraken{} has a depth of 6136 cores whereas \stampede{} has a depth of only 1135 cores. For some resources, typically those with some specialization such as GPU accelerators or large memory, the joint ratio is close to 60:40 indicating more homogeneous core counts from projects. There is only a weak dependency of depth on system total core count, which is most likely due to resource maximum job size policies (see Table \ref{table:user-limits-along-resource-summary} for default maximal job sizes throughout resources).

\begin{table}[H]
\centering
\caption{Changes of depth (by cores) and width over time}
\label{table:depth_and_width_by_years}
\begin{tabular}{R{0.4in}|R{0.4in}|R{0.6in}|R{0.6in}|R{1.0in}|R{1.0in}}
\hline
Year & Joint Ratio & Depth at Joint Ratio & Projects at Joint Ratio & Usage (Core-Hours) at Joint Ratio & Jobs at Joint Ratio\\
\hline
2011 & 75:25 & 1019 & 1,892 & 383,678,799 & 5,889,587\\
\hdashline
2012 & 77:23 & 1020 & 2,231 & 352,896,673 & 3,562,211\\
\hdashline
2013 & 76:24 & 1008 & 2,579 & 369,316,028 & 2,879,818\\
\hdashline
2014 & 72:28 & 755 & 2,249 & 334,853,085 & 2,275,629\\
\hdashline
2015 & 72:28 & 624 & 2,206 & 301,024,614 & 1,425,917\\
\hdashline
2016 & 72:28 & 512 & 2,483 & 328,402,299 & 4,393,705\\
\hdashline
2017 & 73:27 & 538 & 2,688 & 281,268,718 & 3,664,563\\
\hline
\end{tabular}
\end{table}

\begin{table}[H]
\centering
\caption{HPC resources depths (by cores) and widths}
\small
\label{table:depth_and_width_by_resources}
\begin{tabular}{l|R{0.4in}|R{0.4in}|R{0.4in}|R{0.7in}|R{0.6in}|R{0.4in}|R{0.3in}}
\hline
Resource & Joint Ratio & Depth at Joint Ratio & Projects at Joint Ratio & Usage (Core-Hours) at Joint Ratio & Jobs at Joint Ratio & Cores & Nodes\\
\hline
CCT-LSU-SUPERMIC & 64:36 & 313 & 136 & 18,597,422 & 124,411 & 7,200 & 360\\
\hdashline
GATECH-KEENELAND & 63:37 & 380 & 71 & 19,390,100 & 95,117 & 4,224 & 264\\
\hdashline
LONI-QUEENBEE & 70:30 & 202 & 60 & 2,845,034 & 9,376 & 5,344 & 668\\
\hdashline
NICS-DARTER & 63:37 & 2003 & 84 & 43,679,623 & 331,974 & 23,168 & 724\\
\hdashline
NICS-KRAKEN & 71:29 & 6136 & 809 & 700,003,851 & 1,241,083 & 112,896 & 9,408\\
\hdashline
PSC-BRIDGES & 72:28 & 331 & 474 & 44,606,117 & 674,865 & 21,056 & 752\\
\hdashline
PSC-BRIDGES-GPU & 56:44 & 32 & 29 & 737,411 & 2,484 & 1 & 1\\
\hdashline
PSC-GREENFIELD & 62:38 & 51 & 53 & 393,407 & 3,305 & 360 & 3\\
\hdashline
PURDUE-STEELE & 69:31 & 61 & 203 & 16,891,114 & 2,254,942 & 7,144 & 893\\
\hdashline
SDSC-COMET & 69:31 & 452 & 1,251 & 215,885,228 & 5,862,907 & 45,744 & 1,906\\
\hdashline
SDSC-COMET-GPU & 62:38 & 27 & 37 & 1,359,094 & 13,848 & 1,872 & 72\\
\hdashline
SDSC-TRESTLES & 61:39 & 229 & 523 & 90,803,147 & 1,156,484 & 10,368 & 324\\
\hdashline
Stanford-XSTREAM & 63:37 & 14 & 50 & 447,512 & 104,179 & 1,300 & 65\\
\hdashline
TACC-LONESTAR4 & 64:36 & 464 & 520 & 65,905,629 & 370,057 & 22,656 & 1,888\\
\hdashline
TACC-RANGER & 69:31 & 1011 & 678 & 207,779,446 & 382,125 & 62,976 & 3,936\\
\hdashline
TACC-STAMPEDE & 70:30 & 1135 & 2,885 & 789,718,013 & 2,766,606 & 102,400 & 6,400\\
\hdashline
TACC-STAMPEDE2 & 71:29 & 2122 & 371 & 116,417,884 & 80,985 & 285,600 & 4,200\\
\hline
\end{tabular}
\end{table}

\subsection{Summary:Trends in Utilization}

Overall, the allocation utilization of NSF Innovative HPC resources is high, with only 10\% of allocations going unused by researchers, and the percent unused by the top 5\% of users is under 4\%. 
Since 2011, there has been remarkable growth of about two orders of magnitude in utilization of NSF Innovative HPC resources for most  NSF directorates, with Social, Behavioral, and Economic Sciences now consuming as many CPU hours as did the Mathematical and Physical Sciences directorate only 10 years earlier.  Within this time frame, the MPS and Biological Sciences Directorates account for about 70\% of the XD SUs consumed and this percentage has remained relatively constant during this time.  However, Biological Sciences utilization has increased by about 10\% at the expense of MPS.  In terms of parent science, Molecular Biosciences, Physics, and Materials Research account for half of all XD SUs consumed.   Behavioral and Neural Sciences and Integrative Biology and Neuroscience account for over 50\% of all jobs run, with the bulk of those on Open Science Grid.  

Our ``deep and wide'' analysis shows that 27\% of XSEDE projects account for 73\% of the utilization and the remaining 73\% of projects account for only 27\% of the utilization.  Average job size has decreased over time.    Several factors have contributed to the decrease including, the retirement of \kraken{}, the availability of Blue Waters for capability class computing, improved core performance, and resource policies limiting the maximum core count.  However, the introduction of \stampedetwo{}, with its multicore architecture, appears to be reversing this trend.

\newpage
\section{Job Characteristics}\label{sec:jobs}
\textit{
Goals Addressed in Section
\begin{enumerate}\setcounter{enumi}{2}
\item What fraction of the portfolio resources are used for data analytics/data intensive computing (hadoop, spark, etc.)?  Is the trend going up or down? How does it vary among the resources?
\item How much of the resources usage is consumed by high throughput applications (large numbers of loosely-coupled serial, single and small node count jobs) and gateway applications, and is this changing over time?
\item Are jobs using a larger number of cores over time? Are there differences of job size by discipline or application or Innovative HPC Program resource?
\item Is job memory usage increasing/decreasing over time?
\begin{itemize}
\item Are there specific discipline differences in memory usage?
\item Are there specific memory usage differences in the most used applications?
\item Are there memory usage differences among the resources and does this impact throughput (i.e., result in a bottleneck)?
\end{itemize}
\end{enumerate}
}

\subsection{Types of Computing}

\begin{figure}[H]
\centering
\includegraphics{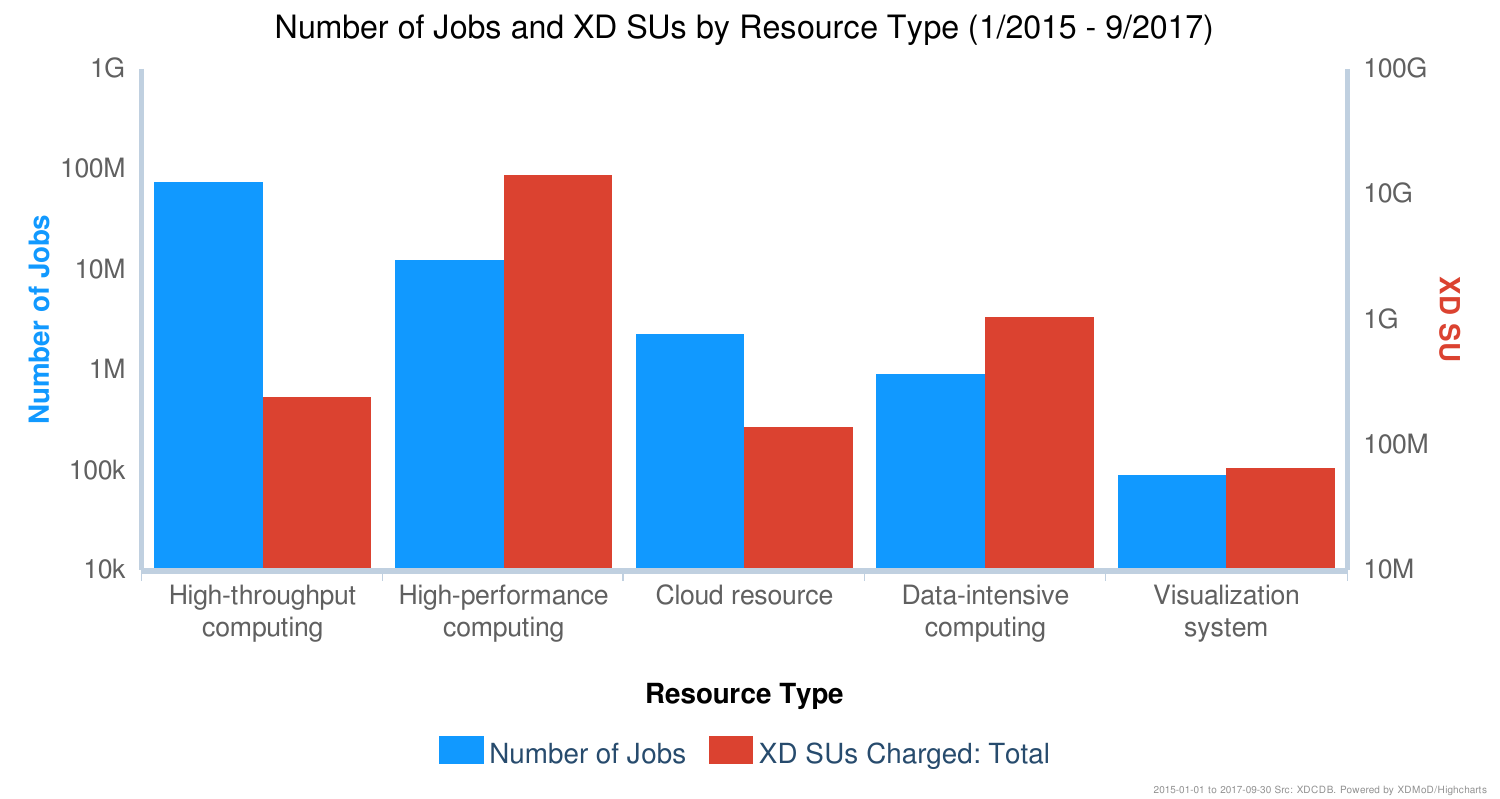}
\caption{\label{fig:NSF_Resource_Type_Utilization}Number of Jobs and XD SUs by Resource Type for the period 2015-01 to 2017-09.  Note the Log Scale.}
\end{figure}

The infrastructure necessary to support research computing, once dominated primarily by high performance computing systems, has diversified over the years and now includes infrastructure to support data intensive applications as well as high-end visualizations. The trend toward diverse infrastructures in support of research is reflected in Figure \ref{fig:NSF_Resource_Type_Utilization} which shows the number of jobs and XD SUs consumed by resource type (HTC, HPC, Cloud, data intensive, or visualization) from January 2015 to October 2017. A log scale was chosen in order to better accommodate the wide range of values among the resource types.  January 2015 was selected as the start date as opposed to July 2011 because prior to 2015, \osg{} (HTC) jobs were not accurately reported in the XSEDE central database.  While the number of HTC jobs is almost 10 times larger then the number of HPC jobs, the HPC jobs consume about 100 times more XD SUs. Data intensive computing is substantial --- almost one tenth the size of HPC in both number of jobs and XD SUs consumed.   As indicated previously, the \osg{} utilization reported here is restricted to community access of \osg{} via the XRAC allocation process (XSEDE).   As discussed in the Introduction, the job type classification scheme employed to generate this figure is based on the resource type classification taken from the Teragrid/XSEDE central database, and not the result of an analysis of individual research awards.    For example, even though all jobs on \gordon{} are classified
as data intensive, we know from the job performance data that
there were community HPC applications (such as \texttt{NAMD},
\texttt{LAMMPS}, and \texttt{GROMACS}) that ran on \gordon{}.
These community HPC applications are likely not
being used for data intensive computing. It is also possible
that data intensive jobs were run on the HPC-type resources in the
NSF portfolio. Unfortunately,
we do not have classification information for each job so
we are not able to address the question about how much data 
intensive application usage is on HPC resources and how
this varies over time.

It is also important to note that the cloud resource utilization in terms of number of jobs is  underreported due to known issues with the accounting data sent to the XSEDE central database.  The XD SUs consumed are correct.  As noted in Appendix \ref{appendix:ResourceCharacteristics} there are two methods by which jobs are submitted to \jetstream{}: via the Openstack API and using the Atmosphere portal developed by CyVerse at the University of Arizona
(\url{http://www.cyverse.org/}) with 31\% and 69\% of jobs submitted using each respective method.
While we have been able to extract accurate data from the XDCDB for jobs submitted via the Openstack
API, accounting data submitted to the XDCDB by Atmosphere is aggregated by user and allocation on a
roughly daily basis and groups together all virtual machines in the given reporting period. Due to
this summarization, we are only able to determine the total number of XD SUs charged to a particular
allocation and unable to determine information such as the number of virtual machines (VMs), the number of cores per
VM, or the times that a given VM was running. We are at present working with 
personnel  to resolve these issues and improve the data reporting.

\subsection{Job Sizes}

\begin{figure}[ht]
\centering
\includegraphics{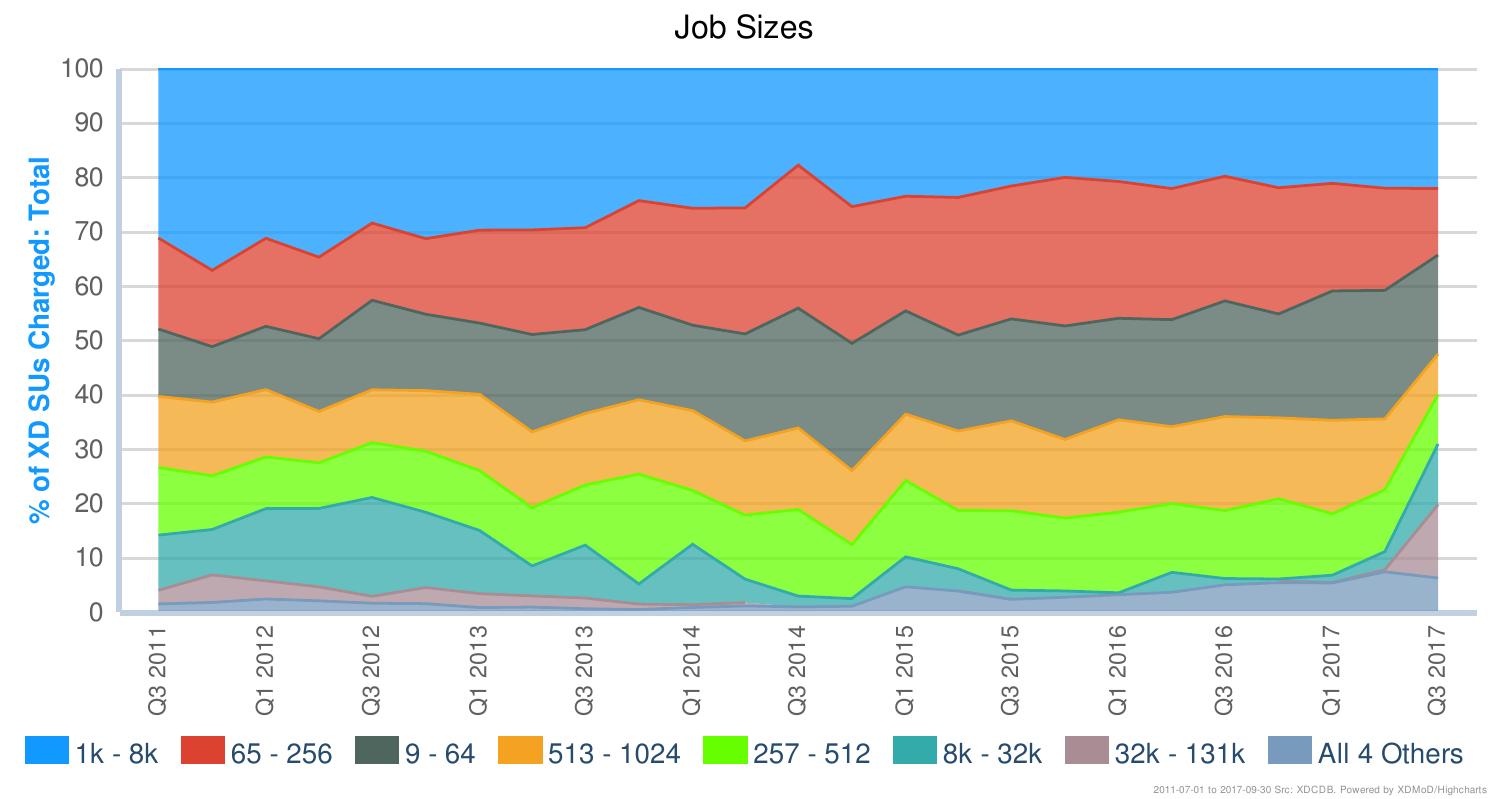}
\caption{\label{fig:job-sizes}Job size distribution for the period 2011-07 to 2017-09.  The job sizes in the ``Others'' category are 1,2-4,5-8, and >131k cores.}
\end{figure}

\begin{figure}[ht]
\centering
\includegraphics[width=0.8\textwidth]{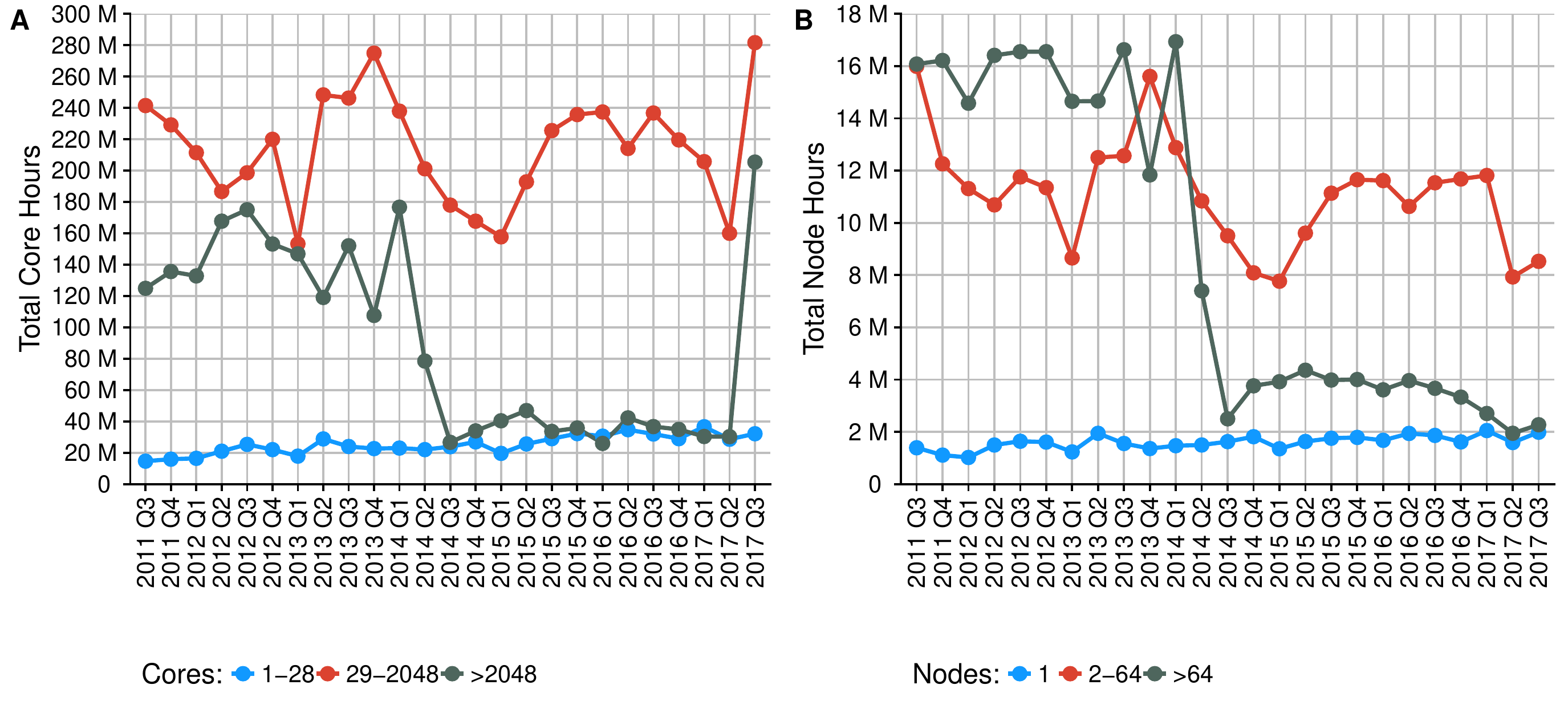} 
\caption{\label{fig:job_sizes_over_time} Changing of HPC job sizes over time (without \osg{} jobs). A - job sizes shown as core count. B - job size shown as node count.}
\end{figure}

\begin{figure}[ht]
\centering
\includegraphics[width=1.0\textwidth]{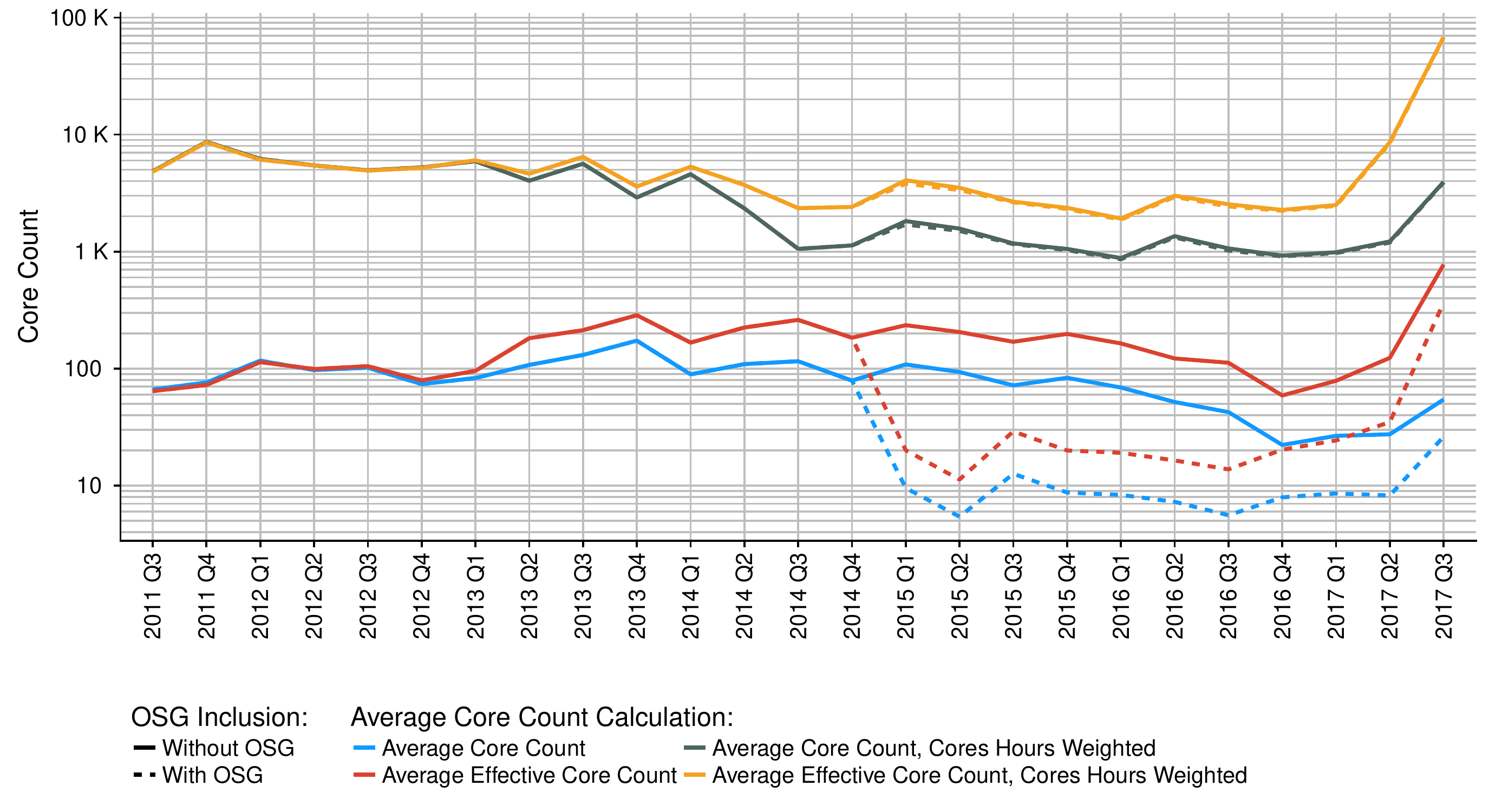}
\caption{\label{fig:ave-core-counts}Average core counts and effective core counts for the period 2011-07 to 2017-09. To account for the difference in CPU core performance, an effective core count was calculated as the actual core count times XD SU conversion factor of the resource where the job ran, divided by the XD SU conversion factor of \kraken{}. The inclusion of \osg{} jobs is shown by dashed lines, while for the solid lines \osg{} jobs are excluded. It should be noted that in the case of \stampedetwo{}, average core counts were computed using 68 physical cores per node as opposed to 272 hardware threads.}
\end{figure}

Figure \ref{fig:job-sizes} shows the job size distribution for the period 2011-07 to 2017-09 by percent of total XD SUs consumed within a given bin size (range of cores).    
While there is substantial fluctuation over the years with respect to the relative contribution that each bin size makes to the total XD SU consumption, the most obvious trend is the decrease in job sizes larger than 1024 cores.   This is supported in our ``deep and wide'' analysis in Section \ref{subsec:depthwidth} and in Figures \ref{fig:job_sizes_over_time} and \ref{fig:ave-core-counts}.   Figure \ref{fig:job_sizes_over_time} shows the distribution of total core and node hours broken out in 3 bin sizes (1-28 cores, 29-2048 cores, and > 2048 cores).   The equivalent distributions in node count are also included.   The precipitous drop in large core jobs in 2014 is clearly evident.  Figure \ref{fig:ave-core-counts} shows the average job size by core count over the same period, both weighted by XD SUs (solid black line) and unweighted (solid blue line).  The unweighted average is fairly steady over the study time period  (dip in 2016), while the weighted average shows a steady decrease from a high of almost 10,000 cores to a plateau of about 1000 cores.   As discussed in Section \ref{subsec:depthwidth}, this decrease coincides with the retirement of \kraken{} in 2014 and with the introduction of Blue Waters, a capability class HPC resource (as well as with resource job submittal policies which limit the maximum number of cores).  Core counts tick up strongly at the end of the study period with the addition of \stampedetwo{} and its multicore architecture.   

Focusing solely on  average core count for the weighted and unweighted jobs without taking into account increases in core performance that occur over time can be misleading.  For this reason, we introduce the concept of an ``effective'' core. The effective core is calculated by taking the actual core count for a given resource and multiplying it by the SU conversion factor for that resource and then dividing the result of the SU conversion factor for \kraken{}. This is shown in Figure \ref{fig:ave-core-counts} which includes time histories of the ``effective'' average weighted (solid gold line) and unweighted (solid red line) core counts. Not surprisingly, ``effective'' cores have the effect of making the decrease in average core count less pronounced.  Also shown in this figure is the effect that \osg{} jobs have on the average weighted (dashed gold and black lines) and unweighted core counts (dashed red and blue lines) for both the ``effective'' and actual average core counts.  Since the \osg{} jobs consume a small amount of XD SUs when compared to the HPC jobs, the weighted curves show no difference when \osg{} jobs are removed. In the case of the unweighted average core counts, inclusion of \osg{} jobs, which were only properly accounted for starting in 2014, results in a substantial drop.

\begin{figure}[ht]
\centering
\subfloat[\label{fig:job-size-1core}Single CPU core jobs 2011-07 to 2017-09.]{%
\includegraphics[width=\textwidth]{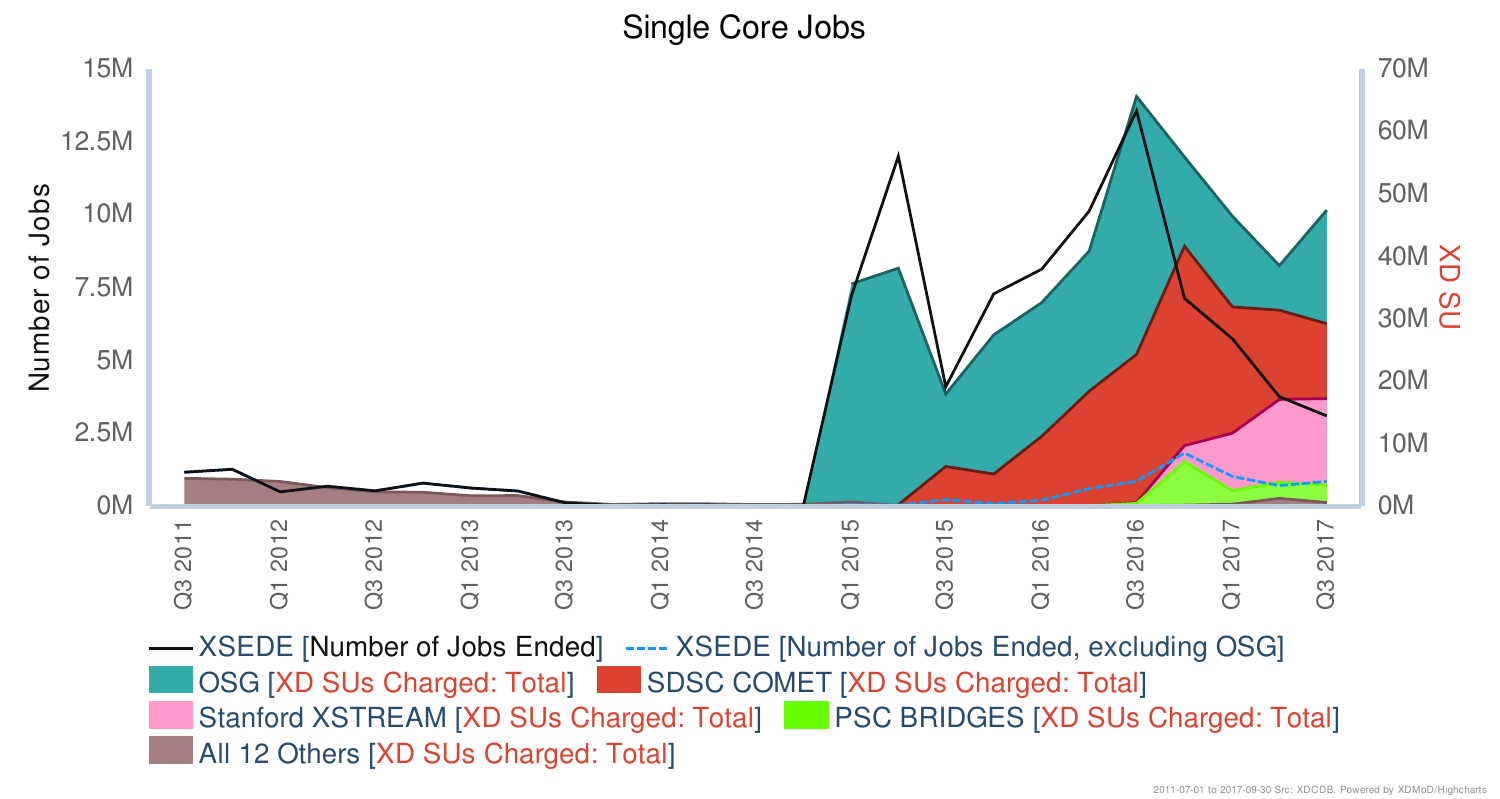}
}
\hfill
\subfloat[\label{fig:job-size-1node}Single node jobs 2011-07 to 2017-09.]{%
\includegraphics[width=\textwidth]{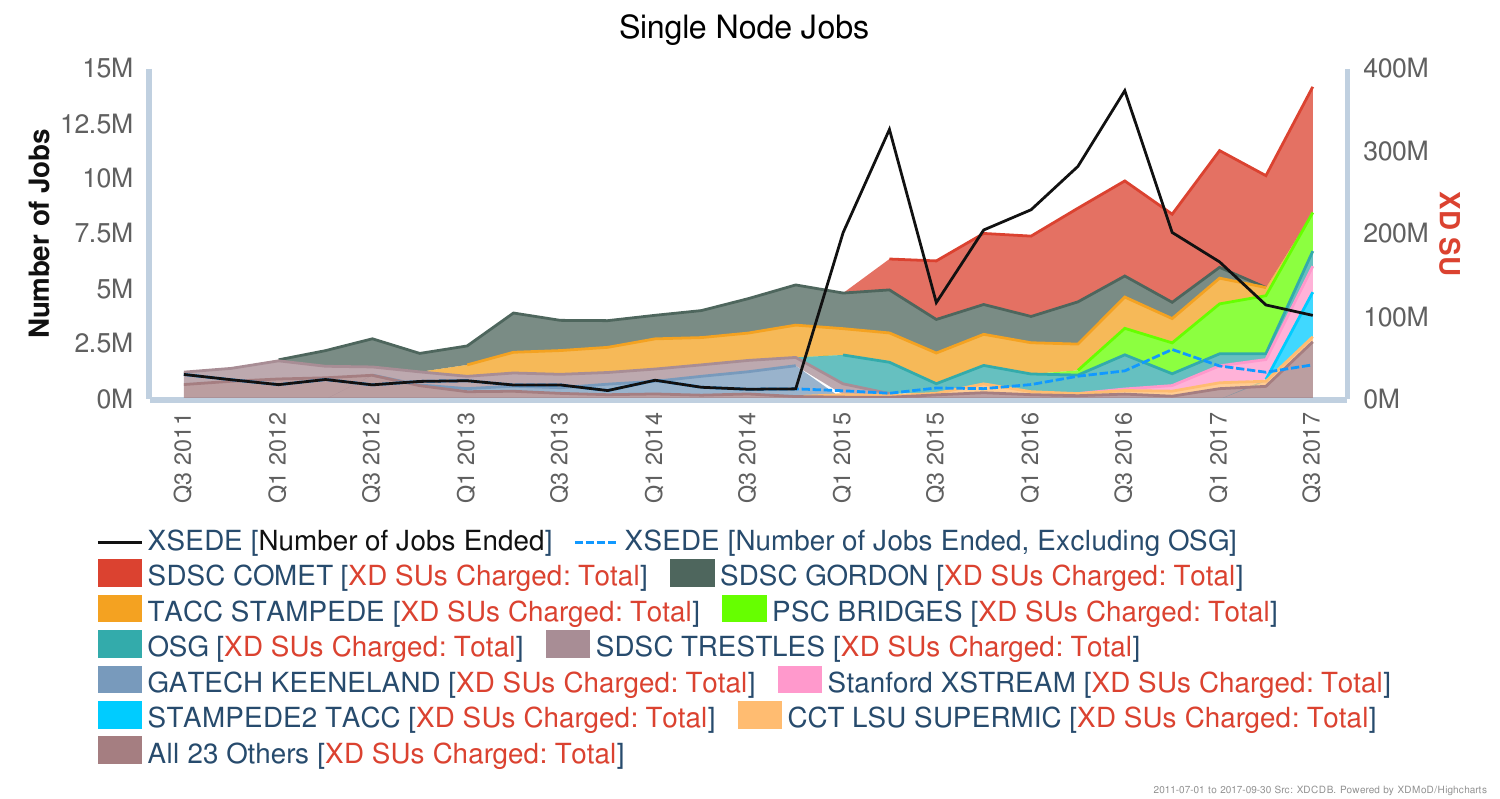}
}
\caption{\label{fig:job-size-1} Single core jobs (a) and single node jobs (b) for the period 2011-07 to 2017-09. Shaded areas are CPU time (in XD SUs).  Solid line indicates job count, dotted line job count excluding \osg{}.}
\end{figure}

Another trend in job sizes is a significant increase in single core and single node jobs as evident in Figure \ref{fig:job_sizes_over_time}. Figures \ref{fig:job-size-1core} and \ref{fig:job-size-1node} look in greater detail at serial and single-node jobs, showing single-core and single-node jobs grouped by resource.  Included in both plots are the total number of jobs run (solid black line) and the total number of jobs run excluding \osg{} jobs (dashed blue line).  The lack of single-core jobs until 2015 is due to exclusive node policies, where an entire node is allocated to single-core requests, and to missing \osg{} data. As discussed previously, prior to 2015, \osg{} jobs were not properly accounted for in the XSEDE central database (they were severely under-counted), and accordingly the large increase in jobs that occurs is the 1st quarter of 2015 is a reflection of these jobs now being properly accounted for.  All \osg{} jobs are serial and although there are a large number of them, they consume a relatively small fraction of the total CPU hours (see also Figure \ref{fig:NSF_Resource_Type_Utilization}). Even excluding \osg{} jobs, serial and single-node jobs have grown over time (dashed blue lines). Much of this can likely be attributed to the introduction of node-sharing policies on \comet{} and \bridges{} as well as increases in single core performance and core count per node.

\begin{figure}[ht]
\centering
\subfloat[\label{fig:xwl_plot_189.pdf}]{%
\includegraphics[width=\textwidth]{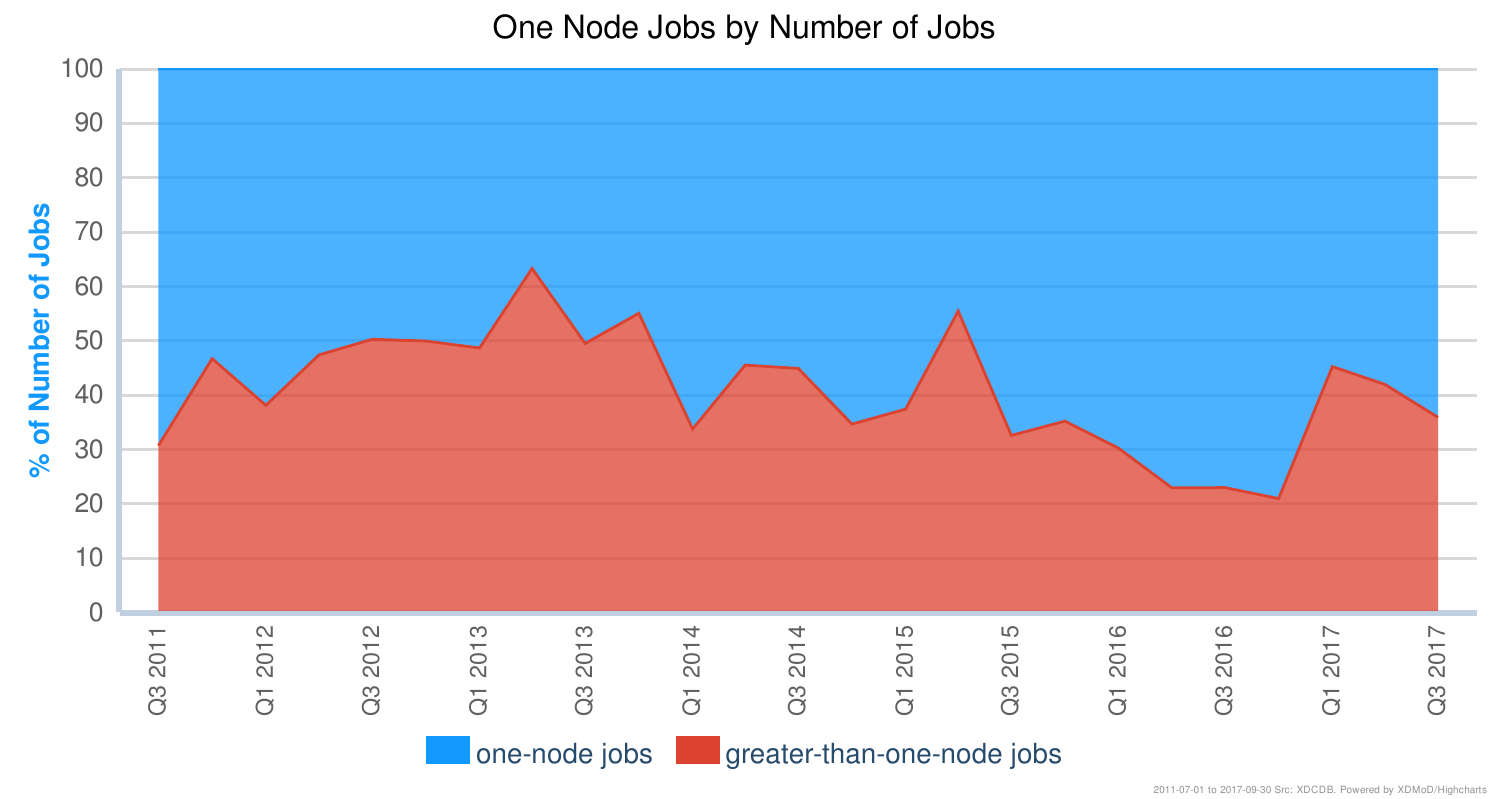}
}\hfil
\subfloat[\label{fig:xwl_plot_190.pdf}]{%
\includegraphics[width=\textwidth]{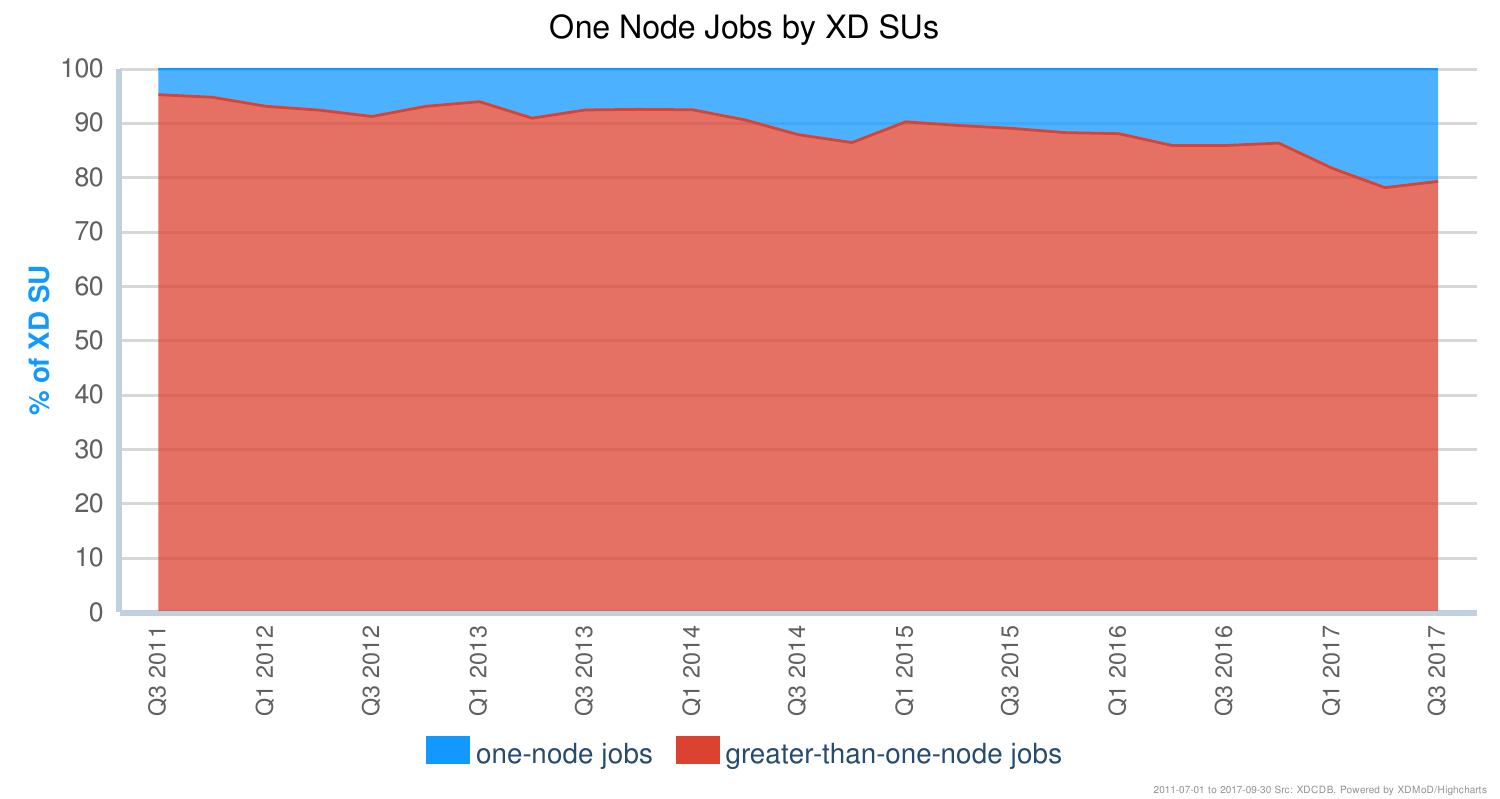}
}
\caption{\label{fig:percentage_single_node} The percent of the job mixture that corresponds to single node jobs by number of jobs (a) and by XD SUs consumed (b).  Note that \osg{} jobs are excluded from these plots.}
\label{fig:one-node}
\end{figure}

In terms of XD SU's consumed, the percentage of single node jobs (excluding \osg{} jobs) increased from 5\% of all XD SUs consumed in the second half of 2011 to 21\% in the first three quarters of 2017 and at least 14\% of the latter was consumed by serial jobs (that is 3\% of all XD SUs consumed). This trend is reflected in Figure \ref{fig:percentage_single_node}, which shows the percent single node jobs in terms of total number of jobs run (Figure \ref{fig:xwl_plot_189.pdf}) and total number of XD SUs consumed (Figure \ref{fig:xwl_plot_190.pdf}).  In the first three quarters of 2017, single node jobs accounted for more than 60\% of the total jobs (excluding OSG jobs).   Subsection \ref{subsec:concurrency} (Concurrency and parallelism) contains additional information on single node and serial jobs utilization.

\clearpage
\subsection{Memory Usage}

This subsection describes the memory usage of the jobs run on the various
facilities. We look at the memory distributions and trends over time and
discuss the large memory jobs in particular.  The memory usage of an HPC job
varies over time but we only sample the memory usage periodically.  
Therefore, we use two different metrics to characterize the job memory usage
--- the mean value of these samples and the maximum value. To facilitate easy
comparison of memory usage between jobs of different sizes we generally
present the memory per core (which is equivalent to the memory used
per process for most parallel HPC jobs).  The memory data presented here
includes the usage by the operating system (O/S) as well
as the HPC user software. Therefore, any observed changes could be due to HPC
user software or the O/S. The O/S buffer cache
and O/S kernel slab cache are not included in the memory usage data, see
Appendix~\ref{appendix:taccstats} for further information.

\begin{figure}[H]
\centering
\includegraphics {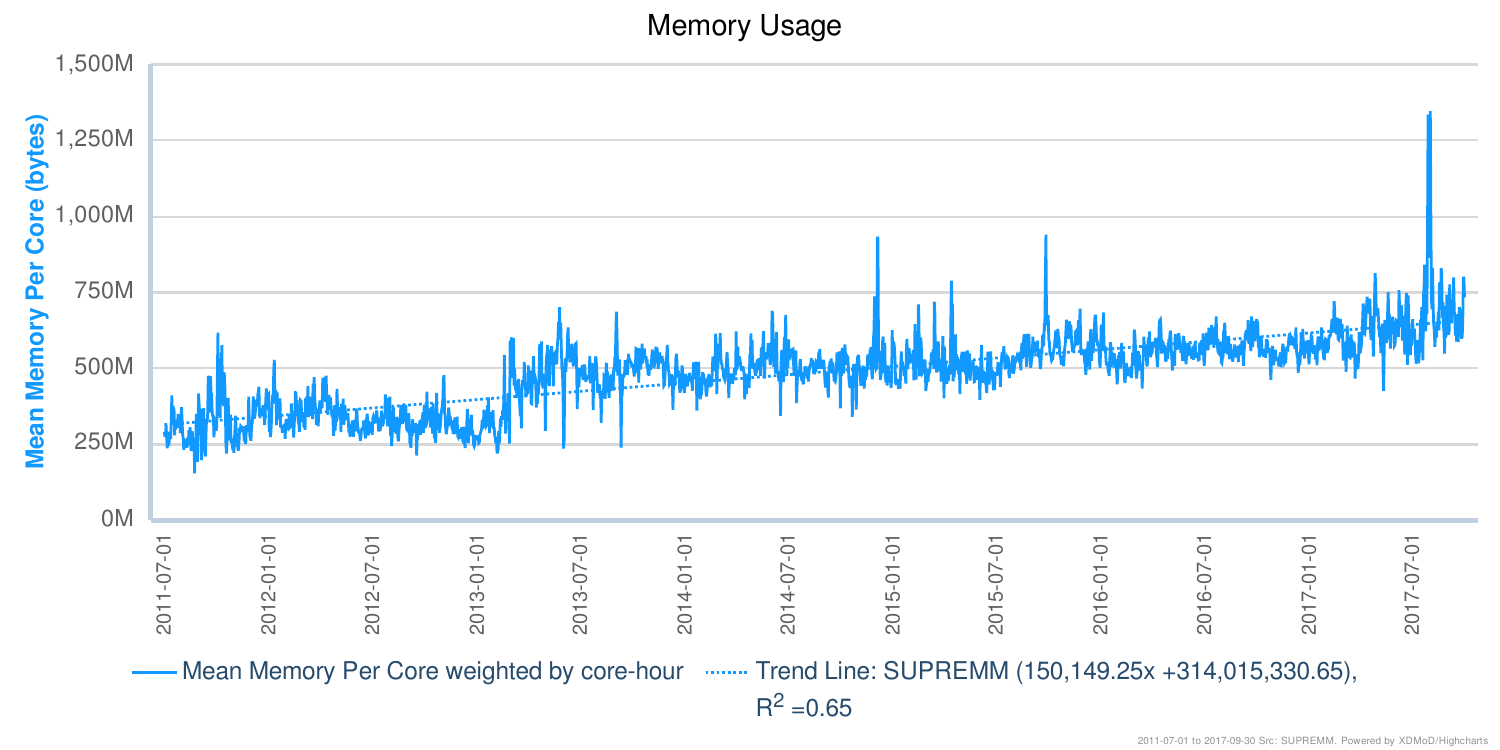}
\caption{\label{fig:xwl_plot_144.pdf}Plot of average memory used per core, weighted by core-hour, for \ranger{}, \lonestar{}, \stampede{}, \gordon{}, \comet{}, \cometgpu{}, \supermic{} and \darter{}.}
\end{figure}

\begin{figure}[H]
\centering
\includegraphics {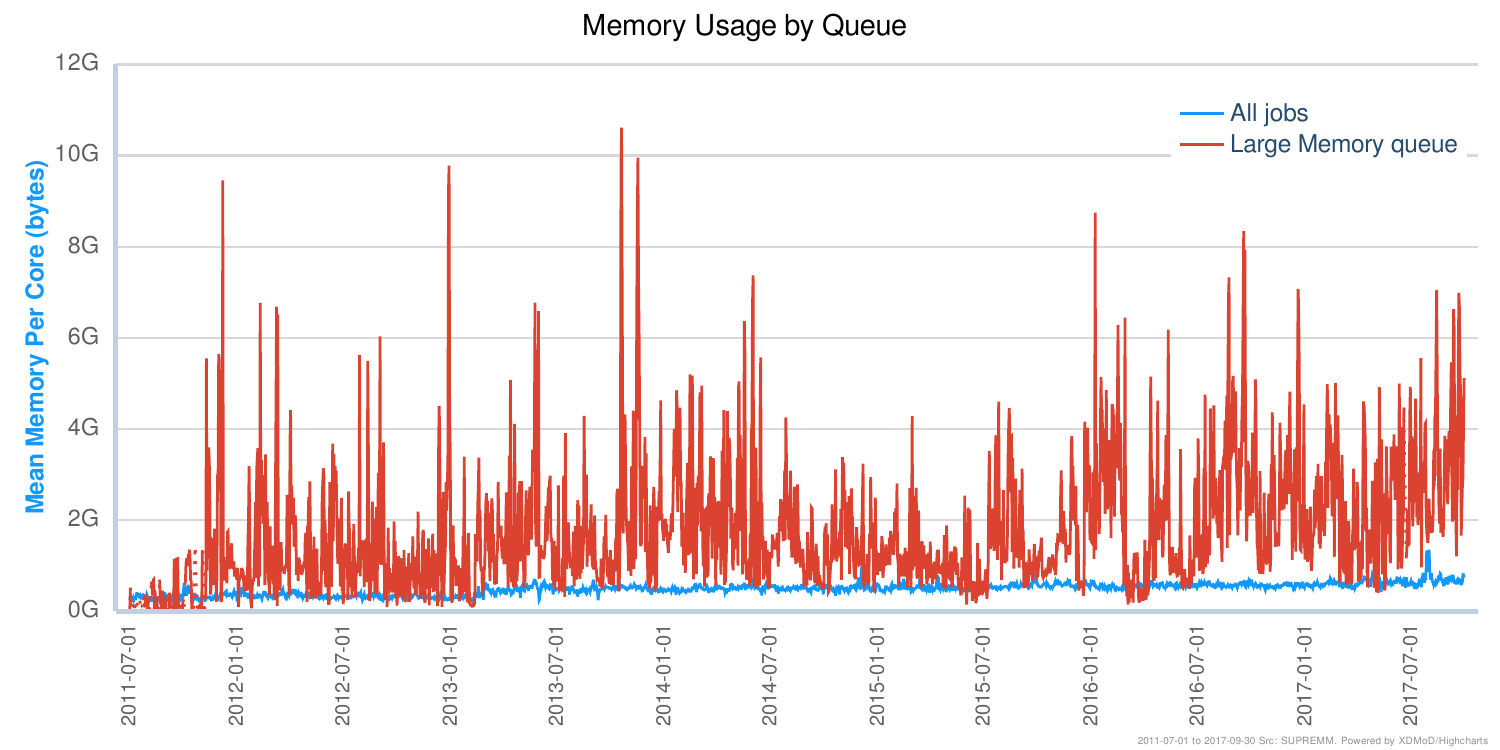}
\caption{\label{fig:xwl_plot_170.pdf}Plot of average memory used per core, weighted by core-hour, for the large memory queues on \ranger{}, \lonestar{}, \stampede{}, \comet{} and \darter{} (red line) and for all queues (blue line) on \ranger{}, \lonestar{}, \stampede{}, \gordon{}, \comet{}, \cometgpu{}, \supermic{} and \darter{}.}
\end{figure}

\begin{figure}[H]
\centering
\includegraphics {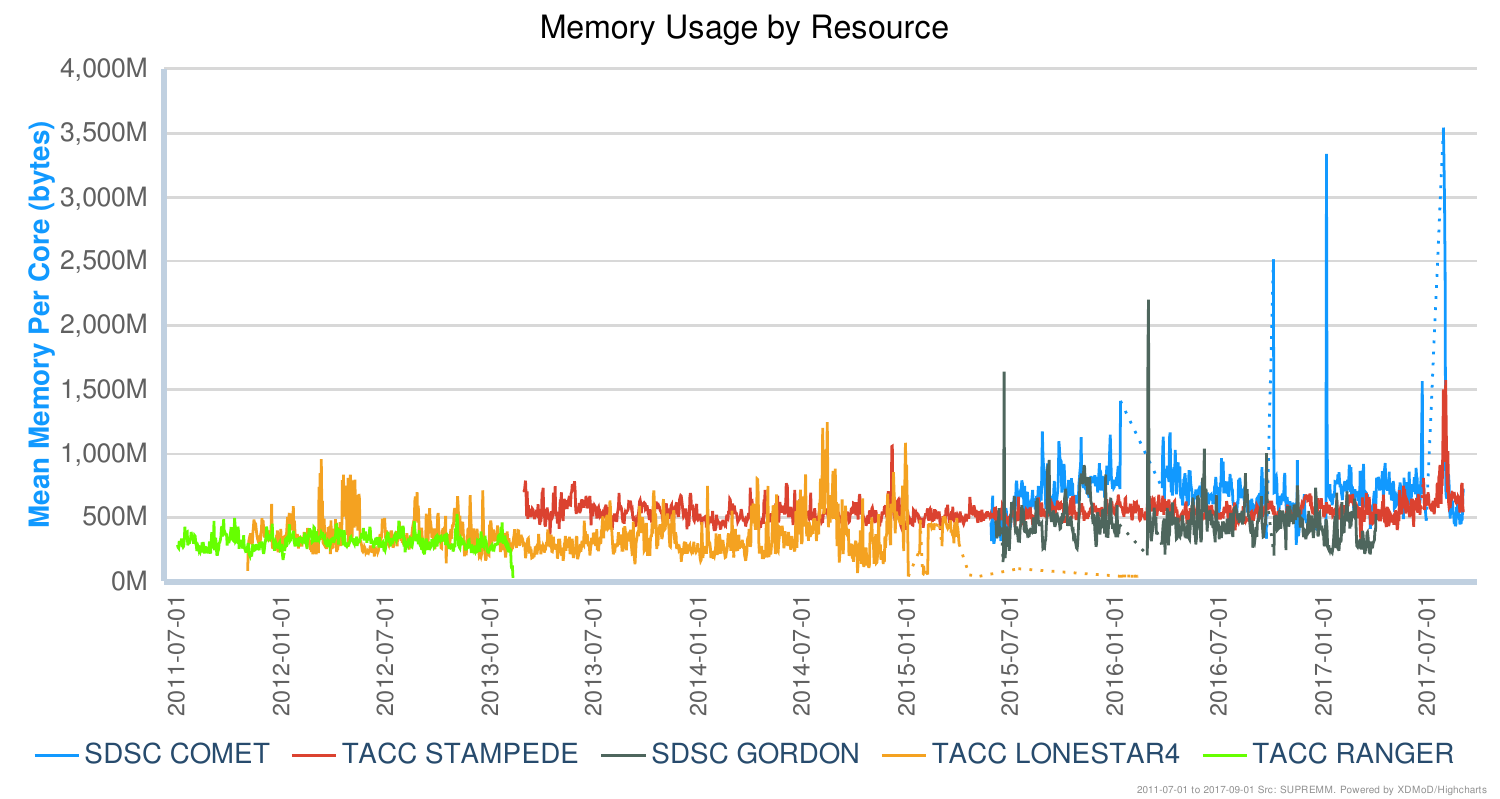}
\caption{\label{fig:xwl_plot_169.pdf}Plot of average memory used per core, weighted by core-hour, analyzed by resource for the ``normal compute'' queues on each resource.}
\end{figure}

The average memory used per core by all resources (for which node level metrics
were available - see note below) weighted by core-hours is shown in  Figure
\ref{fig:xwl_plot_144.pdf}.  There is a modest trend towards increased memory
usage per job, as shown in the figure.  The memory usage by jobs in the large
memory queues is shown in Figure \ref{fig:xwl_plot_170.pdf}.  As expected, the
memory usage per core is substantially higher in the large memory queues but
there is no time dependent trend to higher values.  There is also no trend to
larger usage (that is  no increase in XD SUs) for the large memory queues
relative to overall normal queue usage (not shown). 
The average memory usage per job analyzed by individual resource is shown
in Figure \ref{fig:xwl_plot_169.pdf}. 
Note that all resources have reasonably flat memory usage over time.  Aside
from a few outlying spikes probably caused by a few large memory usage jobs
that occur during low system usage periods, the core-weighted average memory
usage on all resources is less than 1 GB/core. 
The memory usage does vary between different resources, with jobs on newer
resources having a higher average memory usage than older resources.
One of the reasons for the change in memory usage for the different
resources is the O/S memory usage. In general, the O/S 
memory usage  is smaller on older resources. For example, the O/S usage on
 \lonestar{} was approximately 450MB per compute node (37 MB per
core), whereas newer resources such as \stampede{} and \comet{} have approximately
4GB per compute node (\approximately{250MB} and \approximately{170MB} per core, respectively).
A more in-depth picture of the memory usage was obtained by looking
at the entire distribution as a function of time rather than the core weighed average
value (not shown). It confirms that over the time range of the XSEDE program,
that while individual resources have flat memory usage, overall memory usage of
the complete ensemble of machines increases. Figures \ref{fig:xwl_plot_144.pdf}
through \ref{fig:xwl_plot_169.pdf} only include resources for which memory
metrics are available, notably the large memory resource \bridgeslarge{} 
is not included in the analysis due to the absence of data.

\begin{figure}[H]
\centering
\subfloat[\label{fig:xwl_plot_037.pdf}]{%
\includegraphics[width=0.5\textwidth]{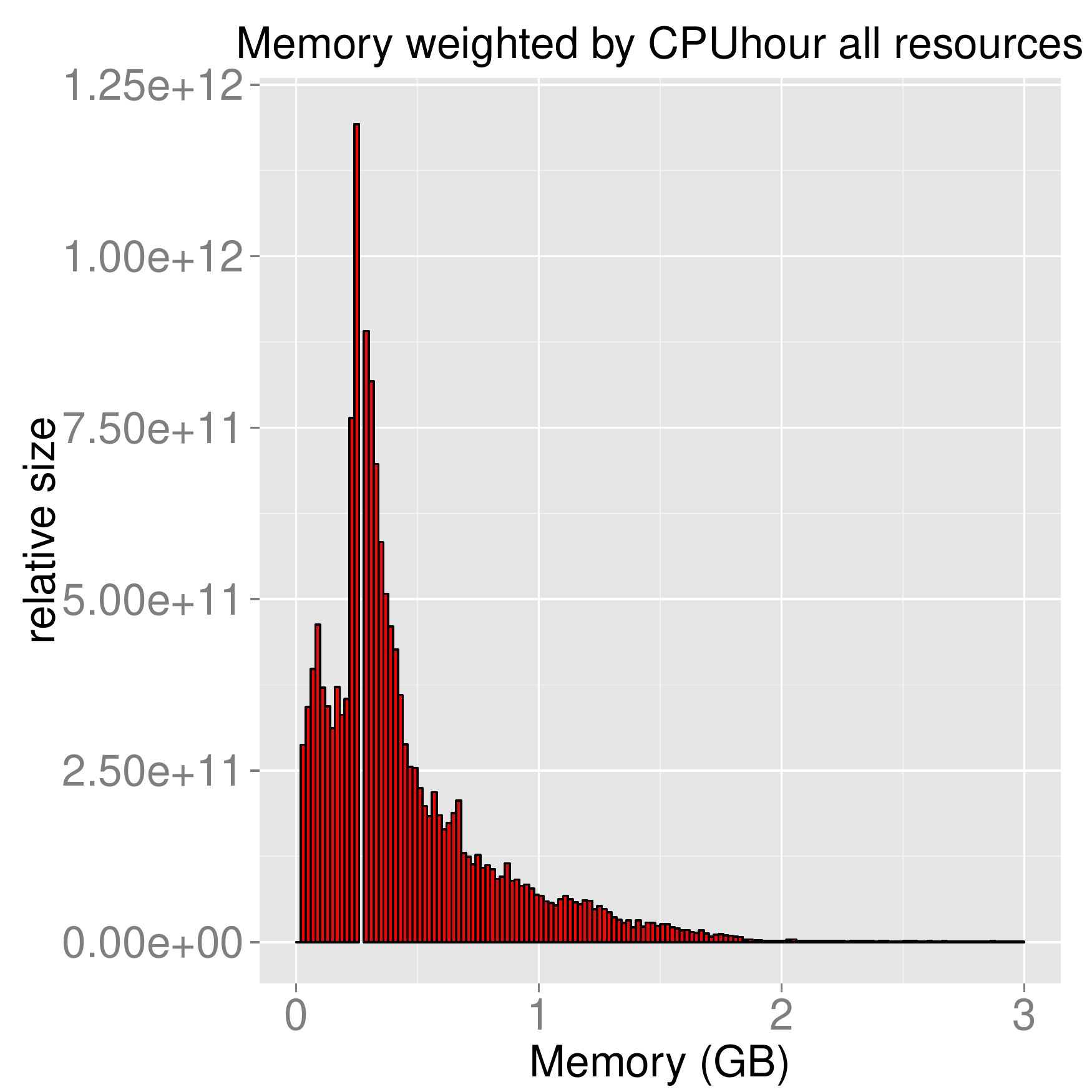}
}
\subfloat[\label{fig:xwl_plot_038.pdf}]{%
\includegraphics[width=0.5\textwidth]{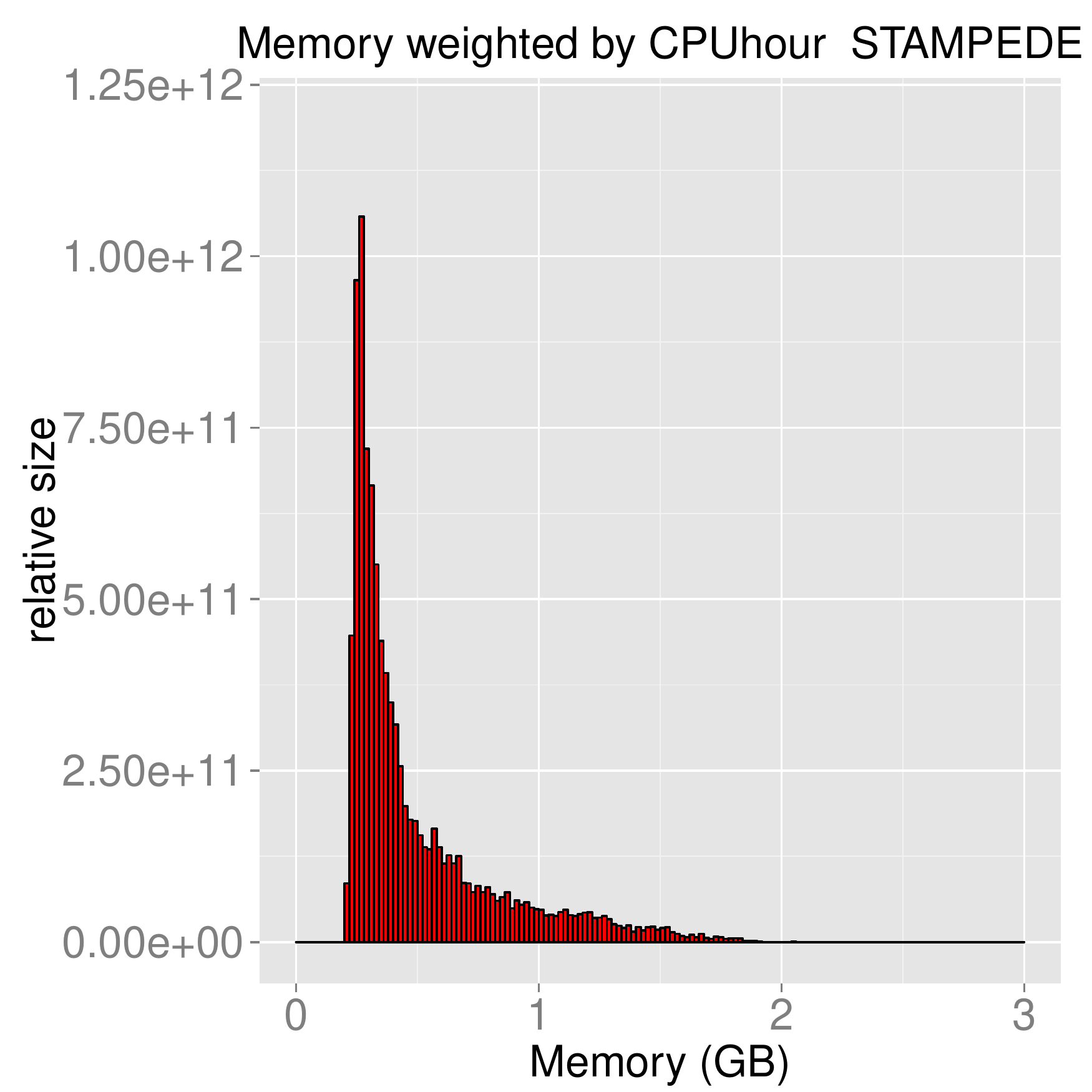}
}
\caption{Histogram of average memory used per core for HPC jobs on (a) all resources that provide memory usage data and (b) \stampede{}.}
\label{fig:cpu-hour-hist}
\end{figure}

\begin{figure}[H]
\centering
\subfloat[\label{fig:xwl_plot_160.pdf}]{%
\includegraphics[width=0.5\textwidth]{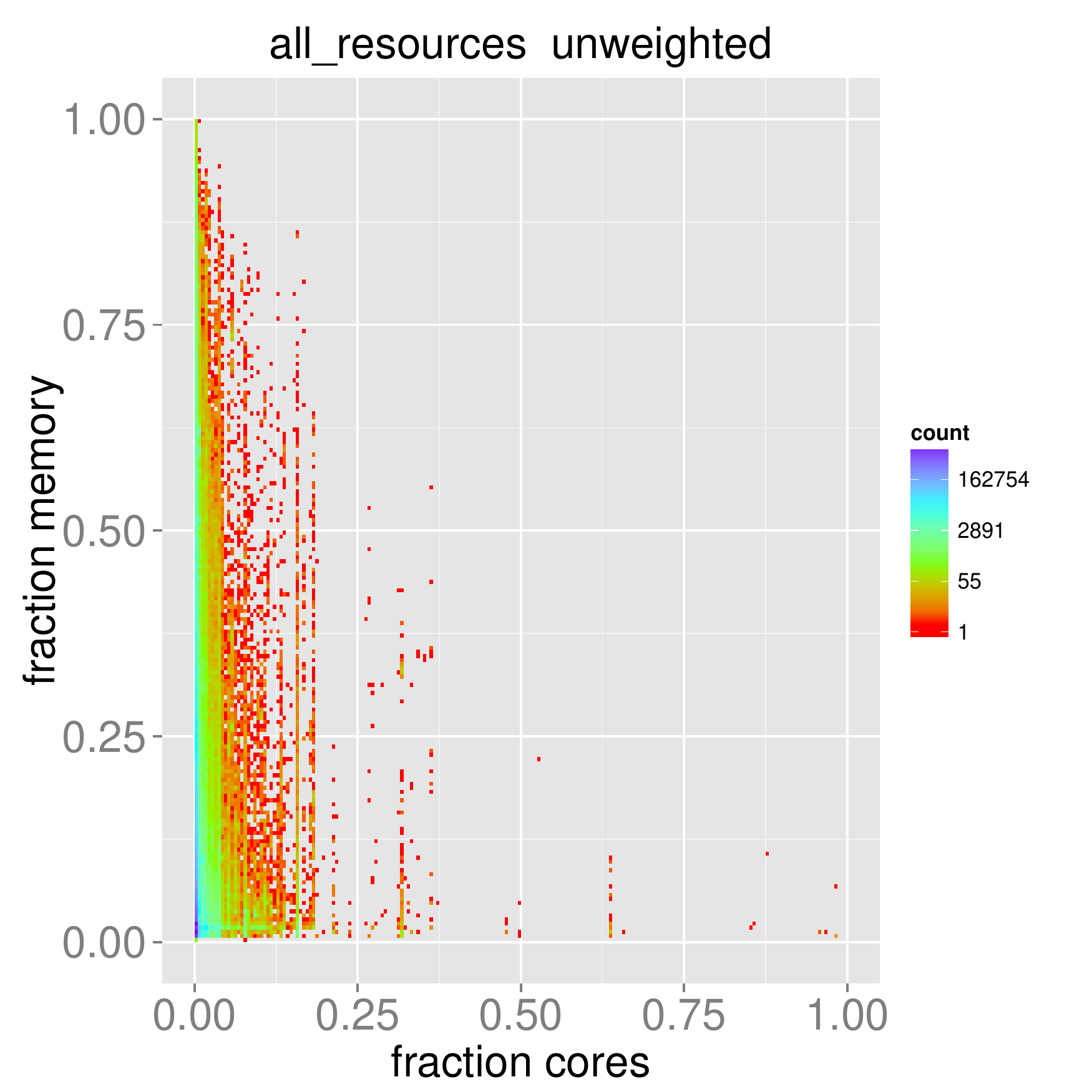}
}
\subfloat[\label{fig:xwl_plot_159.pdf}]{%
\includegraphics[width=0.5\textwidth]{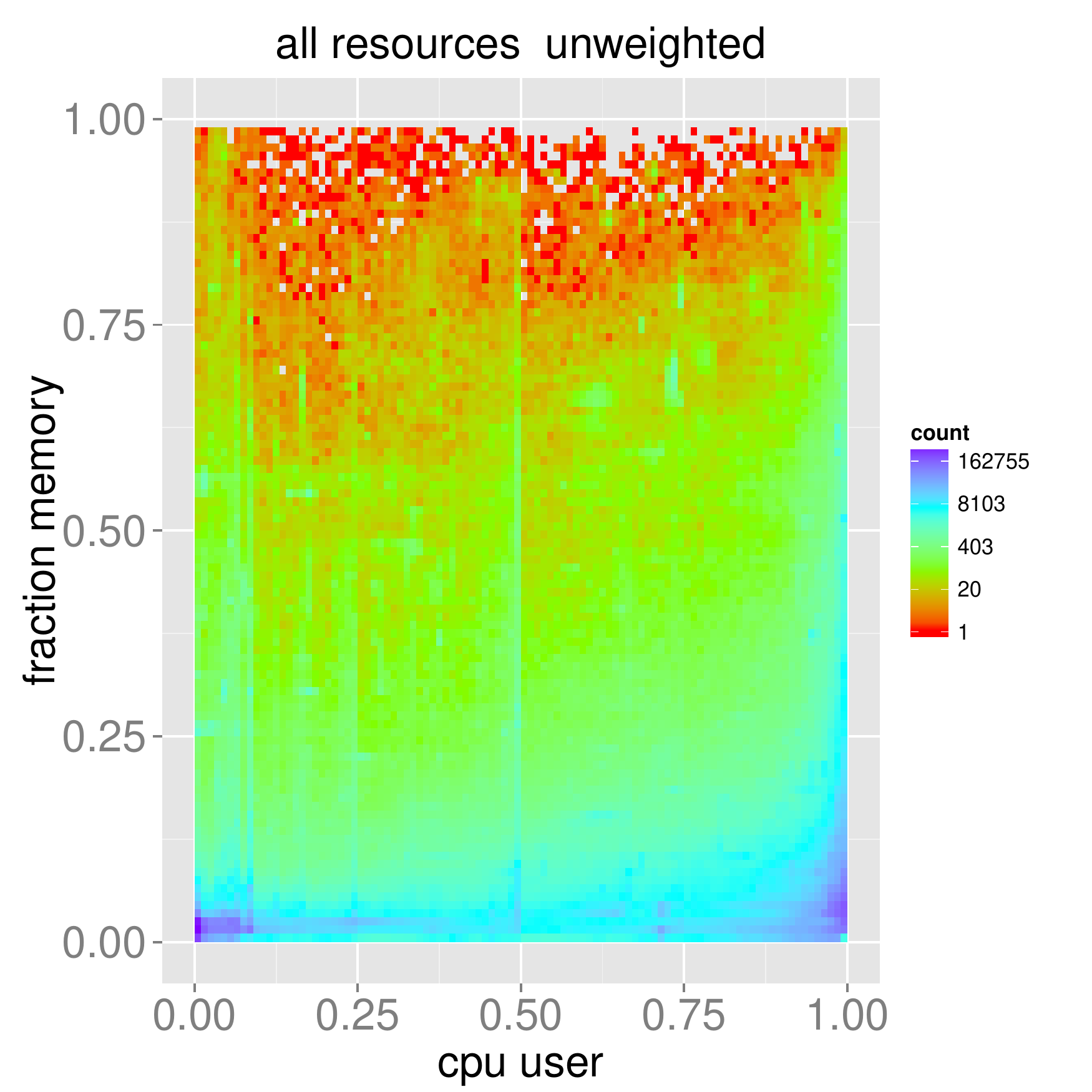}
}
\caption{\label{fig:fractionmem}2-D histograms of available memory used versus fraction of cores used (a) and memory used versus CPU user fraction (b), for all resources on which memory usage data is available, see Appendix~\ref{appendix:ResourceCharacteristics}.}
\end{figure}

A simple histogram can provide the distribution of memory usage by all jobs for
all resources.  Figure \ref{fig:xwl_plot_037.pdf} shows a version with the jobs
weighted by core-hour; the non-weighted version (not shown) is similar.  Most
resources have on the order of 2GB/core or more memory, however, since the
memory per core varies among resources, it is more appropriate to look at
memory utilization for individual resources.  Figure \ref{fig:xwl_plot_038.pdf}
shows the corresponding distribution for \stampede{}; the mean value of memory
used per core is only about 19\% of the available memory. Other resources have
a similar small value of fractional memory used. The 
standard compute nodes in \stampede{} have 2 GB/core,  therefore the tiny tail
beyond 2 GB is due to large-memory nodes which have 32 GB/core.  Note the offset from zero memory
usage is from the O/S memory usage. With the exception of the offset, the
individual resource plots for \comet{}, \gordon{}, and \lonestar{} (not shown)
are similar.  Figure \ref{fig:xwl_plot_160.pdf} shows a 2-D histogram with fraction of memory used
on the $y$-axis and fraction of cores on the $x$-axis.  Fraction of memory is
defined as the fraction of the total memory available that is used by the job;
fraction of cores is defined as the ratio of cores used by the job to the total
available cores on the particular HPC resource on which the job ran. 
As discussed in Section~\ref{subsec:depthwidth} there are resource policies that limit 
the maximum job size allowed for users. Therefore there are very few jobs
that use a sizable fraction of the nodes on any of the resources.
 As far as
the memory usage goes, the largest jobs tend to be very diverse in their memory
usage.  Some use up to half of the available memory; others use only a rather
small fraction of the available memory.  Most of the jobs are relatively small
using much less than 1/10 of the nodes of the resource on which they run.  Here
too the memory use is very diverse.  Only a relatively small tail require more
than 1/2 of the available memory. 

\begin{figure}[h]
\centering
\subfloat[\label{fig:xwl_plot_158.pdf}]{%
\includegraphics[width=0.5\textwidth]{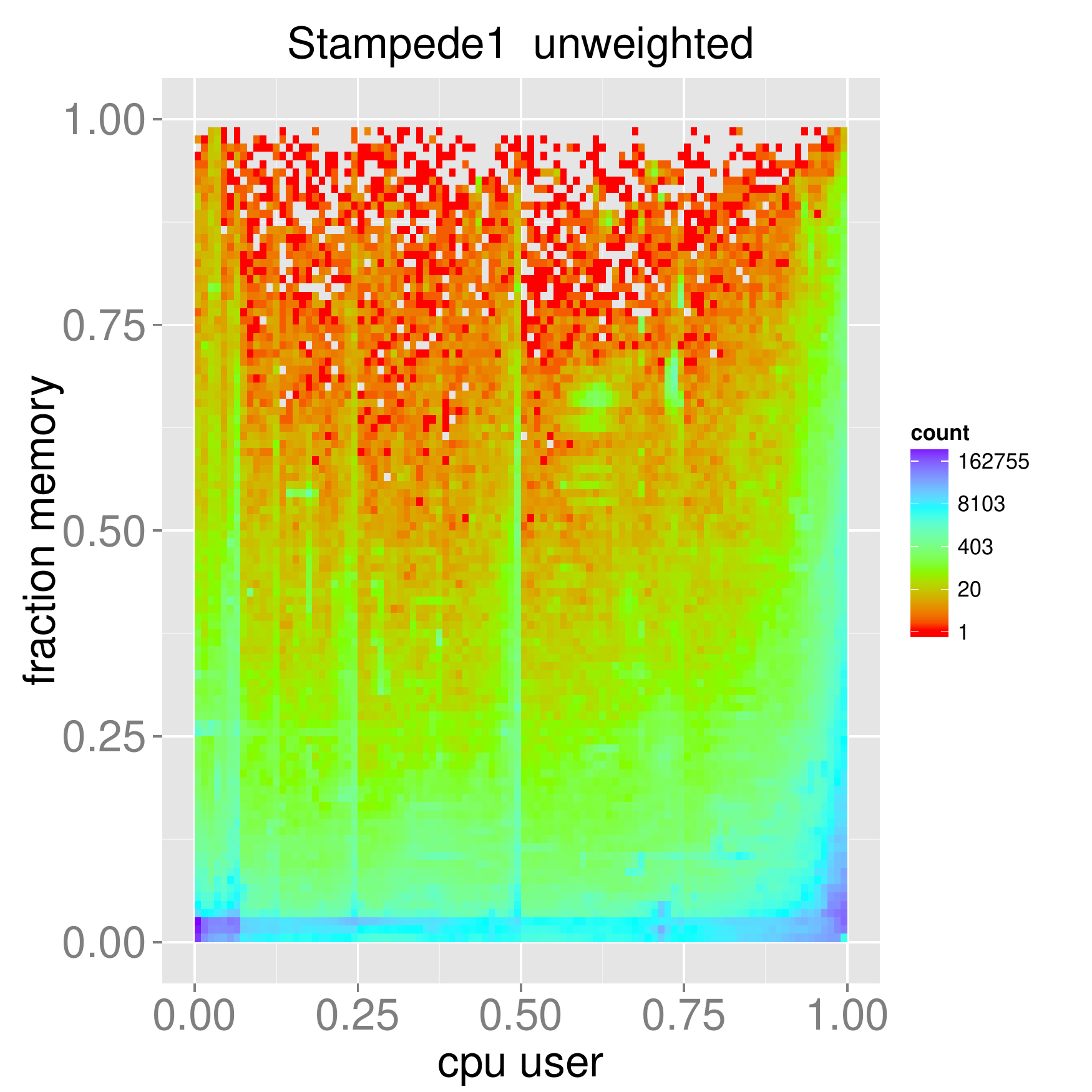}
}
\subfloat[\label{fig:xwl_plot_157.pdf}]{%
\includegraphics[width=0.5\textwidth]{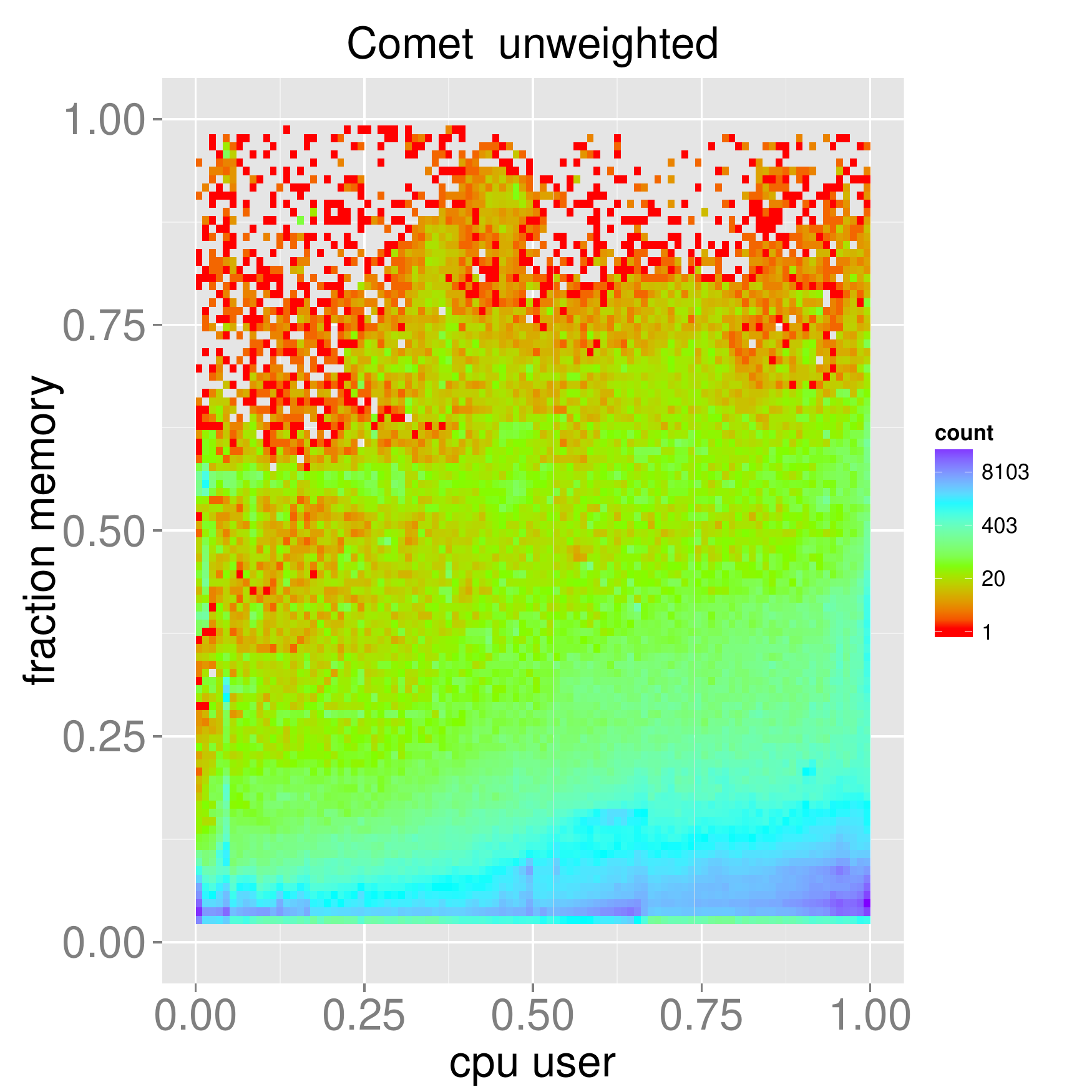}
}
\caption{\label{fig:mem-hist}2-D histogram of memory used vs cpu-user for (a) \stampede{} and (b) \comet{}.}
\end{figure}

A more insightful analysis can be achieved by examining not only the memory usage but simultaneously considering the job CPU usage as measured by the ratio of time spent in user mode to total time (cpu-user).  Figure \ref{fig:mem-hist} shows 2-D histograms of cpu-user on the $x$-axis and fraction of memory on the $y$-axis.  Fraction of memory is defined as the fraction of the total memory available that is used by the job. Figure \ref{fig:xwl_plot_158.pdf} shows this analysis for \stampede{}.  A similar analysis, not shown here, in which the jobs are weighted by cpu-hour has essentially identical features.  The first feature to note is that the jobs with high CPU usage, those with cpu-user near one, also tend to use large fractions of the memory on the order of 0.5 or more.  Conversely, those jobs with low cpu-user values tend to use relatively less memory.  Another feature of the plot is that there are vertical streaks of higher memory usage at fractions of 0.50, 0.25, 0.125 and 0.0625.  \stampede{} had 16 cores per node and these streaks correspond to 8-way, 4-way, 2-way and 1-way jobs, respectively.   In order to determine if these streaks are primarily due to single node jobs that do not utilize all cores on the node, we carried out a similar analysis (not shown) in which single node jobs were excluded.  This analysis produced results similar to Figure \ref{fig:xwl_plot_158.pdf} so we concluded that the observed striping of the memory usage is due predominantly to multi-node jobs.  Many of these jobs that do not use the full 16 cores per compute node may be jobs using fewer cores per node to obtain more memory per process. Interestingly, there is also a faint streak at 0.75 corresponding to 12-way jobs.  This may correspond to jobs ported over from \lonestar{}  in which the users fail to take advantage of \stampede{}'s 16 cores as opposed to \lonestar{}'s 12 cores.  Figure \ref{fig:xwl_plot_157.pdf} shows an analogous analysis for \comet{}.  Here there is almost a linear relationship between cpu-user and memory fraction. Also note that there are no obvious vertical stripes as in \stampede{}.  There are some faint horizontal stripes at \approximately{0.56} corresponding to \approximately{3} GB per core memory usage.   These are too faint to pick up in the simple 1-D histogram of \comet{} memory usage.  Like \stampede{}, \lonestar{} shows vertical streaks at values of cpu-user except that now the strongest are at 2/3 (8-way jobs), 1/3 (4-way jobs), 1/6 (2-way jobs) and 1/12 (1-way jobs) that match Lonestar 4's 12 cores per node.  There is also a streak at cpu-user values of 0.5 corresponding to 12-way jobs which, although not a power of 2, do use exactly half of the cores on each node. Figure \ref{fig:xwl_plot_159.pdf} is an analogous composite plot of all resources for which memory information is available.  Not surprisingly, it most closely resembles the \stampede{} plot since many jobs come from this resource but there are some features from other resources for which memory data is available.  For example, the strong vertical features from \stampede{} indicating jobs using only a fraction of the cores are diluted but there are other faint features corresponding to the same partial core usage on the other resources that have different number of cores per node.

Given that, as shown in Figure \ref{fig:xwl_plot_037.pdf}, the majority of jobs on all resources use less than 2 GB per core, one may be inclined to conclude that there is little need to design HPC architectures with more than 2 GB per core, as long as resources designed specifically for large memory jobs such as \bridgeslarge{} are available. However, figures such as \ref{fig:xwl_plot_037.pdf} only tell part of the story with respect to memory utilization, and making conclusions such as that based on this data is not advisable.  First, for some of the resources, such as \stampede{}, the majority of nodes have only 2 GB per core, and users whose applications require more than this obviously will not run on that resource.  Second, \stampede{}  accounts for the majority of XD SUs consumed since 2014 and therefore averaging the memory use over all resources, as was done in Figure \ref{fig:xwl_plot_037.pdf}, has the effect of masking the (larger) memory use of resources with smaller utilization. Third, many of the scientific applications running on present day HPC systems were designed to run with modest per core memory requirements, however many of today's fastest growing research areas are data intensive requiring modest core counts but large memory
per core. 

\begin{figure}[h]
\centering
\subfloat[\label{fig:xwl_plot_185.pdf}]{%
\includegraphics[width=0.5\textwidth]{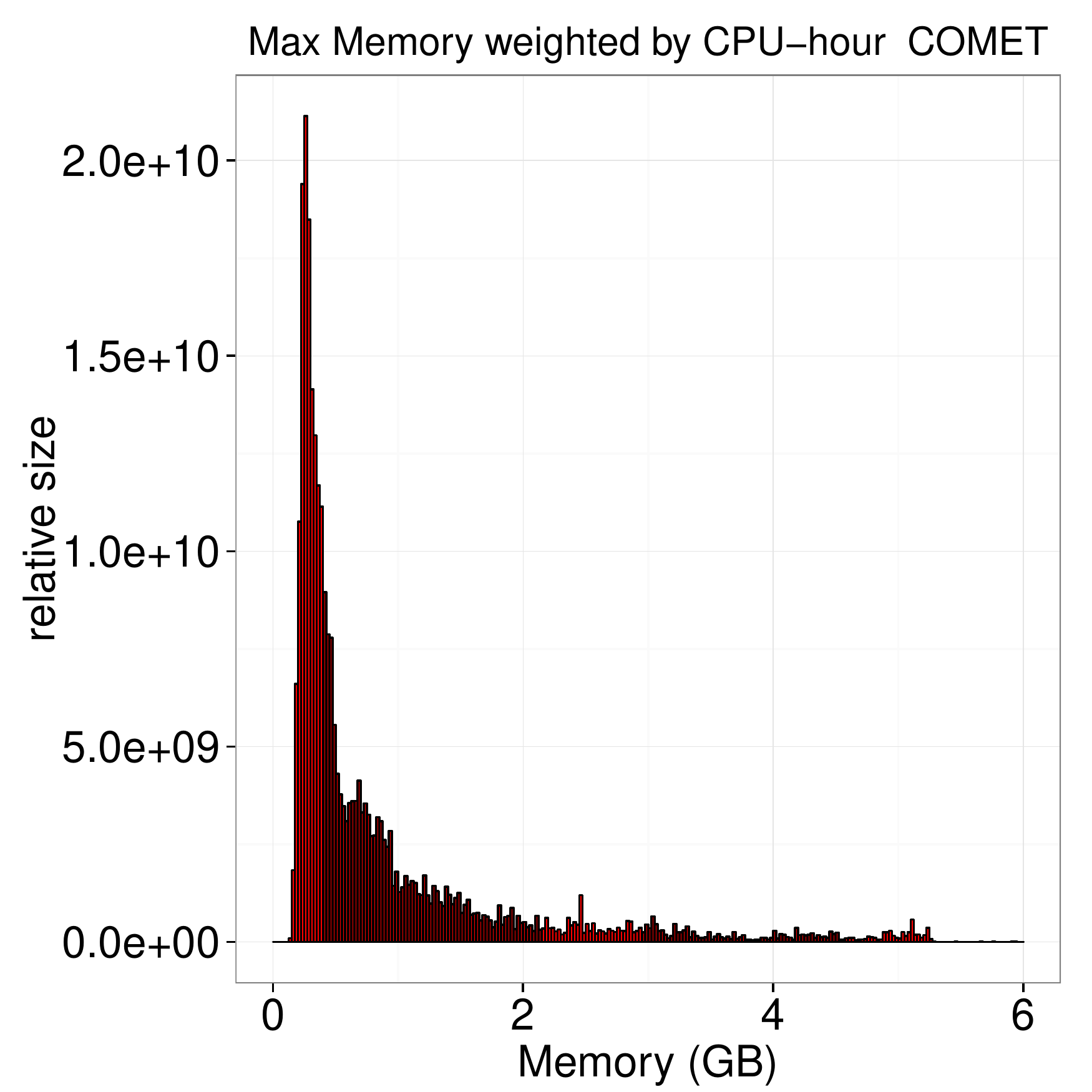}
}
\subfloat[\label{fig:xwl_plot_186.pdf}]{%
\includegraphics[width=0.5\textwidth]{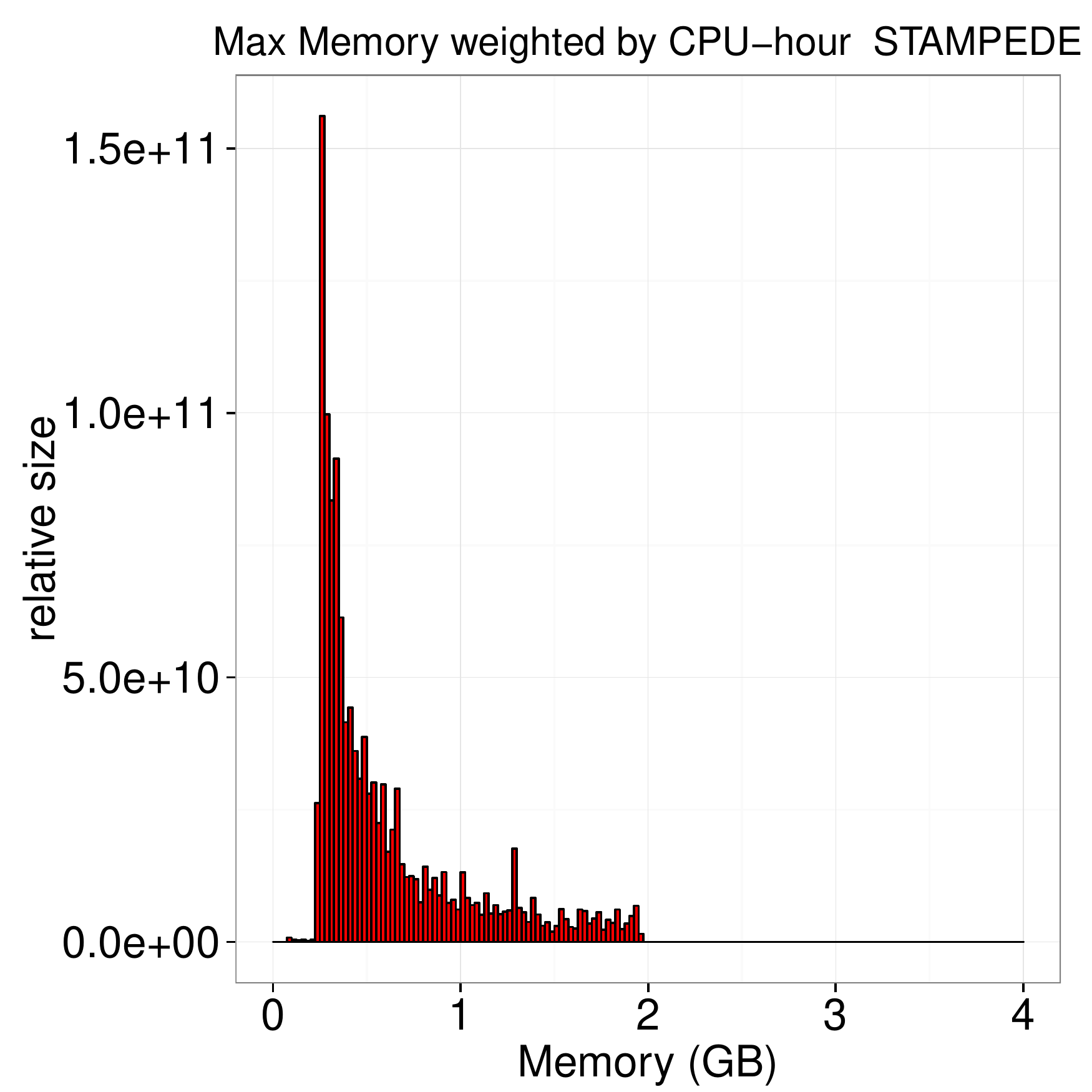}
}
\caption{Maximum memory used per core weighted by core-hour for (a) \comet{} and (b) \stampede{}.}
\label{fig:max-mem}
\end{figure}

As noted above, the usage of jobs that have large per-core memory requirements
is rather small as shown in Figure~\ref{fig:max-mem} which gives the maximum
memory used per core weighted by core-hour for \comet{} and \stampede{}.  
\stampede{} shows few jobs in the large memory queue, those beyond 2 GB/core in
Figure \ref{fig:xwl_plot_186.pdf} and \comet{} shows a long small tail of jobs
in the range of 2--5.3 GB/core and very few jobs in the large memory queue,
Figure  \ref{fig:xwl_plot_185.pdf}.  
This is not surprising since the plots of core-hour weighted data are not able
to show the details of the jobs running on large memory nodes since their usage is very small
compared to the overall ``normal'' queue usage (there are only 16 large memory
nodes out of 6400 on \stampede{} and 4 out of 1944 on \comet{}).

\begin{figure}[h]
\centering
\subfloat[\label{fig:total-mem-per-job-stampede}\stampede{}]{%
    \includegraphics{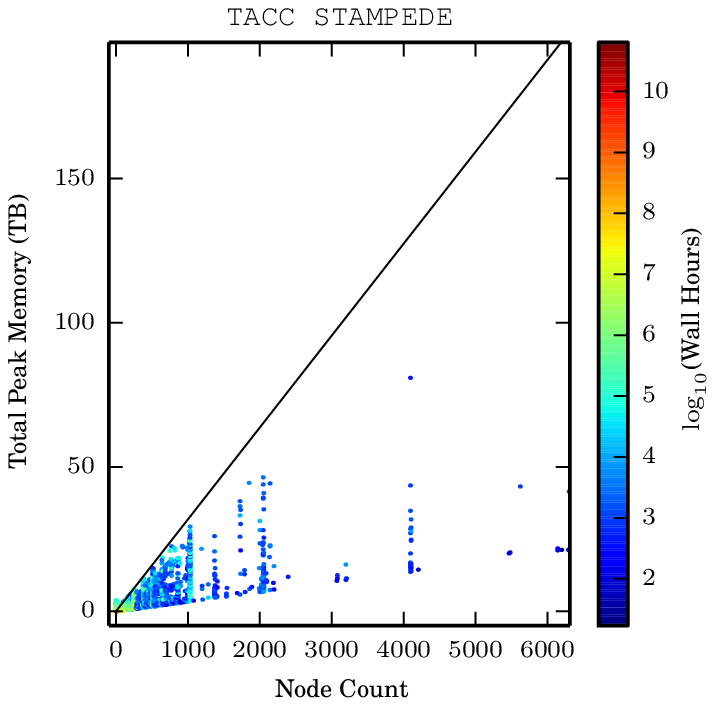}
}
\subfloat[\label{fig:xwl_plot_188.pdf}\comet{}]{%
    \includegraphics{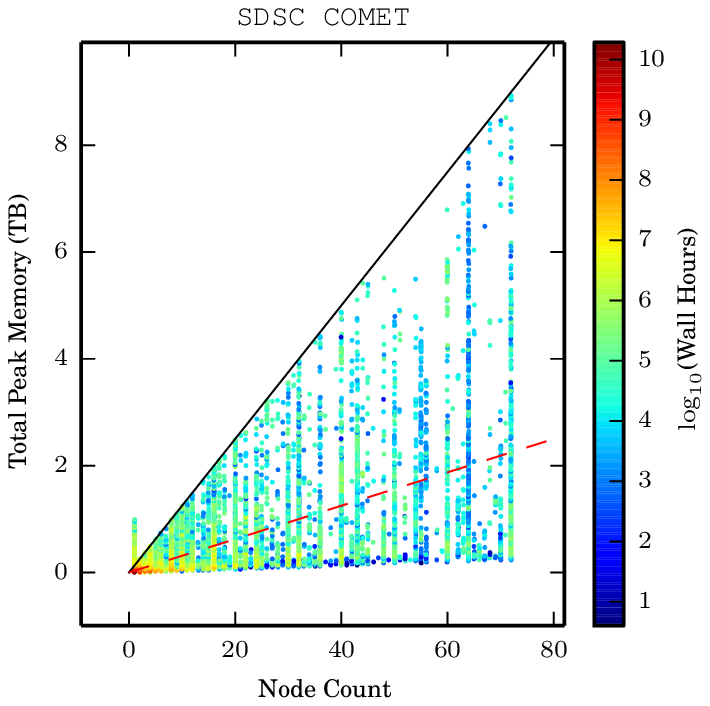}
}
\caption{\label{fig:total-mem-per-job}
    2D binned scatter plot showing the total peak memory usage for jobs on \stampede{} and \comet{}. The
    color coding shows the total wall hours for the jobs in each bin using a log scale. The solid diagonal
    line indicates the points corresponding to the memory available on the normal compute nodes. Note the 
    different $x$ and $y$-axis scales between each plots.  To facilitate comparison between the two charts, the dashed red line has been added to the \comet{} plot; it corresponds to the solid diagonal line in the \stampede{} plot.  That is, the jobs on \comet{} above the dashed red line utilize more memory per node than is available on the standard \stampede{} nodes. There were \approximately{18M} 
node hours of jobs on \stampede{} and \approximately{26M} node hours on \comet{} without memory usage information.
}
\end{figure}

Figure \ref{fig:total-mem-per-job} shows a 2D binned scatter plot of the
total peak memory usage for jobs on \stampede{} and \comet{}. The peak memory usage
is defined as the maximum value of the sampled memory usage for a job. 
These plots show the total usage rather than the per-core value.
Note the logarithmic color coding of the job bins
which is intended to emphasize the distribution of the outlying large memory
jobs. For \stampede{}, Figure \ref{fig:total-mem-per-job-stampede}, the O/S memory usage
is clearly visible as the locus of the lowest memory bins for each node count.
While most of the memory usage points are concentrated in the lower left near
the origin, there are some patterns such as the vertical stripe at 1024 nodes,
reflecting the default maximum permitted job size (without a special request).
Other relatively popular job size choices include 512, 256 and even 2048 nodes.  The actual
memory usage varies widely over these large jobs, although there are no very large (> 2000 node)
jobs that use all of the available memory.  The small vertical line near the origin
corresponds to jobs running on the large memory compute nodes. With the exception
of one outlier all of the jobs use less than 50 TB memory in total. For \comet{}, Figure
\ref{fig:xwl_plot_188.pdf}, as for \stampede{}, most of the jobs are
concentrated near the origin. The O/S memory usage is approximately the
same on \comet{} as \stampede{} but is not as obvious due to the scale on the plot.
The job sizes are dramatically smaller which is expected since the resource
scheduling policy is specifically designed to promote small jobs. The vertical
line at 72 nodes represents the maximum allowed job size.
The jobs running in the large memory queue are clearly visible as the points
at 1 node above the black guide line.
As for \stampede{}, the larger jobs use a
wide range of memory. Note in Figure \ref{fig:xwl_plot_188.pdf} a dashed red
line has been added to the chart.  This line corresponds to the solid black
line in Figure \ref{fig:total-mem-per-job-stampede} for \stampede{}. Comparing the two
charts, one can see that the large memory usage jobs on \comet{} above the
dashed red line utilize more memory per node than is available on the standard
nodes of \stampede{}. Hence, although these jobs are small in terms of node
count they utilize significantly more memory per node.  

\begin{figure}
    \centering
    \includegraphics{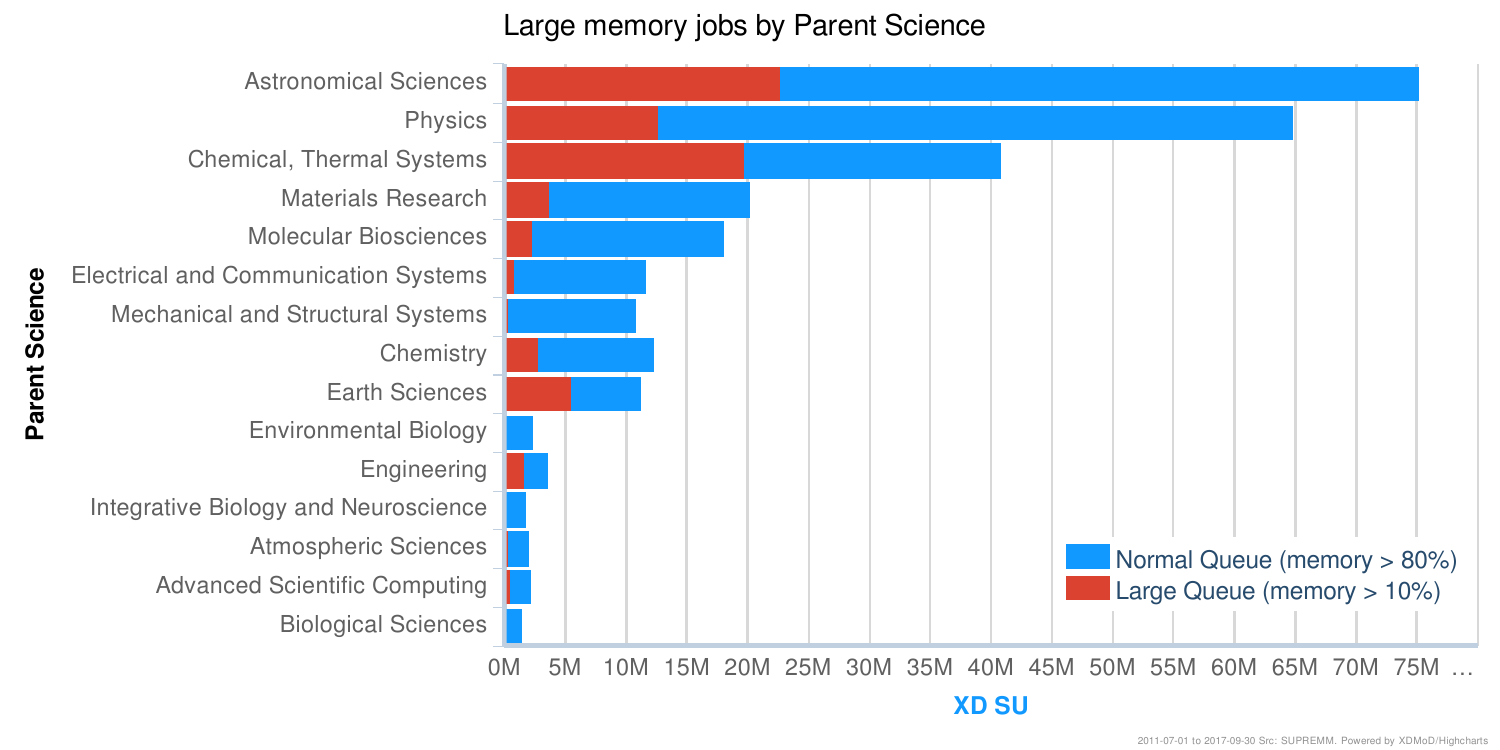}
    \caption{\label{fig:large_mem_by_parent_science}
        XD SUs consumed by jobs that had a high memory usage broken down by
        parent science. The ``Normal Queue'' data
        shows the usage of jobs that used the normal compute nodes and
        used $> \SI{80}{\percent}$ of the available memory on a node
        during the job. The ``Large Queue'' data includes all jobs that used
        $> \SI{10}{\percent}$ of available memory on the large memory compute
        nodes. Of the 20,000M XD SU consumed during this period, the plot cannot represent the 3,400M XD SUs where memory usage information is not available.
    }
\end{figure}

\begin{figure}
    \centering
    \includegraphics{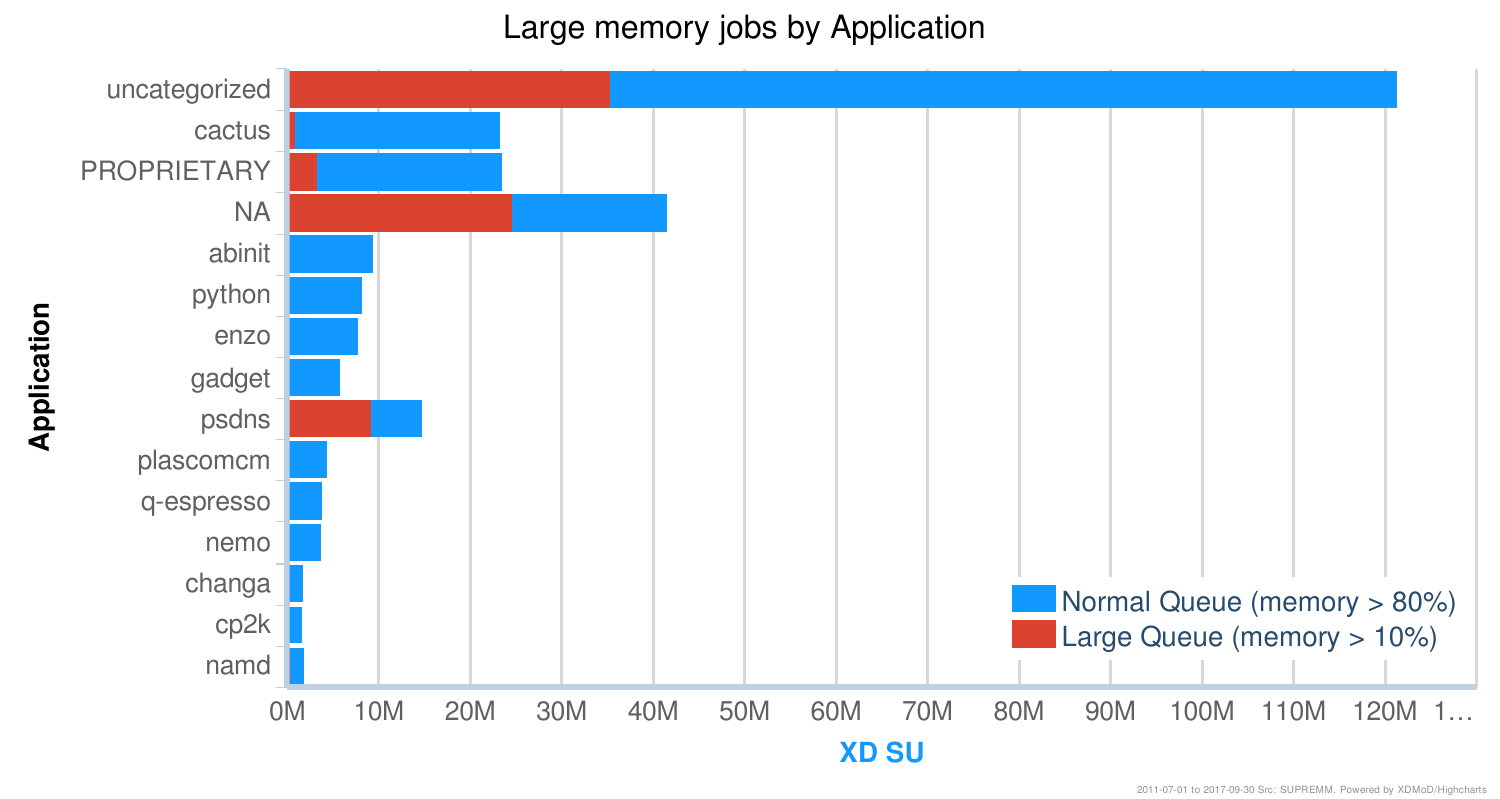}
    \caption{\label{fig:large_mem_by_application}
        XD SUs consumed by jobs that had a high memory usage broken down by application. The ``Normal Queue'' data
        shows the usage of jobs that used the normal compute nodes had memory
        usage was $> \SI{80}{\percent}$ of the available memory on a node
        during the job. The ``Large Queue'' data includes all jobs that used
        $> \SI{10}{\percent}$ of available memory on the large memory compute
        nodes. Of the 20,000M XD SU consumed during this period, the plot cannot represent the 3,400M XD SUs where memory usage information is not available.
    }
\end{figure}

Although we have shown that there are a relatively small number of jobs that utilize greater than 2 GB per core memory, the question remains what type of jobs have large memory usage that would be difficult or impossible to run if available resource memory was substantially reduced.  The Parent Sciences supported by high memory usage jobs is shown in Figure  \ref{fig:large_mem_by_parent_science}.  The figure shows both the large memory queue usage (red) and the high memory tail of the normal compute queue (blue).  The majority of this usage is on \comet{}.  Astronomical Sciences, Physics and Chemical Thermal Systems lead the list of memory intensive disciplines. Biosciences are relatively low on this list considering their overall large usage.  A bit more insight comes from examining the applications that require a large amount of memory, as shown in Figure  \ref{fig:large_mem_by_application}. Although there are a few specific community applications such as \texttt{CACTUS} and \texttt{PSDNS} that are large memory users, by far the largest application on this list is ``uncategorized.'' The classification ``uncategorized'' only occurs when we have the job level data necessary to classify the application but no application is matched. A machine learning based study was done previously on the ``uncategorized'' jobs~\cite{Gallo2015}.  A model was developed that was able to classify jobs with better than 90\% accuracy.  It was found that 80-90\% of the ``uncategorized'' jobs were in fact custom user code and only 10-20\% of the ``uncategorized'' jobs were community applications that had been missed by our regular expression classification scheme.  Hence, the great majority of the jobs in this category are custom user code that requires memory greater than 2 GB per core to run.   Obviously, this is a critical category of innovative usage that needs to be supported.

\clearpage

\subsection{Parallel Filesystem Usage}

\begin{figure}[ht]
\centering
\includegraphics[width=0.9\textwidth]{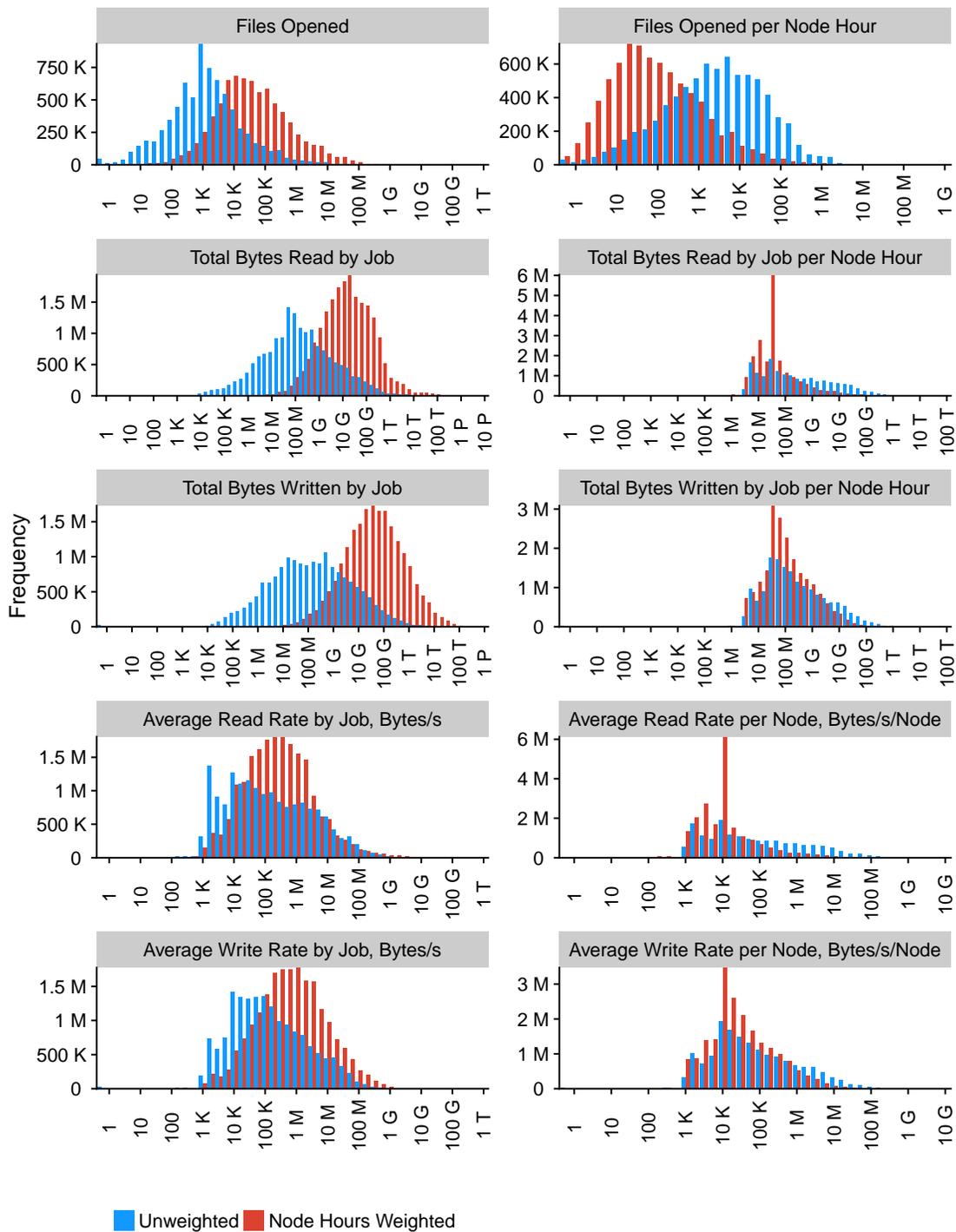} 
\caption{\label{fig:lustre_io_by_jobs} Utilization of Lustre parallel file system. Distribution of files opened, total bytes written and read, write and read rates per job (left) and per job per node hour (right). Write and read rates are shown per node (right). Note that $x$-axis is a log scale. The plots on the left show the distribution of per job characteristics while on the right it is normalized by node-hour and shows the load characterization of compute nodes. The distributions were calculated in two ways: unweighted and weighted by node-hours. This means the unweighted distribution plot ordinates are proportional to the number of jobs and the weighted distribution plot ordinates are proportional to the total node-hours used by the jobs for a particular bin.}
\end{figure}

\begin{figure}[ht]
\centering
\includegraphics[width=0.9\textwidth]{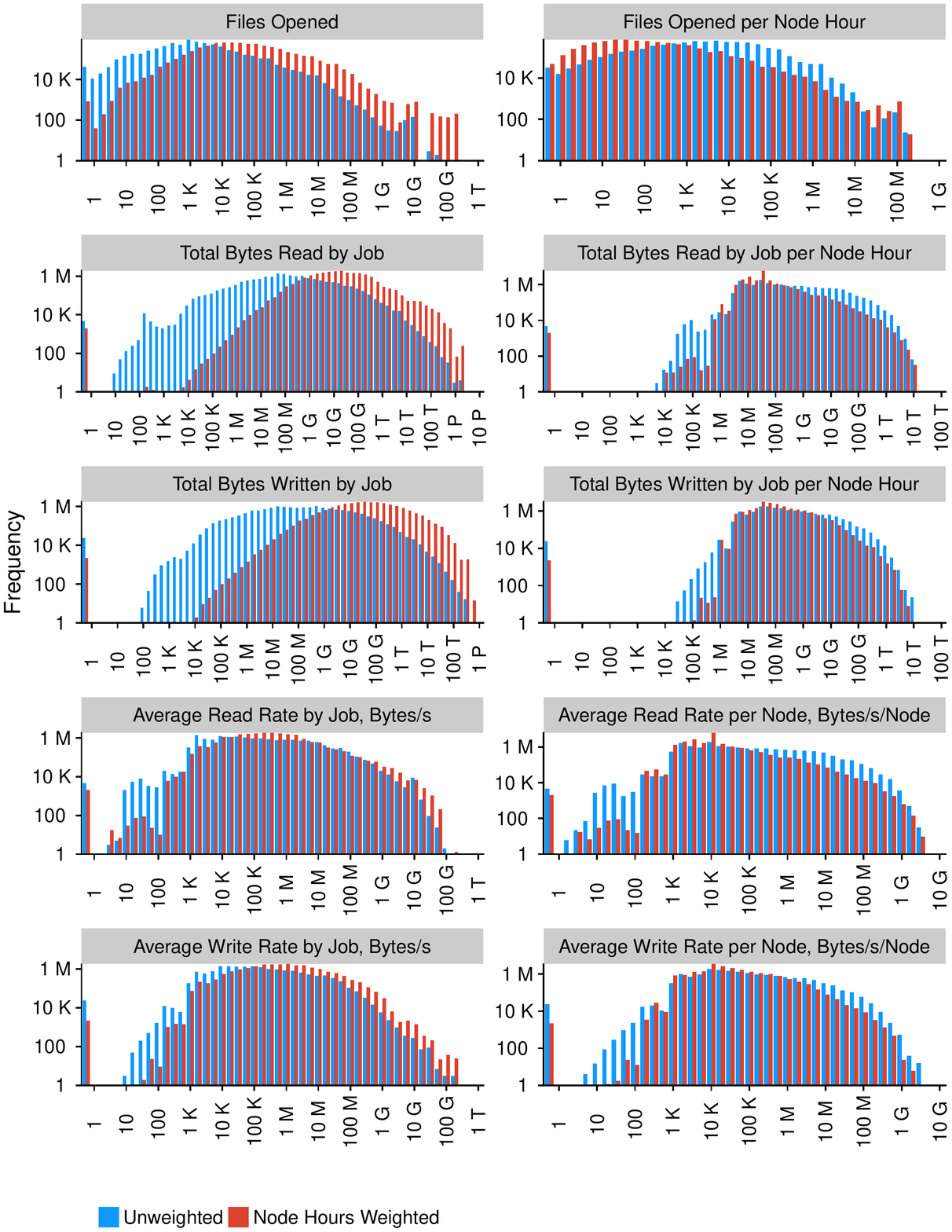} 
\caption{\label{fig:lustre_io_by_jobs_logs} Same as \ref{fig:lustre_io_by_jobs} but frequencies are shown in log scale to highlight extreme usages.}
\end{figure}

All of the resources in the present analysis for which data is available employ the Lustre parallel file system.  These resources include: \comet{}, \gordon{}, \supermic{}, \lonestar{}, \ranger{}, \stampede{} and \stampedetwo{} (see Table \ref{table:parallel_filesystem_resource_data_availability} for metrics availability dates). Based on the resources listed in Table \ref{table:parallel_filesystem_resource_data_availability}, the distributions of several parallel file system characteristics on a per job basis are shown in Figure \ref{fig:lustre_io_by_jobs}, namely file openings, bytes written and read as well as write and read rates. Note that the rates are averaged over the duration of the job and do not show the actual instantaneous network rates. Figure \ref{fig:lustre_io_by_jobs_logs} shows the same distributions but using a log scale for both axes in order to better visualize less frequently occurring but more extreme usage.
The weighted distribution (in red) for files open, read, and write (the three plots on the top left) are shifted to the right of the non-weighted distributions (in blue).  This implies that larger jobs, which are weighted more heavily in the node-hour weighted distribution, do substantially more parallel file system operations.  We can determine why this is so  by considering the average read and write rates shown on the bottom of Figure \ref{fig:lustre_io_by_jobs}.   The average read and write rates and the analogous per node adjusted average read and write rates, show that the weighted and unweighted rate distributions (red and blue distributions) are very similar.  Hence, the greater absolute number of file opens, reads and writes in the large XD SU jobs is primarily due to larger job size (more nodes) rather than such jobs inherently having higher file system usage rates.  In fact for the per node files opened, the (red) weighted distribution is shifted to the left of the (blue) unweighted one indicating the opposite conclusion, namely that the smaller SU jobs do proportionately more file opens. 


\begin{table}[ht]
\centering
\caption{Parallel filesystem metrics availability on resources. \supermic{} is missing information on opened files.}
\label{table:parallel_filesystem_resource_data_availability}
\begin{tabular}{@{}lll@{}}
\toprule
Resource         & Start      & End        \\ \midrule
TACC-RANGER      & 2011-07-01 & 2013-02-12 \\
TACC-LONESTAR4   & 2011-11-01 & 2016-11-02 \\
SDSC-GORDON      & 2015-06-04 & 2017-04-01 \\
TACC-STAMPEDE    & 2013-02-26 & 2017-09-30 \\
CCT-LSU-SUPERMIC & 2015-04-22 & 2017-09-30 \\
SDSC-COMET       & 2015-05-26 & 2017-09-30 \\
TACC-STAMPEDE2   & 2017-06-04 & 2017-09-30 \\
SDSC-COMET-GPU   & 2017-08-01 & 2017-09-30 \\ \bottomrule
\end{tabular}
\end{table}

\begin{figure}[ht]
\centering
\includegraphics[width=0.8\textwidth]{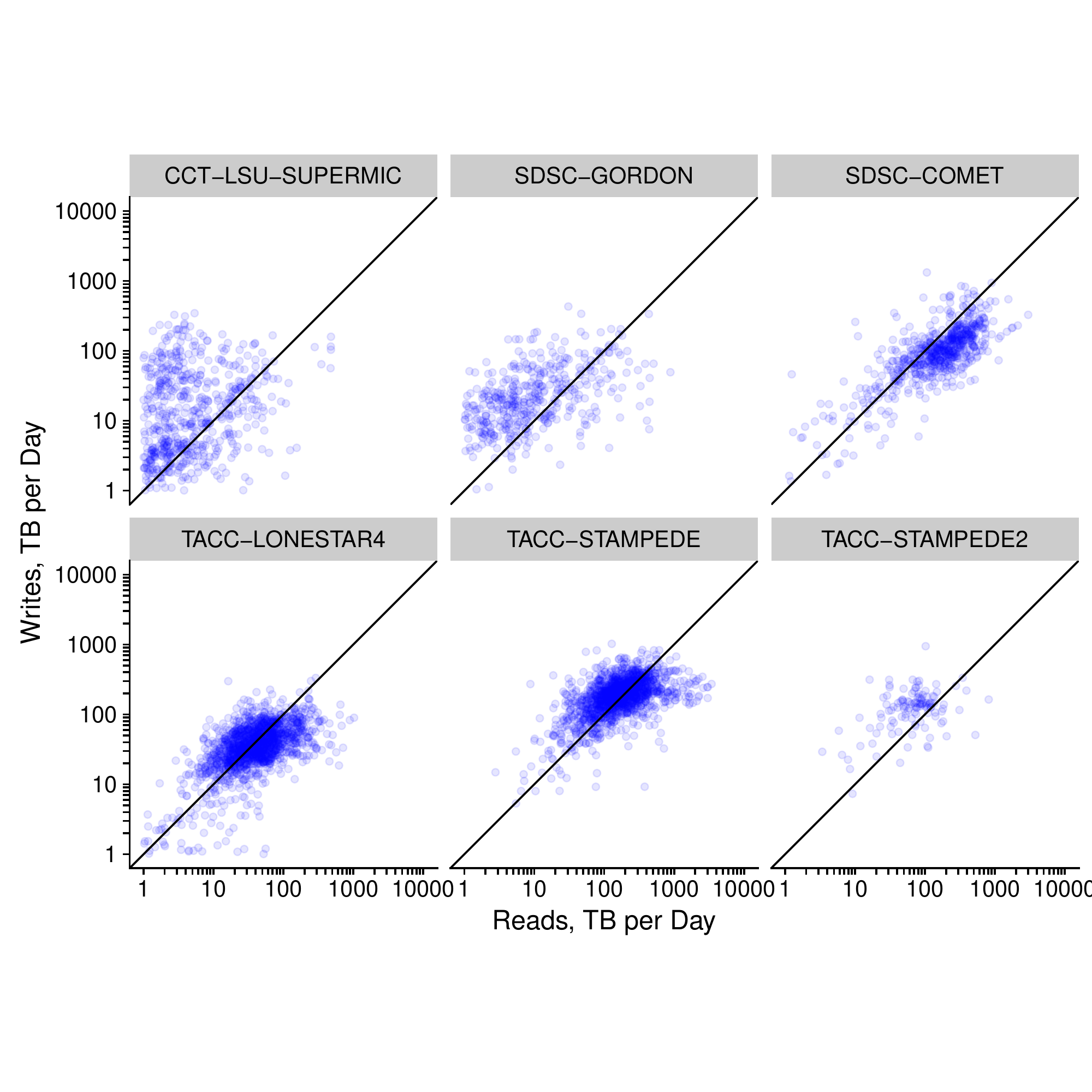} 
\caption{\label{fig:lustre_io_read_vs_write}Total daily reads vs writes to parallel file system from compute nodes for 6 resources. Note the log scale.}
\end{figure}

Figure \ref{fig:lustre_io_read_vs_write} shows daily reads vs writes to the parallel file system for 6 HPC resources. Note that only read and write activity from the compute nodes and not the head nodes is included -  thus data uploaded to the resource and downloaded from the resource are not included.  The diagonal line indicates an equal number of reads and writes and it can be seen that all of the resources show an approximately even balance between reads and writes.  For \lonestar{} and \stampede{} this balance is very even overall between read and write activity to the parallel file-systems, with a fair amount of fluctuations, which is similar to the Blue Waters systems. \texttt{LSU SUPERMIC}, \texttt{SDSC GORDON} and \stampedetwo{} show more writes, which can be due to less data input and more post-analysis on the resource, for example if not all of the generated data is used (e.g. check-files writes, excessively verbose logs ) or the system cache is effective in reducing the actual reads from the file-system. \comet{} on the other hand shows more reads which could be a characteristic of the smaller jobs generally run on this resource.

\clearpage

\subsection{Concurrency and parallelism}
\label{subsec:concurrency}

In this subsection we look at the degree of concurrency and parallelism of the HPC jobs running
on Innovative HPC Program resources. In particular, for multi-process jobs, how
are the processes distributed across the cores on the compute nodes and how has this changed
over time? Does this vary across HPC resources or by queue or application? We also look at the breakdown of different job launch types.  Please note that, at the time of this report, job and node level performance data has not yet been implemented for \bridges{} or \jetstream{} and therefore these resources are only included in analyses that depend on accounting level data (see Appendix \ref{appendix:ResourceCharacteristics} for details). 

\begin{figure}[h]
\centering
\includegraphics[width=1.0\textwidth]{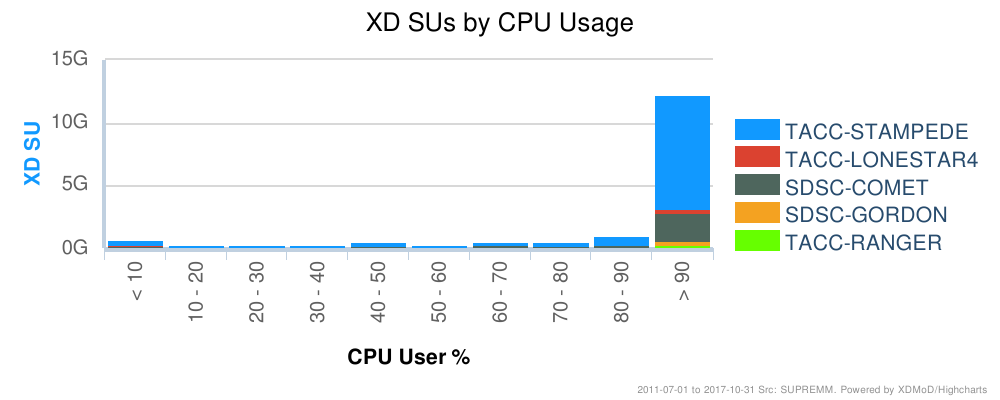}
\caption{\label{fig:overallcpu}Number of XD SUs broken down by CPU usage for
    different resources. There were \approximately{4G} of XD SUs (not shown) where the CPU usage information 
    was not available.
}
\end{figure}

Figure~\ref{fig:overallcpu} shows the XD SUs broken down by CPU usage for \ranger{}, \lonestar{}, \stampede{}, \gordon{} and \comet{}.
The CPU usage data were collected by the  \texttt{tacc\_stats}~\cite{Evans:2014} software.
The CPU usage shown here is the percentage of time that the CPU cores allocated to each HPC 
job spent running user-space
processes. Detailed information about the data collection and processing of the
job performance data is given in Appendix~\ref{appendix:taccstats}.
The overwhelming majority of the jobs have CPU usage over $\SI{90}{\percent}$.
This means that most jobs use all or almost all of their allocated CPU cores.
The  smaller peaks at $\SI{10}{\percent}$ and  $\SI{50}{\percent}$  correspond 
to single core jobs and jobs that use half of the cores on a compute node respectively\footnote{The CPU usage information alone is not sufficient to determine how many 
cores were active since $<\SI{50}{\percent}$ CPU could be half of the cores at full
usage or all of the cores at half usage. This analysis is also based on the process usage data discussed below.}.

\begin{figure}
\subfloat[\label{fig:cpusdifferentqueues_large}Large memory queue]{%
\includegraphics{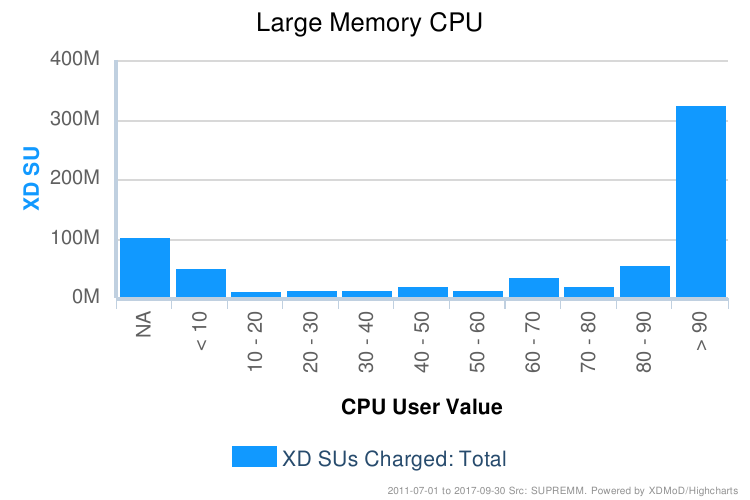}%
}
\subfloat[\label{fig:cpusdifferentqueues_serial}Serial queue]{%
\includegraphics{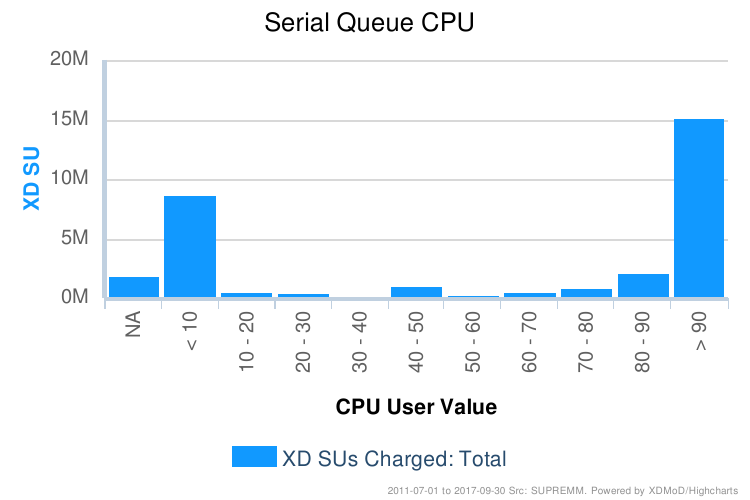}%
}
\caption{\label{fig:cpusdifferentqueues}Number of XD SUs broken down by CPU usage for different queues.
A serial queue was available on \ranger{}, \lonestar{} and \stampede{}. A large
memory queue was available on \ranger{}, \lonestar{}, \stampede{}, \comet{} and \darter{}.}
\end{figure}

Most traditional HPC resources provide a ``serial'' queue that is intended for jobs that do not require the
interconnect and a ``large memory'' queue that provides access to compute nodes with
a large amount of RAM compared to the default compute node. Figure~\ref{fig:cpusdifferentqueues} shows the breakdown of CPU usage for jobs that
were run in the different queues. Large memory compute nodes typically have more CPU sockets (and hence CPU cores) than the ``normal'' queue compute nodes due to
the prevalent NUMA hardware architecture. For example, the large memory nodes
on \stampede{} have 4 sockets (32 cores) rather than the 2 sockets (16 cores) of the 
general compute nodes. The CPU usage of the large memory queues for \ranger{}, \lonestar{}, \stampede{}, \comet{} and \darter{} is shown in Figure~\ref{fig:cpusdifferentqueues_large}. Note that the usage of the large
memory queues is much less than the overall usage. The majority of HPC jobs, by XD SU, 
in the large  memory queues use all of the CPU cores. However, there is a relatively larger peak at 
$\SI{10}{\percent}$ compared to the overall job mix.
The serial queue usage is shown in Figure~\ref{fig:cpusdifferentqueues_serial}.
Many jobs use all of the available cores but nearly one third of the jobs have 
CPU usage less than  ten percent, which corresponds to single or two core
HPC jobs.

\begin{figure}[h]
\centering
\includegraphics{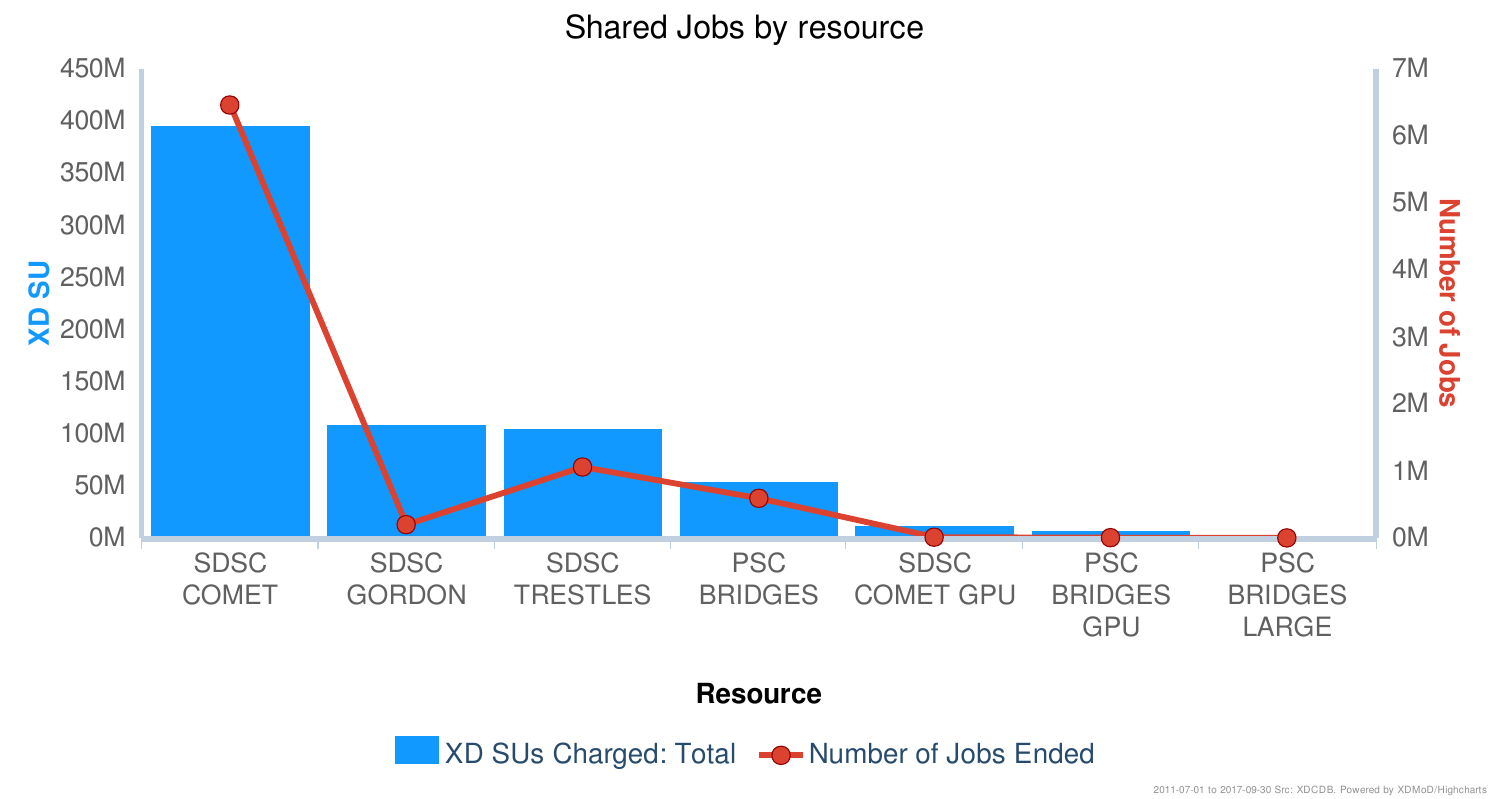}
\caption{\label{fig:shared_jobs_by_resource}Number of jobs and XD SUs for jobs running on a ``shared'' queue for different resources.
}
\end{figure}

The other place to run serial and small (by core count) jobs are the shared queues that are available
on several resources. Jobs running in the shared queue share compute nodes. This improves job throughput
and leads to higher overall resource utilization with only minor impact on individual job performance~\cite{White:2014:ANS:2616498.2616533}.
Figure~\ref{fig:shared_jobs_by_resource} shows the job count and XD SUs for shared queue jobs on the resources that allowed
shared-node jobs. Most of the usage is on \comet{}. Of the \approximately{6.5 M} jobs on \comet{} the majority were single core jobs (5.6 M jobs, 143 M XD SUs).

The HPC resources discussed here so far all have the same order of magnitude of CPU 
cores per compute node. The current trend is to increase the number of
cores per compute node. The Intel Knights Landing compute nodes on \stampedetwo{} have 
68 cores with 4 hardware threads per
core (for a total of 272, which is an order of magnitude more that \stampede{}). 
The recommended usage is to run 64--68 MPI tasks or independent processes per node
with 1--2 threads per process~\cite{Stampede2UserGuide2017}.
Table~\ref{table:stampede2threads} shows the breakdown of usage in XD SUs by 
process/thread count per compute node for \stampedetwo{}. For the jobs that we
have data, the majority are using between 32 and 68 processes per compute node.
There is a significant (198 M XD SUs) usage for jobs that are using less than or equal
to 32 cores. For the jobs in this category the most popular core counts are 16 and 32.
Note that the Intel Skylake compute nodes on \stampedetwo{} came into production after the
end of the workload analysis period so are not considered here.

\begin{table}
\centering
\caption{Process usage on Intel Knights Landing compute nodes on \stampedetwo{} from 2017-06-07 to 2017-10-30.}
\label{table:stampede2threads}
\begin{tabular}{lr}
\toprule
Number of processes/threads & XD SUs \\ \midrule
$\le 32$ & 198 M \\
$ > 32 \le 68$ & 466 M \\
$ > 68  $ & 119 M \\
N/A  & 280 M \\
\bottomrule
\end{tabular}
\end{table}
%

\begin{figure}
\centering
\subfloat[\label{fig:mpivsthreads-normal}XD SU usage by job type (linear scale).]{%
\includegraphics{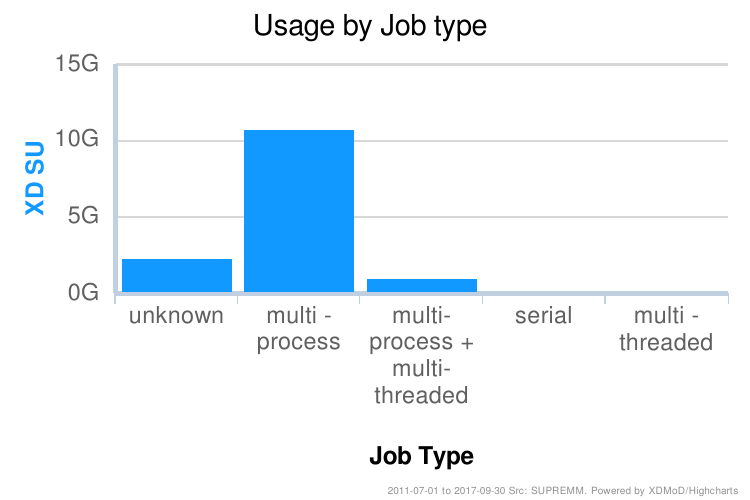}
}
\subfloat[\label{fig:mpivsthreads-log}XD SU usage by job type (log scale).]{%
\includegraphics{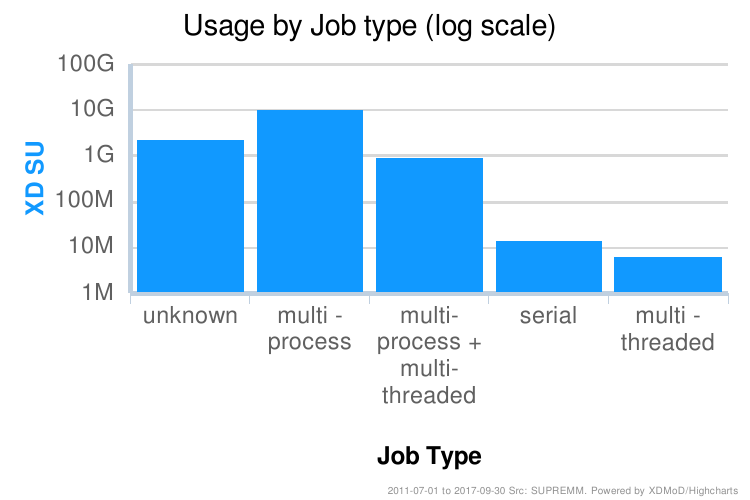}
}
\caption{\label{fig:mpivsthreads}Number of XD SUs broken down by the job launch type. A
type of ``multi-process'' indicates that the job was run using multiple parallel processes.
A type of ``mult-process + multi-threaded'' indicates the job was run using multiple
parallel processes and multiple threads per processes.  ``unknown'' indicates that either the job
was run using a non-instrumented launcher or there was an issue with the data collection
tool. ``serial'' indicates a single process job and ``multi-threaded'' a job with a single process
and multiple threads. These data were collected using the Lariat software
and are only available on \stampede{}, \lonestar{} and \ranger{}.}
\end{figure}

Figure~\ref{fig:mpivsthreads} shows the breakdown of job launch type by XD SUs
for the three TACC resources that ran the Lariat~\cite{Lariat2013} software.
The job launcher
\texttt{ibrun} on these resources is instrumented to collect information about
the job including the number of requested parallel processes and the number of
threads per process. We label a job ``multi-process + multi-threaded'' if there are
multiple parallel processes with multiple threads per process, ``multi-process'' for multiple
parallel proceses and one thread per process, ``serial'' for single process single thread jobs and
``multi-threaded'' for single process multiple thread jobs.  The main
causes for the job launch type to be ``unknown'' are that the job was not launched
using the instrumented \texttt{ibrun} or that there was an error in the data
collection infrastructure.  \ranger{} data before October 2012 had a version of Lariat
that did not collect the thread information but did collect application usage.
There is very little recorded usage of ``serial'' and ``multi-threaded'' jobs
in this data. This is likely an under-estimate of the actual usage because \texttt{ibrun}
may not have been used to launch these classes of job\footnote{The
\stampede{} user guide did not recommended using \texttt{ibrun} to run serial jobs.}.
There are approximately 372 M XD SUs single node jobs in the ``unknown'' category,
which gives an upper bound estimate of the ``serial'' or ``multi-threaded'' job usage.
Note that this analysis does not distinguish between different parallel processing
implementations (mpi, CAF, charm\hspace{-.05em}\raisebox{.4ex}{\tiny\bf ++}, etc.) nor threading models (pthreads, OpenMP)
the concurrency determination is solely inferred from the job configuration.

\begin{figure}
\subfloat[\lonestar{}]{\includegraphics{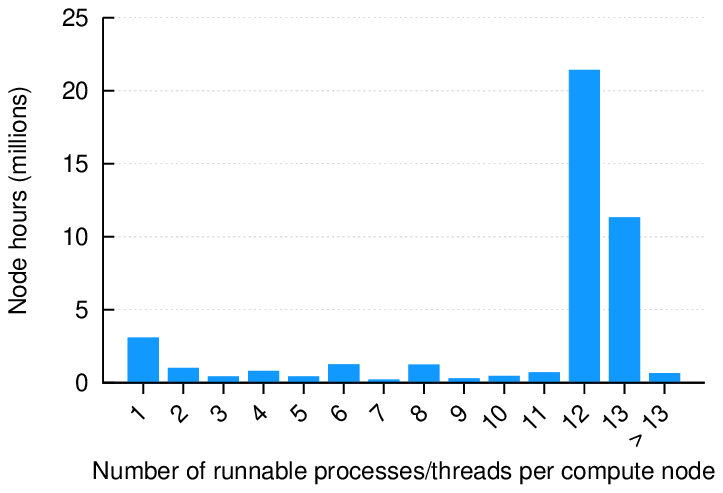}}
\subfloat[\stampede{}]{\includegraphics{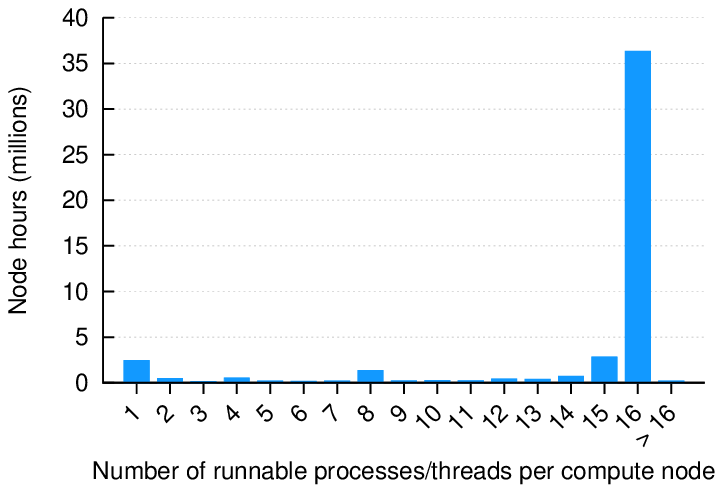}}
\caption{\label{fig:stampedethreads}Number of node hours by the number of 
active kernel scheduling entities (processes and threads) per compute node.}
\end{figure}

Figure~\ref{fig:stampedethreads} shows the amount of node time on \lonestar{}
and \stampede{} by the number of active kernel scheduling entities
(O/S processes) per compute node.  This value is an estimate generated by
recording the number of runnable O/S processes periodically during each HPC job
and then taking the median value see Appendix~\ref{appendix:taccstats} for more
details.  An O/S process could be an MPI task for an
MPI job or a thread for an OpenMP job. The general compute nodes on 
\stampede{} have 16 cores\footnote{Hyperthreading was disabled.}.  The general
compute nodes on \lonestar{} had 12 cores.  For \stampede{}, the vast
majority of the jobs by node hour run at least one O/S process per hardware
core. The next most popular configurations are a single process per compute
node and under-subscription by one half and one quarter (8 and 4 processes
respectively). \lonestar{} has a very similar pattern with peaks at 1/12,
1/2, 1/4 and 2/3.  The plots for the other HPC resources are qualitatively very
similar to the two shown here.  Recall that in Figures \ref{fig:mem-hist} and \ref{fig:xwl_plot_159.pdf} a similar pattern was noted where a minority of jobs use a fraction of the cores per node and since these had relatively high memory usage they were ascribed to memory intensive jobs.
Another typical reason to undersubscribe the cores is for I/O intensive jobs.

The large memory nodes on \stampede{} have 32 cores per compute node and the large memory
jobs that use all cores are in the $> 16$ bin.
Note that \lonestar{} used the Linux kernel version 2.6.18 which used the
$O(1)$ process scheduler, \stampede{} ran the 2.6.32 kernel with the CFS
scheduler. This difference in scheduler software explains why we see 
jobs with $> 12$ runnable threads on \lonestar{} but no jobs with $> 16$ 
runnable threads on \stampede{}. This data does not provide any information
about oversubscription of the cores on the compute nodes since the number of
runnable processes is constrained by the number of cores.

\begin{figure}
\subfloat[\label{fig:taskprocesslonestar}\lonestar{}]{%
    \includegraphics[width=0.45\textwidth]{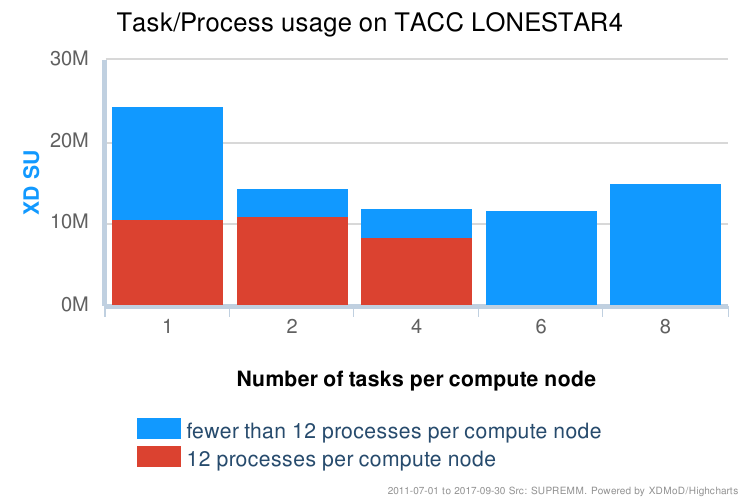}
}%
\hfill
\subfloat[\label{fig:taskprocessranger}\ranger{}]{%
    \includegraphics[width=0.45\textwidth]{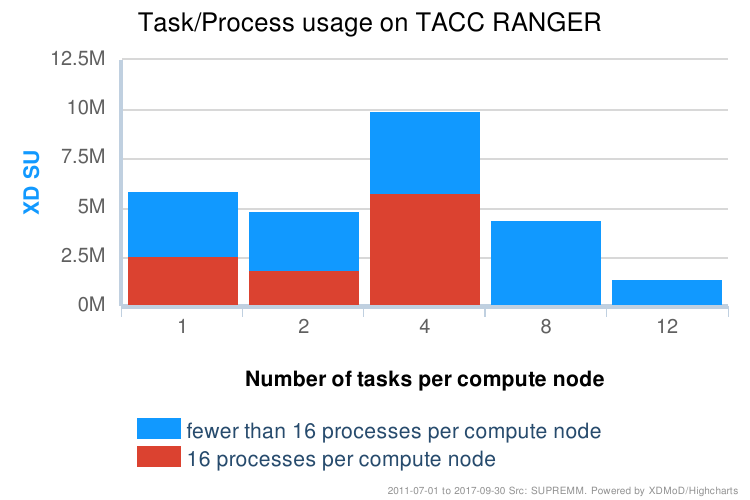}
}%
\caption{\label{fig:taskprocess}Breakdown of O/S process usage for HPC jobs that
under-subscribe the number of processes per compute node. The data for jobs
with 12 processes per node on 
\lonestar{} and 16 processes per node on \ranger{} are not
shown here.}
\end{figure}

The plots in Figure~\ref{fig:taskprocess} illustrate the number of O/S processes 
run on the compute nodes for jobs that undersubscribed the compute nodes (i.e.\ used
fewer processes than available CPU cores). 
Figure~\ref{fig:taskprocesslonestar} on the left
shows the XD SUs consumed by jobs that requested fewer than 12 processes per compute 
node on \lonestar{}. The number of processes per compute node is obtained from
the resource manager log files (both \lonestar{} and \ranger{}
used the Sun Grid Engine (SGE) resource manager). The jobs are split into two 
categories: jobs that ran 12 O/S processes per compute node and those that ran
fewer than 12. The estimate of number of O/S processes per compute node is based
on the median number of runnable processes reported by the Linux kernel as per
the data in Figure~\ref{fig:stampedethreads}. Similar  information is presented for
the 16 cores per compute node \ranger{} in Figure~\ref{fig:taskprocessranger}.


\begin{figure}
\subfloat[\label{fig:comet24coreapps}24 processes per compute node]{%
\includegraphics{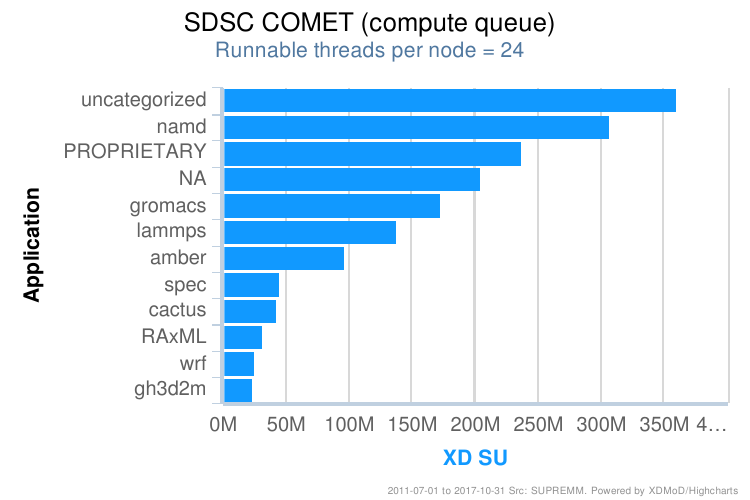}%
}
\hfill
\subfloat[\label{fig:comet16coreapps}16 processes per compute node]{%
\includegraphics{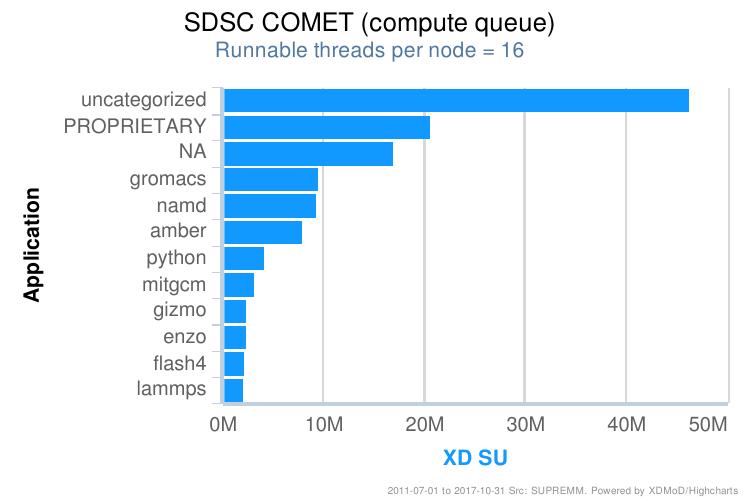}%
}
\caption{\label{fig:cometappsbycores}Application usage by number of O/S processes per compute node for
\texttt{SDSC COMET}. Note the difference in $x$-axis scale between the two plots.}
\end{figure}

To analyze how concurrency varies with application, we plot the application
usage for the general compute queue on \comet{} in Figure~\ref{fig:cometappsbycores}. Figure~\ref{fig:comet24coreapps}
shows the application usage for jobs that used all 24 cores per compute node and
Figure~\ref{fig:comet16coreapps} shows the application usage for the jobs that only used
16 of the 24 cores on the assigned nodes. The determination of the number of active cores
per node was done using the runnable threads metric.
The application mix for the two different cases is quite similar. Note
the difference in scale of the two plots: the $x$-axis is an order of
magnitude larger for the 24 cores per node plot.
One application that appears only in the 16 cores per node list is
\texttt{MITGCM} this code has only been run by one project team
and all instances use 16 cores per node.
If we look at \texttt{NAMD}, the majority of the XD SUs are run at 24 cores per compute
node, but there is a peak at 16 cores per node and at 12 cores per node. The
\texttt{NAMD} usage at 16 is almost entirely due to a single project team. Similarly
the usage a 12 cores per node is predominantly due to a single different project team.

\begin{figure}
\centering
\includegraphics[width=\textwidth]{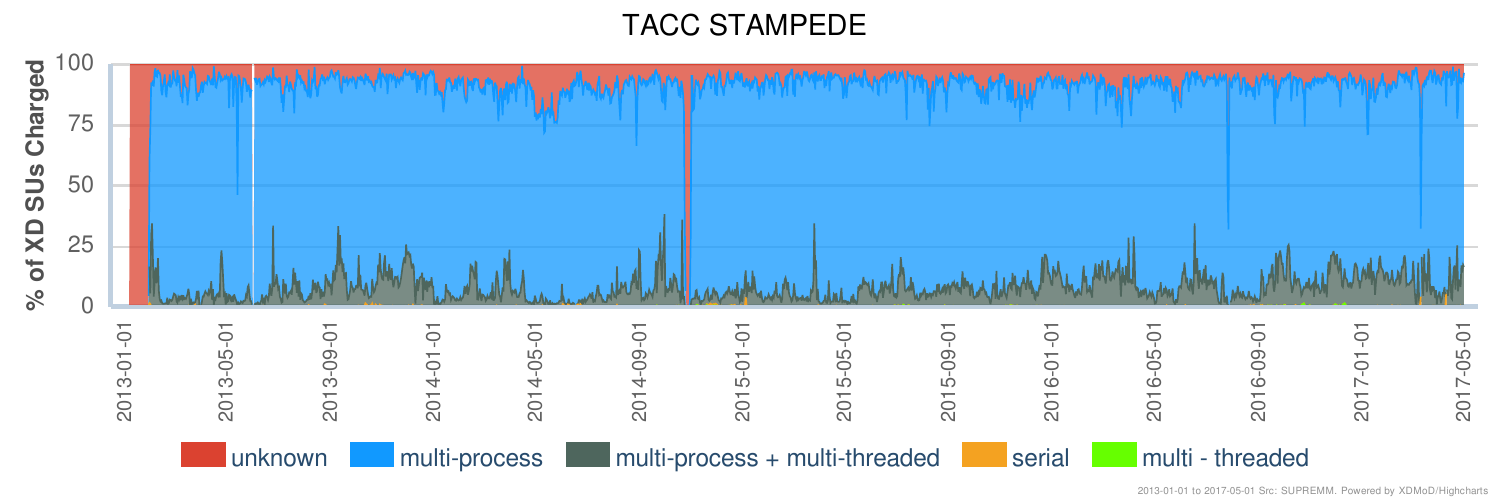}
\caption{\label{fig:stampedeconcurrencybytime}Percentage of XD SUs for \stampede{} by job type over time from 2013-01-01 to 2017-05-01.}
\end{figure}

The time evolution of the job types is illustrated in Figure~\ref{fig:stampedeconcurrencybytime}, which shows the
job types for \stampede{}. The plots for \ranger{} and \lonestar{} are qualitatively similar.
There is very little change in the mix of multiple processes vs multiple processes with multiple threads over time.
The occasional increases in the multiple processes with threads are caused by a project or projects running a using their allocation in a relatively short
space of time. The serial and multi-threaded job usage is so small that it
is barely visible on the plot; the usage of these types is less than $\SI{0.2}{\percent}$ overall.

For the majority of the study period the different HPC resource have
had the same order of magnitude of cores available per compute node (12 -- 24)
with successive generations of hardware having faster cores with increased vector
operations.  With the increase in the number of cores per node in recent
hardware architectures, it will be interesting to see how concurrency changes
in future.  The initial data from \stampedetwo{} suggests that a significant
proportion of jobs are not yet using all of the cores available and are still
using the same number of cores per node as previous resources.
We do not have sufficient data to determine the reason for this, but it 
could be that the users are just reusing existing launch scripts and
simply need to increase the number of processes per node. Alternatively it could
be that the existing software is not well suited to running at large concurrencies
and changes to software may be needed.

\begin{figure}[h]
\centering
\subfloat[\label{fig:jobs-with-Lariat}]{%
\includegraphics[width=0.5\textwidth]{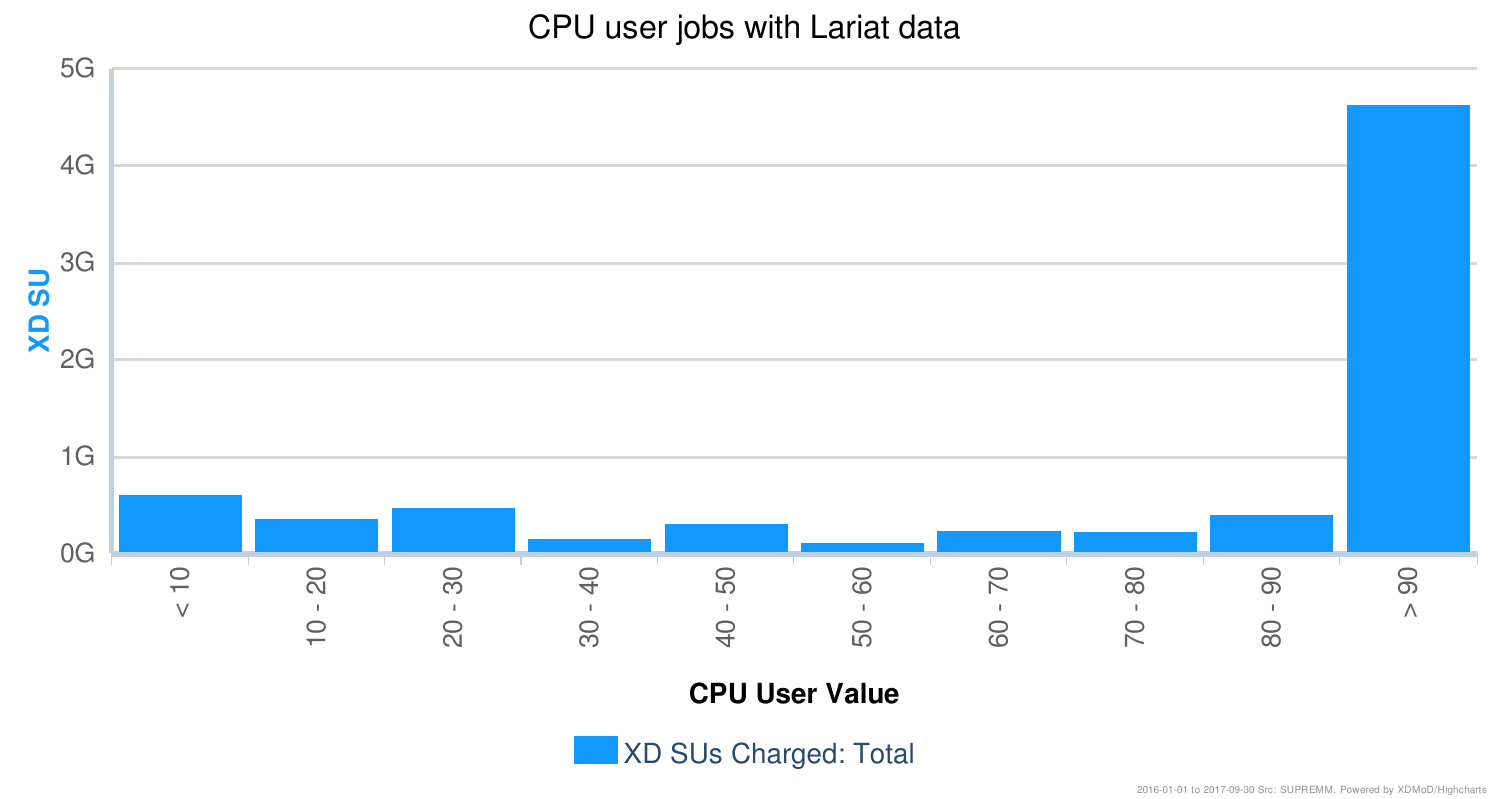}
}
\subfloat[\label{fig:jobs-without-Lariat}]{%
\includegraphics[width=0.5\textwidth]{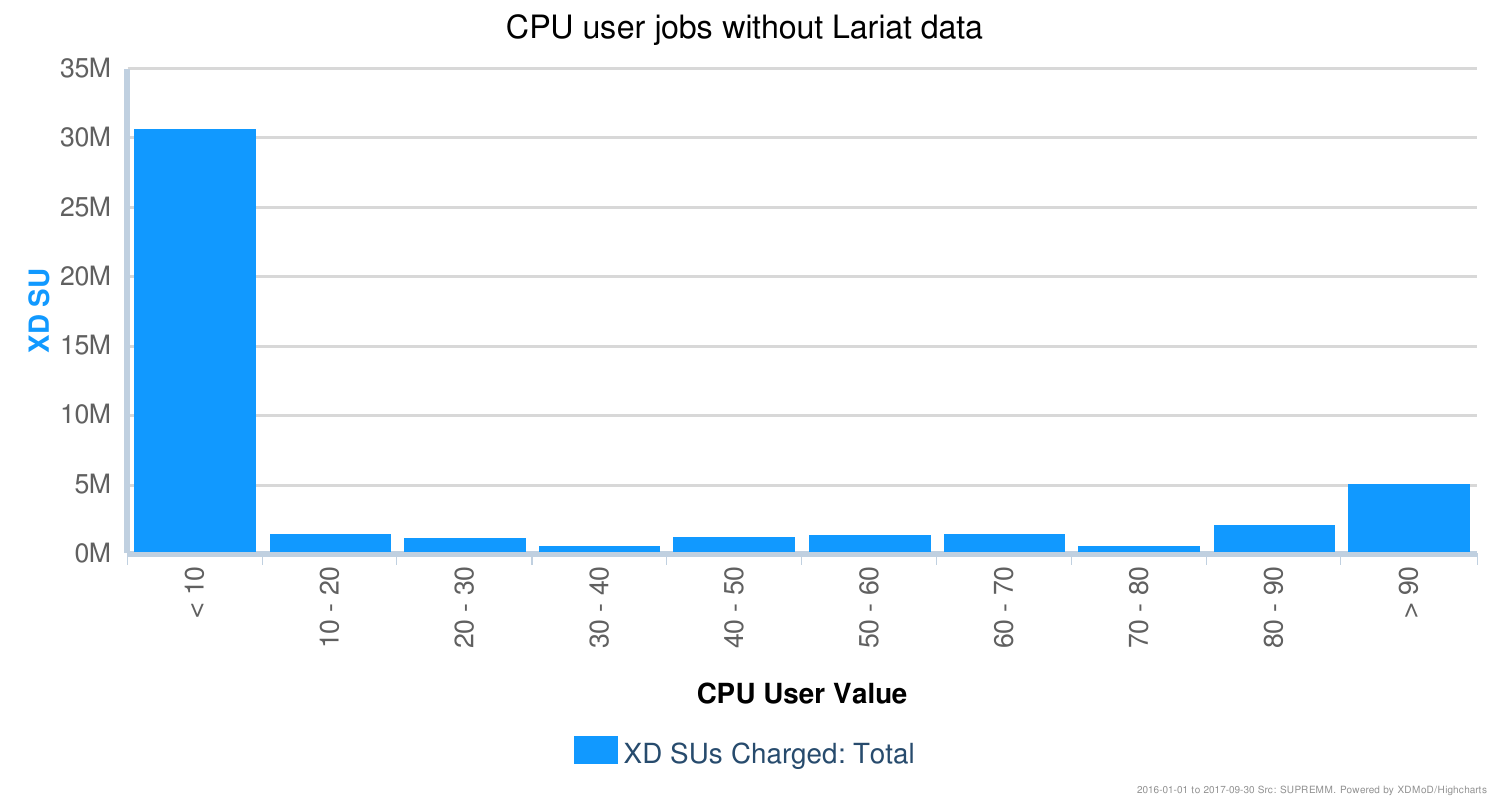}
}
\caption{Comparison the CPU user for (a) jobs with Lariat data (b) jobs without Lariat data.  Jobs with Lariat data are batch (non-interactive) jobs and generally are very efficient with most jobs in the CPU user > 90\% bin.  In contrast, interactive jobs fall into the category of jobs without Lariat data.  These jobs are relatively inefficient with most jobs falling into the lowest CPU user bin, that is, < 10\%. }
\label{fig:with-and-without-Lariat}
\end{figure}

Interactive jobs are expected to have much lower CPU usage than batch jobs
because the compute nodes are likely to be idling when waiting for user input.
Unfortunately, whether a job is interactive or not is not specifically tracked in XDMoD for the
NSF resources as of the date of this report.  We can however, indirectly
illustrate the difference in CPU usage between interactive and non-interactive
(batch) jobs; see Figure \ref{fig:with-and-without-Lariat}.  The distribution
of XD SU by CPU user is shown in Figure \ref{fig:jobs-with-Lariat} for all jobs
for which we have Lariat data in the time frame of 2016-01-01 to 2017-09-01.
Lariat data is only available for jobs that were launched using the $\texttt{ibrun}$
command. The user guide for \stampede{} recommends that users do not use the 
$\texttt{ibrun}$ command to run an interactive job (although the $\texttt{ibrun}$
command could be used to run software with an interactive job).
Note that the
great majority of the jobs have high CPU usage (> 90\%).  Figure \ref{fig:jobs-without-Lariat} shows the CPU user distribution for
jobs for which we do not have Lariat data.  Although there may be other causes
for a job to be missing Lariat data, all jobs that are interactive will be in
this category.  Note the difference in the distribution of CPU user. For this
job category, the great majority of jobs have very low CPU usage (< 10\%).

\clearpage

\subsection{Job Failures}
The exit status of each HPC job is recorded in the job scheduler accounting logs.
The interpetation of the exit status depends on the job scheduler
software used by each resource and most schedulers report the exit status of
the job batch script and also report any failures in the scheduler software or
compute nodes. 
 In this section we investigate the job exit status for \stampede{}
and \comet{}. Both of these resources use the \texttt{slurm} job scheduler.

Table \ref{table:exit-codes} shows a summary of
the job exit codes and the number of jobs that ended with that particular exit
code for \stampede{} and \comet{}.
An exit code of ``completed'' indicates that the job finished
without reporting an error, ``canceled'' indicates that the job was canceled after
submission, ``timeout'' indicates that the job was killed by the scheduler because it
 reached the requested wall time limit, ``node-fail'' indicates that the job ended 
prematurely due to a problem with the scheduler software or a problem with one
of the compute nodes on which the job was assigned. And exit code of ``failed'' 
indicates that the job batch script returned a non-zero exit code.  

\begin{figure}[h]
\centering
\subfloat[\label{fig:xwl_plot_125.pdf}]{%
\includegraphics[width=0.5\textwidth]{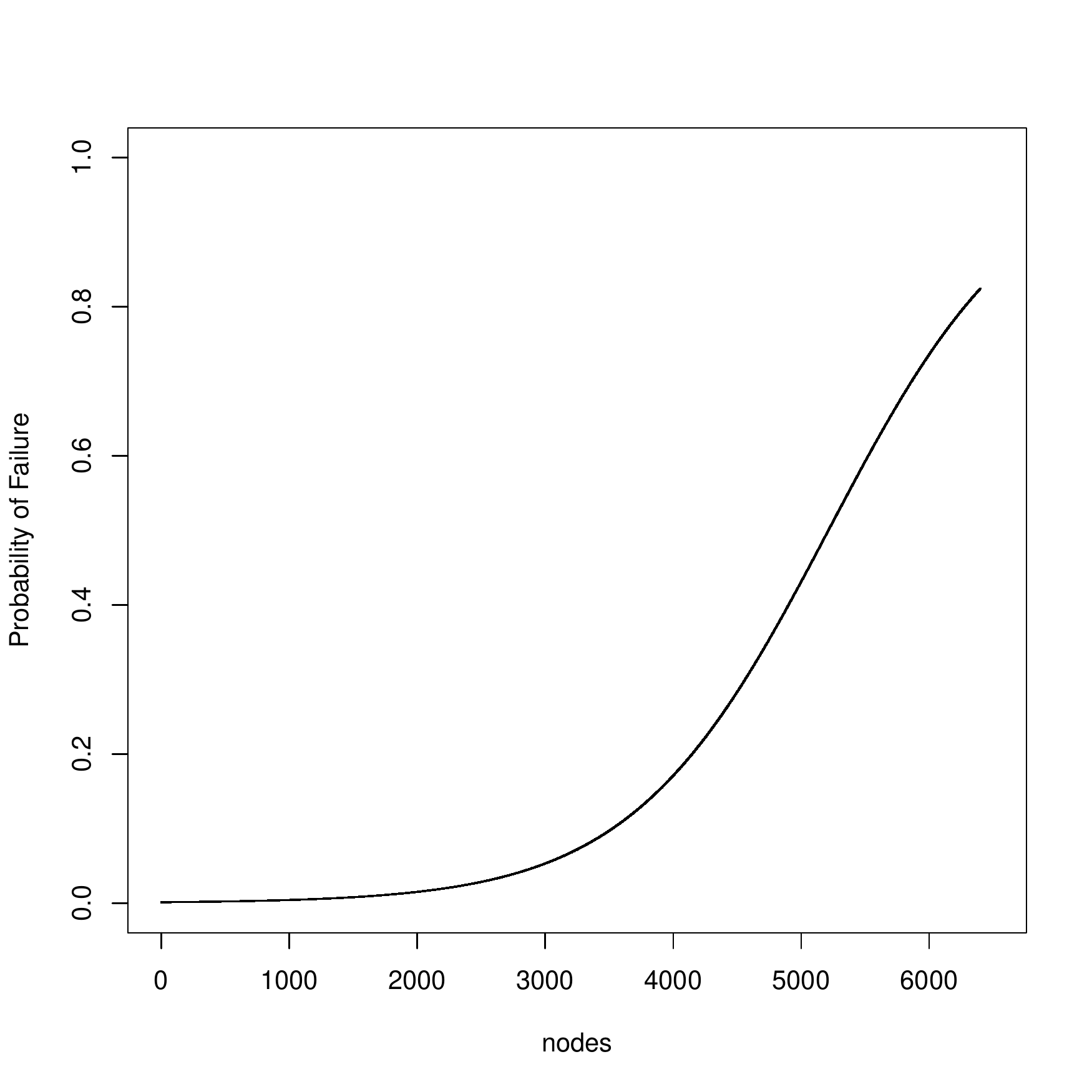}
}
\subfloat[\label{fig:xwl_plot_177.pdf}]{%
\includegraphics[width=0.5\textwidth]{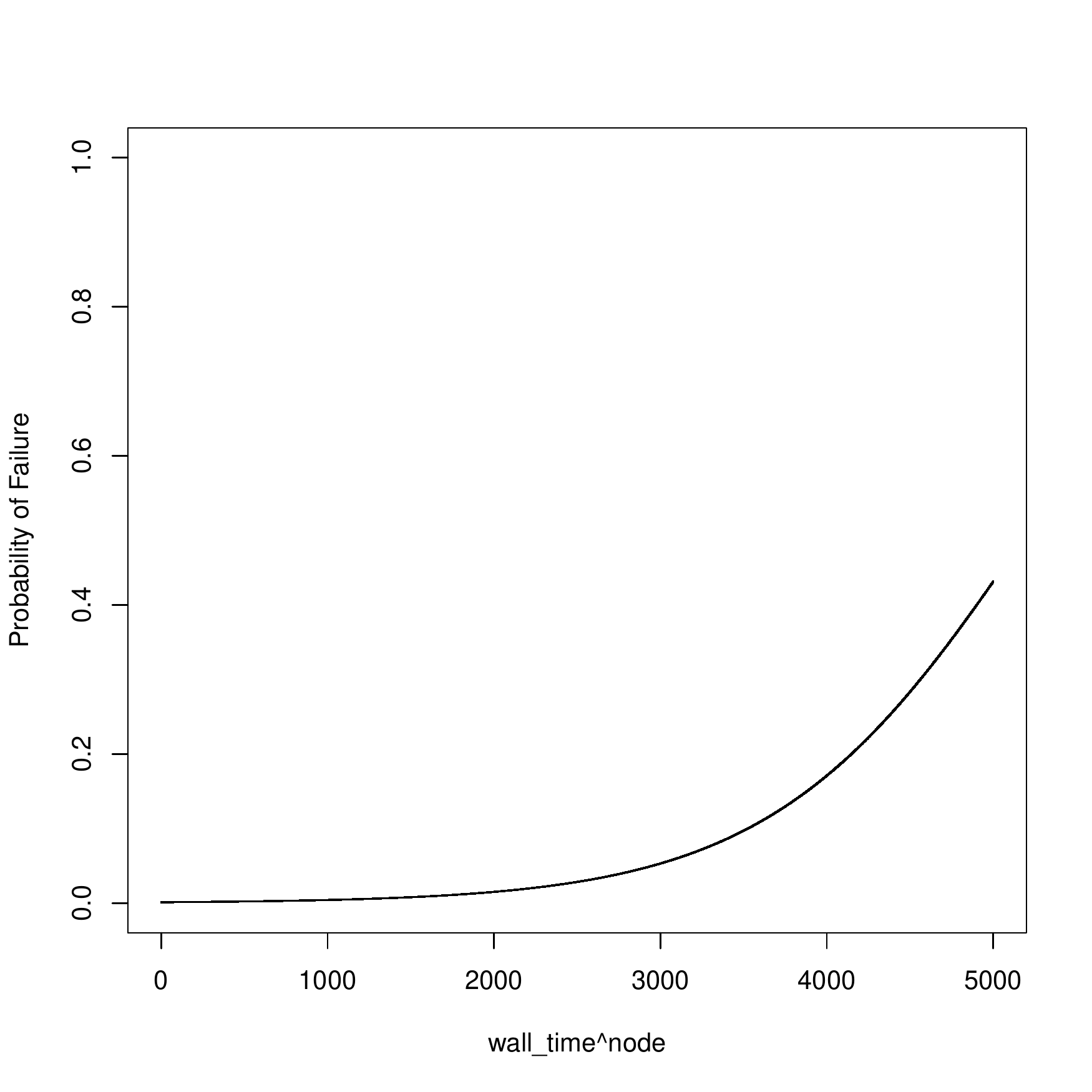}
}
\caption{Logistic regression plot for node-fail jobs for \stampede{} based on (a) a simple model that assumes that job wall time is relatively constant as a function of job size.  The probability of failure is then only a function of job size, and (b) a more sophisticated model where the probability of a job failure is a function of wall time raised to the power of the number of nodes.}
\end{figure}

The job batch scripts are controlled by the end-users so the
reported exit status is unlikely to be consistent between jobs for different
users or even between different jobs for a user.
Despite this known ambiguity around the batch script exit codes, we carried out an analysis
of job exit codes for \stampede{} and \comet{}. The first analysis was carried
out on the node-fail jobs in \stampede{}.  Since this exit status is
a job scheduler status and is not directly controlled by the user supplied
batch script it
should be an unambiguous indicator of a failed job and provides a lower bound.
Since a single node failure on a multi-node job will cause the entire job to
fail we would expect that the probability that a job will fail is strongly
dependent on the number of nodes in the job.  Since a job failure is a binary
process, a job can either fail or complete, a logistic regression will be done
to quantitatively determine the probability of node-failure as a function of
the number of nodes in the job. For a simple initial analysis we will take
advantage of the fact that that the overall distribution of job wall time is
not greatly dependent on job size.  The \stampede{} node-fail data was fit
using logistic regression;  the fit was well behaved and the slope and
intercept were statistically significant as measured by their p-values.  Figure
\ref{fig:xwl_plot_125.pdf} plots the result of this logistic regression;
\stampede{} has 6400 compute nodes.  For large capability jobs using one third
to one half of the nodes, the probability of a node-failure is still small.
Very few jobs exceed this number of nodes on \stampede{} 
as shown in Figure \ref{fig:job-sizes} (the job scheduling policy 
on \stampede{} has a default maximum job size of 1024 nodes).
Hence the data in the knee of the curve in Figure
\ref{fig:xwl_plot_125.pdf} should be the most accurate.  The extrapolation to
full utilization of all 6400 nodes of \stampede{} is somewhat more uncertain
due to the scarcity of the data. We can perform a similar analysis on the
\stampede{} failed jobs if we include both the failed jobs and the node-failed
jobs, however this assumes that all of the jobs with a failed exit code status
are truly failed jobs. This leads to the conclusion that the failure
probability of even small jobs is well above zero (on the order of 0.1) and it
increases nearly linearly with the number of nodes, approaching 1 for jobs that
are the full size of \stampede{}.  A failure rate of this magnitude is
unrealistic and leads to the obvious conclusion that
the job batch script exit status is not a reliable indicator of job failure.
 A similar analysis carried out on \comet{} data showed similar
results. A more sophisticated logistic regression model of the \stampede{} node
failure data can be used where the probability of a job failure is a function
of wall time in years raised to the power of the number of nodes.  This model
is a bit more complex to interpret. Figure \ref{fig:xwl_plot_177.pdf} plots the
result of this more complex logistic regression model.  This model also
produces a well behaved fit with statistically significant parameters.  Once
again the data in the knee of the curve where the failure probability just
starts to rise above zero should be the most accurate.         

\begin{table}[H]
\centering
\caption{Job count grouped by exit code for \stampede{} and \comet{}}
\label{table:exit-codes}
\begin{tabular}{ lrr }
\toprule
Exit Code  & \stampede{}  & \comet{}    \\ \midrule
Completed  &  5575319   &  2244602   \\
Canceled  &   854484   &   176857   \\
Timeout    &   669919   &    81156   \\
Failed     &   889214   &   214634   \\
Not Available         &            &  2715575   \\
Node-fail  &     9151   &      135   \\
\bottomrule
\end{tabular}
\end{table}

\subsection{Summary: Job Characteristics}
  
The average job size, weighted by XD SUs, shows a steady decline from 2011 - 2017 (9,000 cores to 1,000 cores). This is attributable to the retirement of \kraken{}, the introduction of Blue Waters, improved core performance (reflected in higher XD SUs/core-hour), and individual job size limits at the resource level.  However, a very recent trend, with the addition of the \stampedetwo{} Xeon Phi resource, shows increasing percentages for the largest job sizes (8k and larger in core count).

The percentage of single node jobs (excluding OSG jobs) increased significantly from 5\% of all consumed XD SUs in 2011 to 21\% in 2017. In addition, the number of single node jobs accounted for more than 60\% of all jobs in 2017, and while this percentage fluctuates from year to year, it is typically greater than 50\% of all jobs run (excluding OSG jobs).  In 2017, serial jobs comprised at least 14\% of XD SUs consumed by all single node jobs.   
  
Most parallel jobs efficiently use all or almost all of the allocated CPU cores.  However, nearly one-third of single node jobs utilize only one or two cores on \stampede{} and accordingly throughput may be improved through node sharing for single node jobs.

Many of the most heavily utilized HPC community applications have modest per core memory requirements of less than 2 GB per core (primarily because they were designed that way).   However, many custom (user built) codes, which can arguably be associated with innovative usage, requires substantially more than 2 GB per core to run.   

The method of parallelism employed by most HPC jobs is primarily based on multiple single threaded processes as opposed to multi-thread based.   Initial early data from \stampedetwo{} suggests that a significant proportion of jobs are running 32 or fewer processes/threads per compute node and may not yet be making  optimal use of its multicore architecture.

Lustre file system usage as measured by reads, writes and file opens rates is independent of job size.  Total daily aggregate Lustre reads and writes are approximately equal for most resources.

Job batch script exit codes are not a reliable indication of true job failure since there is no
enforced standard for error reporting. The ``node fail'' exit code (indicating a scheduler or compute node
failure) is a reliable indicator of job failure as shown by regression models.

\newpage

\section{Applications}\label{sec:apps}
\subsection{Application Usage}
\begin{figure}[H]
\centering
\subfloat[Top 20 applications.\label{fig:applications_top20}]{%
 \includegraphics[width=0.5\textwidth]{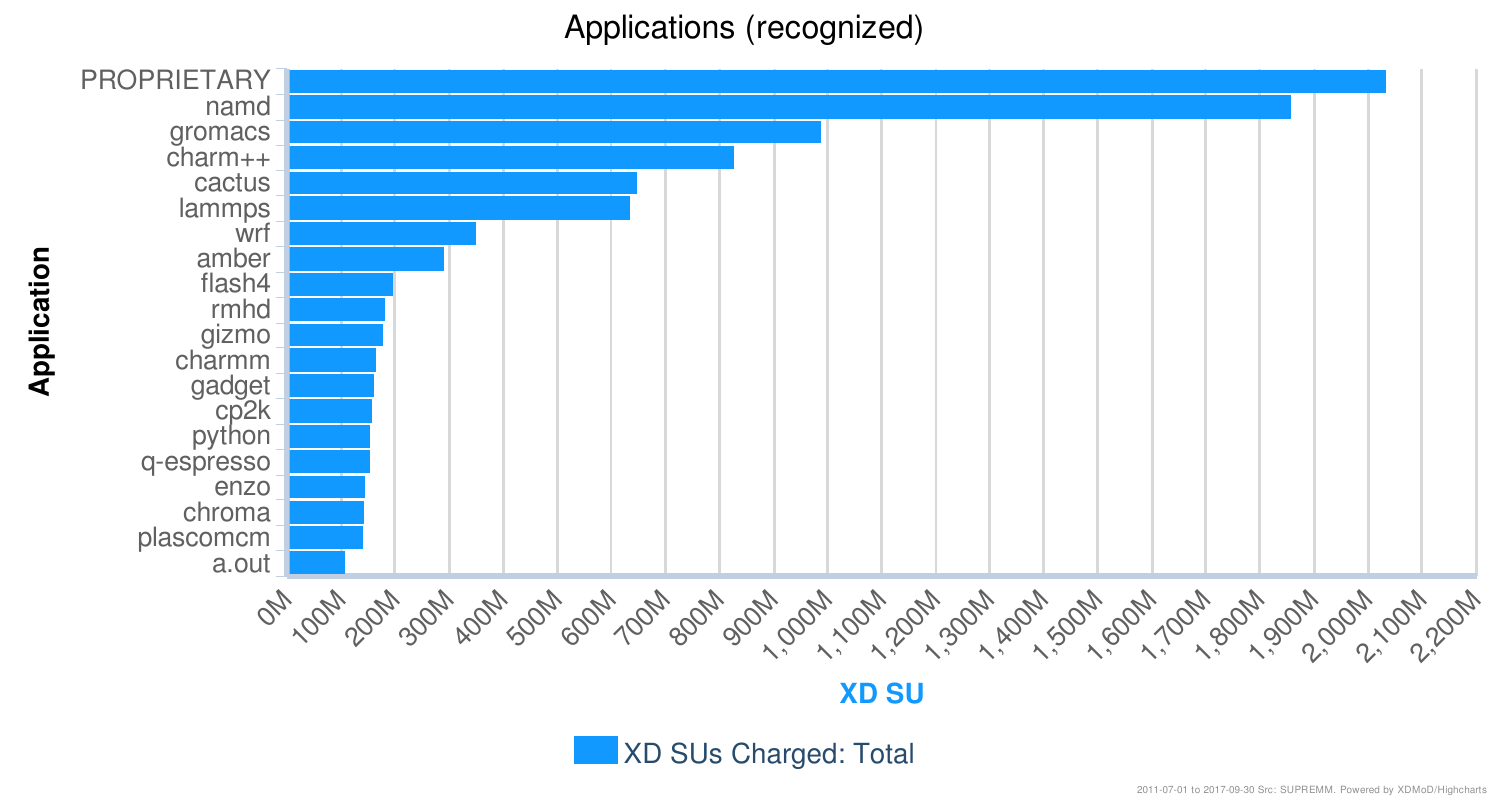}
 }
\subfloat[Top 21-40 applications.\label{fig:applications_next20}]{%
 \includegraphics[width=0.5\textwidth]{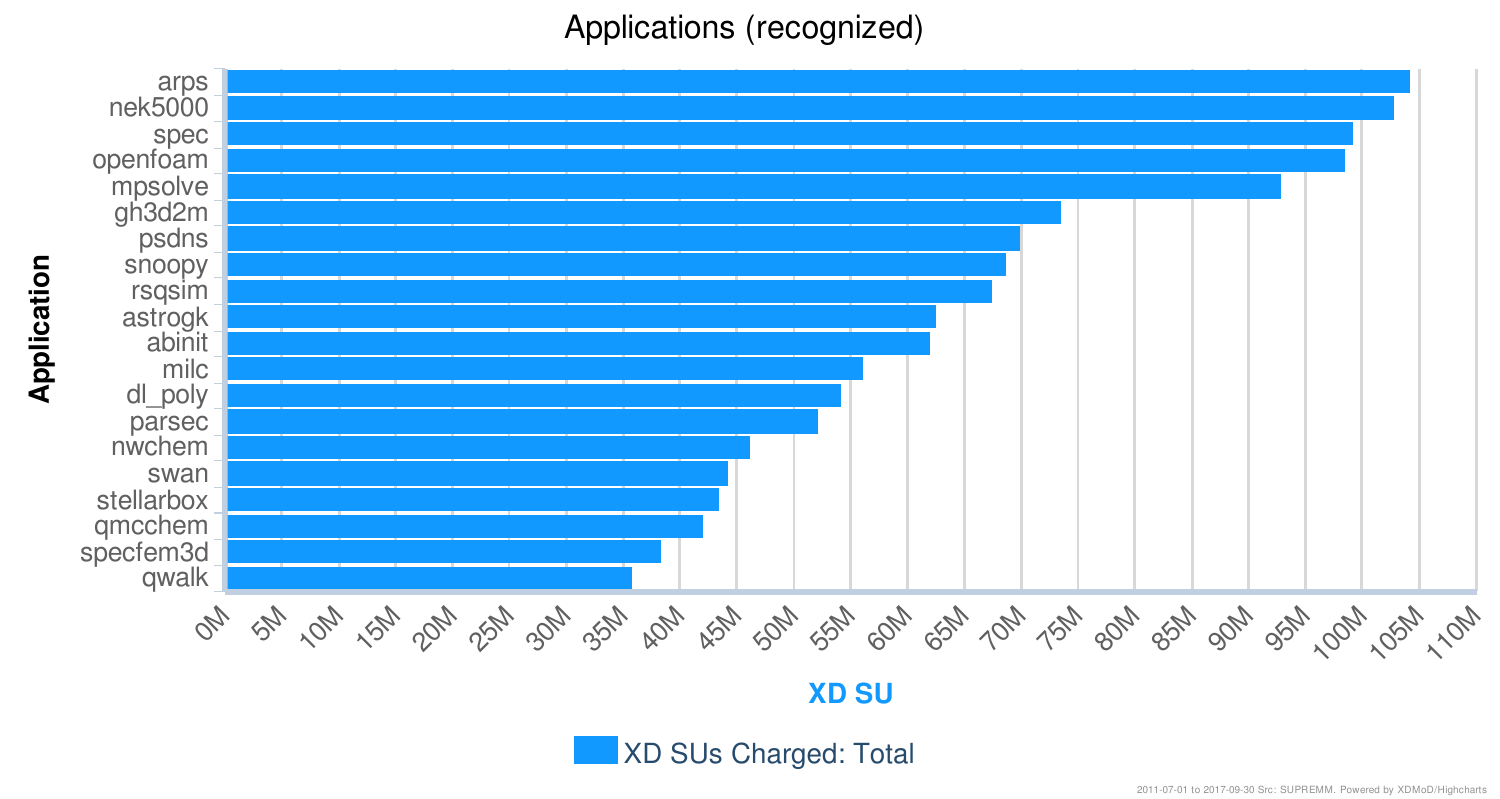}
 }
 \caption{Top 20 (a) and 21-40 (b) applications by XD SUs in the period 2011-07-01 to 2017-09-30.}\label{fig:}
\end{figure}

The most heavily used applications (by XD SU) are shown in Figures \ref{fig:applications_top20}-\ref{fig:applications_next20}.  Note that only resources that record application use are included (see Appendix \ref{appendix:ResourceCharacteristics}). The application classification
uses pattern matching of the job's executable against a reference set of known executables (the methodology is described in detail in Appendix~\ref{appendix:appident}).
Only the most common applications are currently recognized (which are generally open source packages in widespread use)\cite{xdmod-bw2016}.
Note that proprietary applications are intentionally masked (as many have licensing agreements that discourage comparative performance evaluation).
Approximately 39\% of consumed XD SUs were able to be characterized by application name in this study period (an additional 15.6\% were captured but not matched to known applications, and the balance were not collected, mainly due to resources not yet supporting collection).

It should be noted that some of the applications shown in Figures \ref{fig:applications_top20}-\ref{fig:applications_next20} are not ``applications,'' but fall into other categories (see Appendix~\ref{appendix:appident}).  In particular, \texttt{charm++} is a parallel programming system that is used in applications (\texttt{NAMD} is one), and therefore much of its usage should actually be ascribed to other applications, but are not yet recognized as such by our pattern recognition system.  Similarly, \texttt{python} refers to the Python language interpreter, and the underlying application being run by the interpreter is not yet being captured by our system.

\subsection{Application Memory Usage}
\begin{figure}[H]
\centering
\subfloat[\label{fig:xwl_plot_181.pdf}Average memory used per core weighted by core-hour for the top 20 applications by usage.]{%
\includegraphics [width=0.5\textwidth]{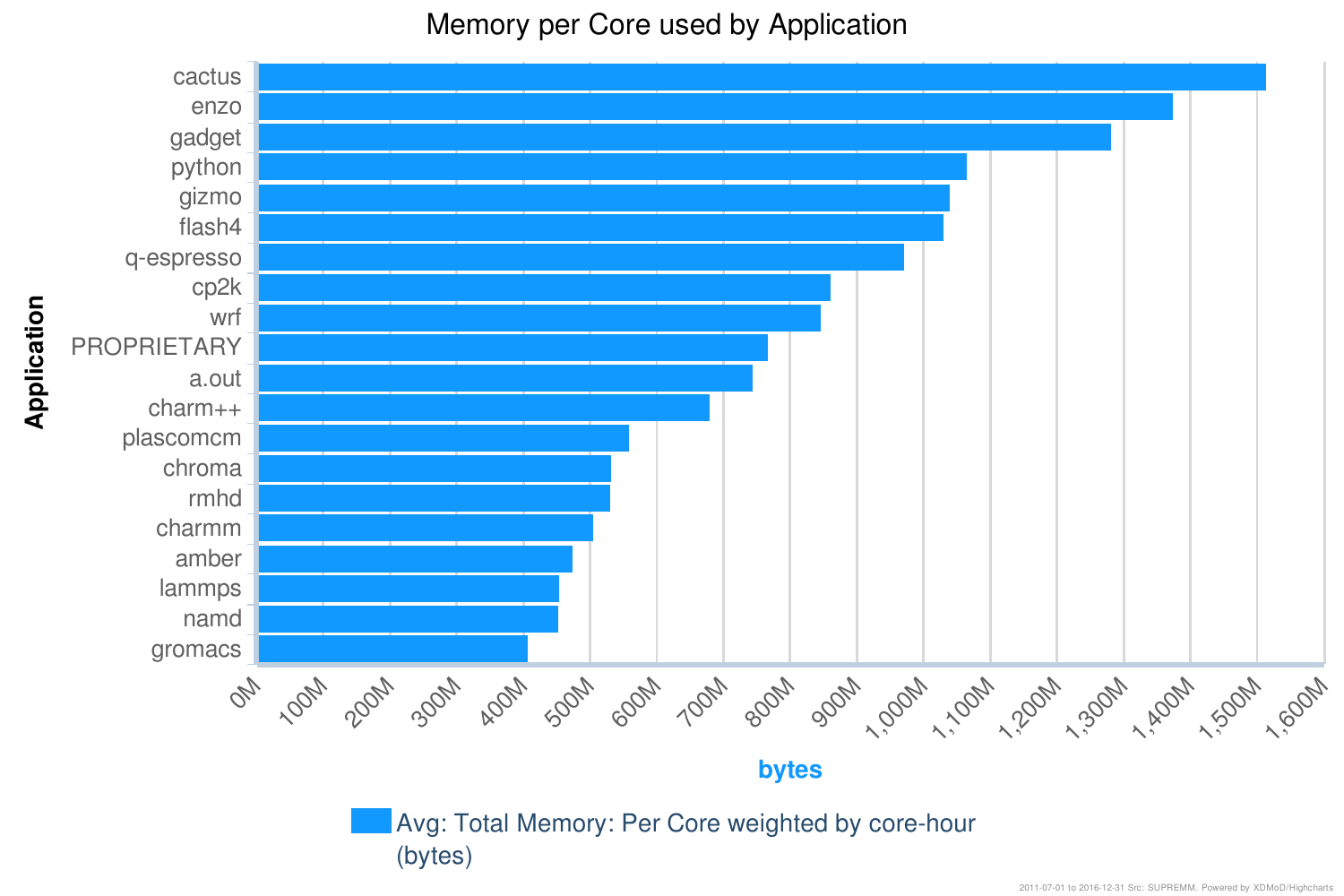}
}
\subfloat[\label{fig:xwl_plot_183.pdf}Average memory used per core weighted by core-hour for the top 20 applications with the largest average memory usage.]{%
\includegraphics[width=0.5\textwidth]{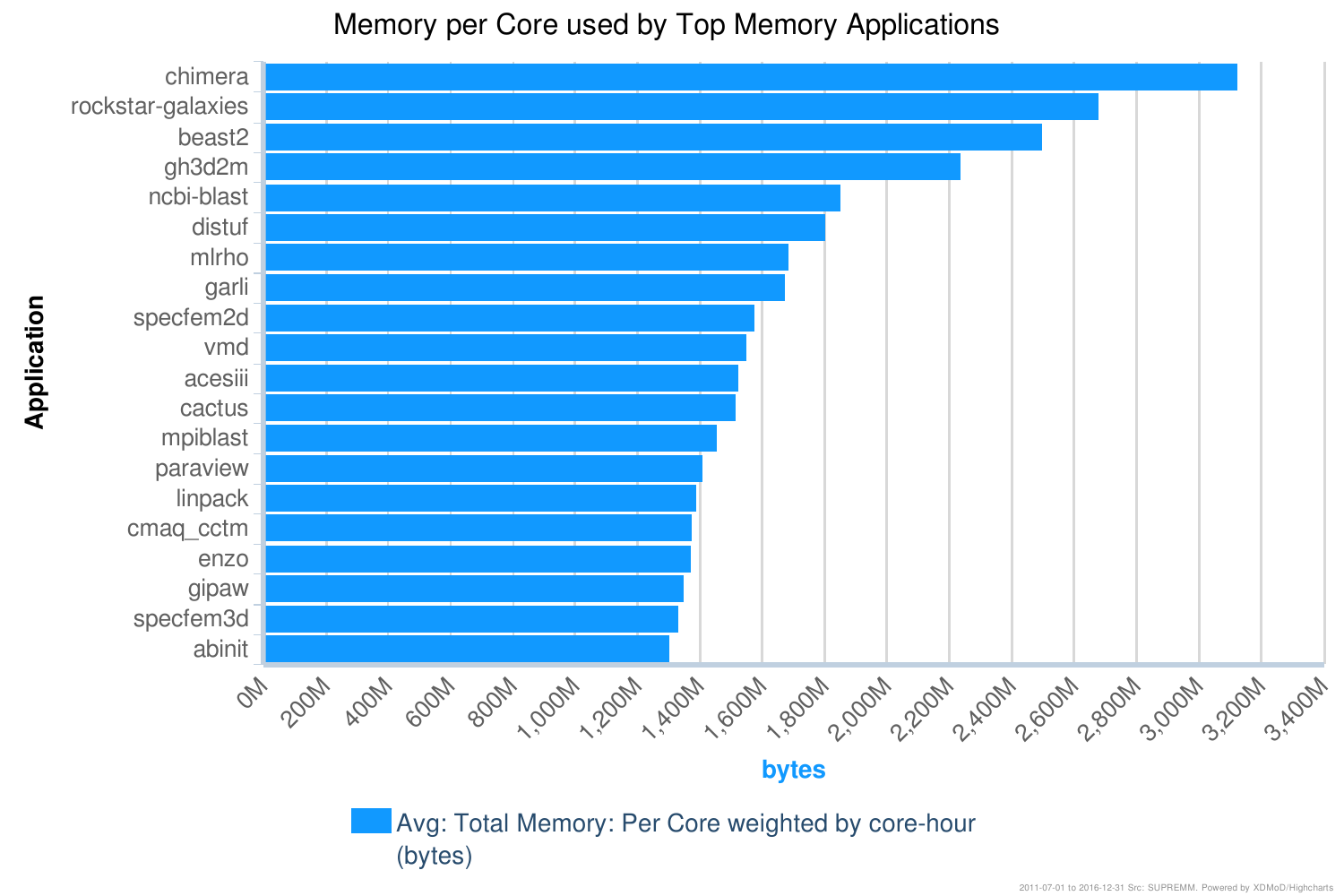}
}
\caption{\label{fig:app_mem}Average memory used per core weighted by core-hour for the top 20 applications by usage (a), and with the largest average memory usage (b) The date range for this plot is from 2011-07-01 to 2017-01-01 (excluding the initial usage on \stampedetwo{}).}
\end{figure}

The average memory used per core weighted by core-hour for the top 20 applications based on usage as given in Figure \ref{fig:applications_top20} is shown in Figure \ref{fig:xwl_plot_181.pdf}.  Only the top three applications substantially exceed 1GB; the remainder range from 0.4GB to 1.05GB per core. Figure \ref{fig:xwl_plot_183.pdf} shows the average memory used per core weighted by core-hour for the top 20 applications with the largest average memory usage. Here the average per core memory lies in the range of 1.3 GB - 3.1 GB per core.  Note that there are only two applications, \texttt{CACTUS} and \texttt{ENZO} that are present on both of these lists, that is a top 20 application by core-hour usage and a top 20 application by average memory usage.  

\begin{figure}[H]
\centering
\includegraphics [width=0.90\textwidth]{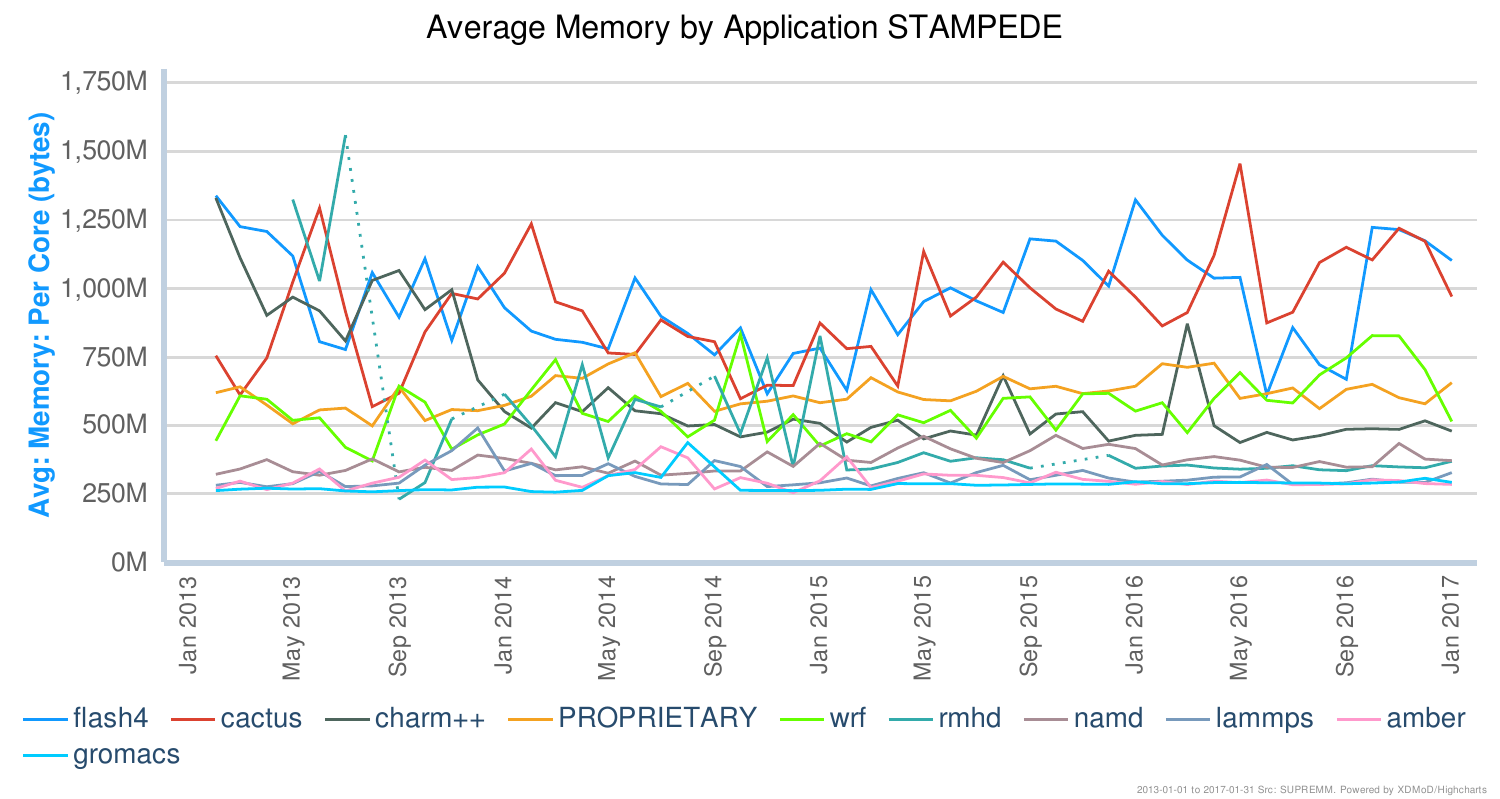}
\caption{\label{fig:xwl_plot_184.pdf}Average memory used per core weighted by core-hour for the top 10 applications with the largest usage on all resources for \stampede{} from 2013-01-01 to 2017-01-01.}
\end{figure}

The average memory usage by application on a given resource over time is remarkably constant throughout the resource lifetime.  Figure  \ref{fig:xwl_plot_184.pdf} shows the average memory usage weighted by core-hour of the top 10 applications by core-hour usage (see Figure \ref{fig:applications_top20}) for \stampede{} throughout its lifetime for which the longest and best memory usage data is available.  Of the 10 applications in the figure, nine of them show no systematic memory trend for \stampede{} throughout its lifetime. Only \texttt{charm++} shows any trend. In the first year, 2013, the memory usage for \texttt{charm++} was higher than for the final three years, 2014-2017, during which is was very flat.  Hence, the individual applications follow the trend already noted for overall memory usage by each resource, see Figure  \ref{fig:xwl_plot_169.pdf}.  

\subsection{Application Interconnect and I/O Use}

\begin{figure}[H]
\centering 
\includegraphics[width=0.8\textwidth]{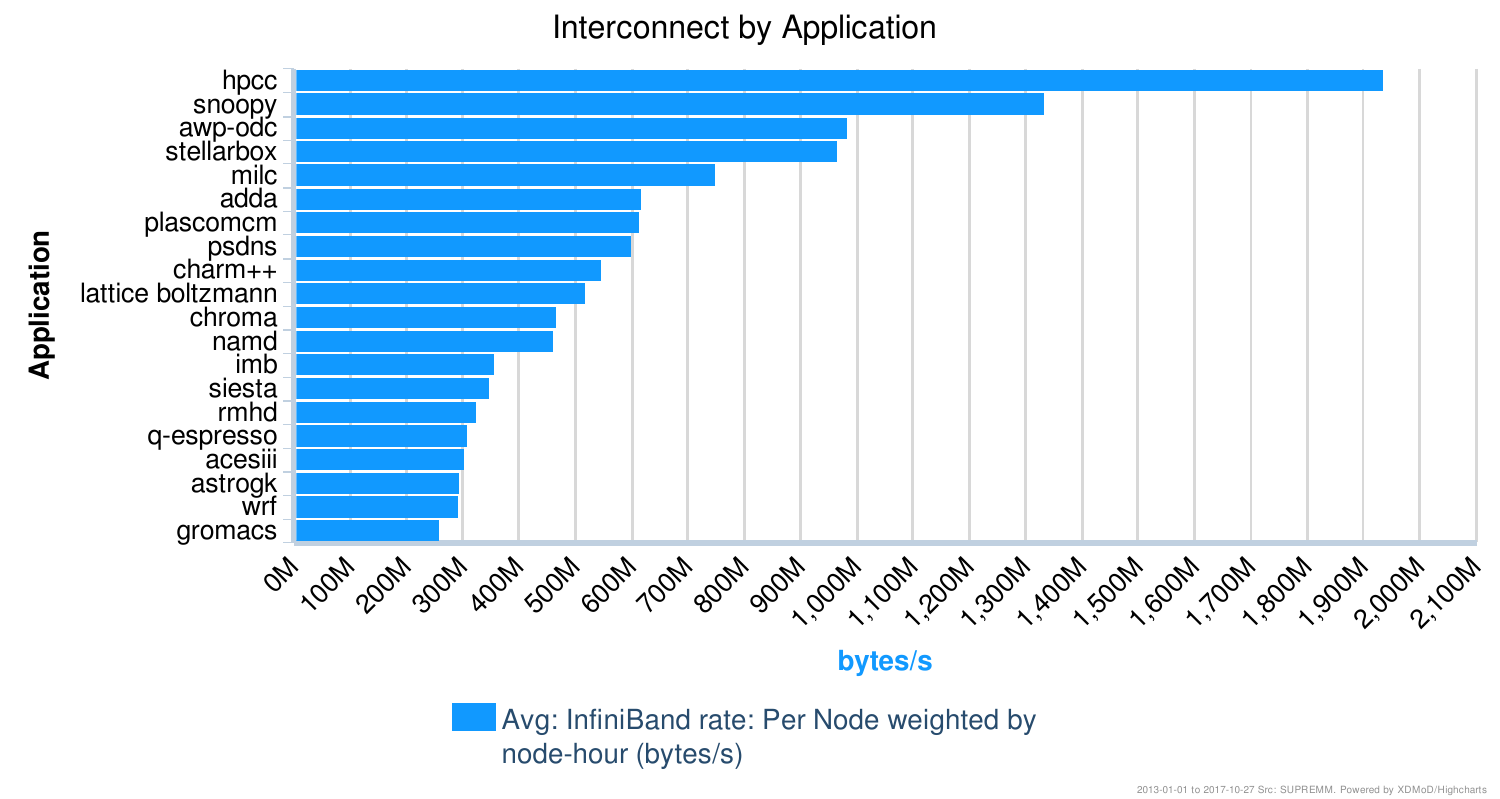}
\caption{Top high-performance interconnect applications (receives only), weighted by node-hours and averaged per job.}
\label{fig:applications_ib-rx}
\end{figure}

\begin{figure}[htb]
\centering 
\subfloat[\label{fig:amber_ib_stampede}]{%
\includegraphics[width=0.25\textwidth]{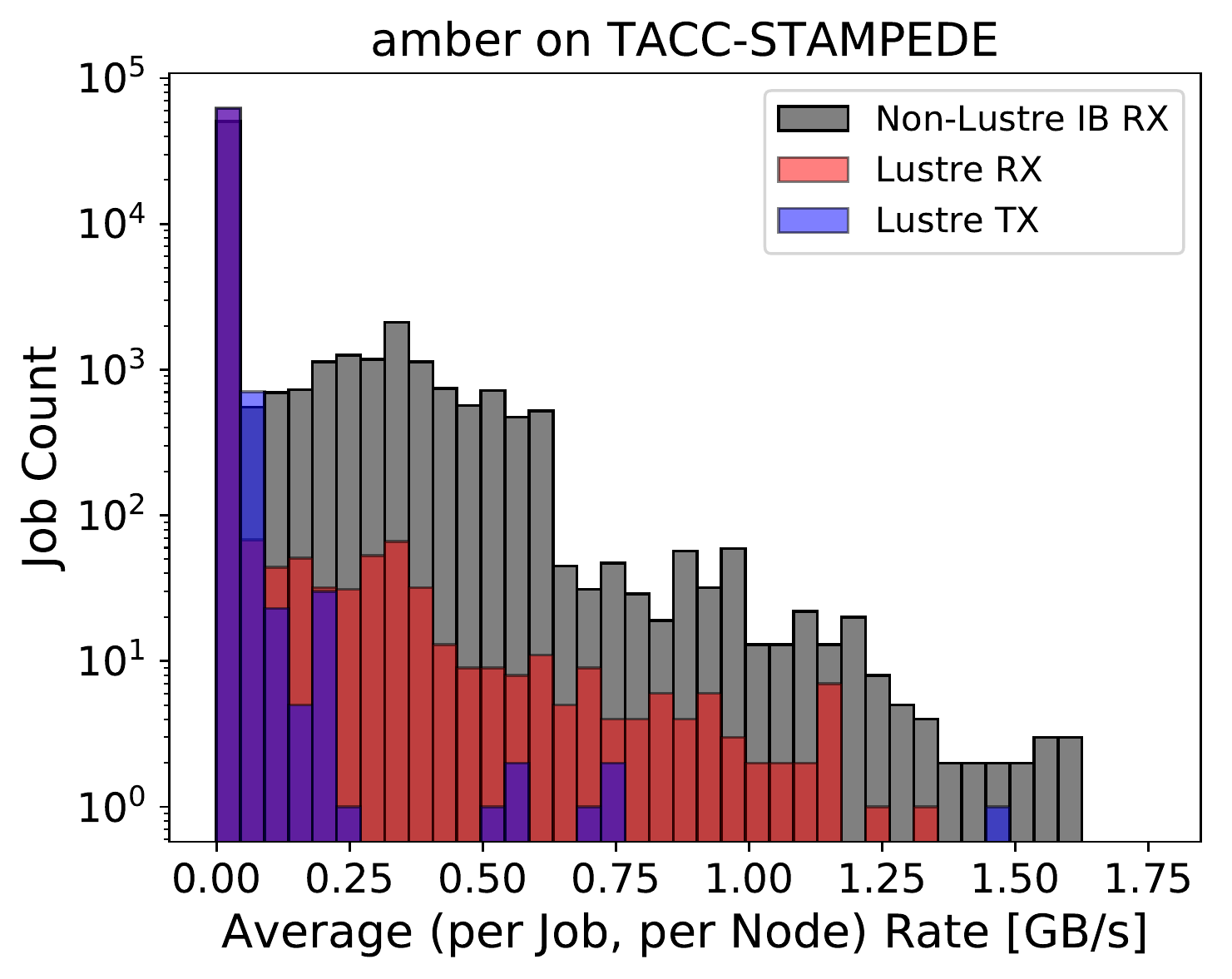}
}
\subfloat[\label{fig:amber_ib_lnet-stampede}]{%
\includegraphics[width=0.25\textwidth]{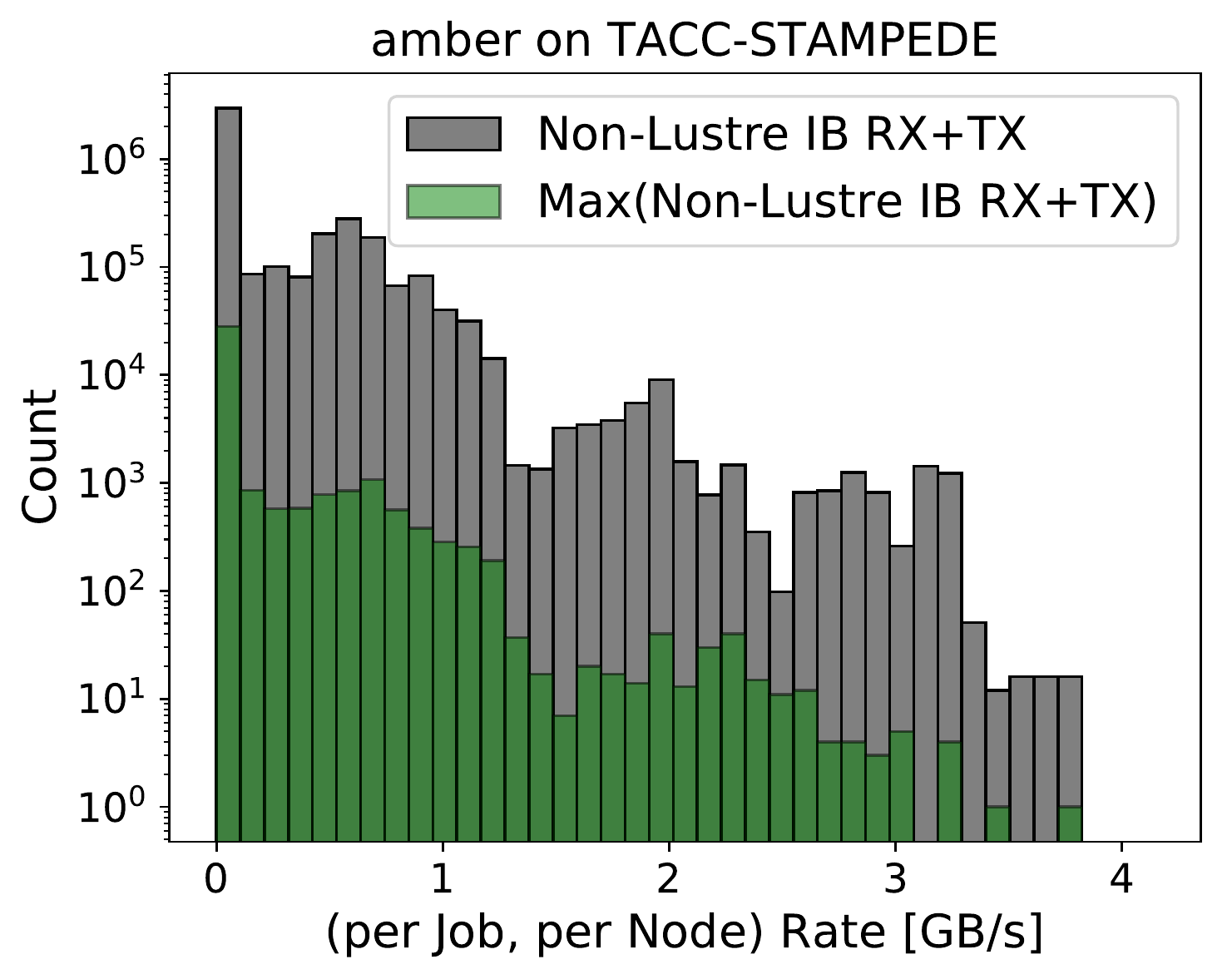}
}
\subfloat[\label{fig:amber_fileopens-stampede}]{%
\includegraphics[width=0.25\textwidth]{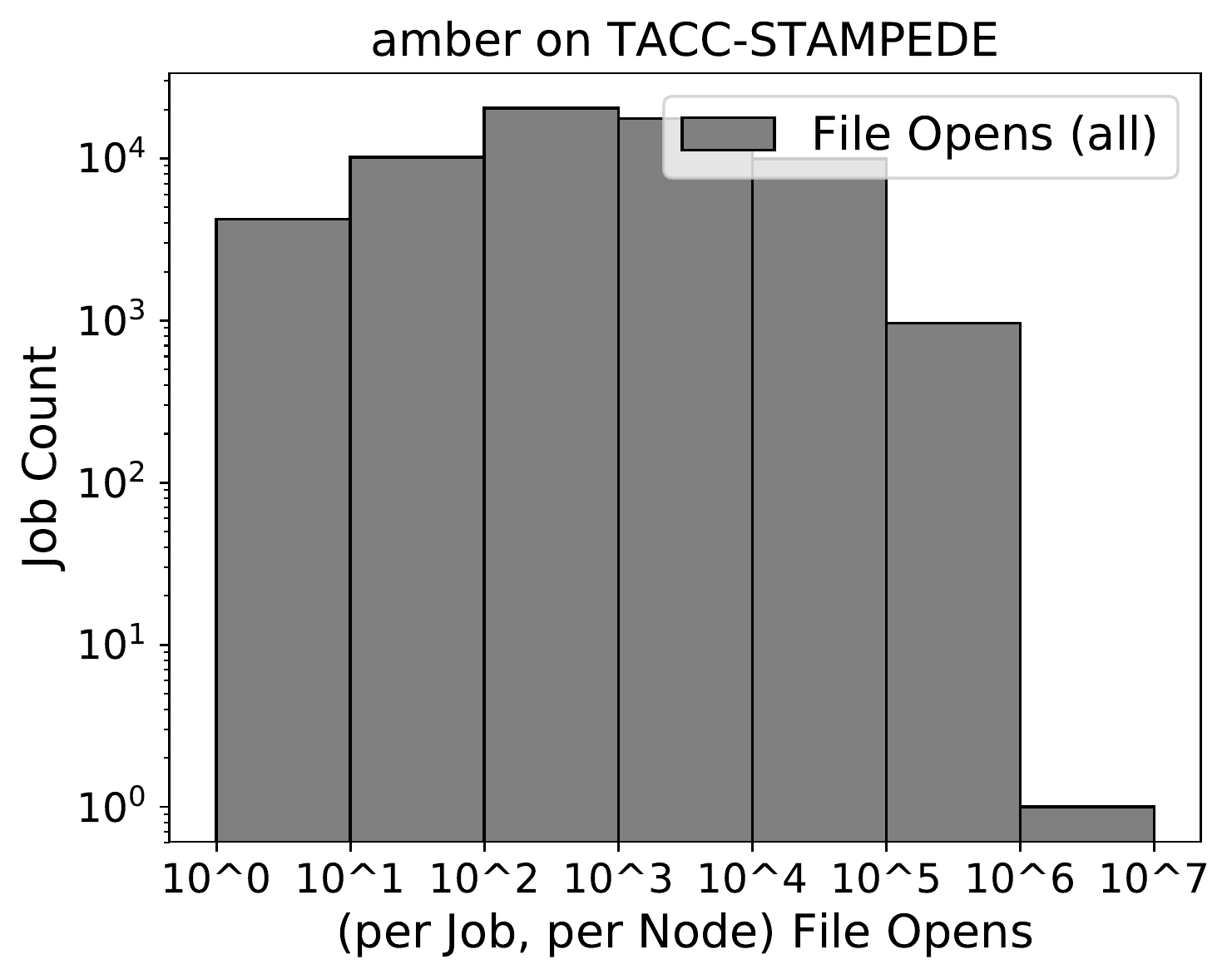}
}
\subfloat[\label{fig:amber_2d_rxtx-stampede}]{%
\includegraphics[width=0.25\textwidth]{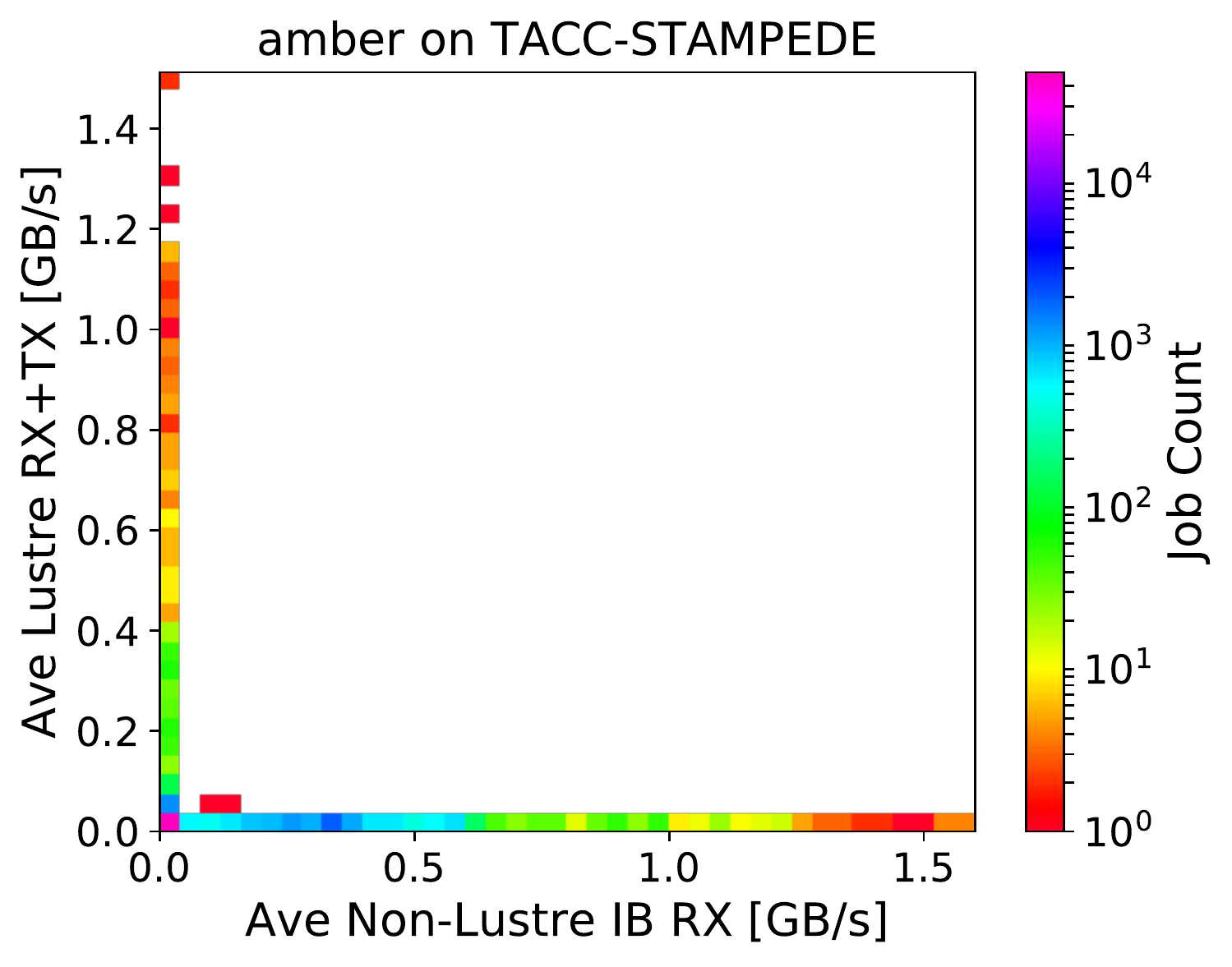}
}\hfil
\subfloat[\label{fig:nwchem_ib_stampede}]{%
\includegraphics[width=0.25\textwidth]{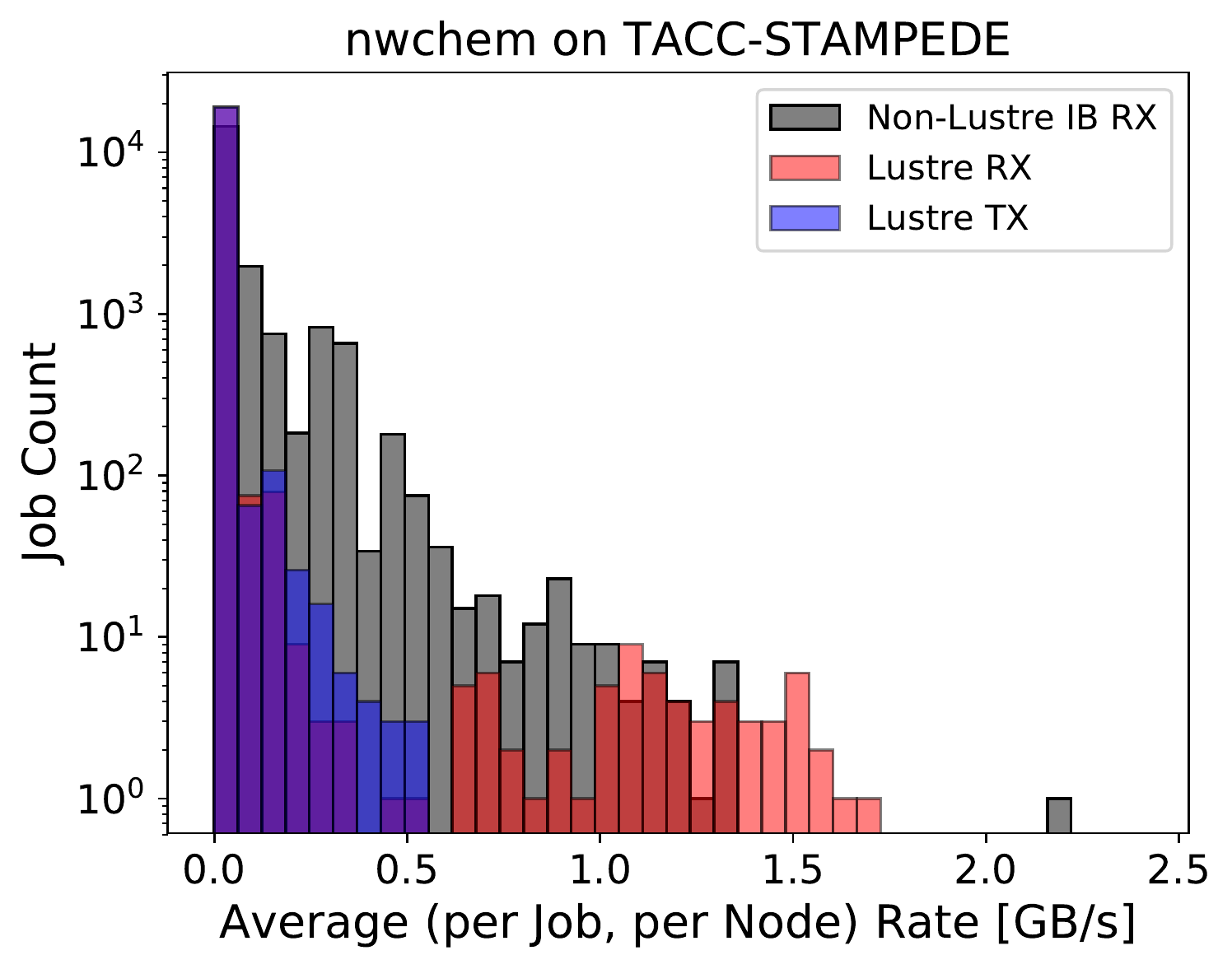}
}
\subfloat[\label{fig:nwchem_ib_lnet-stampede}]{%
\includegraphics[width=0.25\textwidth]{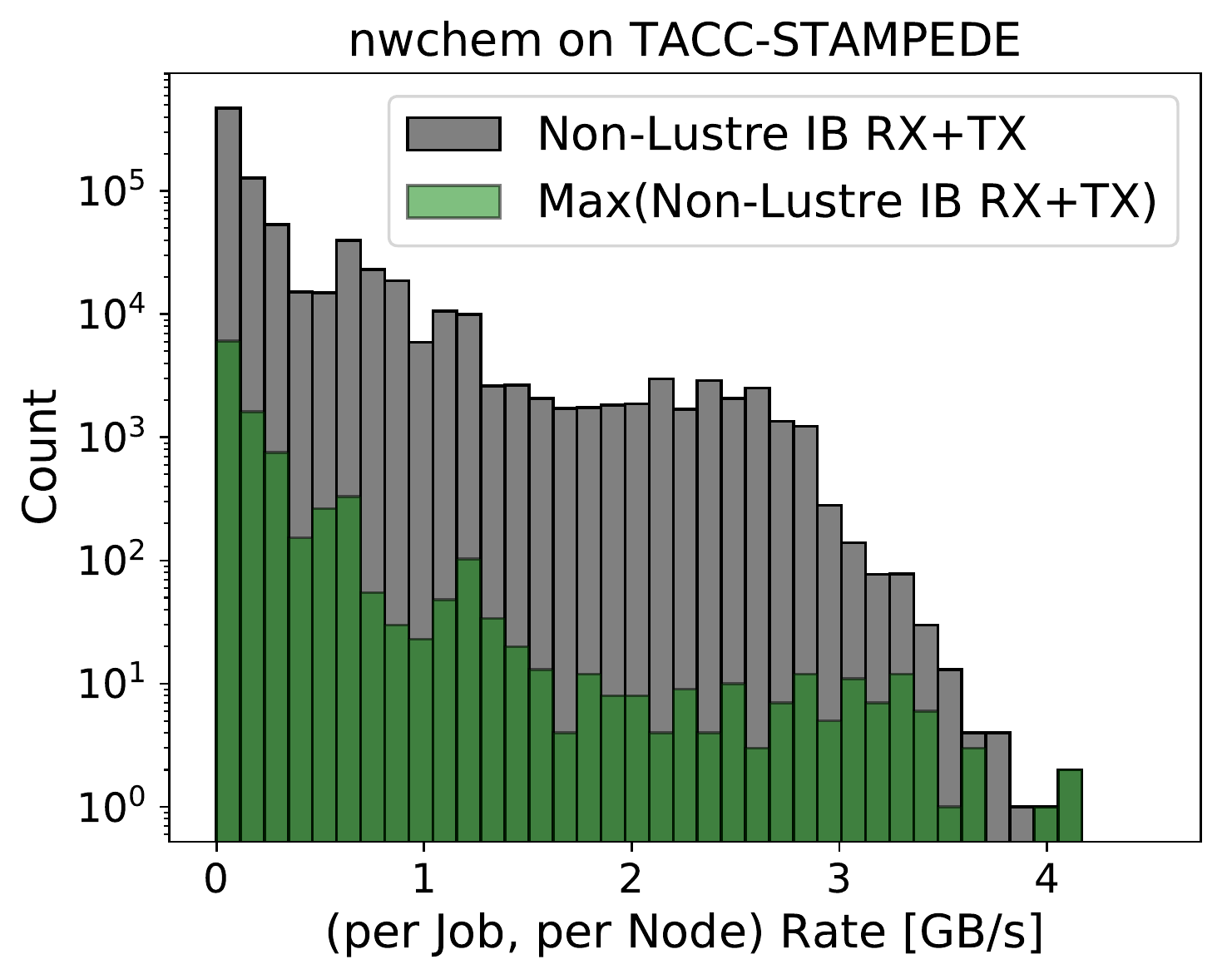}
}
\subfloat[\label{fig:nwchem_fileopens-stampede}]{%
\includegraphics[width=0.25\textwidth]{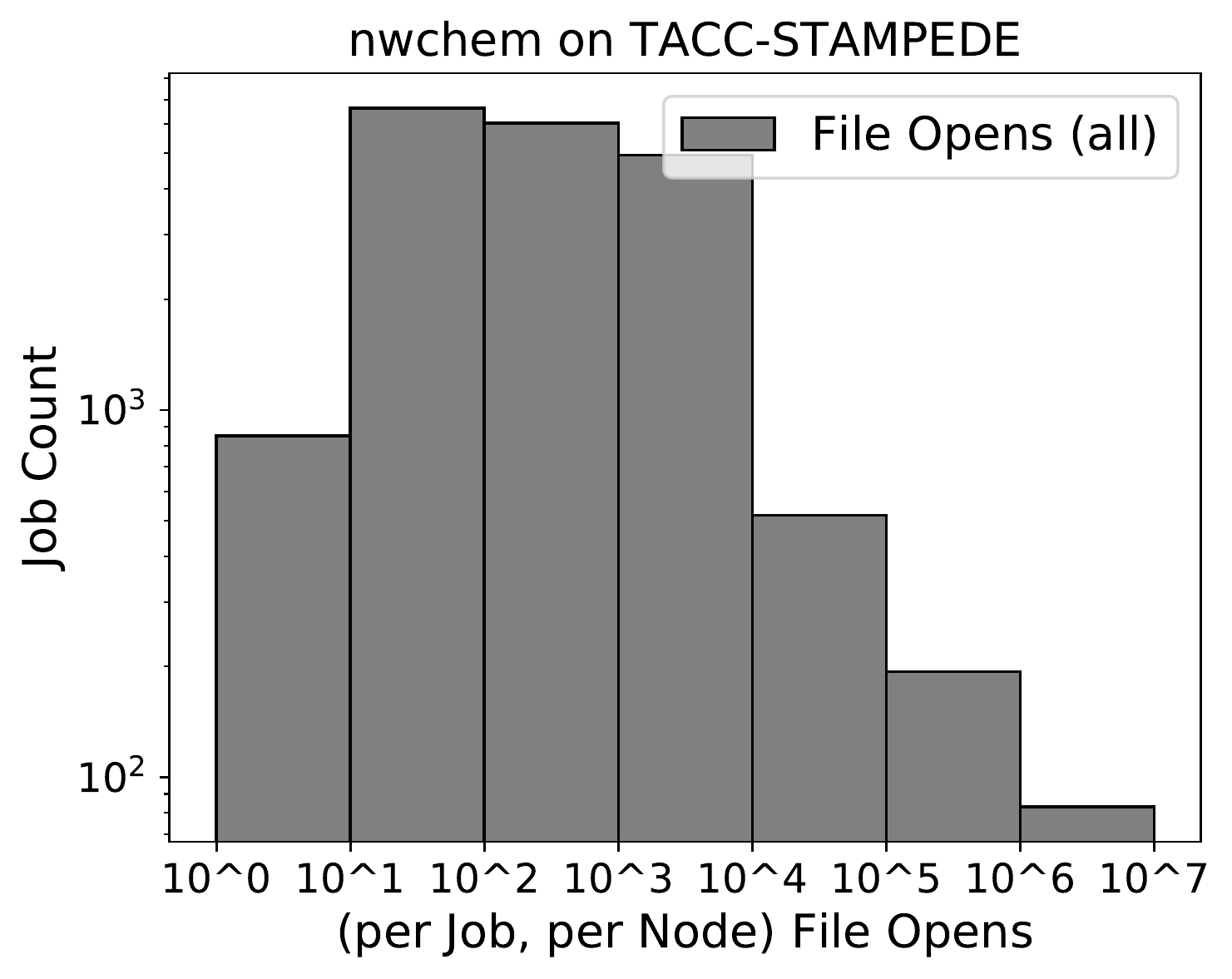}
}
\subfloat[\label{fig:nwchem_2d_rxtx-stampede}]{%
\includegraphics[width=0.25\textwidth]{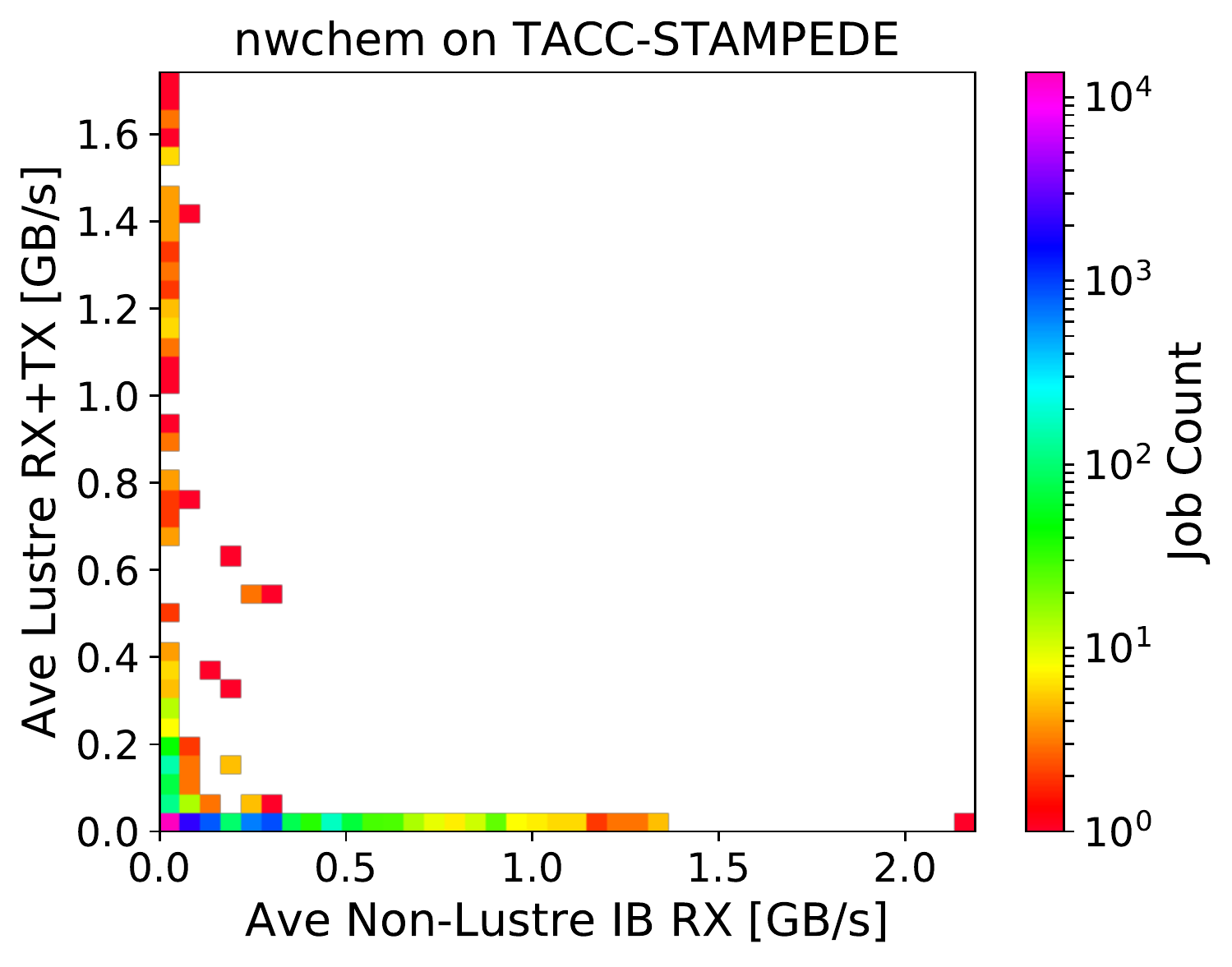}
}\hfil
\subfloat[\label{fig:q-espresso_ib_stampede}]{%
\includegraphics[width=0.25\textwidth]{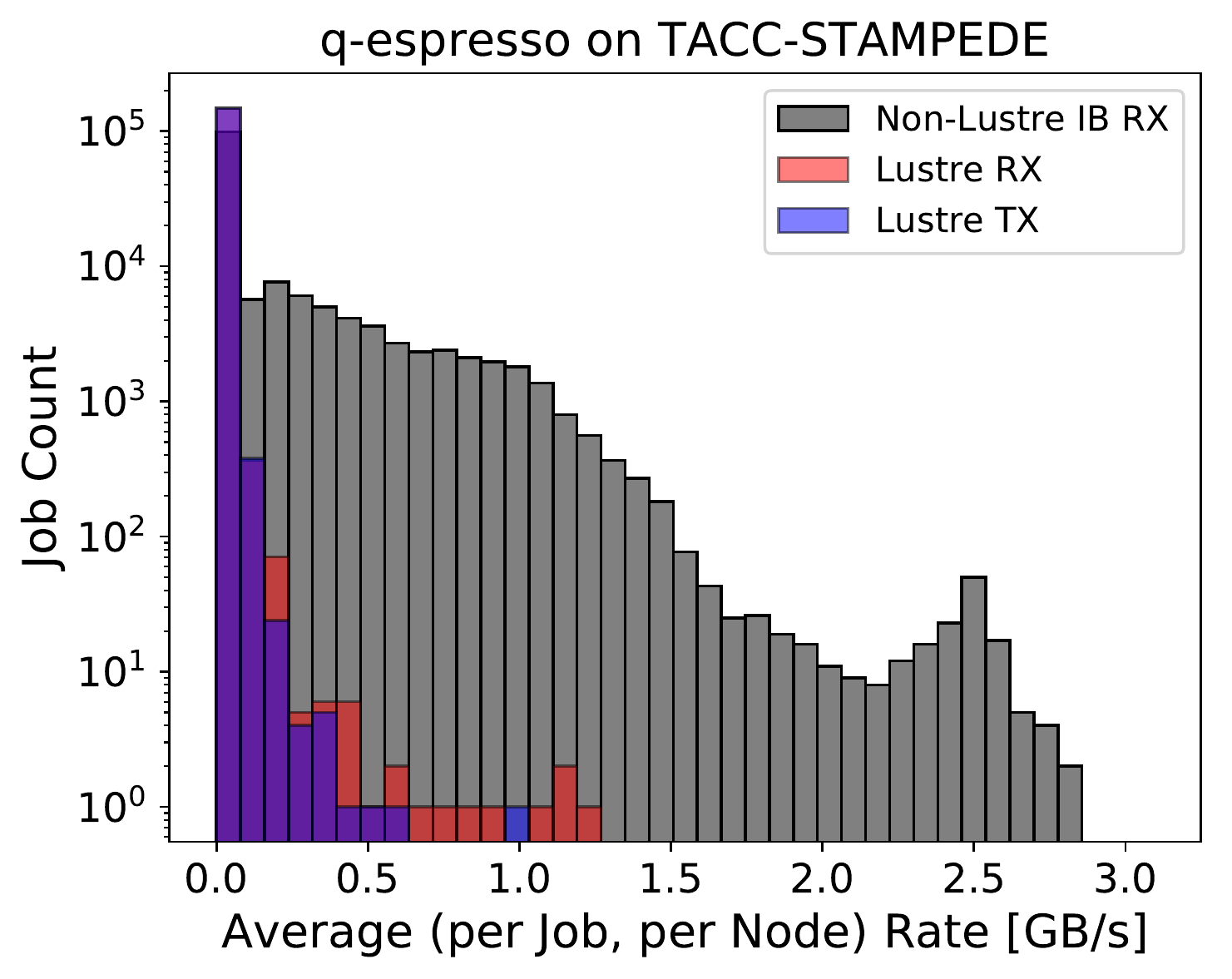}
}
\subfloat[\label{fig:q-espresso_ib_lnet-stampede}]{%
\includegraphics[width=0.25\textwidth]{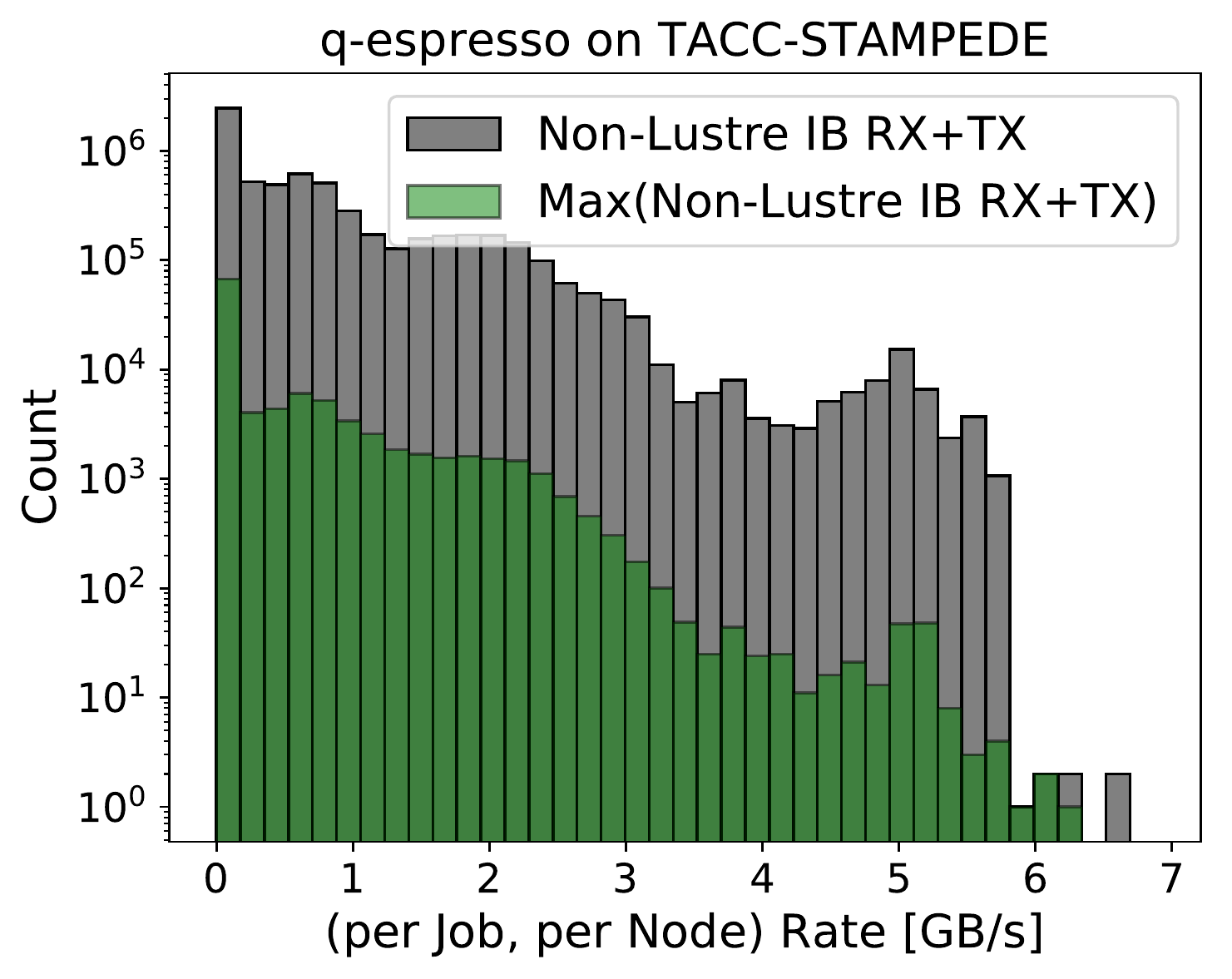}
}
\subfloat[\label{fig:q-espresso_fileopens-stampede}]{%
\includegraphics[width=0.25\textwidth]{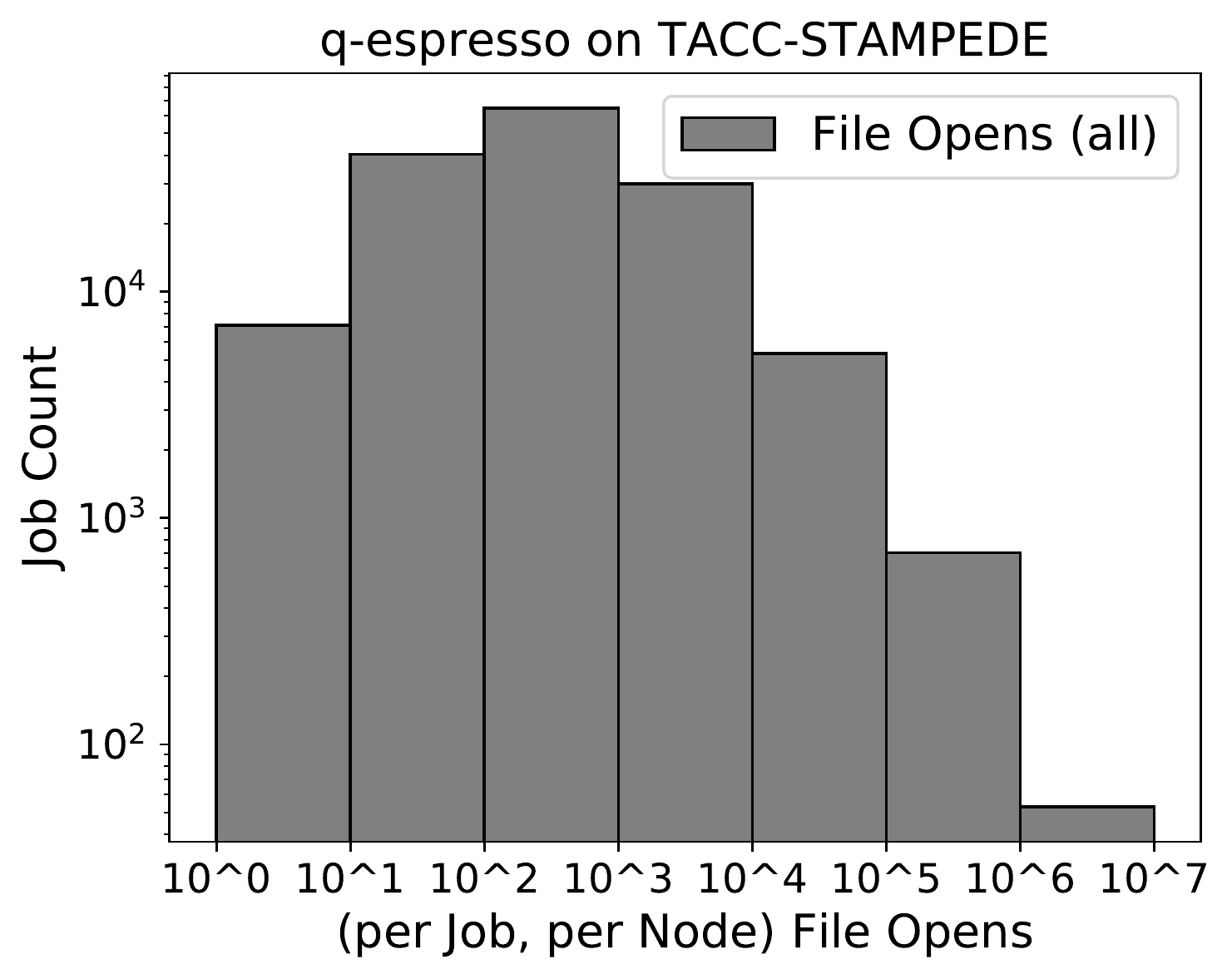}
}
\subfloat[\label{fig:q-espresso_2d_rxtx-stampede}]{%
\includegraphics[width=0.25\textwidth]{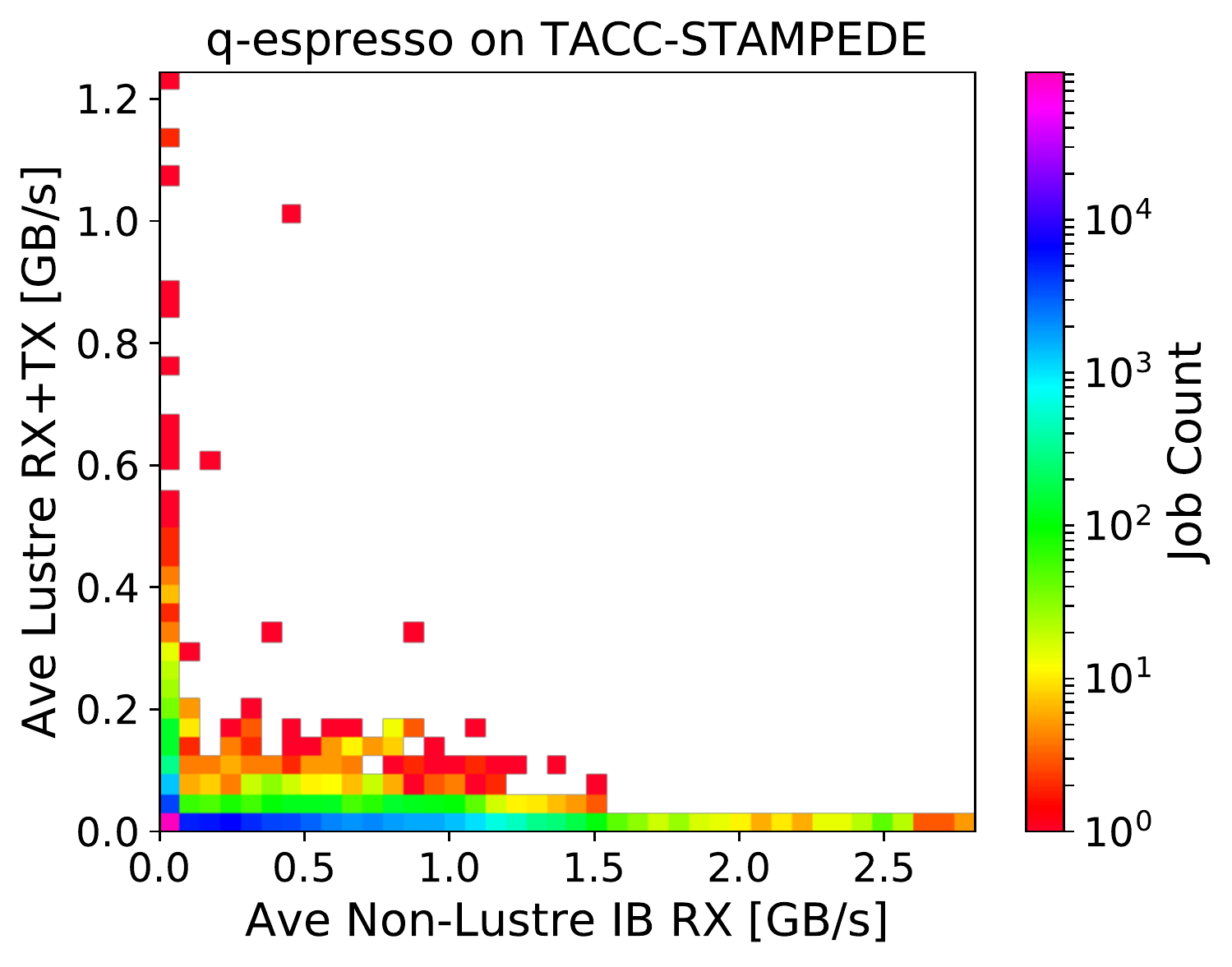}
}\hfil
\subfloat[\label{fig:wrf_ib_stampede}]{%
\includegraphics[width=0.25\textwidth]{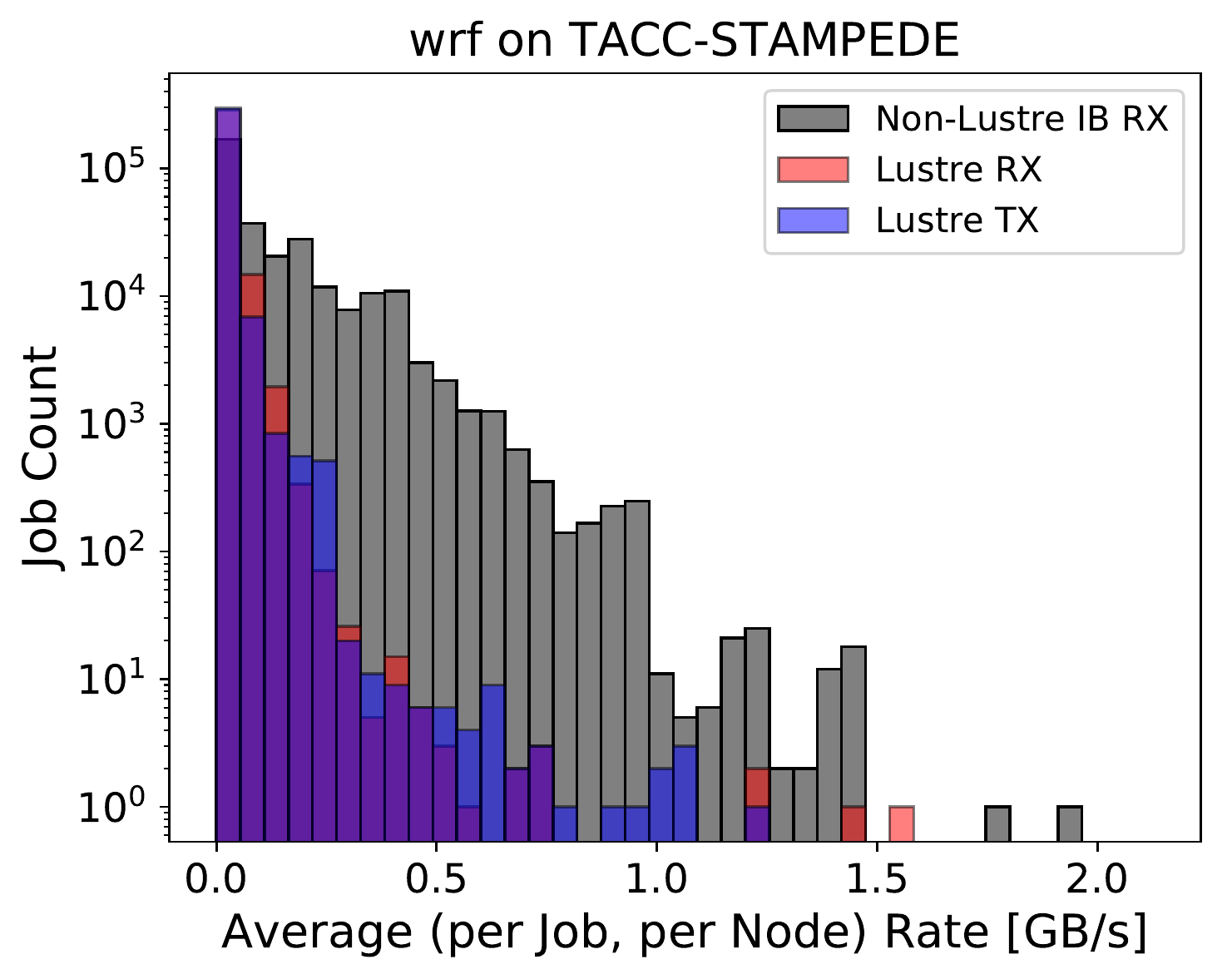}
}
\subfloat[\label{fig:wrf_ib_lnet-stampede}]{%
\includegraphics[width=0.25\textwidth]{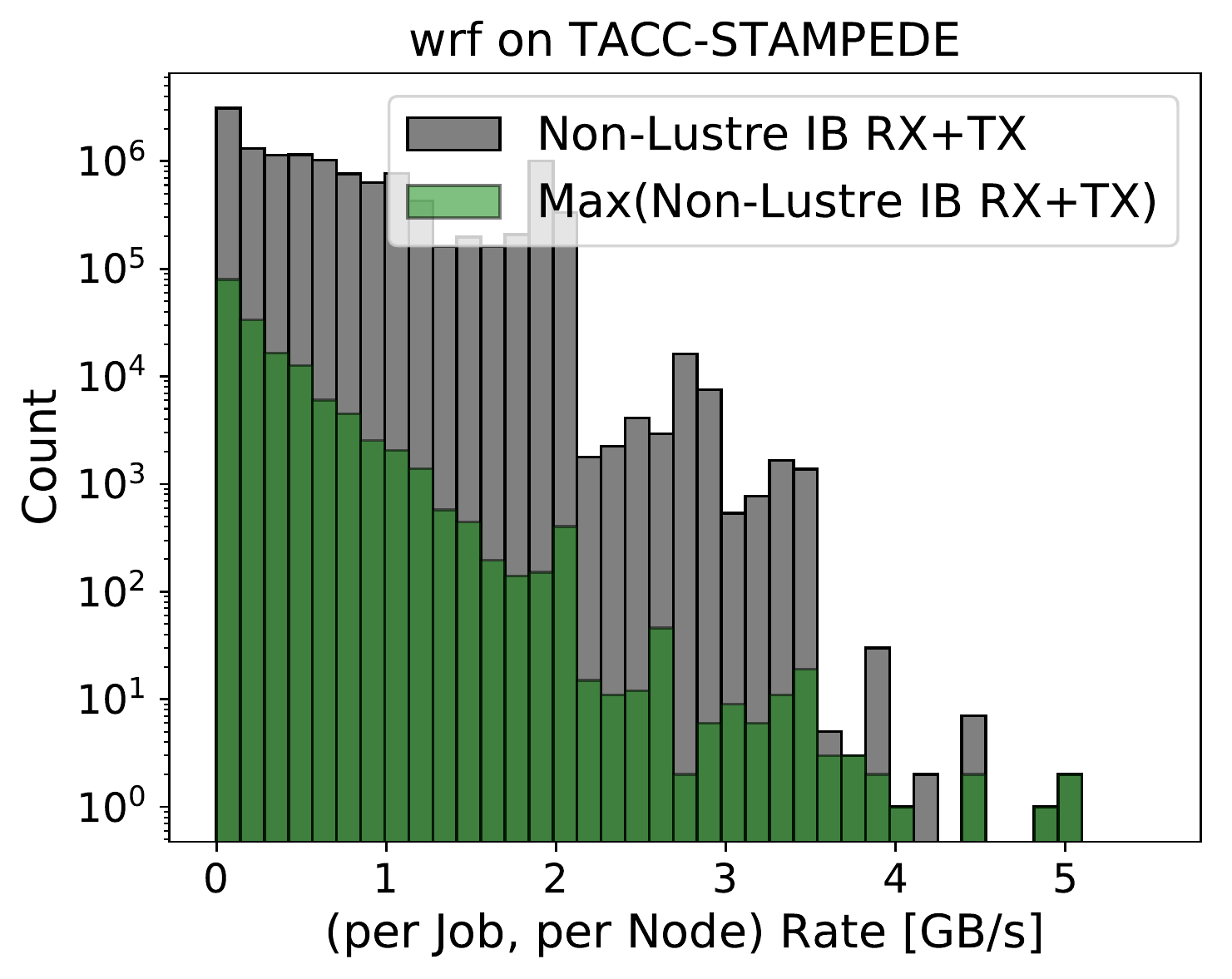}
}
\subfloat[\label{fig:wrf_fileopens-stampede}]{%
\includegraphics[width=0.25\textwidth]{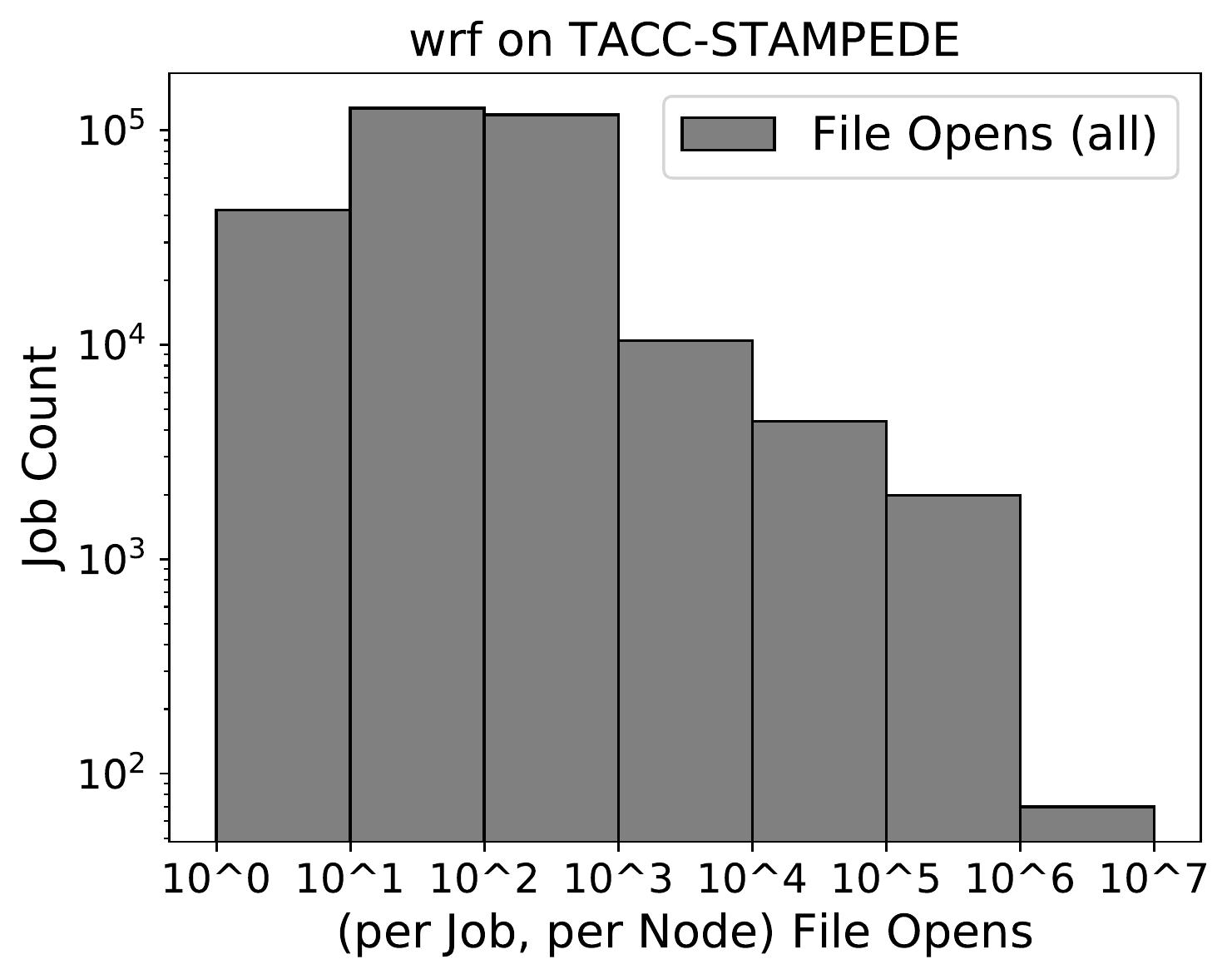}
}
\subfloat[\label{fig:wrf_2d_rxtx-stampede}]{%
\includegraphics[width=0.25\textwidth]{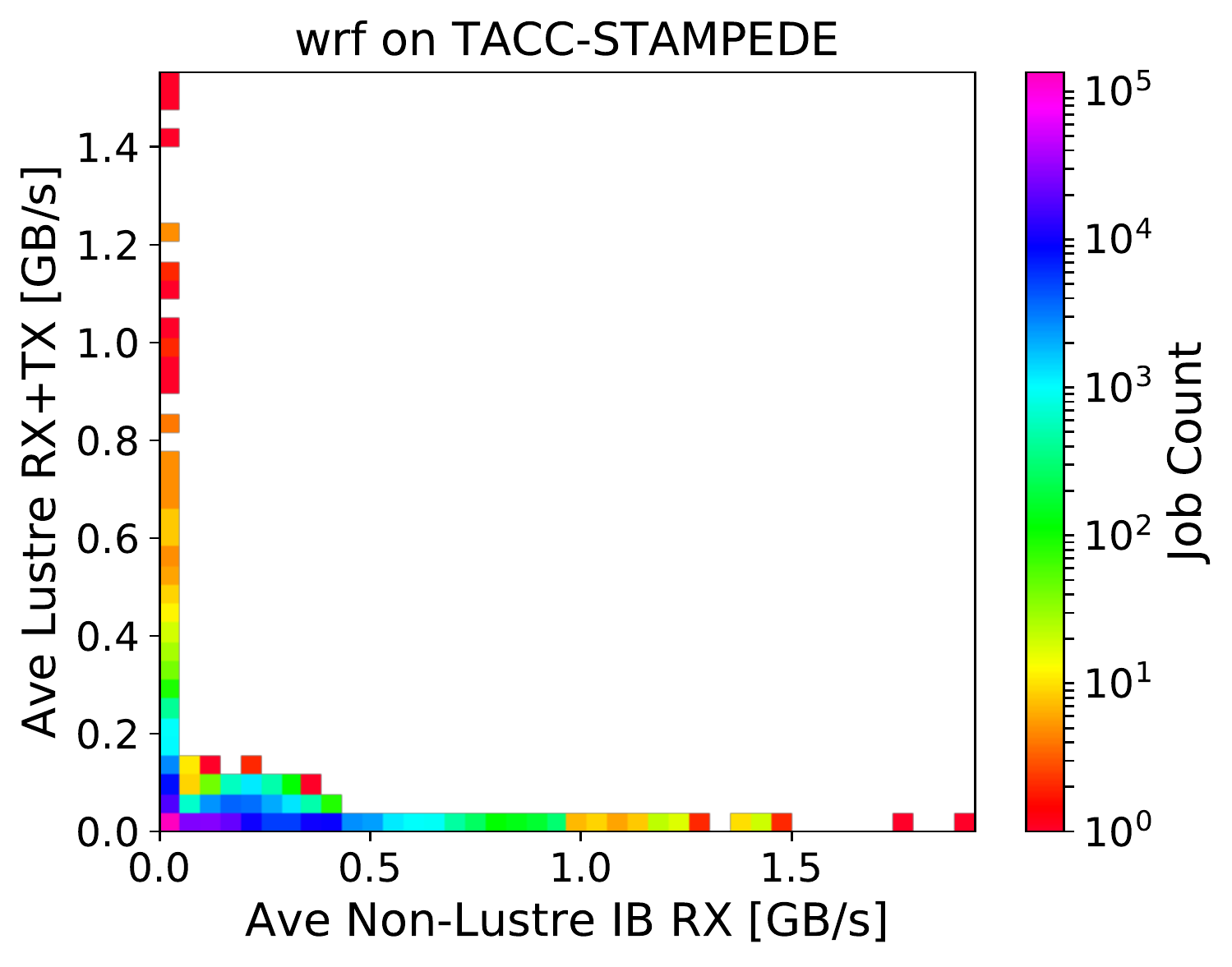}
}
\caption{Application profiles for \texttt{AMBER} (a-d), \texttt{NWCHEM} (e-h), \texttt{Q-ESPRESSO} (i-l),
and \texttt{WRF} (m-p) jobs on \stampede{}.  First column (a,e,i,m) shows the average (per job, per node) Lustre and non-Lustre IB traffic.  Second column (b,f,j,n) shows the per node rate samples for non-Lustre IB as well as the maximum sampled rate.  Third column (c,g,k,o) the distribution of file opens (per job).  Fourth column plots the average Lustre receives and transmits versus non-Lustre IB (per job, per node).}
\label{fig:ib_stampede}
\end{figure}

Figure \ref{fig:applications_ib-rx} shows the average high-performance interconnect use (receives only) by application,
weighted by the node-hours consumed by that application.
The average bandwidth used shown in Figure \ref{fig:applications_ib-rx} is per job and then averaged over all jobs (weighting by node-hours consumed), therefore burst rates are
undoubtedly higher.
Transmits are not shown (mainly due to historical collection issues), therefore one should assume roughly double the bandwidth usage overall, in the case of symmetric bidirectional use (which we examine in more detail below).
The averages shown in Figure \ref{fig:applications_ib-rx} need to be taken in context.  First, the average per job is showing only receives, sampled over relatively coarse time intervals (ten minutes), therefore smoothing over burst rates.  Second, the distribution of averages is not a normal distribution.
In particular, if we examine the distribution of average interconnect use for some popular applications that show relatively high interconnect use, we see highly variable rates.
Figure
\ref{fig:ib_stampede}  shows the distribution of average interconnect use per job for the popular molecular dynamics code \texttt{AMBER}, the quantum chemistry package 
\texttt{NWCHEM}, electronic structure package \texttt{Q-ESPRESSO} and the weather research forecasting package \texttt{WRF} on 
\stampede{}, for several specific usage categories. 
Given that the interconnect sampling interval was ten minutes, the bandwidth burst rates are likely to be
considerably higher.   

Figures~\ref{fig:amber_ib_stampede}, \ref{fig:nwchem_ib_stampede}, \ref{fig:q-espresso_ib_stampede}, and
\ref{fig:wrf_ib_stampede} shows the average interconnect receive rate per job and per node, for the applications \texttt{AMBER}, \texttt{NWCHEM}, \texttt{Q-ESPRESSO}, and \texttt{WRF}, respectively.  
We can also examine the maximum sampled rate per job to see if the maximum value differs significantly from the average.
Figures~\ref{fig:amber_ib_lnet-stampede}, \ref{fig:nwchem_ib_lnet-stampede}, \ref{fig:q-espresso_ib_lnet-stampede}, and
\ref{fig:wrf_ib_lnet-stampede} shows both the maximum sampled rate (in green, now showing the sum of receive and transmit rather than just receives), and also all of the available samples (black), for each job, per node, again for the same applications on \stampede{}.

Note in Figures~\ref{fig:amber_ib_lnet-stampede}, \ref{fig:nwchem_ib_lnet-stampede}, \ref{fig:q-espresso_ib_lnet-stampede} and \ref{fig:wrf_ib_lnet-stampede} that the same trends shown for the averages per job (Figures \ref{fig:amber_ib_stampede}, \ref{fig:nwchem_ib_stampede}, \ref{fig:q-espresso_ib_stampede}, and
\ref{fig:wrf_ib_stampede}) are present, and that indeed the combined receive and transmit values are roughly twice the receives (indicating that the non-Lustre bandwidth use is roughly symmetric).  The maximum bandwidth per job (shown in green in Figures \ref{fig:amber_ib_lnet-stampede}, \ref{fig:nwchem_ib_lnet-stampede}, \ref{fig:q-espresso_ib_lnet-stampede} and \ref{fig:wrf_ib_lnet-stampede}) is only a bit higher than the largest average per job (roughly 20\% for \texttt{WRF}).  One should bear in mind the sample rate is ten minutes, however, so burst rates may indeed be higher still within the sample window.
Also shown in Figures~\ref{fig:amber_ib_stampede}, \ref{fig:nwchem_ib_stampede}, \ref{fig:q-espresso_ib_stampede}, and
\ref{fig:wrf_ib_stampede} are the average Lustre transmits and receives, again showing high variability.  
This variability is also present in file opens, which are shown in Figures \ref{fig:amber_fileopens-stampede}, \ref{fig:nwchem_fileopens-stampede}, \ref{fig:q-espresso_fileopens-stampede}, and
\ref{fig:wrf_fileopens-stampede}.


We compare the parallel filesystem usage to the non-Lustre inter-compute node communication data for selected applications in
Figures~\ref{fig:amber_2d_rxtx-stampede}, \ref{fig:nwchem_2d_rxtx-stampede}, \ref{fig:q-espresso_2d_rxtx-stampede} and \ref{fig:wrf_2d_rxtx-stampede} for \stampede{}. 
\stampede{} and all of the other HPC resources considered here use an InfiniBand high speed interconnect and a Lustre parallel filesystem. The parallel filesystem data is transferred
between the compute nodes and filesystem nodes over the InifiniBand network.
We estimate the inter-compute node communications
by subtracting the Lustre data from the InfiniBand data. This is labeled as ``non-Lustre IB'' in the figures.
The correlation between the non-Lustre InfiniBand network activity and I/O shows only a very modest interaction.  
The correlation is quite dependent on the individual applications as shown by Figures~\ref{fig:amber_2d_rxtx-stampede}, \ref{fig:nwchem_2d_rxtx-stampede}, \ref{fig:q-espresso_2d_rxtx-stampede} and \ref{fig:wrf_2d_rxtx-stampede}.  
\texttt{AMBER} has virtually no correlation, \texttt{NWCHEM} and \texttt{WRF} have some interaction while \texttt{Q-ESPRESSO} has substantial interaction.

\clearpage

\subsection{Summary: Applications}

Of identified, non-proprietary applications, the top five in terms of consumed XD SUs are \texttt{NAMD}, \texttt{GROMACS}, \texttt{CACTUS}, \texttt{LAMMPS}, and \texttt{WRF}.  Most of the identified applications with the greatest utilization use less than 1.5 GB per core, with the majority falling in the range of 0.4--1.1 GB per core.   The average per core memory for the top twenty applications in terms of memory use lies in the range of 1.3 GB--3.1 GB per core.   The average memory usage by application is relatively constant throughout the resource lifetime.  Applications show a wide range of interconnect usage, with the more intensive averaging more than 1GB/s of unidirectional bandwidth, bursting considerably higher.  
Data indicates, at least within the sampling intervals, that applications are not bandwidth limited.
Application I/O usage also shows a wide range of usage patterns.    

\clearpage
\newpage


\section{Science Gateways}\label{sec:gateways}

\textit{
Goals Addressed in Section
\begin{enumerate}\setcounter{enumi}{6}
\item Are there important differences among job types, e.g., interactive jobs, gateway jobs, etc.?
\item What are the characteristics of gateway jobs?
\begin{itemize}
\item What are the characteristics and trends for CPU core hours per job, node counts and types, memory and interconnect usage?
\item Are the parallel jobs simply ensembles (many independent jobs)?
\item  What is the gateway job distribution by resource?
\item How many new (unique) users are using gateways?
\item Are the usage patterns of gateway users significantly different than traditional HPC users?
\item How does gateway utilization and growth differ by discipline
\end{itemize}
\end{enumerate}
}

Science Gateways are defined on the XSEDE User Portal Science Gateway page (\url{https://portal.xsede.org/science-gateways}) as ``Customized portals granting members access to HPC applications, workflows, shared data and other services. XSEDE Gateways unite communities of like-minded members, whether united by discipline or other criteria.'' In this report, we will examine the characteristics of Science Gateways jobs, usage and users showing how they are unique and in what ways they are similar to their traditional HPC analogs. 

\begin{figure}[H]
\centering
\includegraphics [width=0.9\textwidth]{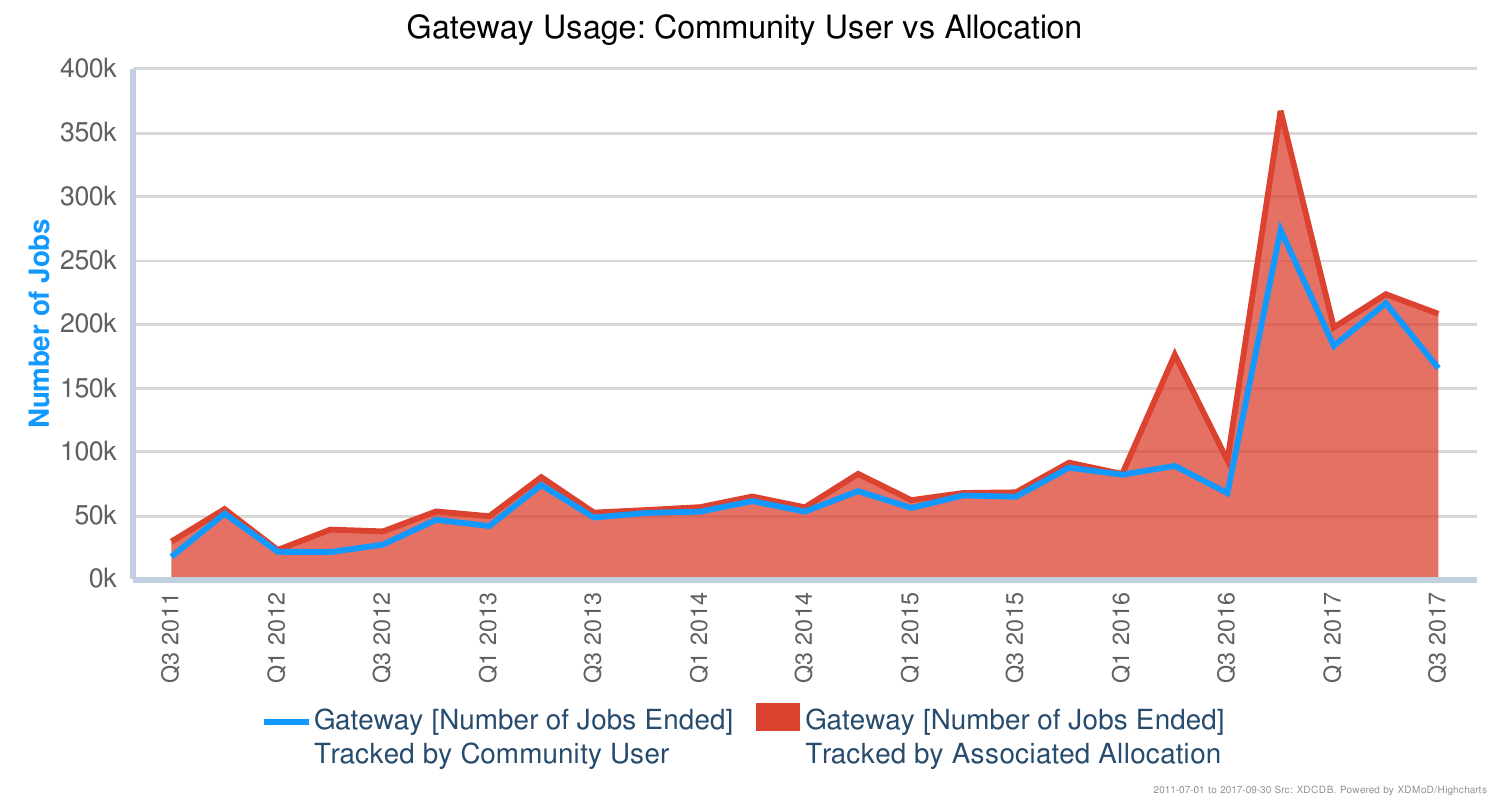}
\caption{\label{fig:Gateway_Usage__Community_User_vs_Allocation_2011-07-01_to_2017-09-30.pdf}Tracking Science Gateway usage via jobs submitted by Community Users versus allocations known to be associated with a Gateway.}
\end{figure}

The methods used for collecting Science Gateway usage data have evolved over the lifetime of XSEDE and there have been multiple methods used to report on Gateway usage as explained in Appendix \ref{appendix:data-science-gateways}. Table \ref{table:gateway-job-attribute-reporting} shows the deployment period (through 2017-09-30) for each Gateway along with other information such as the number of jobs submitted and CPU hours consumed. Gateways shown in \textbf{boldface} are active at the time of writing. For the purposes of this report, we will present Gateway usage via jobs submitted by Community User accounts associated with a particular Gateway where possible as this closely aligns with XSEDE’s definition of a Science Gateway. Figure \ref{fig:Gateway_Usage__Community_User_vs_Allocation_2011-07-01_to_2017-09-30.pdf} shows that, with a few exceptions, the number of Gateway jobs tracked using Community User submissions aligns closely with other methods described in Appendix \ref{appendix:data-science-gateways}.

\subsection{Gateways Usage}

\begin{figure}[H]
\centering
\includegraphics [width=0.9\textwidth]{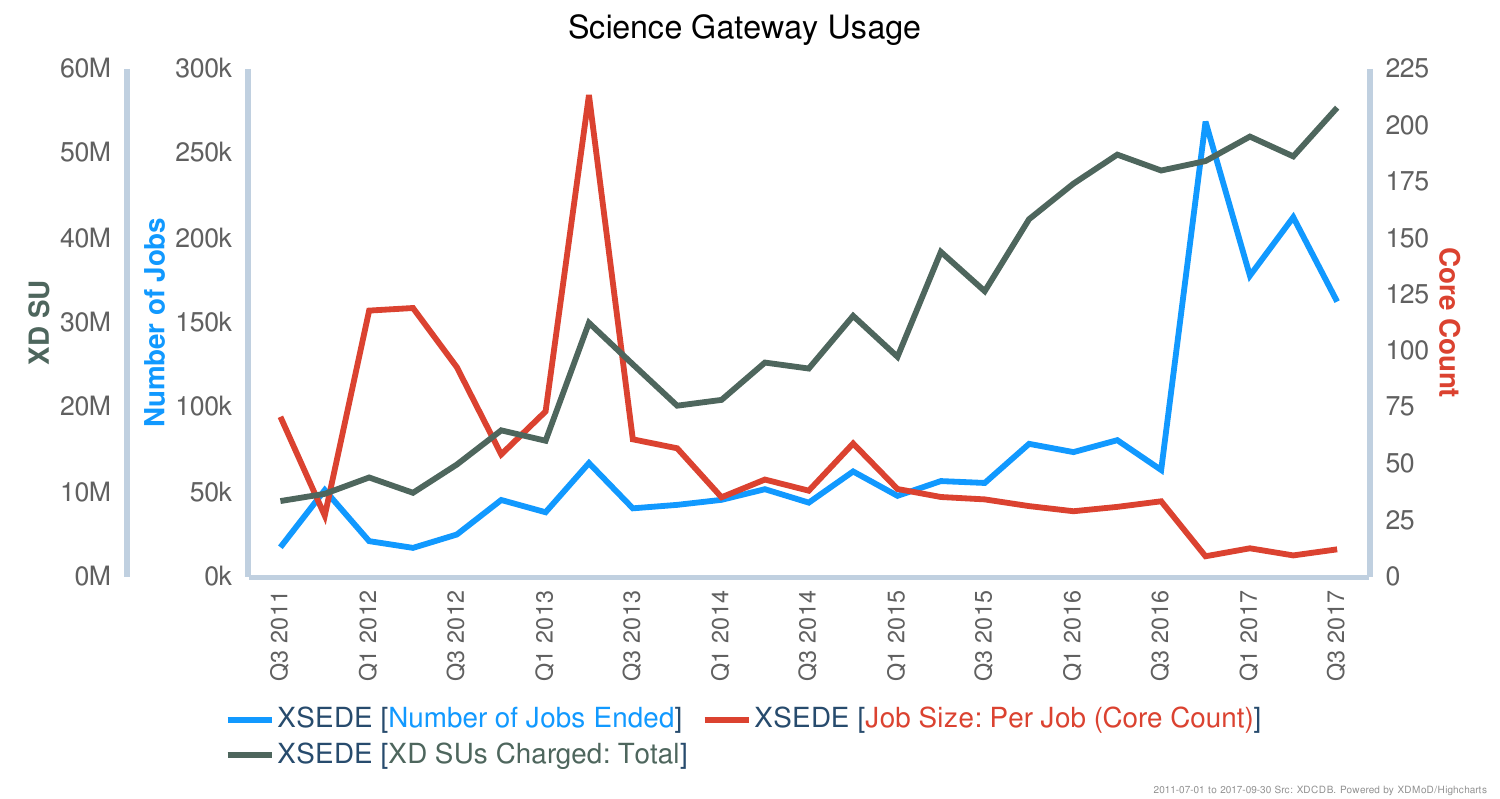}
\caption{\label{fig:Gateway_Usage_2011-07-01_to_2017-09-30.pdf}Science Gateway usage showing number of jobs submitted, core count, and XDSU charged. Note that the drop in core count in 2016-Q3 is due to the large number of single-core jobs submitted by the I-TASSER (Zhanglab) Gateway.}
\end{figure}

In addition to applying for an allocation on NSF Innovative HPC Program resources through XSEDE, researchers can utilize the growing number of Science Gateways to carry out their research.  Indeed, XSEDE usage by Science Gateways has shown a steady increase over the lifetime of XSEDE in the number of jobs run, XD SUs charged, and the number of Gateways in use as shown in Figure \ref{fig:Gateway_Usage_2011-07-01_to_2017-09-30.pdf}. Since the start of XSEDE over 1.983M Gateway jobs have been run while the number of jobs run quarter quarter has increased more than 7-fold from 20,000 in Q3 of 2011 to 150,000 in Q3 of 2017 and the number of XD SUs consumed has increased at a similar rate from less than 10M in Q3 of 2011 to 55M in Q3 of 2017. It is interesting to note that the average job size run by Gateway users has been steadily decreasing since 2014 with a marked drop in 2016-Q3 mainly due to the large number of single-core jobs submitted by the I-TASSER (Zhanglab) Gateway. If we discount the effect of this Gateway, the average job size has shown only a slight decrease between 2014 and 2017.

\begin{figure}[H]
\centering
\includegraphics [width=0.9\textwidth]{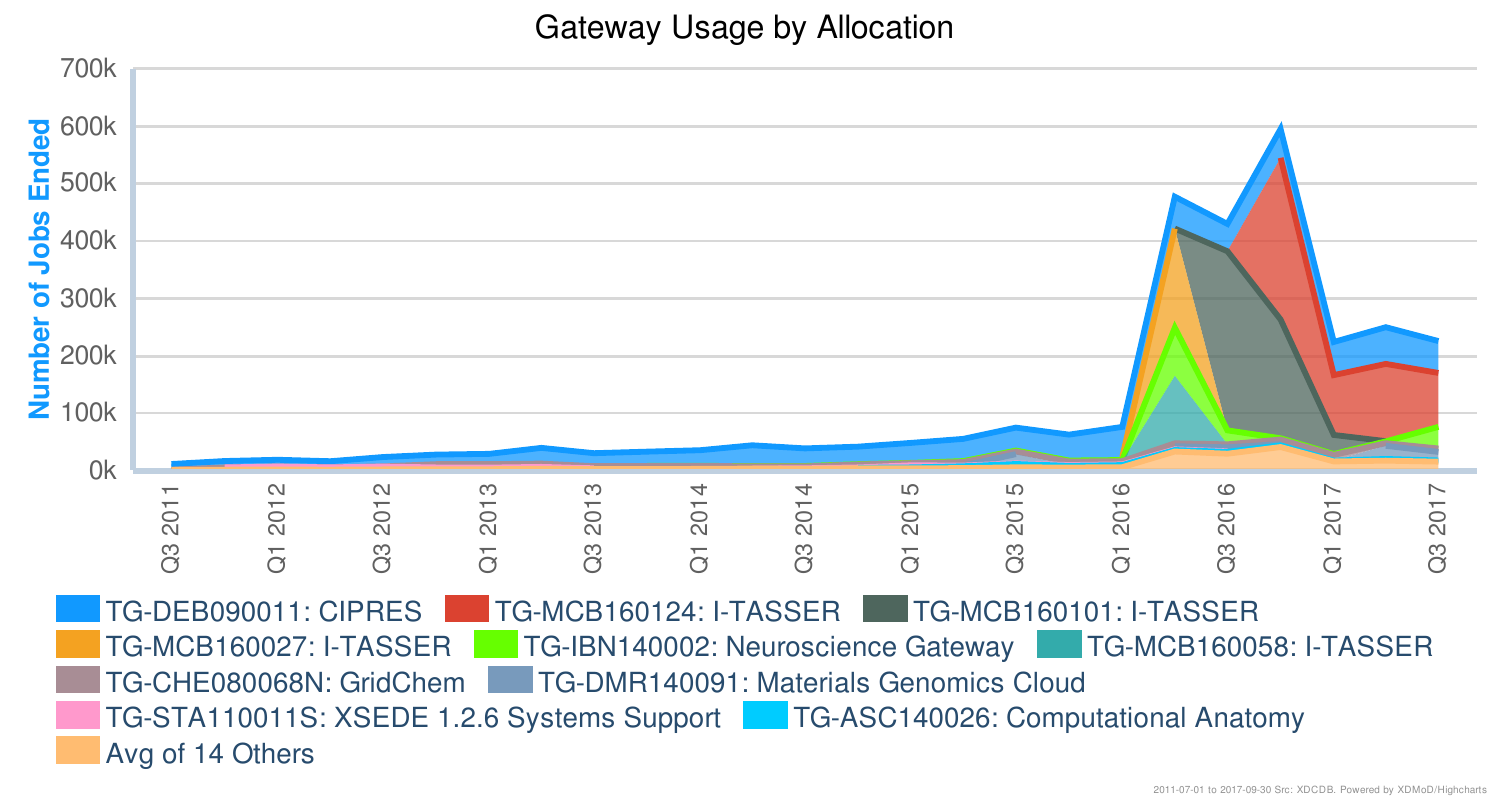}
\caption{\label{fig:Gateway_Usage_by_Allocation_2011-07-01_to_2017-09-30.pdf}Science Gateway usage by allocation for the period 2011-07 to 2017-09. Note the increase in number of jobs in March 2016 as additional Gateways come online.}
\end{figure}

If we look at the allocations known to be associated with Science Gateways (Figure \ref{fig:Gateway_Usage_by_Allocation_2011-07-01_to_2017-09-30.pdf}), CIPRES, which supports phylogenetic research, and GridChem, which provides access to a wide variety of computational chemistry programs, were the only active gateways from 2011-07-01 through October 2013. At the end of 2013, several other Science Gateways came online bringing the total number of active gateways to seven in February 2016 although the CIPRES Gateway accounted for most of the jobs run up to this point. Starting in March 2016 the I-TASSER and Neuroscience gateways came online and we see a marked increase in the number of jobs submitted. As of 2017-Q3 there were 14 active Science Gateways submitting a total of 975,000 jobs up to that point.

\begin{figure}[H]
\centering
\includegraphics{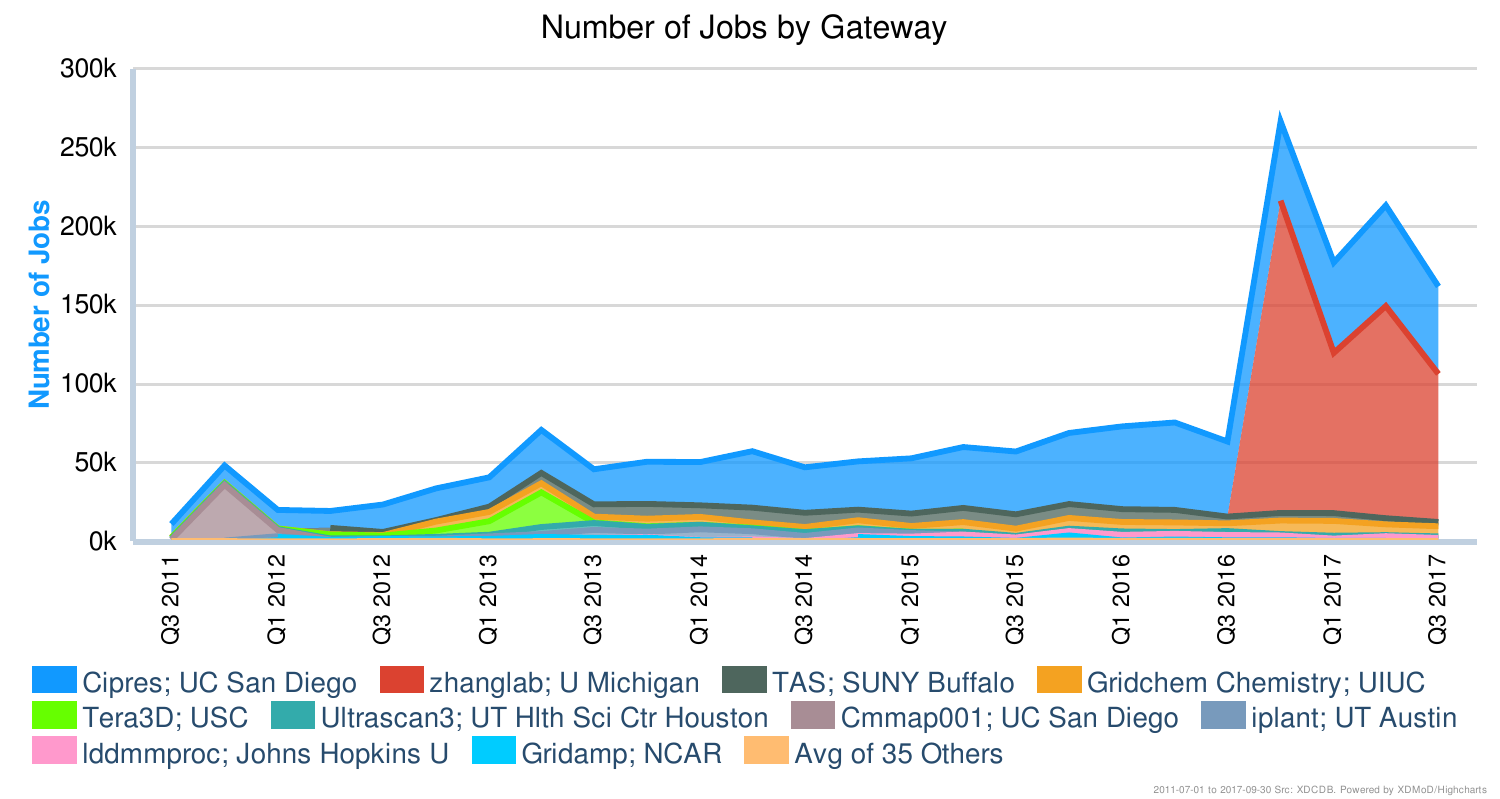}
\caption{\label{fig:Number_Jobs_Gateways}Number of jobs run by Gateways for the period 2011-07 to 2017-09.}
\end{figure}

\begin{figure}[H]
\centering
\includegraphics{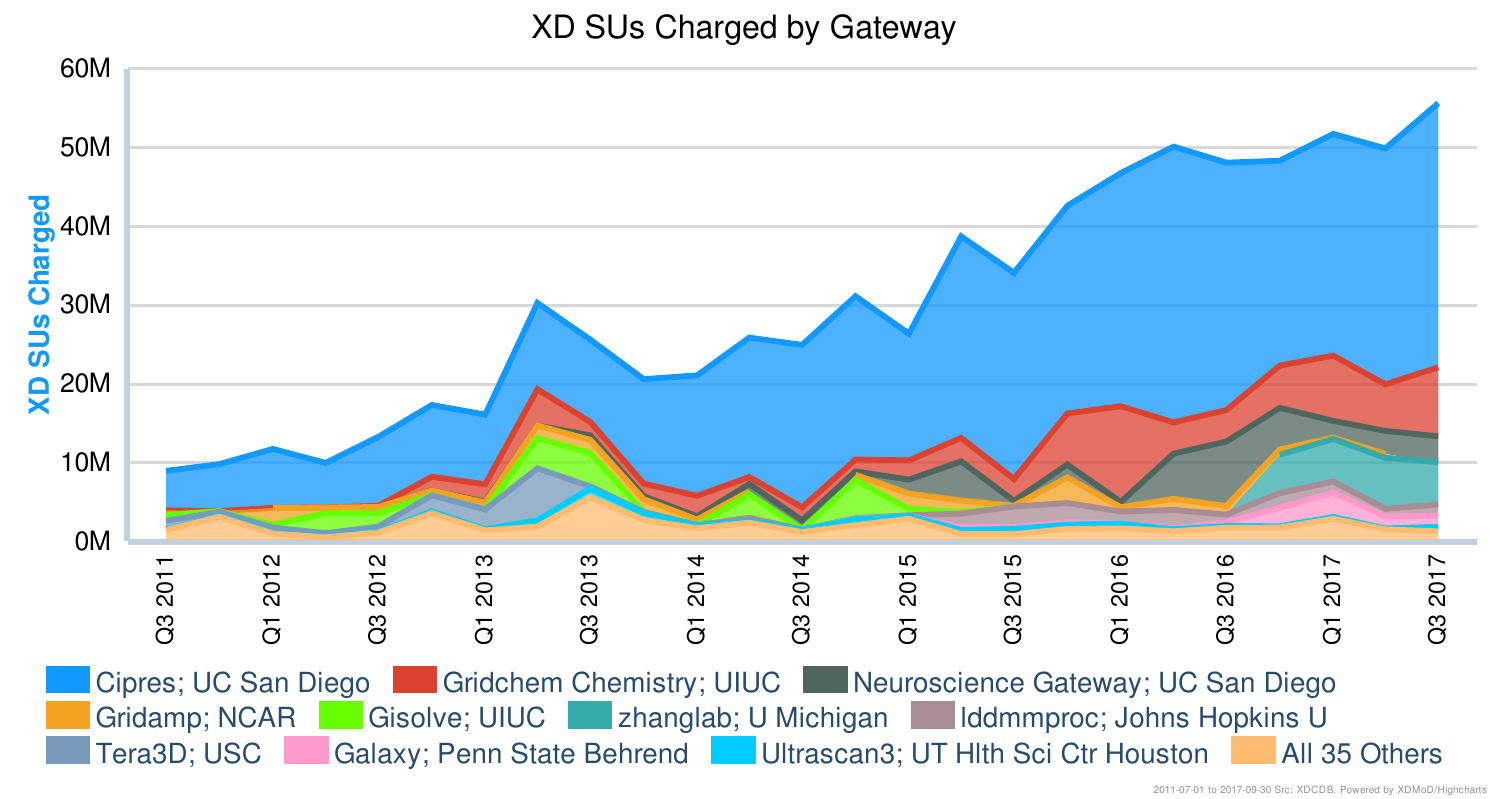}
\caption{\label{fig:XD_SUs_Gateways}XD SUs Charged by Gateway for the period 2011-07 to 2017-09.}
\end{figure}

Figure \ref{fig:Number_Jobs_Gateways} is similar to Figure \ref{fig:Gateway_Usage_by_Allocation_2011-07-01_to_2017-09-30.pdf} but shows the historical trend as a function of Science Gateway rather than individual allocations. Prior to 2016, CIPRES was the dominant gateway based on number of jobs.  However, since late 2016, the I-TASSER (ZhangLab) Gateway, which conducts protein folding predictions, is now the leading provider of Science Gateway jobs.  Third in this figure is the University Buffalo's TAS gateway which runs application kernels to provide provide quality of service metrics for XSEDE \cite{simakov2015application}.  Since the application kernels are designed to be computationally lightweight, it is reassuring to note that the TAS Gateway does not appear in Figure \ref{fig:XD_SUs_Gateways} which shows the top ten  Gateways in terms of XD SUs consumed. UC San Diego's CIPRES Gateway is the largest consumer of CPU cycles followed by Gridchem. 

\begin{figure}[H]
\centering
\includegraphics{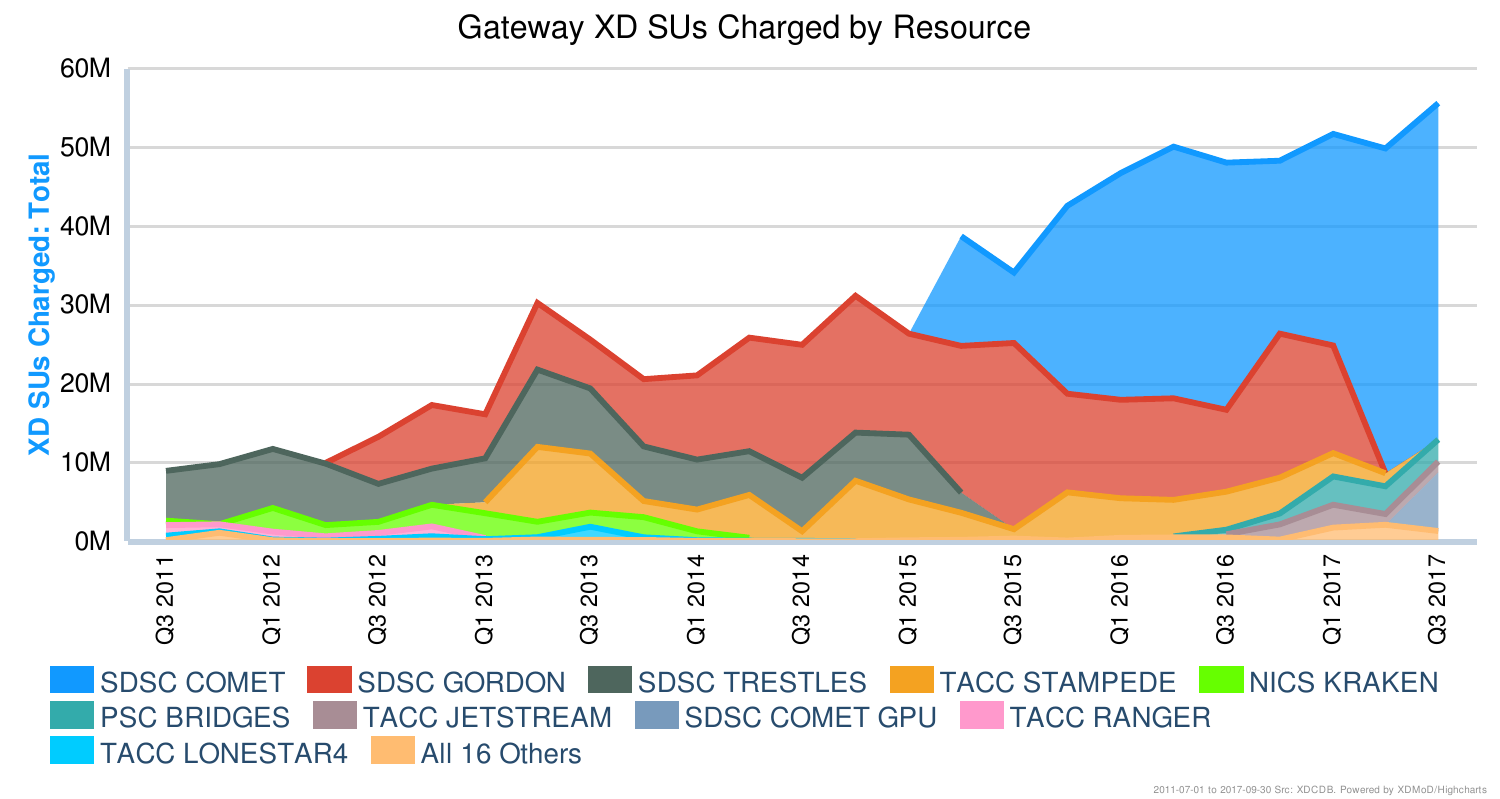}
\caption{\label{fig:Number_SUs_Gateways_by_Resource}Number of Gateway SUs Charged by Resource for the period 2011-07 to 2017-09.}
\end{figure}

\begin{figure}[H]
\centering
\includegraphics [width=0.9\textwidth]{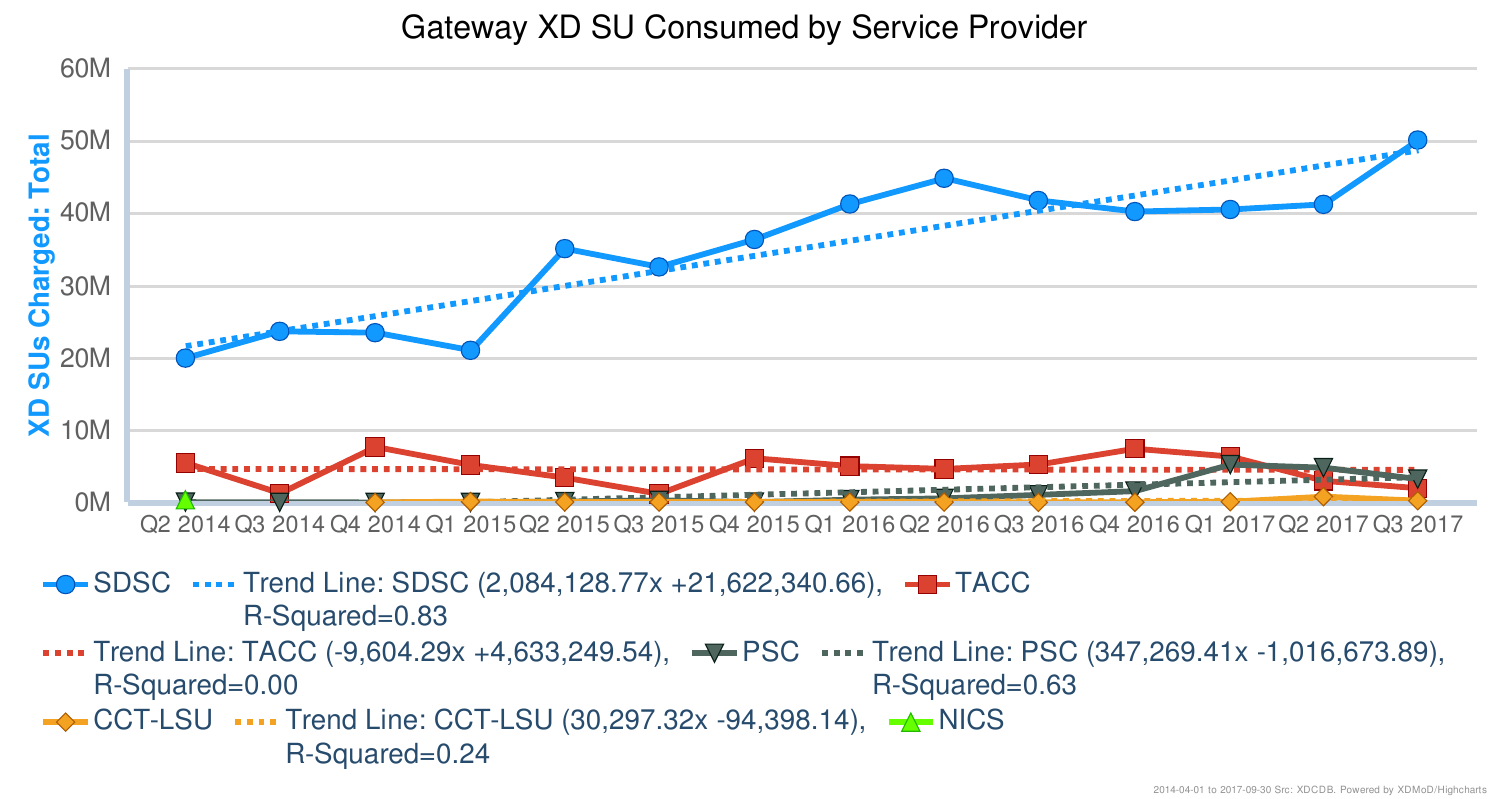}
\caption{\label{fig:figures/Gateway_XD_SU_Consumed_by_Service_Provider_2014-05-01_to_2017-09-30.pdf}XD SU consumed by Science Gateway jobs run at each XSEDE service provider between 2014-05 and 2017-09. Note the increase in XD SU provided by SDSC.}
\end{figure}

\begin{figure}[H]
\centering
\includegraphics [width=0.9\textwidth]{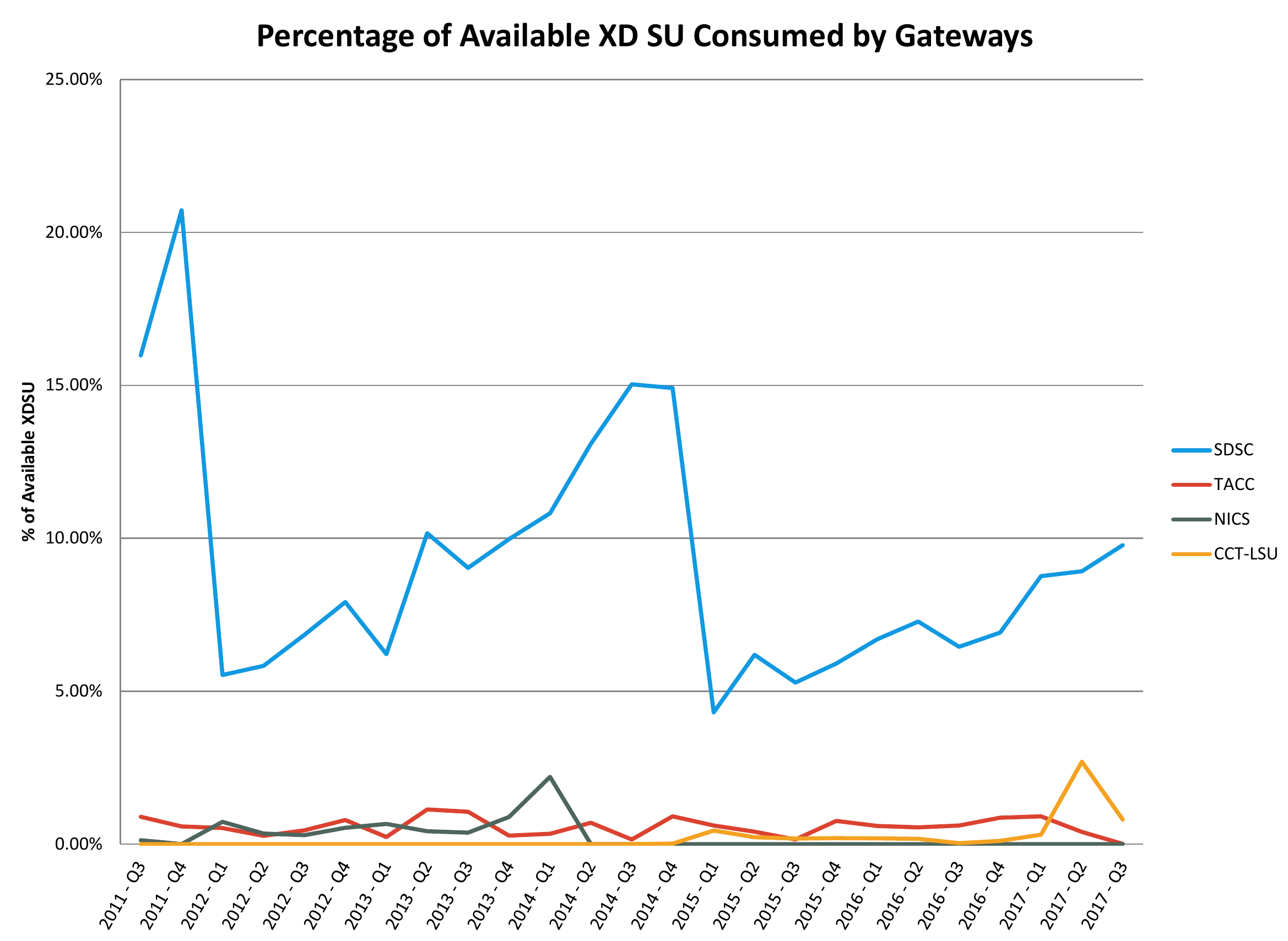}
\caption{\label{fig:xsede_gw_xdsu_as_pct_available}Percentage of the total available XD SU consumed by Science Gateway jobs at SDSC, TACC, CCT-LSU, and TACC for the period 2014-05 and 2017-09.}
\end{figure}

As shown in Figure \ref{fig:Number_SUs_Gateways_by_Resource} Science Gateways have submitted jobs to all of the major XSEDE compute resources over the time period of this report with the exception of Keeneland. However, SDSC's HPC systems have been the primary resources utilized by Science Gateways. Indeed, as the trend lines in Figure \ref{fig:figures/Gateway_XD_SU_Consumed_by_Service_Provider_2014-05-01_to_2017-09-30.pdf} show, Gateway usage on SDSC resources is growing steadily while usage is actually decreasing on TACC resources with the retirement of \stampede{} and increasing only slightly at PSC and LSU. Out of these, SDSC Gateway usage has averaged 7.6\% of total available XD SUs while usage at each of the other service providers averages less than 0.5\% as shown in Figure \ref{fig:xsede_gw_xdsu_as_pct_available}.

\begin{figure}[H]
\centering
\includegraphics{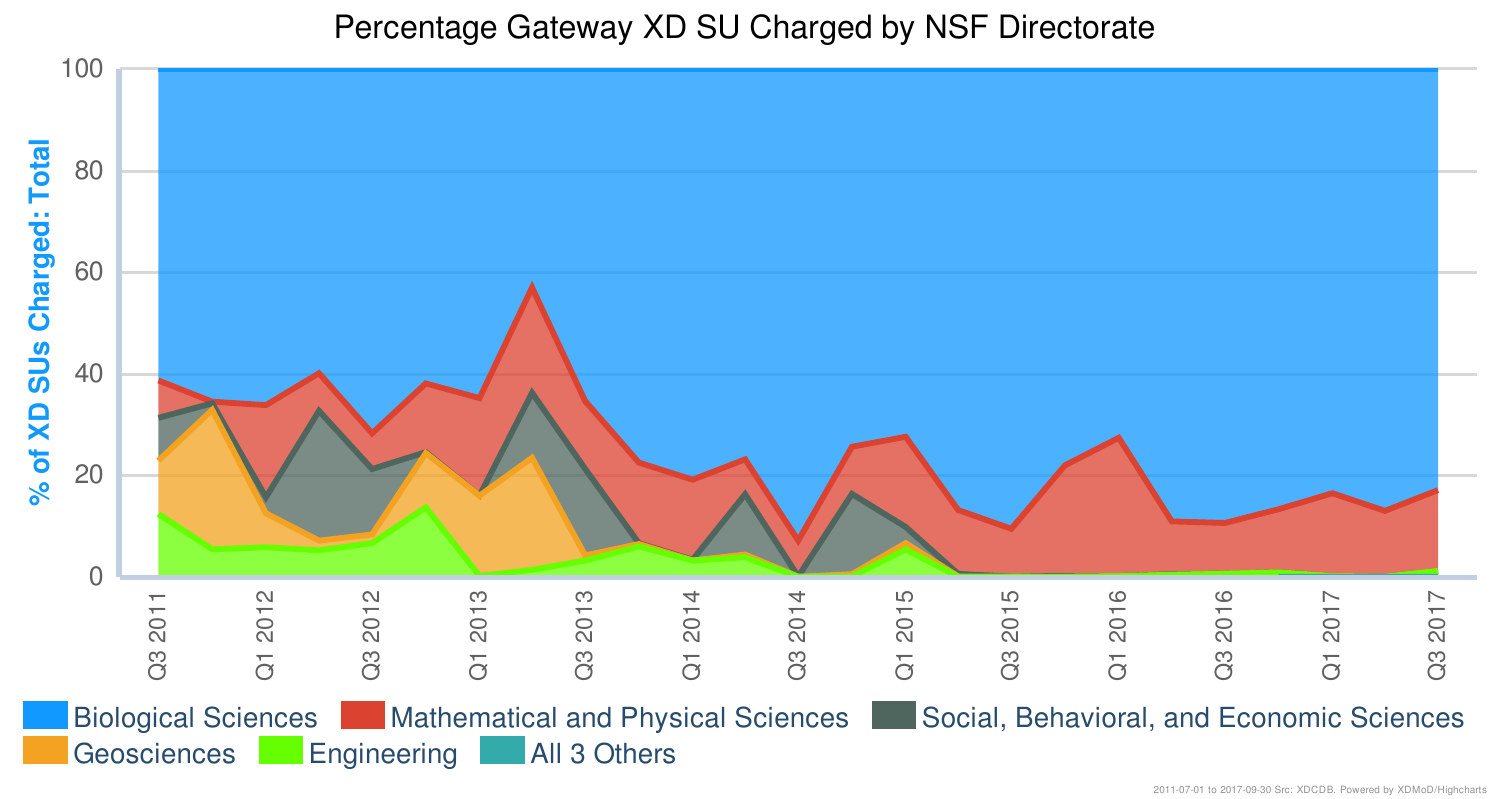}
\caption{\label{fig:Percentage_Gateway_XD_SU_Charged_by_NSF_Directorate_2011-07-01_to_2017-09-30}Number of Gateway SUs Charged by Resource for the period 2011-07 to 2017-09.}
\end{figure}

Individual Science Gateways are typically designed to support a particular discipline such as Population Biology (Cipres Gateway), Chemistry (Gridchem Gateway), or Neuroscience (Neuroscience Gateway) with many XSEDE Science Gateways servicing users in the Biological Sciences.  From 2011 - 2014 XSEDE Science Gateways supported research in five NSF Directorates (Biological, Mathematical \& Physical, Social, Behavioral \& Economic, Geosciences, and Engineering) with Biological Sciences accounting for over 60\% of the total XD SU consumed. However, as shown in Figure \ref{fig:Percentage_Gateway_XD_SU_Charged_by_NSF_Directorate_2011-07-01_to_2017-09-30} between 2015 and 2017 the number of directorates serviced has declined with the Biological and Mathematical \& Physical Sciences now accounting for almost 98\% of all Gateway XD SUs consumed.

\subsection{Gateways Users}

\begin{figure}[H]
\centering
\includegraphics[width=\textwidth]{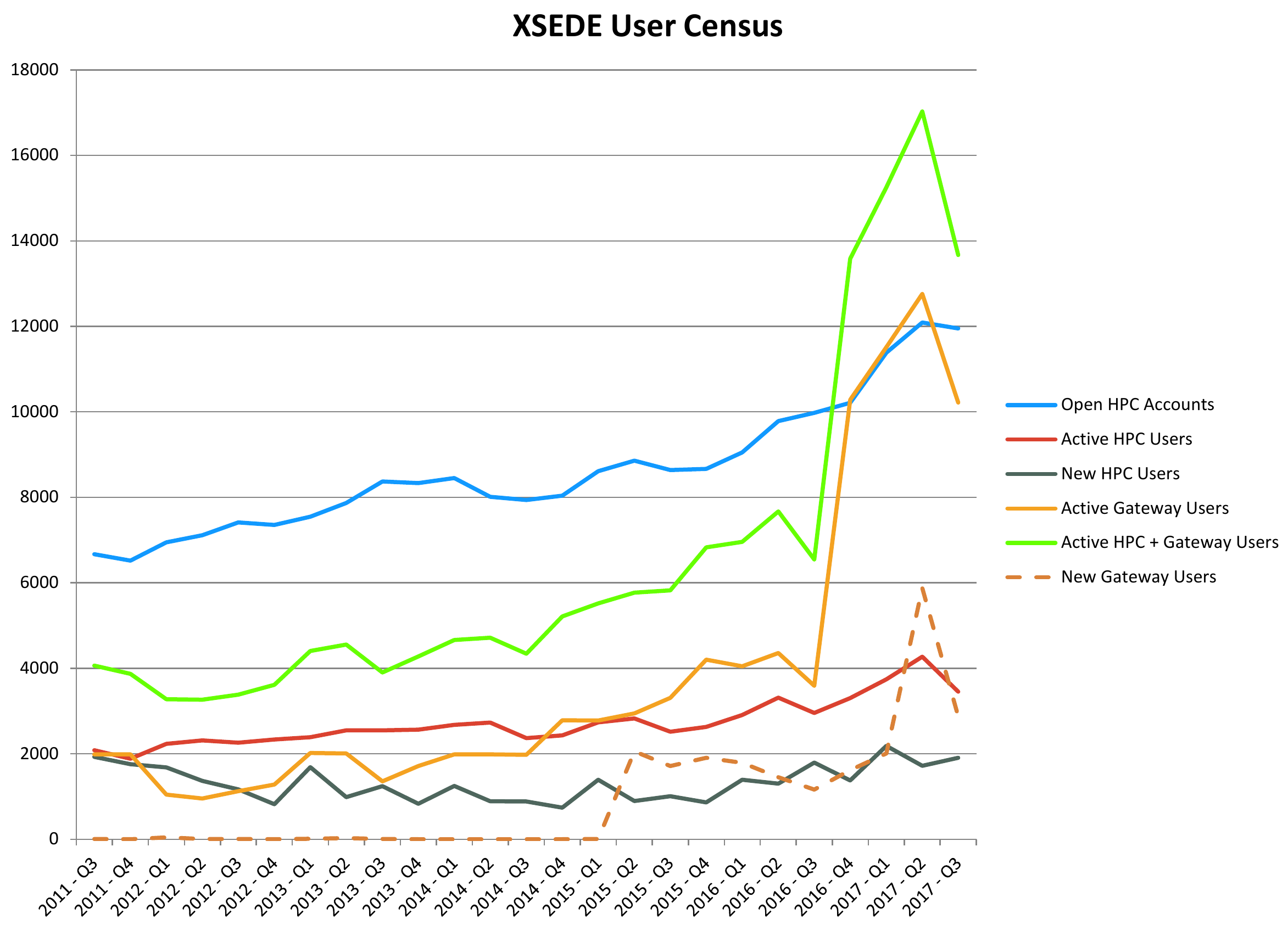}
\caption{\label{fig:xsede_user_census_updated}XSEDE User Census showing the number of open HPC user accounts, active HPC users, HPC users newly added to allocations, active Science Gateway users, and new Science Gateway users. The dashed line for new Gateway users represents a lower bound due to incomplete data reporting prior to 2015-Q2.}
\end{figure}

A census comparing XSEDE HPC and Science Gateway users, compiled from the data listed in Table \ref{table:xsede_user_census}, is shown in Figure \ref{fig:xsede_user_census_updated}. The data shows that, while the overall XD SU consumed by Gateway users is only \~3\% of the total XSEDE XD SUs, the number of active Gateway users has been steadily increasing at a faster rate than active HPC users. In 2015, the number of active Gateway users surpassed the number of active XSEDE HPC users and has been steadily rising.

The number of HPC users actively running jobs each quarter has steadily increased from 2,075 at the start of XSEDE to 3,432 in 2017-Q3. However, the number of new HPC users added to allocations each quarter dropped between 2012 and 2016, until finally regaining original 2011 levels in 2017. During this period it is significant to note that the number of active Science Gateway users has grown steadily - increasing from 1,983 to 3,590 users - and has shown significant growth between 2016-Q3 and 2017-Q3 topping out at 12,757 users in 2017-Q2. This increase is mainly a result of the I-TASSER Gateway with over 6,000 active users in 2016-Q4 and over 7,000 active users in 2017-Q1 and is also visible in the increased number of jobs ended shown in \ref{fig:Gateway_Usage_2011-07-01_to_2017-09-30.pdf}.

Gateway users are identified by a unique user name, typically an email address, that is unique to each Gateway and is specified by the gateway upon job submission. The same user name on two different Gateways could potentially refer to two different people. It is important to note that data for newly created Gateway users was not reliably collected prior to 2015-Q2 as explained in Appendix \ref{appendix:data-science-gateways}, so the new Gateway user values presented here typically represent a lower bound.

\begin{table}[H]
\centering
\caption{Data sources for the XSEDE User Census shown in Figure \ref{fig:xsede_user_census_updated}}
\label{table:xsede_user_census}
\begin{tabular}{ *3l }
\toprule
Metric & Source & Definition \\ \midrule
Open HPC Accounts & XDCDB & The number of HPC accounts (i.e., people) \\
& & that have access to XSEDE resources. They \\
& & may or may not have run any jobs. \\
Active HPC Accounts & XDCDB & The number of HPC accounts that have run \\
& & at least one job during the specified period.  \\
New HPC Accounts & XDCDB & Newly created HPC accounts \\
Active Gateway Users & XDCDB/PI polling & Science gateway users who have run at least \\
& & one XSEDE job via a Gateway in the specified period. \\
Active HPC + & XDCDB/PI polling & Sum of Active HPC Users and Active Gateway users. \\
Gateway Users & & \\
New Gateway Users & XDCDB & Unique \textbf{new} Gateway users created in the specified period. \\
& & Note that this is a lower bound as gateway reporting is \\
& & incomplete prior to 2015-Q2 as discussed in Appendix \ref{appendix:data-science-gateways} \\
\bottomrule
\end{tabular}
\end{table}

In an attempt to determine if Gateway users eventually converted to XSEDE users with their own allocations, we attempted to cross-reference XSEDE users and Gateway users by using their email address, limiting our analysis to users who ran more than 10 XSEDE jobs. As shown in Table \ref{table:gateway_to_xsede_user_conversion}, a very limited number of users started off using a Science Gateway and then later ran jobs on XSEDE. Indeed, only six users started as Gateway users and went on to run more than 100 XSEDE jobs.

\begin{table}[H]
\centering
\caption{The number of gateway users (anonymized) who later ran jobs on XSEDE.}
\label{table:gateway_to_xsede_user_conversion}
\begin{tabular}{ llrlrl }
\toprule                 
Gateway  & Gateway User             & \# GW Jobs & First GW Job & \# XSEDE Jobs & First XSEDE Job\\ \midrule
Cipres   & user01@uta.edu      & 43 & 2015-06-17 & 3841 & 2017-03-02\\
Cipres   & user02@gmail.com    & 2  & 2015-10-04 & 47   & 2017-04-12\\
Cipres   & user03@uta.edu      & 3  & 2015-04-28 & 273  & 2017-03-09\\
Cipres   & user04@gmail.com    & 1  & 2016-03-10 & 33   & 2017-06-26\\
Cipres   & user05@indiana.edu  & 17 & 2016-10-17 & 904  & 2017-02-09\\
Cipres   & user07@berkeley.edu & 63 & 2015-04-29 & 48   & 2015-12-08\\
Cipres   & user08@uni-koeln.de & 5  & 2015-09-28 & 28   & 2017-03-03\\
Cipres   & user09@berkeley.edu & 24 & 2016-02-17 & 17   & 2016-06-07\\
Cipres   & user10@upr.edu      & 14 & 2016-08-31 & 27   & 2017-01-27\\
Cipres   & user11@uark.edu     & 1  & 2016-03-04 & 5243 & 2016-09-14\\
Galaxy   & user12@gmail.com    & 26 & 2016-09-10 & 17   & 2017-01-05\\
Galaxy   & user13@upr.edu      & 1  & 2015-10-11 & 25   & 2016-09-28\\
Galaxy   & user14@sdstate.edu  & 2  & 2015-10-17 & 20   & 2016-10-05\\
Galaxy   & user15@umsl.edu     & 6  & 2015-12-24 & 2642 & 2016-08-15\\
I-TASSER & user16@syr.edu      & 3  & 2017-04-17 & 112  & 2017-07-06\\
I-TASSER & user18@hawk.iit.edu & 8  & 2017-05-14 & 23   & 2017-06-17\\
I-TASSER & user18@fsu.edu      & 1  & 2017-05-18 & 20   & 2017-08-10\\
\bottomrule
\end{tabular}
\end{table}

\subsection{Gateways Job Mix and Characteristics}

\begin{figure}[H]
\centering
\includegraphics{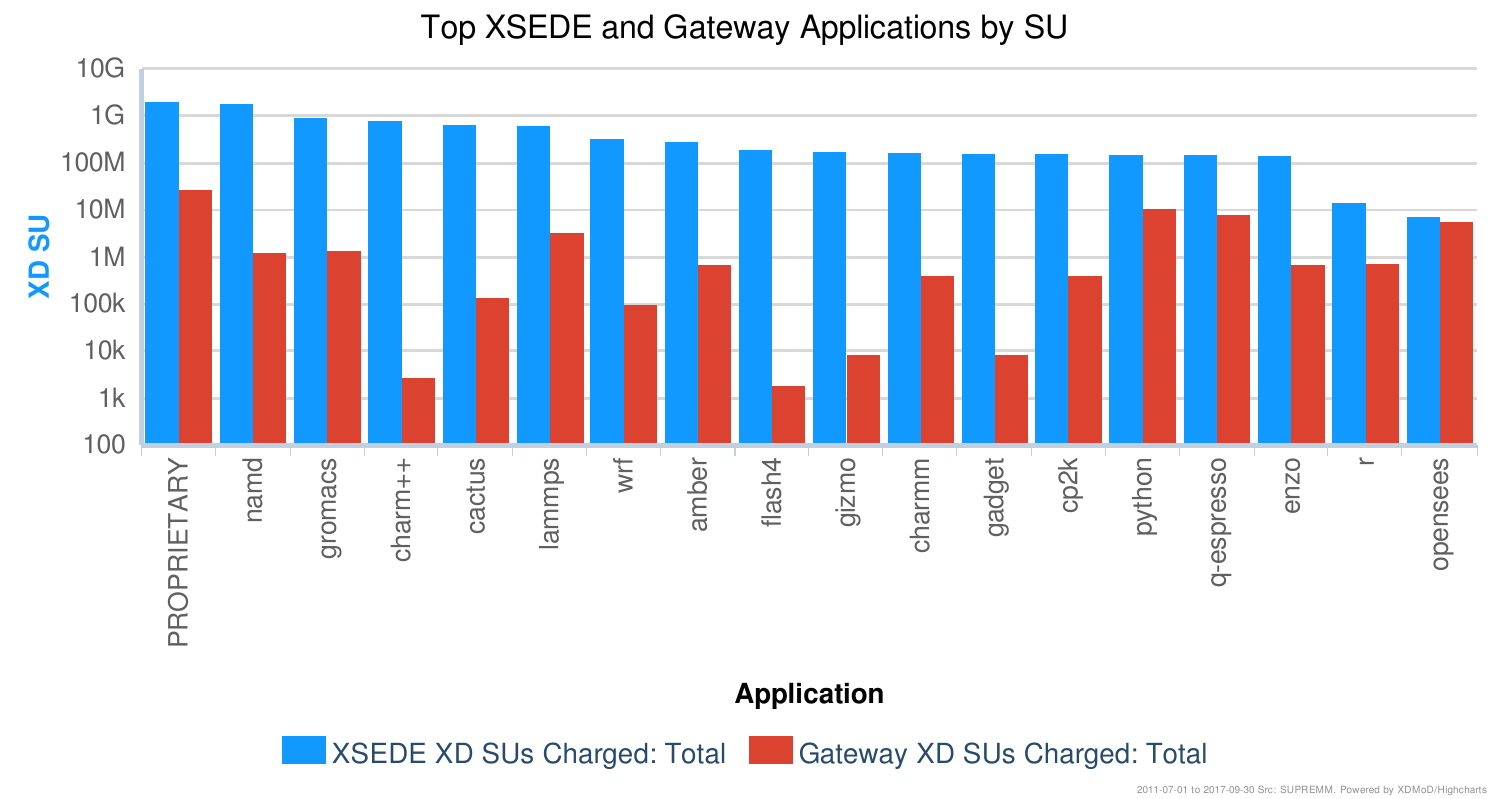}
\caption{\label{fig:TopXSEDEandGatewayApplicationsbySU.pdf}Number of SU used by the top applications run on XSEDE resources compared to those run specifically via Science Gateways for the period 2011-07 to 2017-09. The set of applications is the union of the top 12 applications by SU for each category. Note that the XSEDE values include all XSEDE jobs, including those submitted via Science Gateways.} 
\end{figure}

\begin{figure}[H]
\centering
\subfloat[\label{fig:aps}]{%
\includegraphics[width=0.5\textwidth]{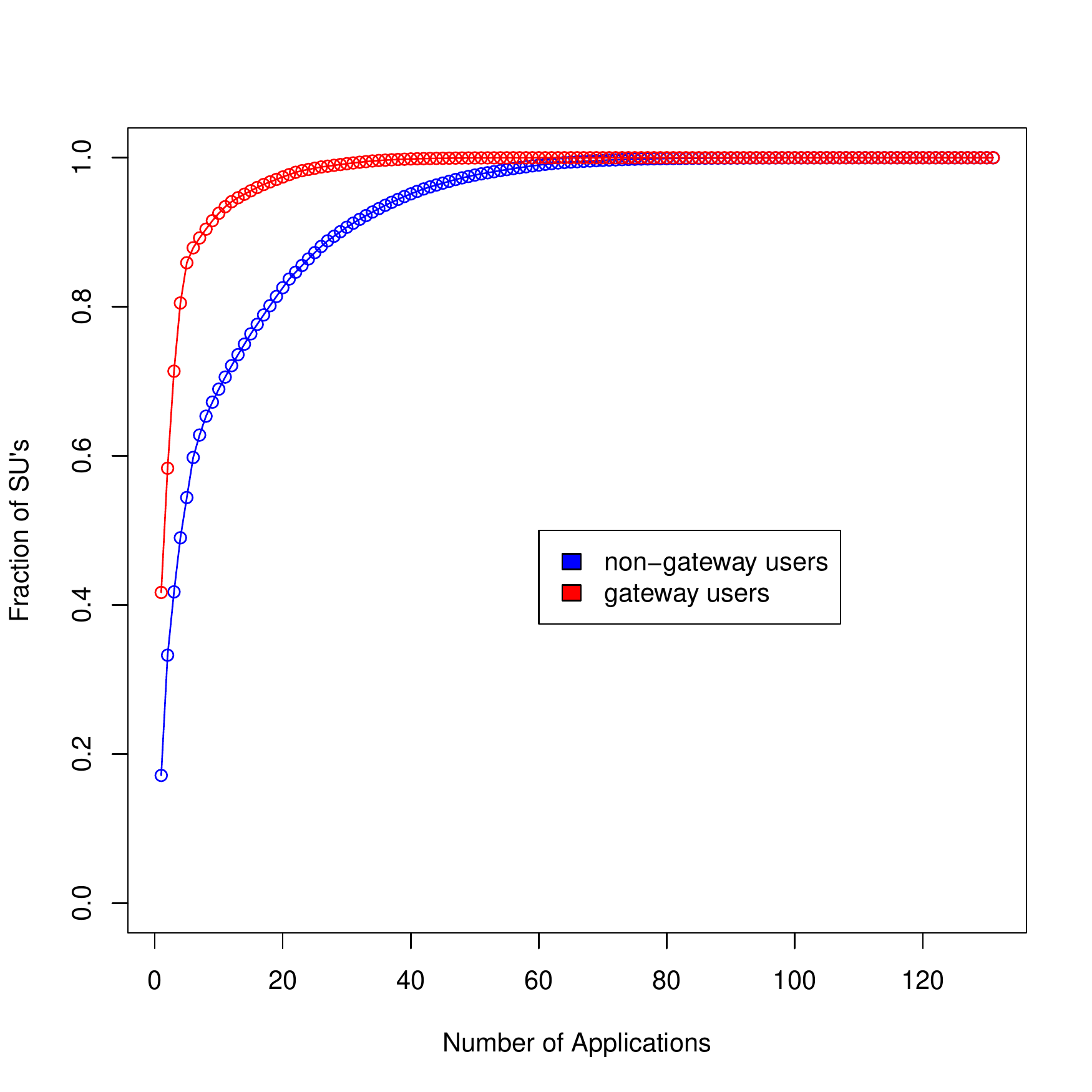}
}
\subfloat[\label{fig:fos}]{%
\includegraphics[width=0.5\textwidth]{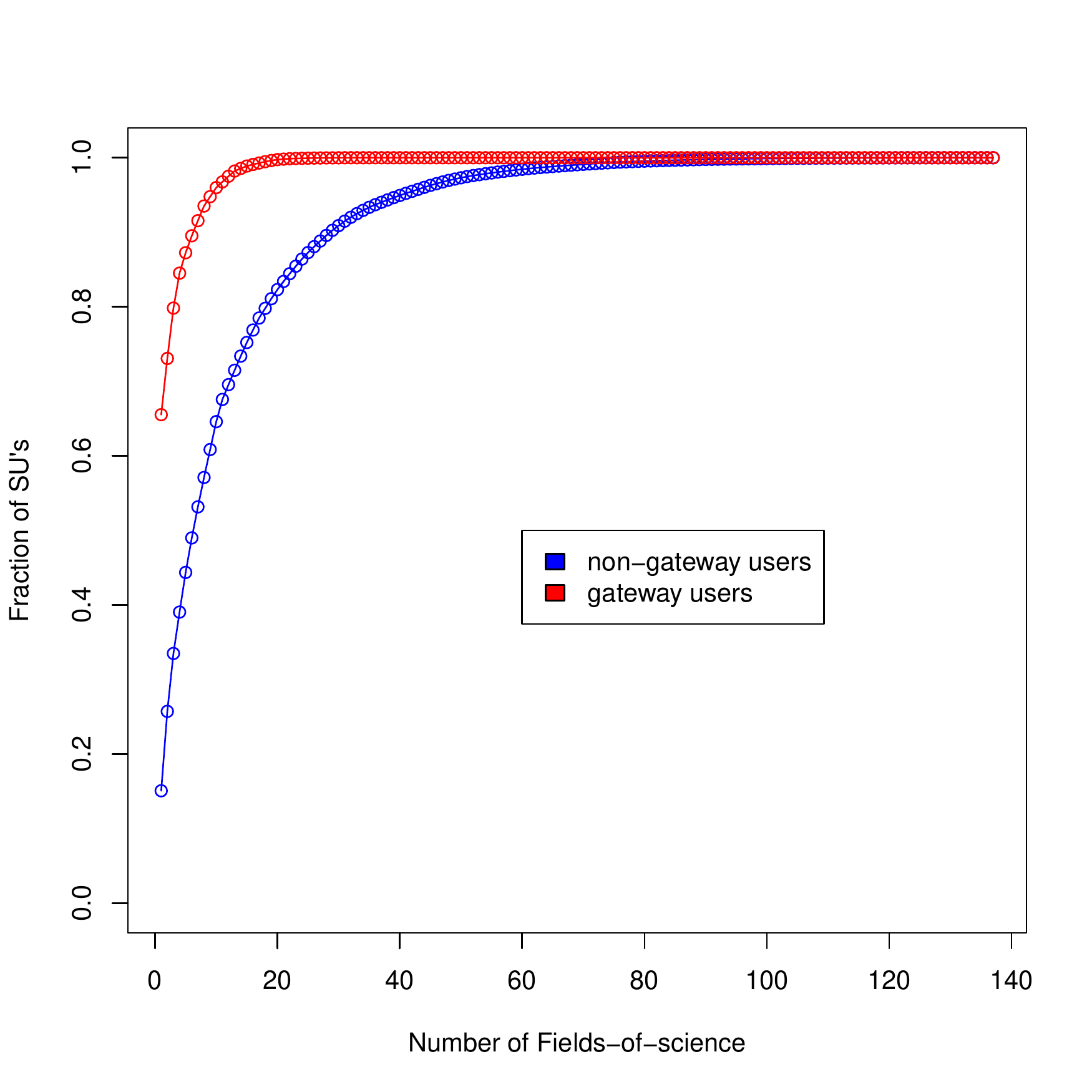}
}
\caption{Cumulative usage comparison of gateway and non-gateway usage by (a) application (b) fields-of-science for the period 2011-07 to 2017-09.  Both by the number of applications used and the number of fields-of-science spanned the gateway usage is much more limited than the non-gateway usage.}
\label{fig:aps and fos}
\end{figure}

In order to determine whether or not Gateway users utilize HPC resources differently than traditional XSEDE HPC users, either in the types of applications used or the properties of jobs run, we examined the mix of applications submitted by these two categories of users. The top 12 applications by XD SUs consumed by traditional HPC and Science Gateway users were selected and are shown in Figure \ref{fig:TopXSEDEandGatewayApplicationsbySU.pdf};  XD SU usage is on a log scale. Note that this includes only jobs for which we were able to identify the application that was run, which is 55\% of the total XD SU for XSEDE and 19\% of the total XD SU for Science Gateways. The top applications for Gateway users clearly overlaps those of traditional HPC users however, looking at the total SU consumed for XSEDE and Gateways, 99\% of the XSEDE HPC XD SUs are consumed by 64 categorized applications while 99\% of the Gateway XD SUs are consumed by 27 categorized applications.  Figure \ref{fig:aps} shows a cumulative plot of the usage of applications for Gateway and non-Gateway users.  Note that the Gateway usage is much more limited on the number of applications employed.  Since gateways typically target a particular field of science, a narrower application scope is expected. Note that the XSEDE values include Gateway usage. Similarly, Figure \ref{fig:fos} shows that the Gateway usage encompasses many fewer fields-of-science.


\begin{figure}[H]
\centering
\includegraphics[width=\textwidth]{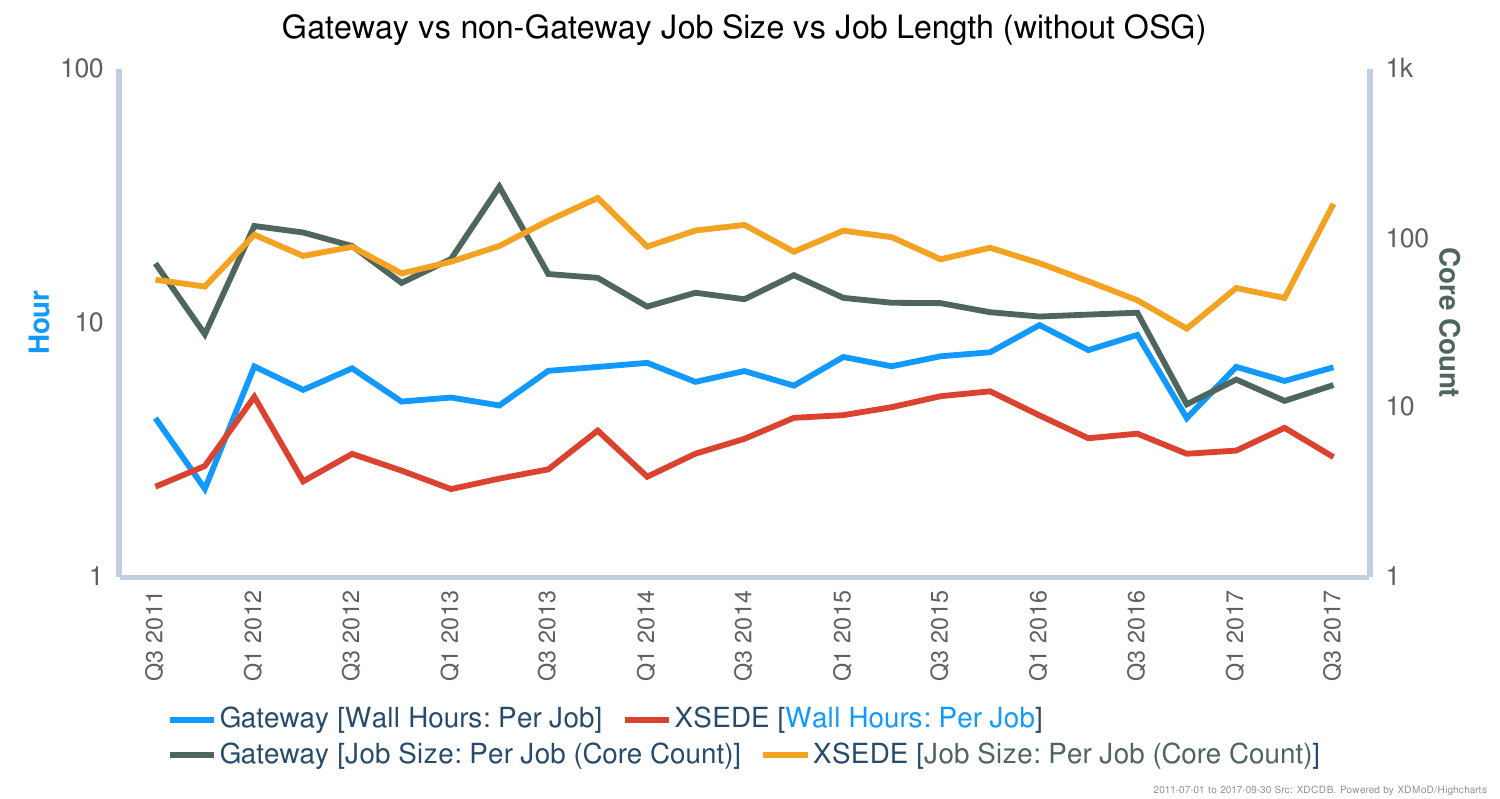}
\caption{\label{fig:Gateway_vs_non-Gateway_Job_Size_vs_Job_Length__without_OSG__2011-07-01_to_2017-09-30}Comparison of Gateway vs non-Gateway Job Size by Core Count and Job Length in Wall Hours for the period 2011-07 - 2017-09 in log scale. Note that \osg{} jobs are not included.}
\end{figure}

In addition to the applications utilized by Science Gateways users, we examined the average job size by Core Count Per Job and the average run time per job. Note that we have not included jobs run on \osg{} in this analysis as they are exclusively single core jobs at a rate of 30M jobs per year which causes a steep drop in the average job size starting in 2015. We have also excluded jobs run on \jetstream{} as many of these are very long running, low core-count jobs. Figure \ref{fig:Gateway_vs_non-Gateway_Job_Size_vs_Job_Length__without_OSG__2011-07-01_to_2017-09-30} (log scale) shows that the average size of non-Gateway jobs (yellow line) is roughly twice that of Gateway jobs (black line) but the gateway jobs tend to run for longer periods of time (blue line vs red line), roughly a factor of two longer through most of the time span.  Starting in Q3 of 2016 when the Zanglab gateway started both the gateway per job core count and the run time dropped sharply. 

\subsection{Summary: Science Gateways}
Science Gateway usage has increased steadily over the lifetime of the XSEDE program, with a 5-fold increase in the number of jobs run in 2017 relative to 2011.  Today, the Biological (82\%) and Mathematical and Physical Sciences (16\%) directorates account for almost 98\% of Gateway XD SUs consumed, up from 70\% in 2011.  While Science Gateways consume only about 3\% of the total XD SUs, the number of active Gateway users is growing more rapidly than the active HPC users, surpassing the number of active HPC users in 2015. The Gateway job mix in terms of applications run and fields of science is more narrow than that of HPC users, as expected due to the targeted nature of Science Gateways. The average size of non-Gateway jobs is roughly twice that of Gateway jobs.  However,  gateway jobs tend to run longer than the non-gateway jobs.

\newpage

\section{Job Submission Patterns \& Over-Subscription}\label{sec:patterns}
\textit{
Goals Addressed in Section
\begin{enumerate}\setcounter{enumi}{8}
\item Are jobs constrained by resource policy limits such as queue length, user limits or node sharing?
\begin{itemize}
\item How does this vary by resource?
\item Do these limits affect the analysis?
\end{itemize}
\item Are there differences in the job mixes among the resources and if so, how does this impact job throughput? 
\begin{itemize}
\item What is the relative proportion of these jobs between systems?
\item Is this the result of allocation decisions, or something else we can determine? 
\end{itemize}
\item How do wait times, throughput and queue length vary among the resources?  Has this changed over time?
\item What is the run-time over-subscription? What is the breakdown by resource and resource type?
\end{enumerate}
}
\subsection{Over-subscription}

\begin{figure}[H]
\centering
\includegraphics[width=1.0\textwidth]{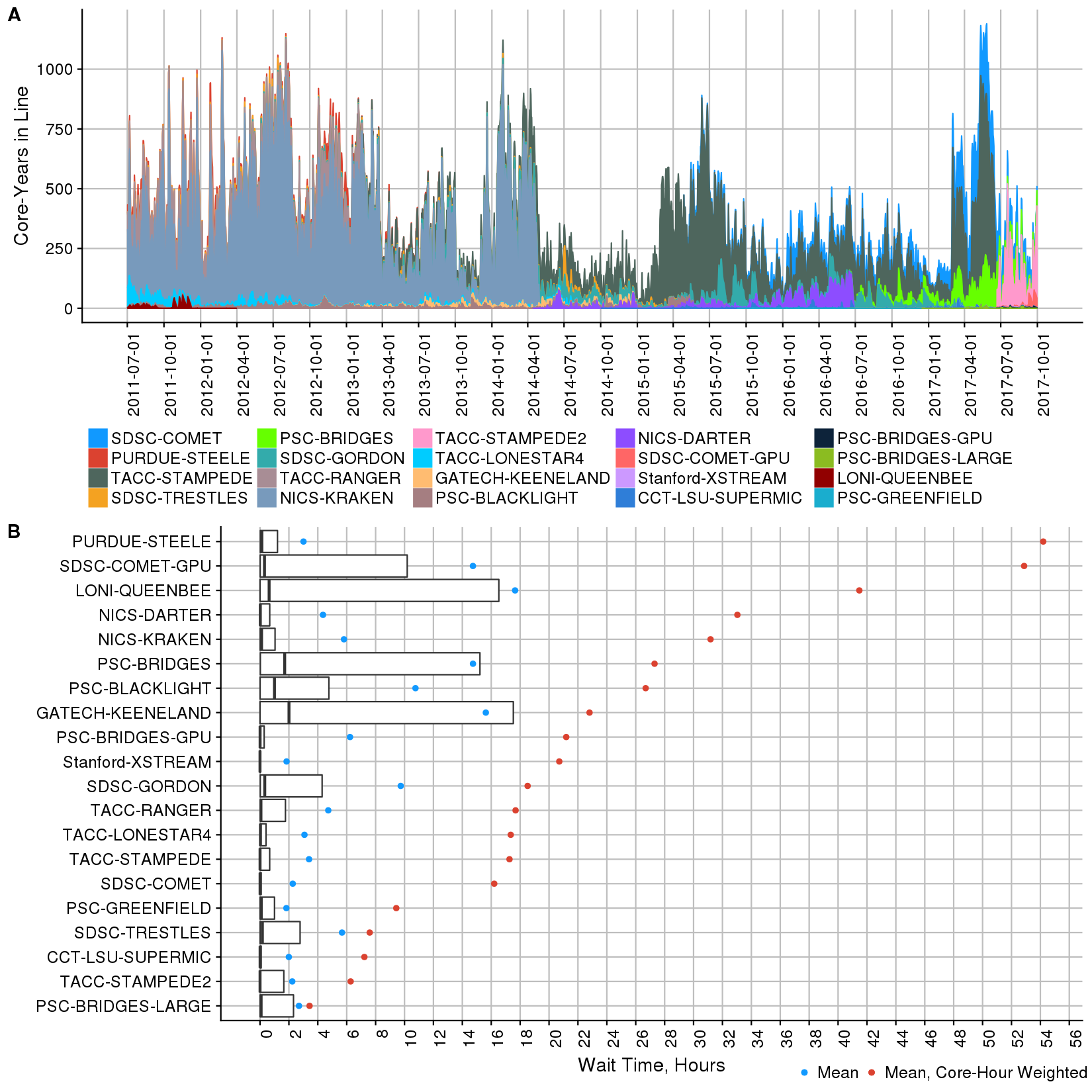} 
\caption{\label{fig:corehours_in_line_waiting_time_hpc_dic} \textbf{A} -Cumulative plot of core years of all jobs waiting execution across all resources over time. \textbf{B} - Wait time of queued jobs: horizontal box-plot shows quartile (left side), median (middle notch) and third quartile (right side) values.  Also included is the mean wait time (blue dots) and the mean wait time weighted by core hours (red dots).   Time period is over the production lifetime of each resource}
\end{figure}

In this subsection we address the issue of over subscription in which the total number of requests for resources exceeds the capacity of the existing resources and therefore user jobs are queued waiting for other jobs to end. Figure 
\ref{fig:corehours_in_line_waiting_time_hpc_dic} characterizes the backlog of jobs by looking at the time history of queued jobs in terms of core years (Figure \ref{fig:corehours_in_line_waiting_time_hpc_dic}.A) as well as analyzing the wait times (mean, median, etc) of the queued jobs broken out by resource (Figure \ref{fig:corehours_in_line_waiting_time_hpc_dic}.B).  While the backlog of jobs shows a substantial fluctuation over time, the demand for resources clearly outstrips capacity as is readily evident in Figure \ref{fig:corehours_in_line_waiting_time_hpc_dic}.A. In terms of wait times, the horizontal box plots in Figure \ref{fig:corehours_in_line_waiting_time_hpc_dic}.B shows that there is wide variation in the wait times among the resources.  Particularly noteworthy is the large difference on most resources between the mean wait time (blue dots) and the mean wait time weighted by job core hours (red dots), indicating that large jobs have much longer wait times then small jobs.

\begin{figure}[H]
\centering
\includegraphics[width=1.0\textwidth]{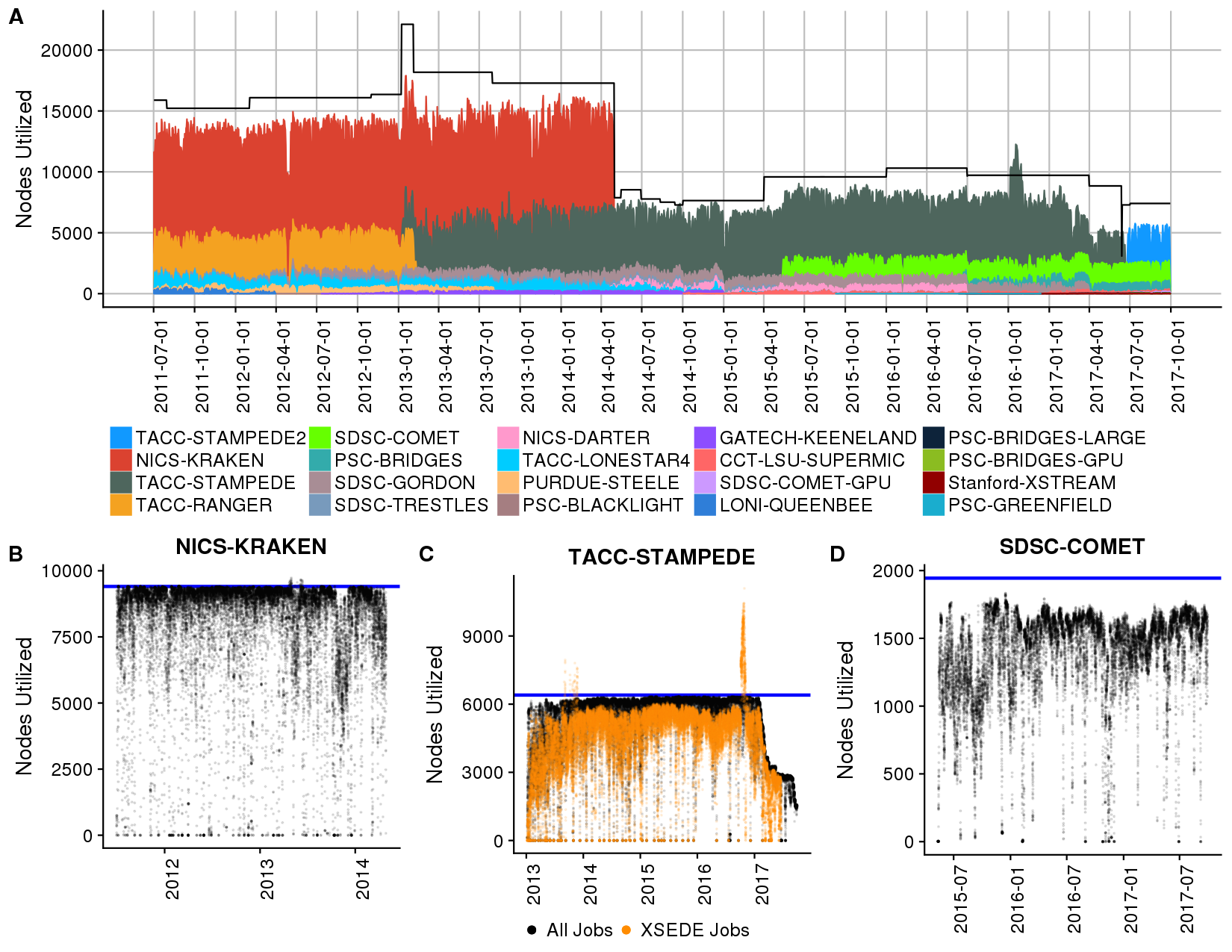} 
\caption{\label{fig:nodes_utilized_hpc_dic} \textbf{A} - Cumulative utilization of NSF Innovative HPC Program resources. Solid black line shows total nodes available to XSEDE program. The spike for \stampede{} around November, 2016 is most likely due to the addition of early \stampedetwo{} nodes to \stampede{}. Node utilization for \kraken{} (\textbf{B}), \stampede{} (\textbf{C}) and \comet{} (\textbf{D}). Blue line shows total number of nodes for the system. Black dots shows utilization from all jobs and orange from only XSEDE jobs, note that \kraken{} and \comet{} are 100\% dedicated to XSEDE allocated jobs}
\end{figure}

To address the question of how much larger NSF Innovative HPC Program resources would have to be in order to dramatically reduce wait times, we analyzed the utilization time history and queued job time history of these resources.  Figure \ref{fig:nodes_utilized_hpc_dic}.A shows the node-based utilization time history of each HPC resource along with the maximum nodes available (solid black line).   The utilization shown (80-90\%) is typical of HPC systems that support large parallel jobs in which the job scheduling software holds nodes idle while waiting for a sufficient number of nodes to become available to allow a job to run. Most job schedulers backfill jobs whenever possible in order to maximize resource utilization while at the same time facilitating the throughput of large parallel jobs.  Figure \ref{fig:nodes_utilized_hpc_dic}.B shows the node utilization for 3 specific resources, namely \kraken{}, \stampede{}, and \comet{}.  Both \kraken{} and \comet{} are entirely dedicated to XSEDE jobs (black dots), while \stampede{} is 90\% allocated to XSEDE (orange dots).  

\begin{figure}[H]
\centering
\includegraphics[width=1.0\textwidth]{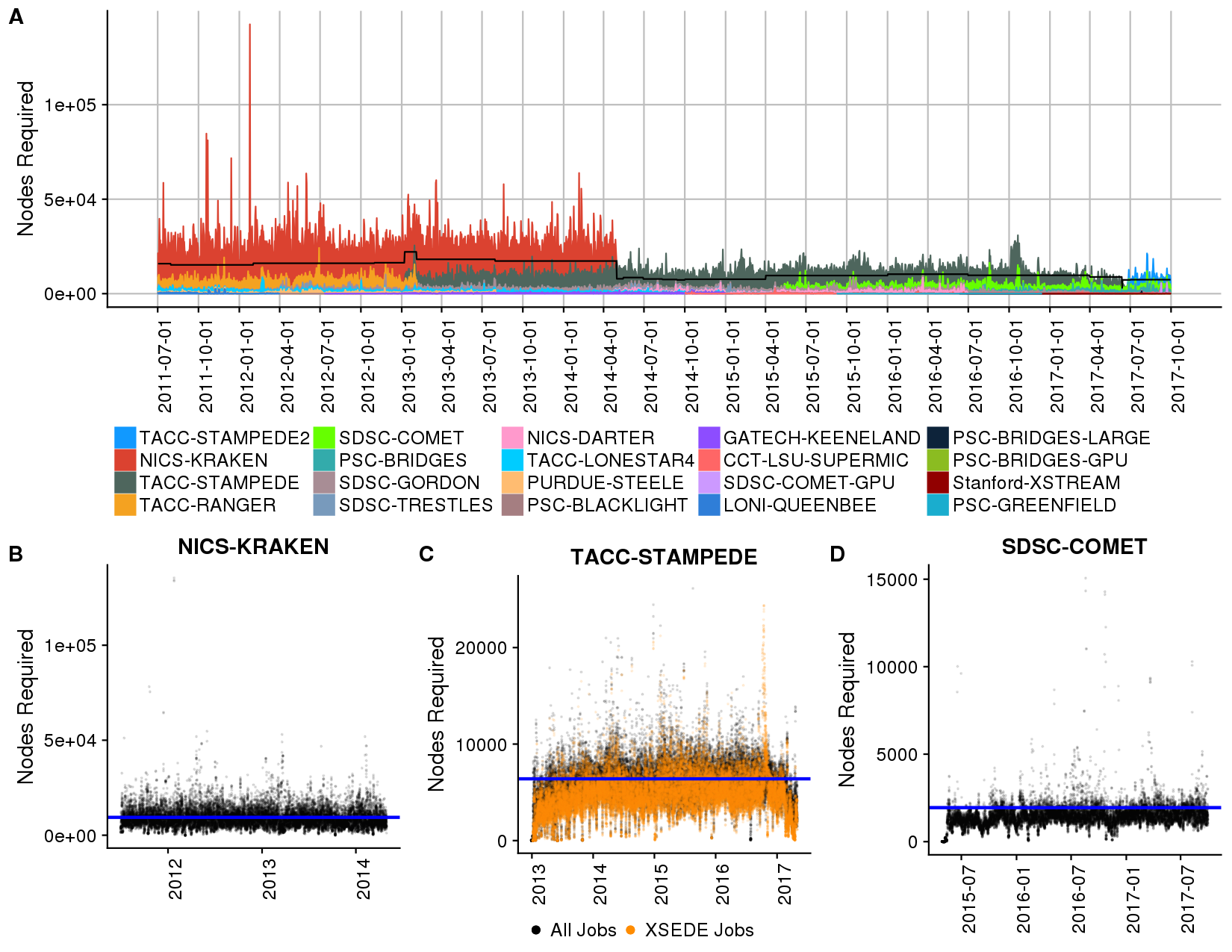} 
\caption{\label{fig:nodes_req_hpc_dic2} \textbf{A} - Number of nodes that would be required to run all queued jobs immediately over time.  Solid black line represents the available nodes.  Nodes required for: \kraken{} (\textbf{B}), \stampede{} (\textbf{C}) and \comet{} (\textbf{D}). Blue line shows total number of nodes for the system. Black dots show required nodes for all jobs and orange for only XSEDE jobs, note that \kraken{} and \comet{} are 100\% allocated to XSEDE jobs}
\end{figure}

Since we know the time history of the node utilization and the attributes of the queued jobs (nodes requested, etc), we can project the number of nodes that are required to run all queued jobs immediately,    This result is shown in Figure \ref{fig:nodes_req_hpc_dic2}.A where the solid black line represents the cumulative number of nodes that are available and anything over this line indicates needed additional nodes.      Figure \ref{fig:nodes_req_hpc_dic2}.B shows the same thing but this time for specific resources, namely  \kraken{}, \stampede{}, and \comet{}.  

Providing a system with no wait time is not realistic.  However, we can also frame our analysis in terms of what size would each of the HPC systems need to be to immediately run a given percentage of the queued jobs.  This is presented in Table \ref{table:resource-nodes-needed-to-run-immediately}  in which we show the number of additional nodes and cores needed for 95\% or 99\% percent of the queued jobs on each resource to run immediately.  At the 95\% level, many of the systems would require about a 10 - 20\% increase in size.   Of course, this analysis assumes that the allocations are not increased to take advantage of the increased resource size.


\begin{table}[H]
\begin{threeparttable}
\centering
\caption{\label{table:resource-nodes-needed-to-run-immediately}Number of nodes required to run 95\% or 99\% jobs immediately.}
\small
\begin{tabular}{@{}l|R{0.4in} R{0.8in} R{0.8in}|R{0.4in} R{0.8in} R{0.8in}@{}}
\hline
resource          & Nodes, Actual & 
Nodes to run immediately 95\% of jobs\tnote{1}   & 
Nodes to run immediately 99\% of jobs\tnote{1}    & 
Cores, Actual & Cores to run immediately 95\% of jobs\tnote{1}     & 
Cores to run immediately 99\% of jobs\tnote{1}     \\
\hline
CCT-LSU-SUPERMIC  & 360           & 267 (0.74)  & 433 (1.20)    & 7200          & 5300 (0.74)   & 8640 (1.20)    \\ \hdashline
GATECH-KEENELAND  & 264           & 424 (1.61)  & 769 (2.91)   & 4224          & 6784 (1.61)   & 12304 (2.91)  \\ \hdashline
LONI-QUEENBEE     & 668           & 519 (0.78)  & 909 (1.36)   & 5344          & 4152 (0.78)   & 7272 (1.36)   \\ \hdashline
NICS-DARTER       & 724           & 1073 (1.48) & 1884 (2.60)   & 11968         & 17168 (1.43)  & 30144 (2.52)  \\ \hdashline
NICS-KRAKEN       & 9408          & 15042 (1.60) & 22665 (2.41) & 112896        & 179388 (1.59) & 270924 (2.4)  \\ \hdashline
PSC-BLACKLIGHT    & 2             & 2.2 (1.08)  & 3 (1.63)     & 4096          & 4432 (1.08)   & 6672 (1.63)   \\ \hdashline
PSC-BRIDGES       & 752           & 1034 (1.38) & 1579 (2.10)   & 21056         & 28953 (1.38)  & 44200 (2.10)   \\ \hdashline
PSC-BRIDGES-GPU   & 48            & 54 (1.13)   & 82 (1.70)     & 1472          & 1662 (1.13)   & 2509 (1.70)    \\ \hdashline
PSC-BRIDGES-LARGE & 46            & 34 (0.73)   & 39 (0.85)    & 4512          & 3294 (0.73)   & 3851 (0.85)   \\ \hdashline
PSC-GREENFIELD    & 3             & 2.5 (0.83)  & 3 (1.08)     & 360           & 300 (0.83)    & 390 (1.08)    \\ \hdashline
PURDUE-STEELE     & 893           & 654 (0.73)  & 1062 (1.19)  & 7144          & 5232 (0.73)   & 8496 (1.19)   \\ \hdashline
SDSC-COMET        & 1944          & 2088 (1.07) & 2677 (1.38)  & 46656         & 50113 (1.07)  & 64256 (1.38)  \\ \hdashline
SDSC-COMET-GPU    & 72            & 116 (1.61)  & 177 (2.46)   & 1872          & 3014 (1.61)   & 4600 (2.46)   \\ \hdashline
SDSC-GORDON       & 1024          & 1192 (1.16) & 1818 (1.78)  & 16384         & 18336 (1.12)  & 28160 (1.72)  \\ \hdashline
SDSC-TRESTLES     & 324           & 352 (1.09)  & 534 (1.65)   & 10368         & 11255 (1.09)  & 17094 (1.65)  \\ \hdashline
Stanford-XSTREAM  & 65            & 20 (0.30)    & 35 (0.54)    & 1300          & 394 (0.30)     & 696 (0.54)    \\ \hdashline
TACC-LONESTAR4    & 1888          & 1080 (0.57) & 1578 (0.84)  & 22656         & 12960 (0.57)  & 18936 (0.84)  \\ \hdashline
TACC-RANGER       & 3936          & 4674 (1.19) & 6096 (1.55)  & 62976         & 74784 (1.19)  & 97536 (1.55)  \\ \hdashline
TACC-STAMPEDE     & 6400          & 7344 (1.15) & 9435 (1.47)  & 102400        & 117504 (1.15) & 150960 (1.47) \\ \hdashline
TACC-STAMPEDE2    & 4200          & 3803 (0.91) & 4917 (1.17)  & 285600        & 258604 (0.91) & 334352 (1.17) \\ \hline
\end{tabular}
\begin{tablenotes}
\item[1] The value in brackets shows fraction of actual number of nodes or cores.
\end{tablenotes}
\end{threeparttable}
\end{table}


\subsection{Allocations Impact}

\begin{figure}[ht]
\centering
\includegraphics[width=1.0\textwidth]{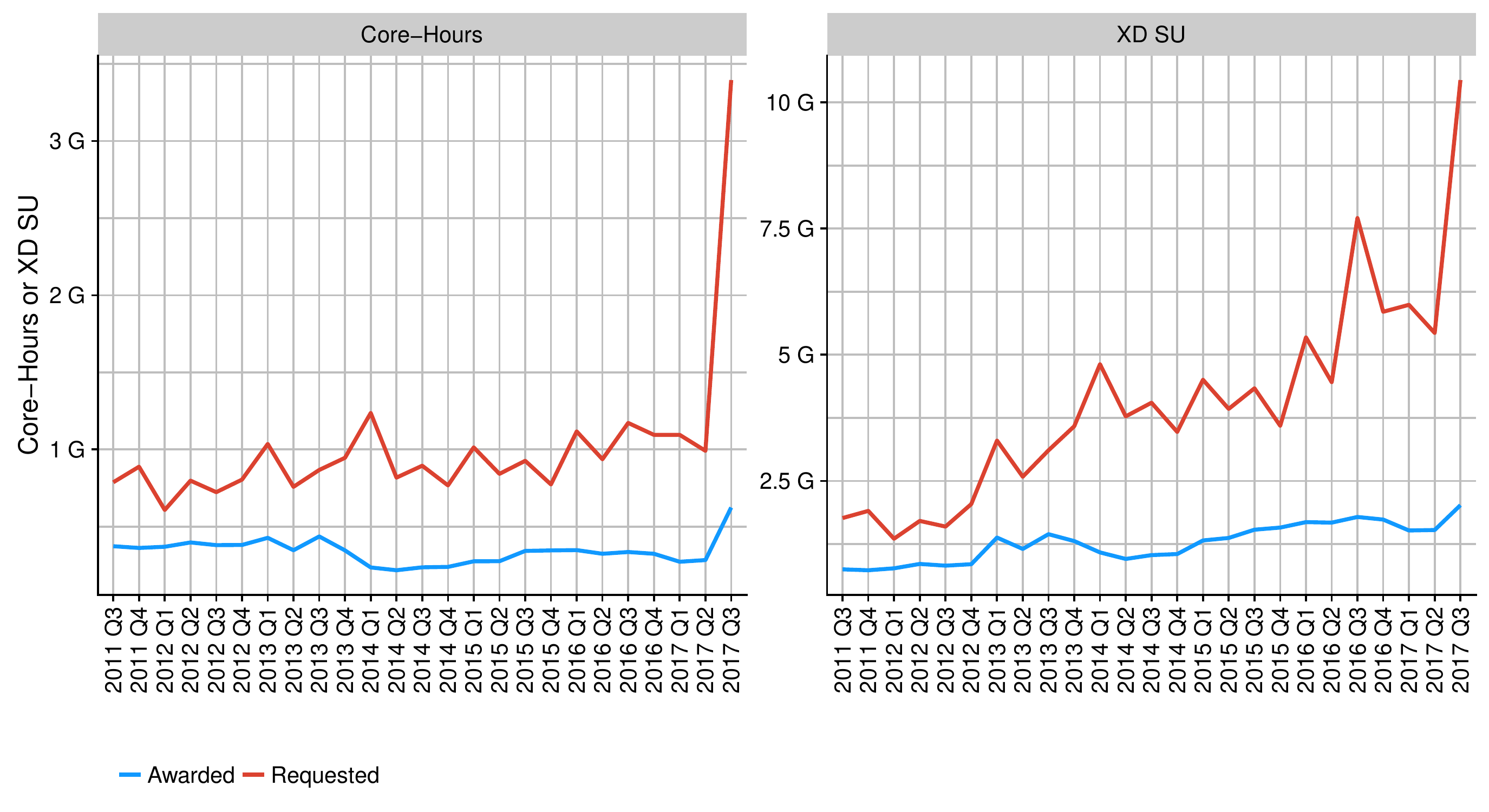} 
\caption{\label{fig:requested_vs_awarded} Time history of requested and awarded allocations. Left plot shows requested and awarded allocation measured by core hours and right plot shows same but measured by XD SU (which includes performance based scaling)}
\end{figure}

The previous subsection addressed the question of how much larger a given HPC resource would need to be in order for 95\% or 99\% of queued jobs to run immediately. Another associated question is how would increasing the awarded allocations influence hypothetical system sizes. The answer to this question could help inform future system requirements. This is especially useful given the decrease in percentage of awarded SU from the 40's in 2011 to the 20's in 2017 as indicated in Figure \ref{fig:requested_vs_awarded} which shows the requested and awarded allocations over the time. Prior to the introduction of \stampedetwo{}, the requested allocations were steadily increasing in terms of both absolute core hours and performance scaled XD SUs (red lines)  while the total core count for all available XSEDE systems has remained about the same and the available XD SUs has increased only moderately (blue lines).   

Given the pent up demand for CPU cycles shown in this figure, it is natural to ask how big would HPC resources need to be to accommodate a given percentage increase in allocation awards.   The trivial answer is that system size should be increased proportional to the increase in the allocation. However, there are several factors that can affect the scaling. First, not all projects use their entire allocation.  As shown in Table \ref{table:allocation_stats_summary}, in aggregate, projects utilized about 89\% of their allocation. Second, larger HPC systems have more opportunities to schedule jobs, and therefore improve job throughput. 
Accordingly, the increase in system size needed to support a targeted increase in allocations might be somewhat smaller.

Whatever new systems of increased size are commissioned, it is always worthwhile planning to operate them optimally to serve the computational community.   There are potential ways to increase allocations by  optimizing the utilization of existing systems and trade-offs to be considered between throughput and wait time. One  possibility is to institute over-allocation, a practice common in other industries such as internet service providers. If properly balanced over-allocation can improve utilization without a significant negative effect on users. This can lead to a compromise solution in which increased system utilization can be accompanied by a modest increase in job wait times.  There are tools available to study the impact of over subscription on system utilization and wait times~\cite{Simakov2018_SlurmSim}.  Another area that can be exploited in order to allow for a more effective use of allocations is node sharing for small core jobs, which has been shown in most cases not to impact  performance for the jobs sharing a node~\cite{White:2014:ANS:2616498.2616533,Simakov2018_SlurmSim}.

\subsection{Temporal and Spatial Resource Usage Patterns}

\begin{figure}[htb]
\centering
\includegraphics[width=0.8\textwidth]{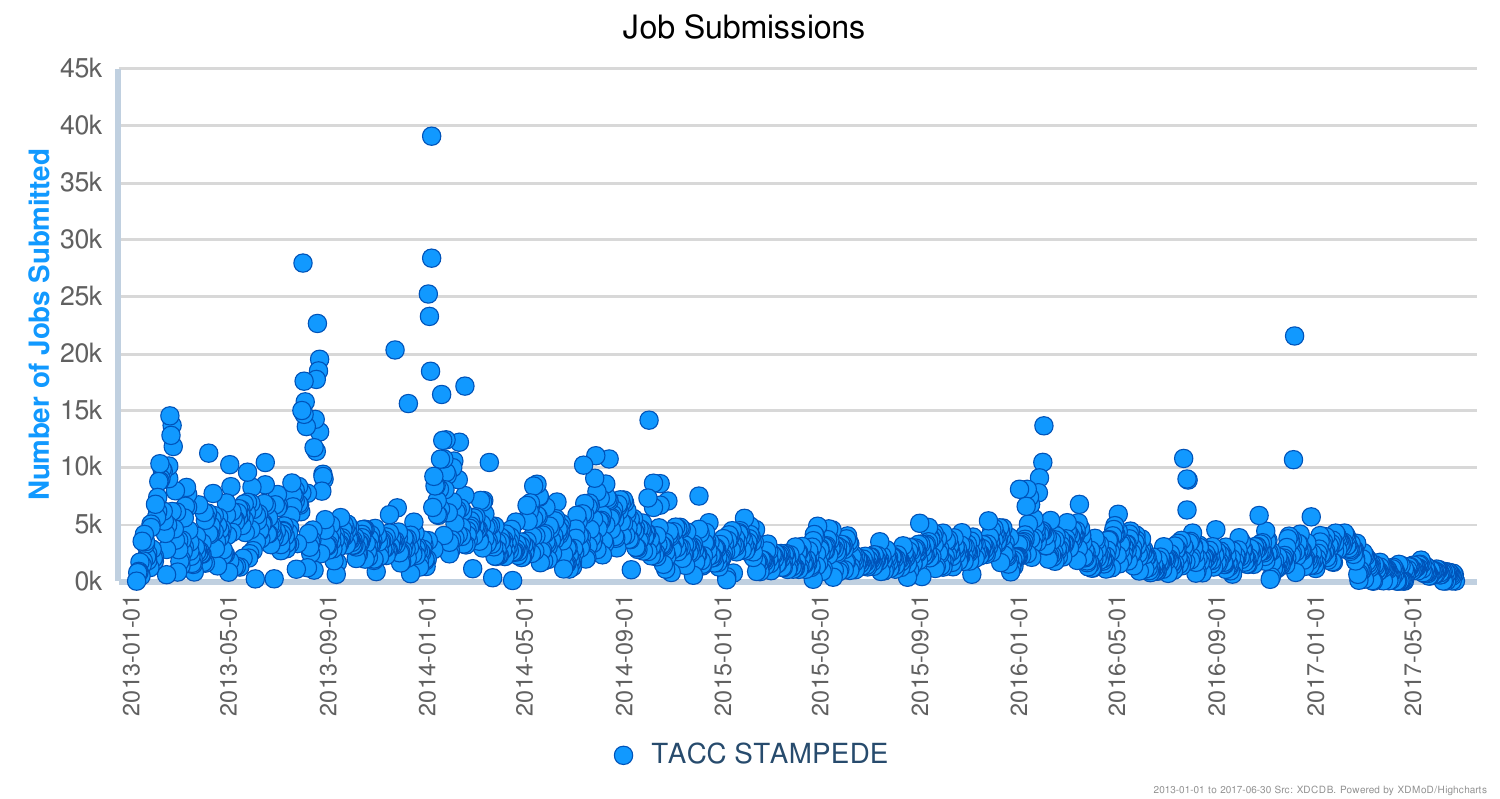}
\caption{Time history of the number of jobs submitted to \stampede{} over the time period 2013-01-01 to 2017-06-30.}\label{fig:xwl_plot_143.pdf}
\end{figure}

We can gain insight into temporal resource usage patterns by analyzing the variation in the job submission queues over time.  Here we include only the results for \stampede{} as the other resources showed similar patterns. Figure \ref{fig:xwl_plot_143.pdf} is a plot of the time history of the number of jobs submitted to \stampede{} over the 2013-2017 time range. 

\begin{figure}[H]
\centering
\subfloat[\label{fig:xwl_plot_091.pdf}Lomb-Scargle periodogram for \stampede{}.]{%
\includegraphics[width=0.4\textwidth]{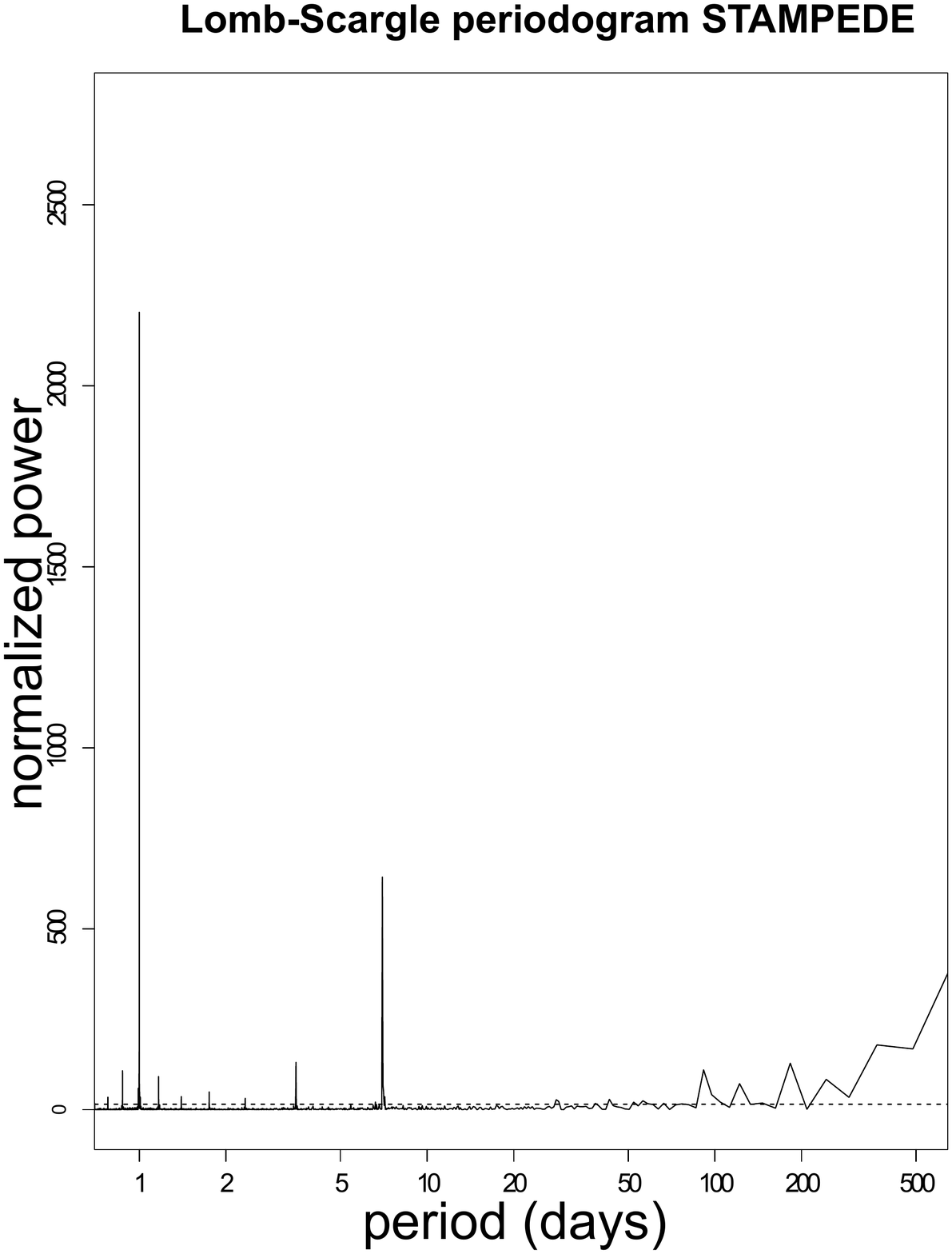}
}
\subfloat[\label{fig:xwl_plot_093.pdf}Histogram of \stampede{} job submissions by hour of the day.]{%
\includegraphics[width=0.4\textwidth]{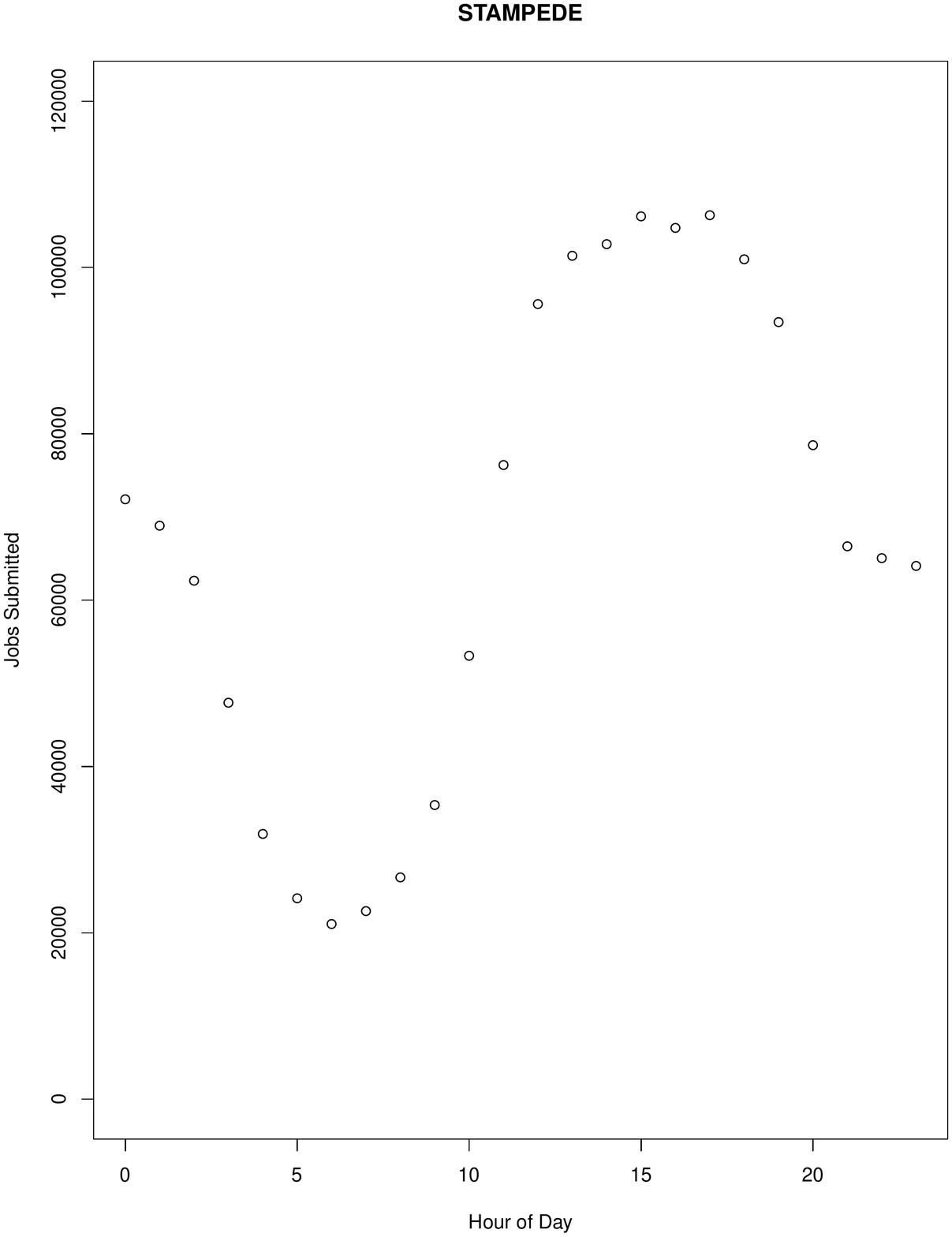}
}
\caption{(a) Lomb-Scargle periodogram of the TACC STAMPEDE job submission analysis over the time period 2013-2017.  Note the strong periodic behavior at 1 day and 1 week and the very weak periodicity over 1 year.  (b) Histogram of \stampede{} job submissions by hour of the day.}
\end{figure}

In order to detect periodic trends in the data we carried out a Lomb-Scargle analysis of the queue data. A Lomb-Scargle analysis is similar to a FFT analysis except the exact time of each point can be specified and there is no requirement for uniformly sampled points.  Figure \ref{fig:xwl_plot_091.pdf} shows a computation of the Lomb-Scargle periodogram for the \stampede{} job submission data.  There are three main periodic trends detected, namely 1 day,  7 days (1 week), and \~365 days (1 year).  Note that there are some other small peaks which are overtones (artifacts) as established by a comparison to the auto correlation function, not shown.  The daily variation of \stampede{} job submission over the 2013-2017 time range is shown in Figure \ref{fig:xwl_plot_093.pdf}.  The queue is at its minimum in the early hours of the morning.  As users go to work they submit more jobs and the queue grows to a maximum at the end of normal working hours.  The submission then decreases back down until the users return to work.  The effect is such that the minimum submission early in the morning is only a small fraction of the peak value.

\begin{figure}[H]
\centering
\subfloat[\label{fig:xwl_plot_092.pdf}]{%
\includegraphics[width=0.5\textwidth]{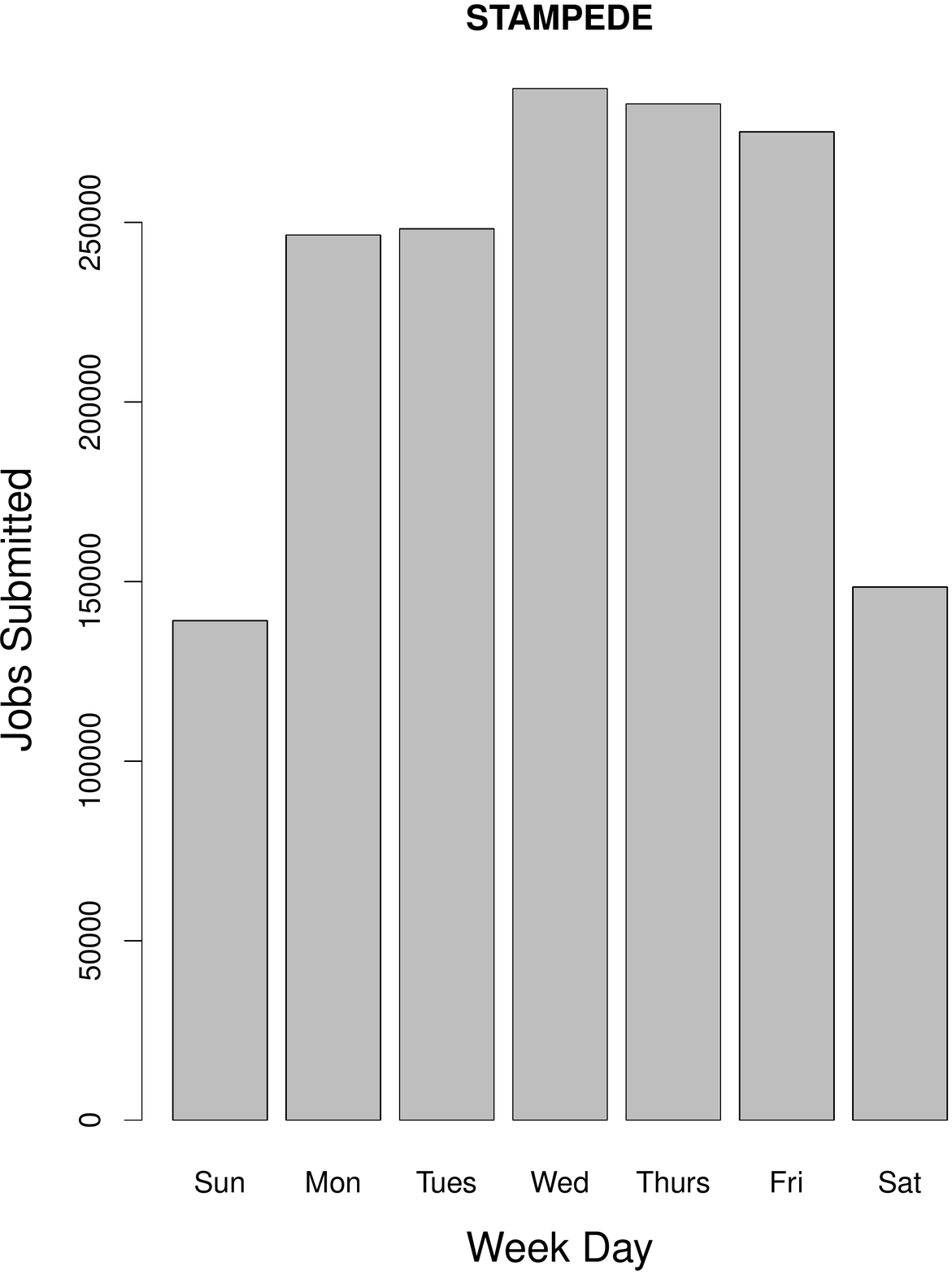}
}
\subfloat[\label{fig:xwl_plot_095.pdf}]{%
\includegraphics[width=0.5\textwidth]{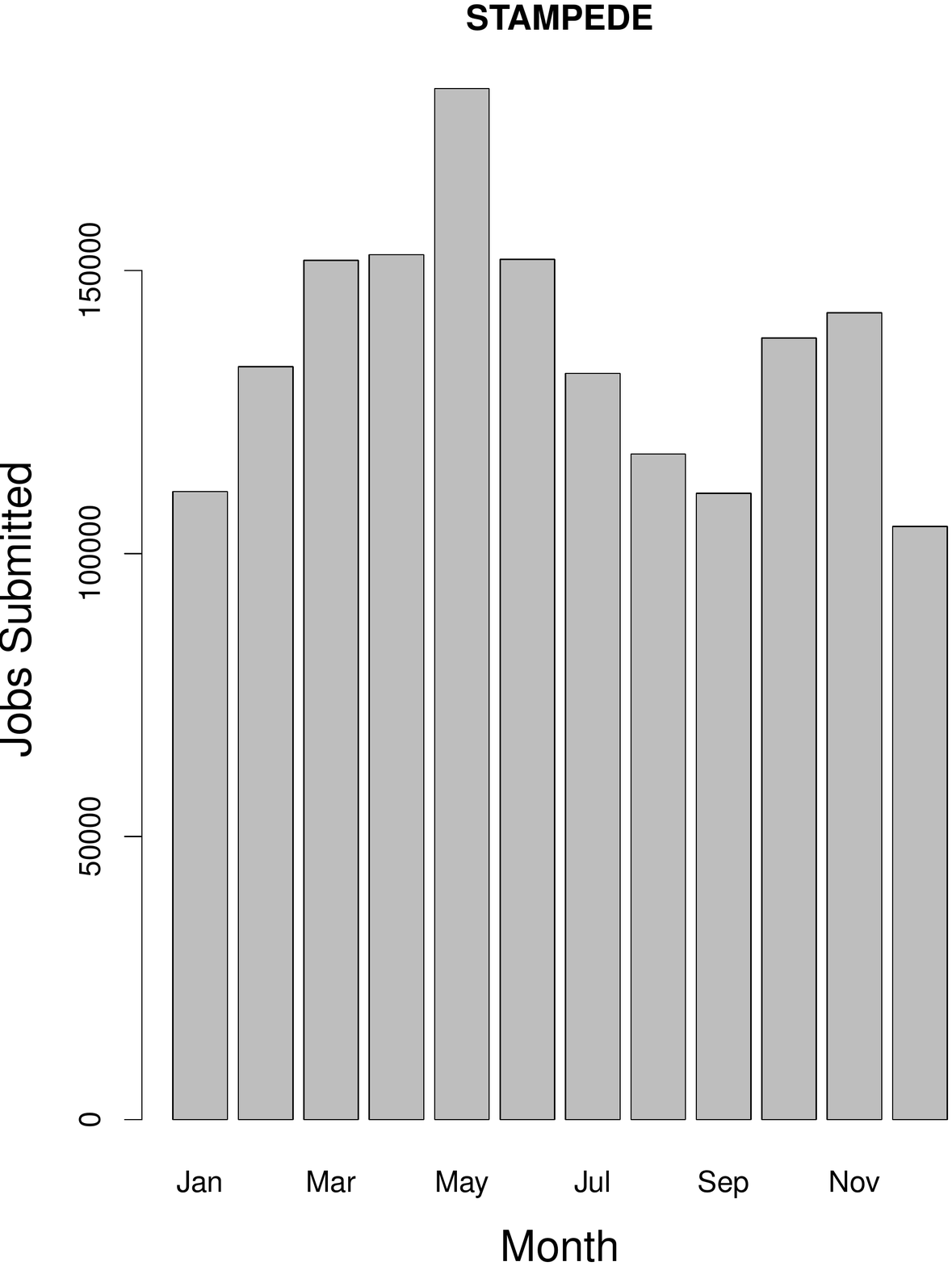}
}
\caption{Histogram showing the distribution of the number of jobs submitted to \stampede{} by (a) day of the week and (b) month of the year.  The time range is 2013-2017.}
\end{figure}

The weekly variation of \stampede{} job submission over the 2013-2017 time range is shown in Figure \ref{fig:xwl_plot_092.pdf}.  Not surprisingly, the job submission is smallest during the weekend (both Saturday and Sunday) relative to the week days.  The week-end value is about 1/2 of the peak week-day values.   The monthly variation is shown in Figure \ref{fig:xwl_plot_095.pdf}. The most noticeable trend is that the queue during the holiday months of December and January is substantially less than its value during peak months.  We have repeated a similar analysis for the other resources. The same general trends are evident but with a bit more noise in the plot. Note that in reference ~\cite{Hart2011b} a closely related study of periodicity in job submissions also detected similar trends for the TeraGrid resources. 

\begin{figure}[H]
\centering
\includegraphics[width=0.75\textwidth]{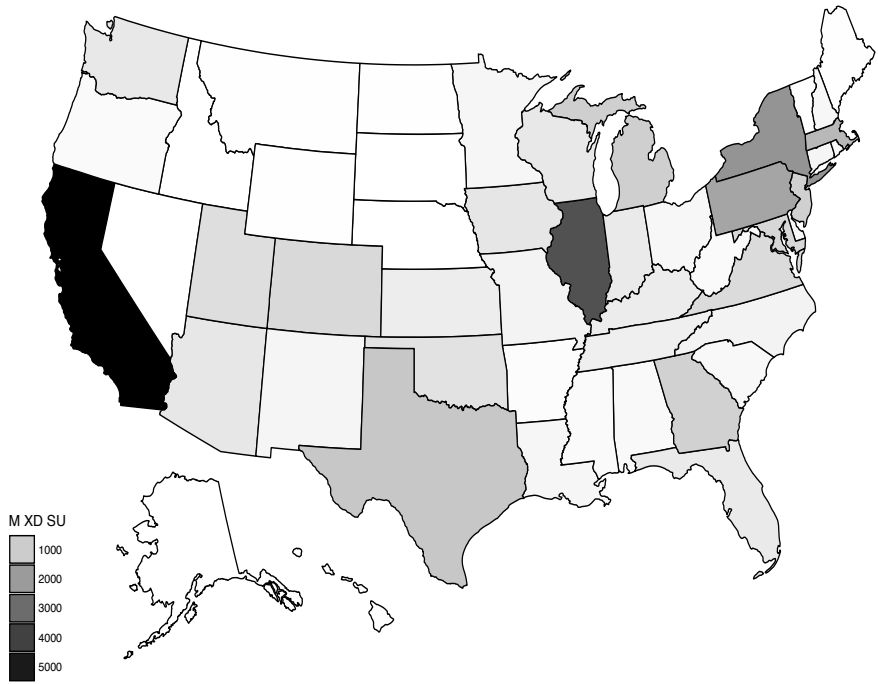}
\caption{\label{fig:xwl_plot_175.pdf}Usage of the NSF facilities by state in millions of XD SUs.}
\end{figure}

\begin{figure}[H]
\centering
\includegraphics[width=0.75\textwidth]{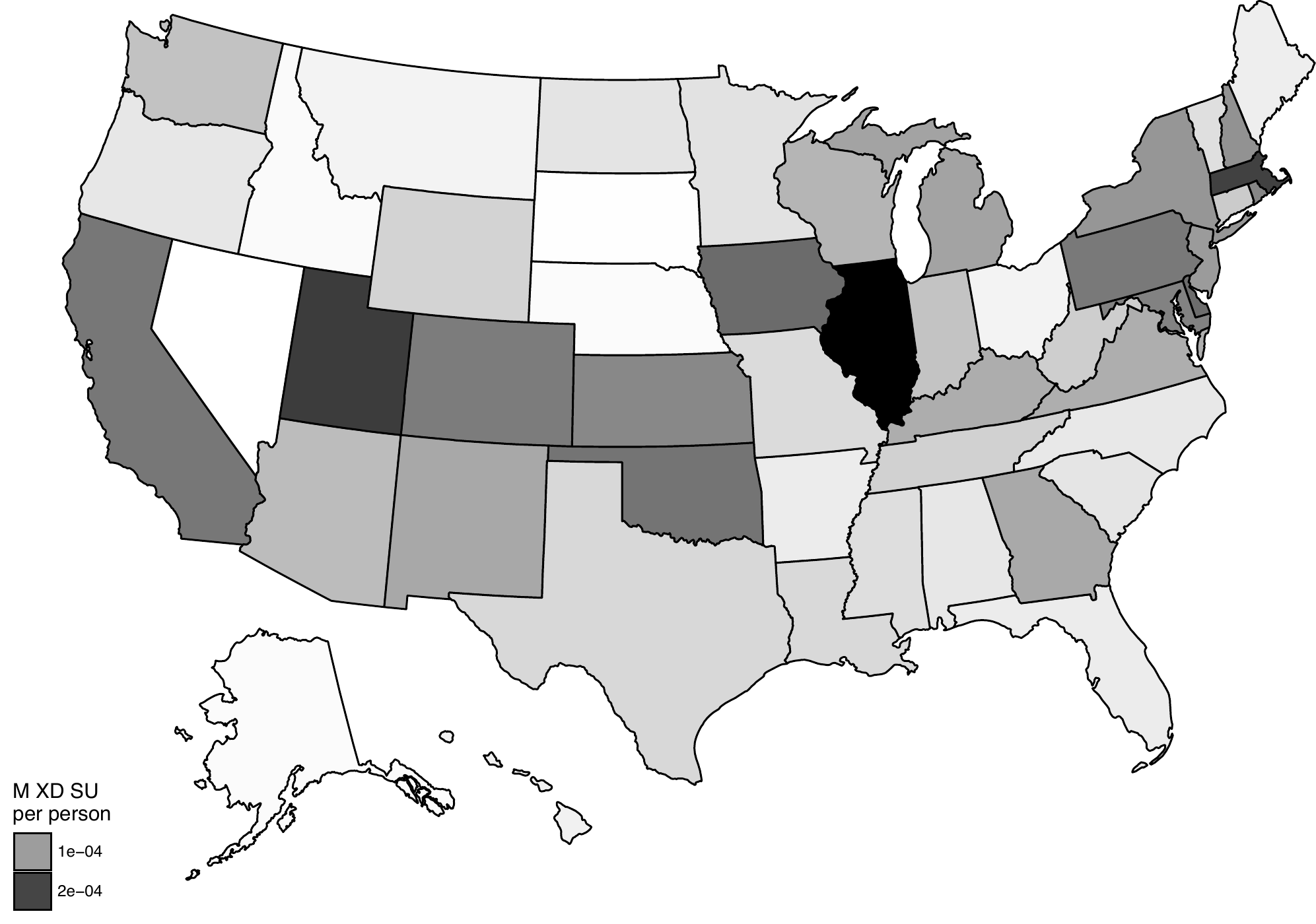}
\caption{\label{fig:xwl_plot_176.pdf}Usage of the NSF facilities by state in millions of XD SUs normalized by state population, that is the per capita usage.}
\end{figure}

\begin{figure}[H]
\centering
\includegraphics[width=0.75\textwidth]{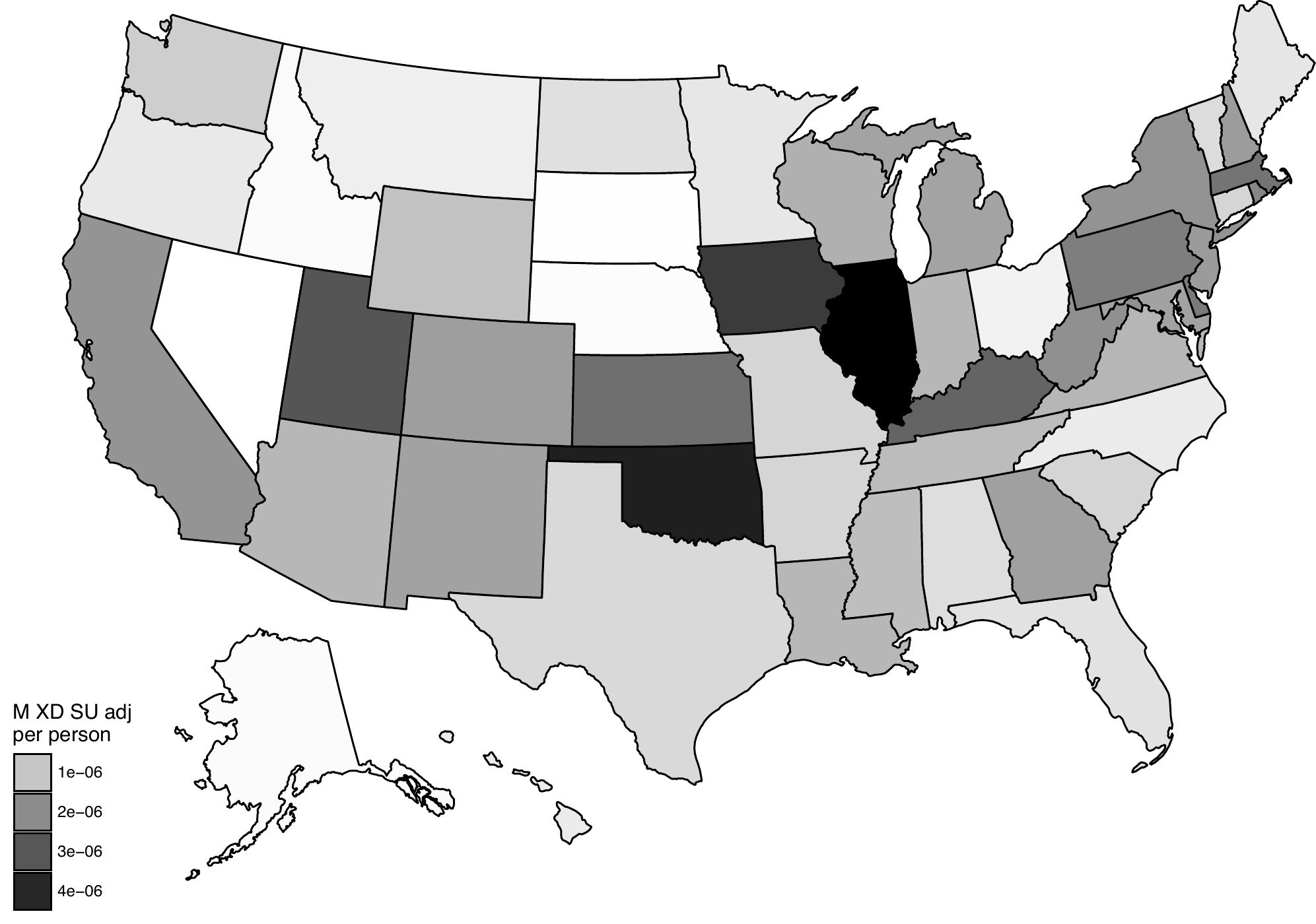}
\caption{\label{fig:xwl_plot_178.pdf}Usage of the NSF facilities by state in millions of XD SUs normalized by state population and adjusted by the state technology index.}
\end{figure}

The resources on which this analysis is based are intended to serve the entire country as a whole. Figure \ref{fig:xwl_plot_175.pdf} shows a map of the facility usage in millions of XD SUs by state of origin.  Not surprisingly, the usage tends to be concentrated in a few of the most populous states. Figure \ref{fig:xwl_plot_176.pdf} shows a map of per capita usage, that is, the usage has been normalized by state population. The per capita usage is substantially more uniform although there are still some high and low usage states.  We can make a second adjustment based on the individual state economies and and their reliance on high technology.  We would expect that states that have economies that are more dependent on high technology would be more able to take advantage of the facilities that NSF provides.  When adjusted for the state technology index the usage map appears as in Figure \ref{fig:xwl_plot_178.pdf}.  The adjusted usage seems to be rather evenly distributed over various regions of the country. Interestingly, the main feature common to the three maps is that Illinois stands out as relatively bright (high usage).      

\subsection{User Limits \& Throughput}

\begin{figure}[H]
\centering
\includegraphics[width=0.8\textwidth]{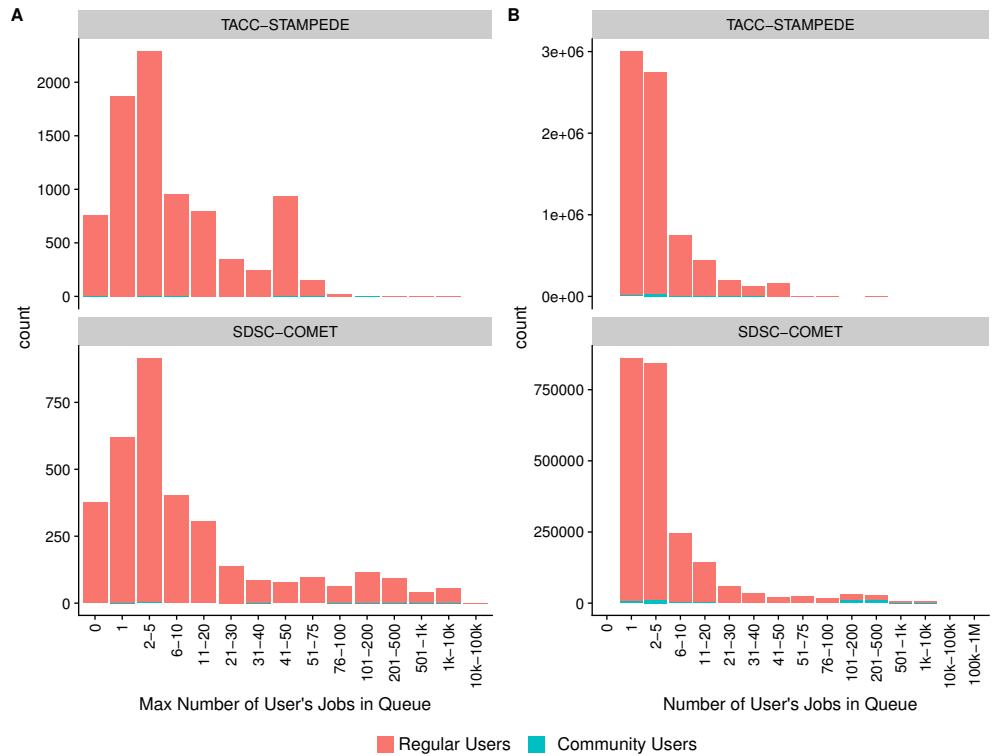}
\caption{\label{fig:jobs_per_user__stampede_comet} \textbf{A}: Distribution of maximum jobs per user for \stampede{}, \comet{}. \textbf{B}: Distribution of jobs per user.  Time period is over the production lifetime of each resource. Users are split in two categories: regular users, where each user corresponds to single person, and community user, where multiple people can use the same system user account. Gateways typically use community user accounts for their jobs execution.}
\end{figure}

In this subsection we study the effect that queue limits have on job throughput. 
Table \ref{table:user-limits-along-resource-summary} shows the maximum number of queued jobs allowed per user among different resources.  Only TACC and \xstream{} impose limits.  In order to determine what effect, if any, imposing a limit on the number of queued jobs a user can have,  we analyzed the distribution in the maximum number of queued jobs per user and the number of jobs per user on the resources \stampede{} and \comet{}; one which has a queue limit (\stampede{}) and one which does not (\comet{}).    The results are shown in Figure \ref{fig:jobs_per_user__stampede_comet}.  The time period is over the production lifetime of each resource.  Figure \ref{fig:jobs_per_user__stampede_comet}.A shows the distribution of maximum queued jobs per user for both resources.  While \stampede{} has a limit of 50 jobs, the vast majority of users do not reach this limit.  However, a moderate peak at 50 queued jobs is apparent, which confirms the imposition of \stampede{}'s queue limit. Interestingly, there is also a faint tail beyond the limit of 50 jobs which must be attributable to a relaxed constraint for some users. \comet{}, which has no limit has a much more pronounced tail beyond 50 jobs but otherwise the two distributions are similar.  Figure \ref{fig:jobs_per_user__stampede_comet}.B, which shows the distribution in the number of jobs run by users over the same time period, addresses the issue of potential impact of the queue limit.  Based on this figure, we see little if any impact since  most of the users for both resources run fewer than 50 jobs.  \comet{} exhibits a more pronounced tail however.

\begin{table}[H]
\centering
\caption{User limits along XSEDE resources.}
\label{table:user-limits-along-resource-summary}
\begin{tabular}{ lrrrr }
\toprule
\textbf{Resource} & \textbf{Max Duration, Hours} & \textbf{Max Jobs per User} & \textbf{Max Nodes} & \textbf{Max Cores} \\ \midrule
TACC Stampede2    & 48                  & 50                & 2048      & 139264    \\ \hdashline 
TACC Stampede     & 48                  & 50                & 1024      & 16384     \\ \hdashline 
TACC Jetstream    &                     &                   & 1         &           \\ \hdashline 
TACC Maverick2    & 12                  & 32                & 640       & 20        \\ \hdashline 
TACC Lonestar4    & 24                  & 50                & 342       & 4104      \\ \hdashline 
TACC Ranger       & 6                   & 50                & 128       & 2048      \\ \hdashline 
SDSC Comet        & 48                  &                   & 72        & 864       \\ \hdashline 
SDSC Comet GPU    & 48                  &                   & 8         & 96        \\ \hdashline 
SDSC Gordon       & 48                  &                   & 64        & 1024      \\ \hdashline 
SDSC Trestles     &                     &                   &           &           \\ \hdashline 
Stanford XSTREAM  & 168                 & 512               & 16        & 320       \\ \hdashline
Purdue Steele     &                     &                   &           &           \\ \hdashline 
Purdue Condor     &                     &                   &           &           \\ \hdashline 
PSC Bridges       & 48                  &                   & 42        &           \\ \hdashline 
PSC Bridges Large & 96                  &                   &           & 250       \\ \hdashline 
PSC Bridges GPU   & 48                  &                   & 16        &           \\ \hdashline 
PSC Greenfield    & 168                 &                   &           &           \\ \hdashline
PSC Blacklight    & 168                 &                   &           &           \\ \hdashline
NICS Kraken       &                     &                   &           &           \\ \hdashline
NICS Darter       & 24                  &                   & 748       & 11968     \\ \hdashline
NICS Nautilus     & 48                  &                   &           &           \\ \hdashline
LSU SuperMIC      & 72                  &                   & 200       & 4000      \\ \hdashline
LONI QueenBee     & 48                  &                   & 256       &           \\ \hdashline
GATech Keeneland  & 48                  &                   & 132       & 2112      \\ \hdashline
OSG               & 12                  &                   & 1         & 1         \\ \bottomrule
\end{tabular}
\end{table}

\subsection{Summary: Job Submission Patterns \& Over-Subscription}

Given current XSEDE allocations, in order for 95\% of the backlog of queued jobs to run immediately, current HPC resources would need to be 10-20\% larger.  Queue limits in terms of the total number of jobs a user is allowed to have in a queue are imposed only by TACC and \xstream{} resource and were found to have no impact on job throughput.  The majority of users do not reach the maximum queue limit in their work flow.

\newpage
\section{Acknowledgments}

We gratefully acknowledge the support of NSF awards OCI 1025159, ACI 1445806, ACI 1566393, ACI 1763033, and OCI 1203560.  
This analysis would not have been possible without the support and expertise of the XDMoD development team,
including Cynthia Cornelius, Martins Innus, Ryan Rathsam, Jeanette Sperhac, and former members Thomas Yearke, Amin Ghadersohi and Ryan Gentner. 

\newpage

\appendix

\section{Supplementary Data}

\subsection{Data Sources}\label{appendix:data-sources}

XDMoD draws information from a number of data repositories including the XSEDE Central Database (XDCDB), the XSEDE Resource Allocation System (XRAS), the XSEDE Resource Description Repository (RDR), \texttt{tacc\_stats} software, resource manager accounting logs, and Cray RUR software~\cite{Barry:2013}. In many cases, additional data is required that is not available in one of these existing repositories and is provided to XDMoD directly from the source. For example, the XDCDB does not track the hosts that an individual job was allocated so this information comes directly from the resource manager accounting data at the service providers. Open Science Grid (\osg{}) processes over 30M jobs each year, which is a higher volume than the XSEDE accounting workflow can currently handle so \osg{} provides this data directly to XDMoD. Metrics for cloud and innovative resources such as \jetstream{} require detailed event information so this data is provided data directly to XDMoD from the service providers. Data from these various repositories is brought into XDMoD through an Extract/Transform/Load (ETL) process where it is cleaned, normalized, and cross-referenced in the XDMoD data warehouse.

\begin{table}[H]
\centering
\caption{Data sources utilized for this report.}
\label{table:data-sources}
\begin{tabular}{ *4l S }
\toprule
\textbf{Source}      & \textbf{Description} \\ \midrule
XDCDB                & Primary source of job accounting data as well as XSEDE users, organizations, \\
                     & system usernames, allocation charge numbers, gateway usernames, etc. \\
XRAS                 & Source of XD SUs requested and allocated. \\
RDR                  & Resource configuration information (nodes, cores, production dates, etc.) \\
\texttt{tacc\_stats} & Job performance data for \texttt{TACC RANGER}, \lonestar{}, \stampede{}, \\
                     & \stampedetwo{}, \texttt{LSU SUPERMIC}, \texttt{SDSC GORDON} and \texttt{SDSC COMET} \\
Cray RUR & Job performance data for \darter{} \\
Resource Mgr Log Files & Source for the hosts allocated to individual jobs \\
\osg{} Job Accounting & Detailed accounting data for jobs submitted to \osg{} \\
\bottomrule
\end{tabular}
\end{table}

\subsection{Resource Characteristics}
\label{appendix:ResourceCharacteristics}

This appendix contains architectural and operational characteristics of the resources studied in this report, including configuration, dates of operation, and the data available for analysis. Table \ref{table:resource-charachteristics} lists resources and their composition, and Table \ref{table:resources-service-dates} includes resource service dates and data collected. 

It is important to note that there are two methods by which jobs are submitted to \jetstream{}: via the Openstack API and using the Atmosphere portal developed by CyVerse at the University of Arizona (\url{http://www.cyverse.org/}) with 31\% and 69\% of XD SUs submitted using each respective method. While we have been able to extract accurate data from the XDCDB for jobs submitted via the Openstack API, accounting data submitted to the XDCDB by Atmosphere is aggregated by user and allocation on a roughly daily basis and groups together all virtual machines (VMs) in the given reporting period. Due to this summarization, we are only able to determine the total number of XD SU charged to a particular allocation and unable to determine information such as the number of VMs, the number of cores per VM, or the times that a given VM was running. We have been in contact with the Atmosphere team to obtain more detailed accounting records going forward, but the data is not currently available at the time of this report.

\begin{figure}[H]
\centering
\includegraphics[width=1.0\textwidth]{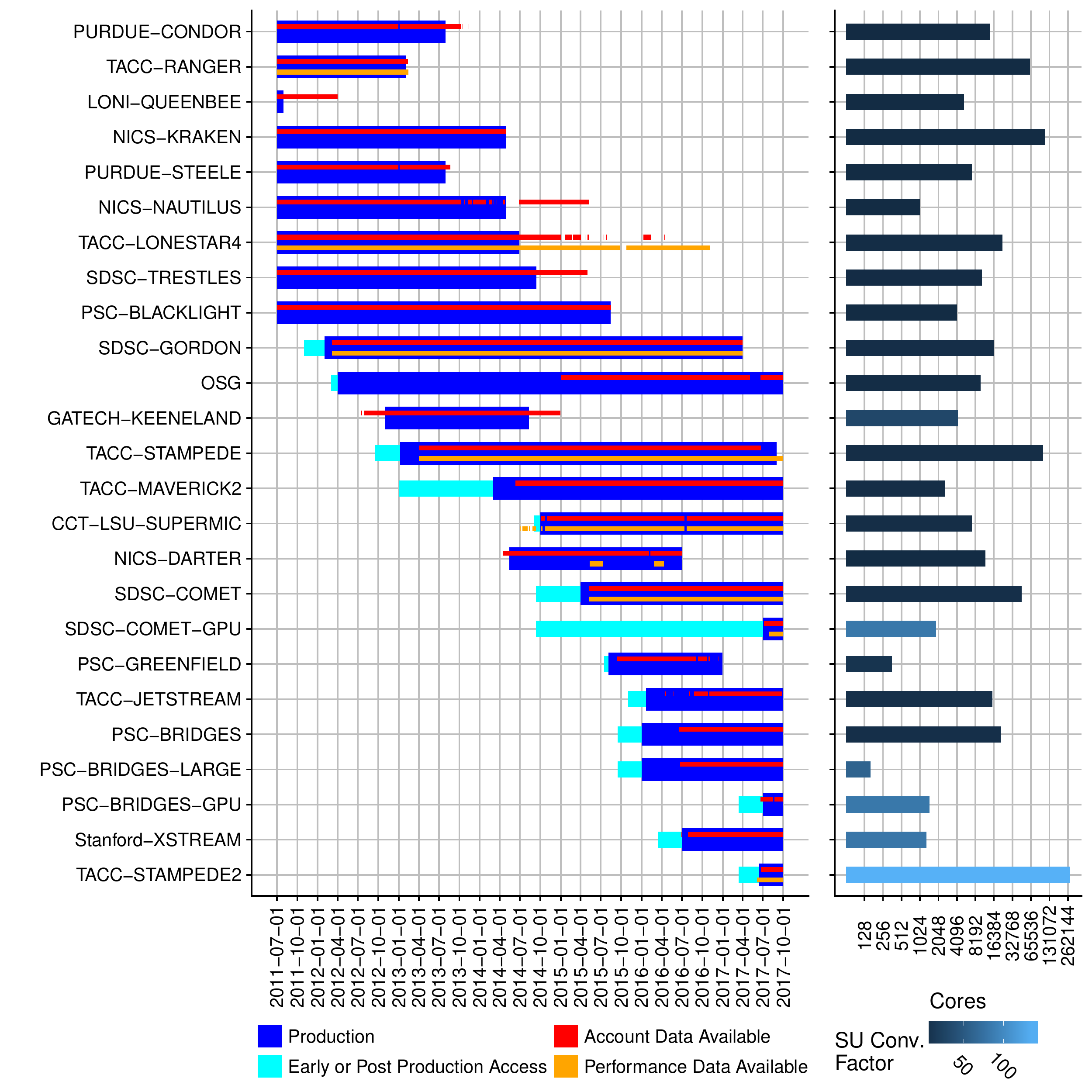}
\vspace*{-0.4in}
\caption{\label{fig:resources_serving_time} Production time along with pre and post production time for XSEDE resources.  Also included is an indication of those resources for which performance data (orange line) and account data (red line) is available.  SU conversion factors for each resource are shown in the bar chart on the right.}
\end{figure}

Figure \ref{fig:resources_serving_time} is a detailed plot showing the dates each resource was in production, along with the time period for which accounting and job level performance data is available.   The size of each resource in cores is also given.  In addition to job accounting data for almost all of the production systems during the period covered by this report, job level performance data was available for \ranger{}, \lonestar{}, \gordon{}, \stampede{}, \supermic{}, \darter{}, \comet{}, and \stampedetwo{}, as indicated in this figure.  At the time of this report, the collection of job level performance data has not yet been implemented for the current production systems \bridges{} and \jetstream{}.  Neither accounting or job level performance data is available for \wrangler{}.

\begin{table}[H]
\begin{threeparttable}
\centering
\caption{XSEDE resources characteristics.}
\footnotesize
\label{table:resource-charachteristics}
\begin{tabular}{l l r r R{0.6in} H R{0.65in} R{0.7in} R{0.55in} H}
\hline
Resource          & Type\tnote{1}  & Cores       & Nodes       & Cores per Node & CPU                                               & CPU Clock Rate, GHz & RAM, GiB & GPU/MIC per Node & GPU                           \\
 \hline
PURDUE-CONDOR     & HTC   & 14000       & 1750        & 8              & Intel Misc                                        & \textgreater3.0     & 0.5-32    &              &                               \\ \hdashline
TACC-RANGER       & HPC   & 62976       & 3936        & 16             & AMD Opteron                                       & 2.3                 & 32        &              &                               \\ \hdashline
LONI-QUEENBEE     & HPC   & 5344        & 668         & 8              & Intel                                             & 2.33                & 8         &              &                               \\ \hdashline
NICS-KRAKEN       & HPC   & 112896      & 9408        & 12             & AMD Opterons                                      & 2.6                 & 16        &              &                               \\ \hdashline
PURDUE-STEELE     & HPC   & 7144        & 893         & 8              & Intel E5410                                       & 2.33                & 16, 32    &              &                               \\ \hdashline
NICS-NAUTILUS     & Vis   & 1024        & 1           & 1024           & Intel Nehalem EX                                  &                     & 4096      & 8            &                               \\ \hdashline
TACC-LONESTAR4    & HPC   & 22656       & 1888        & 12             & Intel Westmere                                    & 3.3                 & 24        &              &                               \\ \hdashline
SDSC-TRESTLES     & HPC   & 10368       & 324         & 32             & AMD Magny-Cours                                   & 2.4                 & 64        &              &                               \\ \hdashline
PSC-BLACKLIGHT    & DIC   & 4096        & 2           & 2048           & Intel, X7560, Nehalem)                            & 2.27                & 16384     &              &                               \\ \hdashline
SDSC-GORDON       & DIC   & 16384       & 1024        & 16             & Intel, Sandy Bridge                               & 2.6                 & 64        &              &                               \\ \hdashline
OSG               & HTC   & $\sim$10000 & $\sim$10000 & 1              &                                                   &                     &           &              &                               \\ \hdashline
GATECH-KEENELAND  & HPC   & 4224        & 264         & 16             & Intel, Sandy Bridge                               &                     & 32        & 3            & NVIDIA M2090 \\ \hdashline
TACC-STAMPEDE     & HPC   & 102400      & 6400        & 16             & Intel, E5-2680, Sandy Bridge                      & 2.7                 & 32        & 1, 2        & Intel MIC                     \\ \hdashline
TACC-MAVERICK2    & Vis   & 2640        & 132         & 20             & Intel, E5-2680 v2                                 & 2.8                 & 256       & 1            & NVIDIA Tesla K40              \\ \hdashline
CCT-LSU-SUPERMIC  & HPC   & 7200        & 360         & 20             & Intel, Ivybridge                                  & 2.8                 & 64        & 2            & Intel Xeon Phi 7120P          \\ \hdashline
NICS-DARTER       & HPC   & 11968       & 724         & 16             & Intel, E5-2600, Sandy Bridge                      & 2.6                 & 32        &              &                               \\ \hdashline
SDSC-COMET        & HPC   & 46656       & 1944        & 24             & Intel, E5-2680v3, Haswell                         & 2.5                 & 128       &              &                               \\ \hdashline
SDSC-COMET-GPU    & HPC   & 1872        & 72          & 24, 18         & Intel, Broadwell and haswell                      & 2.5                 & 128       & 4            & K80 or K100                   \\ \hdashline
PSC-GREENFIELD    & HPC   & 360         & 3           & 240, 60, 60    &                                                   &                     &           &              &                               \\ \hdashline
TACC-JETSTREAM    & Cloud & 15360       & 640         & 24             & Intel E5-2680v3 Haswell                           & 2.5                 & 128       &              &                               \\ \hdashline
PSC-BRIDGES       & HPC   & 21056       & 752         & 28             & Intel, E5-2695 v3                                 & 2.3                 & 128       &              &                               \\ \hdashline
PSC-BRIDGES-LARGE & DIC   & 160         & 10          & $\sim$98       & Intel                                             &                     & 3TB, 12TB &              &                               \\ \hdashline
PSC-BRIDGES-GPU   & HPC   & 1472        & 48          & 1              & Intel                                             &                     & 128       & 4, 2         & K80/ P100                     \\ \hdashline
Stanford-XSTREAM  & HPC   & 1300        & 65          & 20             & Intel Ivy-Bridge                                  &                     & 256       & 8 (16-logical)        & K80                           \\ \hdashline
TACC-STAMPEDE2    & HPC   & 285600      & 4200        & 68             & Intel KNL                                         & 1.4                 & 192       &              &                              \\ \hdashline
\hline
\end{tabular}
\begin{tablenotes}
\item[1] Resource types: 
HPC - High-performance computing, 
HTC - High-throughput computing,
DIC - Data-intensive computing,
Cloud - Cloud resource and
Vis - Visualization system.
\end{tablenotes}
\end{threeparttable}
\end{table}

\begin{table}[H]
\begin{threeparttable}
\centering
\caption{Resource types, service dates, and data collected.  For service dates, early denotes availability of early access, start and end the beginning and end of the production phase, and retired the end of service.  In the Data column, \textbf{A} is accounting (resource manager, data provided to the TeraGrid/XSEDE central database), \textbf{J} is job-level metrics (applications, from \texttt{tacc\_stats} and/or Lariat), \textbf{N} is node-level metrics (cpu,i/o,network,memory, all from \texttt{tacc\_stats}.)}
\label{table:resources-service-dates}
\begin{tabular}{lllllll}
\toprule
Resource          & Type\tnote{1} & Early & Start & End & Retired & Data \\ 
\midrule
PURDUE-CONDOR     & HTC           &               & 2008-01-01       & 2013-07-30     & 2013-07-31  & \textbf{A}   \\ \hdashline
TACC-RANGER       & HPC           &               & 2008-02-04       & 2013-02-03     & 2013-02-04  & \textbf{A,J,N}   \\ \hdashline
LONI-QUEENBEE     & HPC           &               & 2008-02-01       & 2011-07-30     & 2011-07-31  & \textbf{A}   \\ \hdashline
NICS-KRAKEN       & HPC           & 2009-01-01    & 2009-10-05       & 2014-04-30     & 2014-05-01  & \textbf{A}   \\ \hdashline
PURDUE-STEELE     & HPC           &               & 2008-01-01       & 2013-07-30     & 2013-07-31  & \textbf{A}   \\ \hdashline
NICS-NAUTILUS     & Vis           & 2010-04-07    & 2010-10-01       & 2014-04-30     & 2015-04-30  & \textbf{A}   \\ \hdashline
TACC-LONESTAR4    & HPC           &               & 2011-02-01       & 2014-06-29     & 2014-12-31  & \textbf{A,J,N}   \\ \hdashline
SDSC-TRESTLES     & HPC           &               & 2011-01-01       & 2014-09-13     & 2015-05-01  & \textbf{A}   \\ \hdashline
PSC-BLACKLIGHT    & DIC           &               & 2011-01-17       & 2015-08-15     & 2015-08-16  & \textbf{A}   \\ \hdashline
SDSC-GORDON       & DIC           & 2011-11-01    & 2012-02-01       & 2017-03-31     & 2017-04-01  & \textbf{A,J,N}   \\ \hdashline
OSG               & HTC           & 2012-03-01    & 2012-04-01       & 2019-05-31     &                & \textbf{A} \\ \hdashline
GATECH-KEENELAND  & HPC           & 2012-07-01    & 2012-10-31       & 2014-08-11     & 2014-12-31  & \textbf{A}   \\ \hdashline
TACC-STAMPEDE     & HPC           & 2012-09-15    & 2013-01-07       & 2017-08-31     &                & \textbf{A,J,N} \\ \hdashline
TACC-MAVERICK2    & Vis           & 2013-12-15    & 2014-03-03       & 2018-03-31     &                & \textbf{A} \\ \hdashline
CCT-LSU-SUPERMIC  & HPC           & 2014-09-01    & 2014-10-01       &                &                & \textbf{A,J,N} \\ \hdashline
NICS-DARTER       & HPC           & 2014-04-14    & 2014-05-15       & 2016-06-30     & 2016-07-01 & \textbf{A,J,N}    \\ \hdashline
SDSC-COMET        & HPC           & 2014-09-14    & 2017-07-01       & 2019-01-30     &                & \textbf{A,J,N} \\ \hdashline
SDSC-COMET-GPU    & HPC           & 2014-09-14    & 2017-07-01       & 2019-01-30     &                & \textbf{A,J,N} \\ \hdashline
PSC-GREENFIELD    & HPC           & 2015-07-15    & 2015-08-05       & 2016-12-31     &                & \textbf{A} \\ \hdashline
TACC-WRANGLER    & DIC            &               & 2015-01-31       &                          
&                & \\ \hdashline
TACC-JETSTREAM    & Cloud         & 2015-11-01    & 2016-01-20       & 2019-11-30     &                & \textbf{A} \\ \hdashline
PSC-BRIDGES       & HPC           & 2016-01-01    & 2016-01-01       & 2019-11-30     &                & \textbf{A} \\ \hdashline
PSC-BRIDGES-LARGE & DIC           & 2016-01-01    & 2016-01-01       & 2019-11-30     &                & \textbf{A} \\ \hdashline
PSC-BRIDGES-GPU   & HPC           & 2017-03-13    & 2017-07-01       & 2019-11-30     &                & \textbf{A} \\ \hdashline
Stanford-XSTREAM  & HPC           & 2016-03-01    & 2016-07-01       &                &                & \textbf{A} \\ \hdashline
TACC-STAMPEDE2    & HPC           & 2017-03-14    & 2017-06-13       & 2021-09-30     &                & \textbf{A,J,N} \\  \hdashline
\bottomrule
\end{tabular}
\begin{tablenotes}
\item[1] HPC resource types: 
HPC - High-performance computing, 
HTC - High-throughput computing,
DIC - Data-intensive computing,
Cloud - Cloud resource and 
Vis - Visualization system
\end{tablenotes}
\end{threeparttable}
\end{table}

\newpage

\subsection{XD SU Conversions}
\label{appendix:SU Conversions}
Table~\ref{table:xdsu-conversion-factors} shows the conversion factors from local service units (local SU) to
XSEDE SU (XD SU). Although local SUs are intended to be costing units they are generally proportional to CPU core-hours (or node-hours in the case of \stampedetwo{} and \wrangler{}) but can be more complex formulas depending on the type of resource as described by \href{https://portal.xsede.org/allocations/policies}{XSEDE allocation policy}. For example, \jetstream{} defines a local SU as 1 virtual CPU core-hour plus 2 GB of memory.

\begin{table}[H]
\centering
\caption{Conversion Factors used to translate Local SU to XD SU.}
\label{table:xdsu-conversion-factors}
\begin{tabular}{ *4l S }
\toprule
Resource         & Type  & Factor Start & Factor End & {Conversion Factor}\\ \midrule
TACC STAMPEDE2   & HPC   & 2017-03-14   & 2017-06-06 & 73.584\\
                 &       & 2017-06-07   & Current    & 143.719\\
TACC STAMPEDE    & HPC   & 2012-09-15   & Current    & 4.599\\
IU Jetstream   & Cloud & 2016-01-01   & Current    & 6.856\\
TACC Maverick2   & Viz   & 2014-02-10   & Current    & 4.8375\\
TACC LONESTAR4   & HPC   & 2010-12-01   & 2011-04-14 & 2.9\\
                 &       & 2011-04-15   & 2014-12-31 & 3.288\\
TACC RANGER      & HPC   & 2008-01-01   & 2008-06-15 & 1.256\\
                 &       & 2008-06-16   & 2013-02-07 & 1.533\\
TACC WRANGLER    & DIC   & 2015-01-31   & Current    & 7.0\\
SDSC COMET       & HPC   & 2015-01-31   & Current    & 6.165\\
SDSC COMET GPU   & HPC   & 2017-04-12   & Current    & 86.31\\
SDSC GORDON      & DIC   & 2011-09-15   & Current    & 4.932\\
SDSC Trestles    & HPC   & 2011-01-01   & 2011-04-14 & 1.615\\
                 &       & 2011-04-15   & 2015-04-30 & 2.3\\
Stanford XSTREAM & HPC   & 2016-03-01   & Current    & 85\\
Purdue Steele    & HPC   & 2008-05-01   & 2013-07-31 & 1.606\\
Purdue Condor    & HTC   & 2005-10-01   & 2013-07-31 & 0.849\\
PSC Bridges      & HPC   & 2016-01-01   & Current    & 6.165\\
PSC Bridges Large & HPC  & 2016-01-01   & 2016-04-28 & 11.4944\\
                  &      & 2016-04-29   & Current    & 64.7325\\
PSC Bridges GPU  & HPC   & 2017-03-14   & 2017-04-24 & 1\\
                 &       & 2017-04-25   & Current    & 86.31\\
PSC Greenfield   & HPC   & 2015-07-15   & Current    & 11.4944\\
PSC Blacklight   & DIC   & 2011-01-01   & 2015-08-15 & 1.796\\
NICS Kraken      & HPC   & 2008-03-31   & 2008-08-03 & 1.075\\
                 &       & 2008-08-04   & 2009-02-01 & 1.691\\
                 &       & 2009-02-02   & 2009-10-04 & 1.623\\
                 &       & 2009-10-05   & 2014-05-01 & 2.04\\
NICS Darter      & HPC   & 2014-04-14   & 2016-07-01 & 2.028\\
NICS Nautilus    & Viz   & 2010-07-01   & 2015-04-20 & 1.572\\
LSU SuperMIC     & HPC   & 2014-04-01   & Current    & 4.599\\
LONI QueenBee    & HPC   & 2008-01-01   & 2011-07-31 & 1.606\\
GATech Keeneland & HPC   & 2012-07-01   & 2014-12-31 & 34\\
OSG              & HTC   & 2012-03-01   & 2015-02-11 & 1\\
                 &       & 2015-02-12   & Current    & 3.147\\
\bottomrule
\end{tabular}
\end{table}

\subsection{Science Gateways}
\label{appendix:data-science-gateways}

We define Science Gateways as described on the XSEDE User Portal Science Gateway page
(\url{https://portal.xsede.org/science-gateways}) which states that Science Gateways are ``Customized portals granting members access to HPC applications, workflows, shared data and other services. XSEDE Gateways unite
communities of like-minded members, whether united by discipline or other criteria.''

Historically, Science Gateway usage has been tracked in multiple ways, with varying degrees of completeness, including
\begin{enumerate}
\item Verbal communication with Science Gateway PIs,
\item Jobs run under a shared Community User account (one for each Gateway),
\item Jobs charged to an allocation that a Community User has utilized at some point, and
\item Jobs run under an allocation that is tagged as a "Science Gateway" grant
\end{enumerate}
The number of active and unique Gateway users has been tracked using two methods
\begin{enumerate}
\item Verbal communication with Science Gateway PIs and
\item Job attributes that link a Gateway user to a particular job that they submitted (supplied by the Gateway to the XDCDB starting in 2015-Q2)
\end{enumerate}

When examining Gateway jobs, we will primarily look at jobs that were run under a Gateway's shared Community User account as these appear to most closely match both the definition of a Science Gateway and also best captures Gateway Usage by end users. This will be supplemented by data collected through verbal communication with Gateway PIs, which is especially useful for examining active and unique Gateway users prior to 2015-Q2.

Jobs run under an allocation that a Community User has utilized at some point casts a larger net and includes jobs not submitted by Community User accounts including development and preliminary jobs. Comparing jobs submitted via Community User accounts versus jobs submitted against an associated allocation that the Community User utilized at some point we see that, with a few exceptions such as SciGaP and Tera3D, usage through the Community User accounts aligns fairly closely with that through associated allocations as shown in Table \ref{table:gateway-community-user-vs-allocation-jobs}.

Allocations that are tagged as a ``Science Gateway'' in the XDCDB appear to comprise a much broader definition of gateways and are not restricted to portals. In addition, jobs submitted via a Community User account are not always charged to an allocation tagged as a ``Science Gateway,'' making this method unreliable.  For example, the ChemCompute portal (\url{https://chemcompute.org/}) charged usage in 2017 to charge numbers TG-CDA160009 and TG-CDA170003 but only TG-CDA160009 is tagged as a "Science Gateway" allocation. Also of note is that many ``Science Gateway'' allocations are not listed as a Science Gateway on the XSEDE User Portal such as TG-PHY140033 ``Cloud Computing on Jetstream for the ATLAS Experiment at the Large Hadron Collider'' which appears to have grown out of the ATLAS Connect (\url{https://connect.usatlas.org/}) project.

\begin{table}[H]
\centering
\caption{Gateway jobs submitted via Community User accounts versus jobs submitted against allocations that the Community Users had access to. Data is limited to Gateways submitting more than 1,000 jobs.}
\label{table:gateway-community-user-vs-allocation-jobs}
\begin{tabular}{ l *4r }
\toprule 
\textbf{Gateway} & \multicolumn{2}{c}{\textbf{via Community User}} & \multicolumn{2}{c}{\textbf{via Associated Allocation}}\\
                 & \textbf{\# Jobs} & \textbf{SUs} & \textbf{\# Jobs}  & \textbf{SU}  \\
\midrule
Cipres                     & 840,082 & 102,700,214 & 892,827 & 111,083,931 \\
I-TASSER (zhanglab)        & 592,351 & 3,868,193   & 620,430 & 3,954,523 \\
Galaxy                     & 168,998 & 1,759,180   & 198,437 & 1,929,403 \\
Gridchem Chemistry         & 112,714 & 18,961,818  & 206,265 & 21,324,347 \\
SciGaP                     & 49,353  & 111,856     & 208,266 & 19,129,439 \\
Tera3D                     & 41,986  & 5,716,028   & 73,014  & 49,060,022 \\
Ultrascan3                 & 40,527  & 2,475,550   & 45,927  & 3,928,918 \\
Cmmap001                   & 39,586  & 1,208,171   & 47,931  & 6,976,294 \\
iplant                     & 34,343  & 2,221,089   & 54,707  & 8,208,360 \\
lddmmproc                  & 31,650  & 3,754,803   & 52,733  & 4,572,435 \\
Gridamp                    & 27,019  & 8,333,392   & 27,101  & 8,432,809 \\
UCI Social Science Gateway & 24,906  & 25,191      & 25,076  & 58,427 \\
OGCE                       & 23,887  & 298,587     & 229,031 & 23,998,196 \\
ROSIE                      & 21,766  & 1,946,403   & 24,043  & 3,093,687 \\
Neuroscience Gateway       & 18,091  & 8,527,274   & 225,785 & 24,803,875 \\
olam                       & 12,754  & 1,478,205   & 12,775  & 1,478,489 \\
ChemCompute                & 7,184   & 51,182      & 17,508  & 188,138 \\
Gisolve                    & 6,665   & 6,833,523   & 13,595  & 8,272,114 \\
ATLAS Connect              & 6,350   & 273,038     & 68,886  & 1,836,002 \\
C4e4                       & 5,312   & 1,669       & 5,684   & 36,473 \\
NEES                       & 4,577   & 1,956,594   & 20,517  & 2,567,117 \\
Nanohub                    & 2,387   & 483         & 6,732   & 17,224,963 \\
Ultrascan                  & 2,306   & 1,145,519   & 45,927  & 3,928,918 \\
DESDM                      & 1,527   & 40,845      & 3,937   & 520,599 \\
diagrid                    & 1,362   & 5,971       & 143,398 & 16,422,155 \\
\bottomrule
\end{tabular}
\end{table}

In addition to submitting jobs on behalf of its users using a shared Community User account, a gateway may also provide job attributes to the XDCDB to associate individual Gateway users (typically via an email address) with the jobs run on their behalf. These users are unique to a gateway and cannot be assumed to be unique across multiple gateways. Also note that not all Science Gateways provide job attributes, as shown in Table \ref{table:gateway-job-attribute-reporting}, and that no Gateway has complete coverage in this respect. To improve reporting by science gateways and facilitate the storing of job attributes in the XDCDB, the XSEDE Cyberinfrastructure Integration (XCI) and  XDCDB teams developed the \texttt{gateway-submit-attributes} tool. This tool allows a gateway to easily associate the local gateway user who submitted a job with that job's record in the XDCDB. The first record of job attributes being recorded using this method is 2015-03-05. What this means is that any information presented on the number of unique or new Gateway users is a lower bound and that prior to 2015-Q2 the tracking of unique Gateway users was far less reliable.

It should be noted that in the early stages of a Science Gateway's implementation, job attribute data may not be submitted to the XDCDB. For example, it may take some time for a Gateway to set up the infrastructure for submitting job attributes to track individual users and a community user may not have been set up for a gateway. In these cases, data has been supplemented by verbal communication with gateway PIs by the XSEDE Science Gateways Team which has provided this data to XMS.

\begin{table}[H]
\centering
\caption{Science Gateway jobs showing the first and last dates that a job was run and the number of jobs that provided attributes such as gateway user name during that period. Note that the two largest Gateways by CPU hours have less than 60\% coverage. Gateways in \textbf{boldface} are active as of the writing of this report.}
\label{table:gateway-job-attribute-reporting}
\begin{tabular}{ *3l *3r }
\toprule
\textbf{Gateway} & \multicolumn{2}{c}{\textbf{Gateway Deployment}} & \textbf{\# Jobs} & \textbf{CPU Hrs} & \textbf{w/Attrib}(\%)\\
 & \textbf{First Job} & \textbf{Last Job} &  &  & \\ \midrule
ATLAS Connect                & 2014-03-31 & 2015-05-14 & 6,350   & 187,409     & 0(0\%) \\
C4e4                         & 2011-07-02 & 2013-08-19 & 5,312   & 1,679       & 0(0\%) \\
Ccsmuser                     & 2011-07-01 & 2014-03-14 & 371     & 34,798      & 0(0\%) \\
\textbf{ChemCompute}         & 2017-01-03 & 2017-09-30 & 9,768   & 64,840      & 1,126(12\%) \\
CIGportal                    & 2011-07-01 & 2014-07-31 & 60      & 14,625      & 0(0\%) \\
\textbf{Cipres}              & 2011-07-01 & 2017-09-30 & 859,902 & 102,645,960 & 507,146(59\%) \\
Cmmap001                     & 2011-07-07 & 2012-05-10 & 39,586  & 1,208,171   & 0(0\%) \\
\textbf{COSMIC2}             & 2016-11-17 & 2017-09-30 & 72      & 502         & 61(85\%) \\
DES                          & 2011-07-26 & 2011-09-24 & 132     & 838         & 0(0\%) \\
DESDM                        & 2011-07-05 & 2012-06-09 & 1,527   & 40,806      & 0(0\%) \\
\textbf{diagrid}             & 2014-05-28 & 2017-09-30 & 1,377   & 6,672       & 104(8\%) \\
\textbf{dREG}                & 2016-10-11 & 2017-09-30 & 311     & 48,053      & 0(0\%) \\
GAAMP                        & 2013-03-28 & 2014-02-05 & 110     & 2,157       & 0(0\%) \\
\textbf{Galaxy}              & 2014-09-15 & 2017-09-30 & 176,968 & 1,888,196   & 5,979(3\%) \\
Gisolve                      & 2011-07-01 & 2015-09-16 & 6,665   & 6,831,836   & 70(1\%) \\
Gridamp                      & 2011-07-01 & 2017-06-19 & 27,019  & 8,759,981   & 14,742(55\%) \\
\textbf{Gridchem Chemistry}  & 2011-07-08 & 2017-09-30 & 114,991 & 19,891,823  & 28,588(25\%) \\
iplant                       & 2011-10-27 & 2015-04-01 & 34,343  & 790,854     & 0(0\%) \\
\textbf{lddmmproc}           & 2013-09-30 & 2017-09-30 & 33,139  & 3,777,986   & 22,842(69\%) \\
Medici Gateway               & 2015-07-15 & 2015-09-17 & 740     & 1,695       & 702(95\%) \\
mgcloud                      & 2015-02-03 & 2015-07-17 & 782     & 4,505       & 0(0\%) \\
Nanohub                      & 2011-07-01 & 2012-06-29 & 2,387   & 482         & 17(1\%) \\
\textbf{NEES}                & 2012-05-25 & 2017-09-30 & 4,628   & 3,076,664   & 318(7\%) \\
\textbf{Neuroscience Gateway} & 2012-09-19 & 2017-09-30 & 18,588  & 8,753,898   & 11,956(64\%) \\
OGCE                         & 2011-07-11 & 2017-05-21 & 23,887  & 1,534,618   & 4(0\%) \\
olam                         & 2011-07-01 & 2012-11-27 & 12,754  & 1,478,320   & 0(0\%) \\
Parametrization Gateway      & 2013-08-12 & 2014-01-07 & 350     & 44          & 0(0\%) \\
PhastaCommunity              & 2017-01-17 & 2017-04-25 & 25      & 2,216       & 0(0\%) \\
ROSIE                        & 2013-10-08 & 2017-02-02 & 21,766  & 1,946,413   & 10,486(48\%) \\
sbwp                         & 2012-07-09 & 2014-01-17 & 499     & 548,653     & 86(17\%) \\
\textbf{SciGaP}              & 2015-06-04 & 2017-09-30 & 50,770  & 112,603     & 27(0\%) \\
Sidgrid                      & 2011-07-23 & 2012-01-05 & 428     & 31,394      & 0(0\%) \\
Simpleg                      & 2011-07-13 & 2014-06-27 & 128     & 2,526       & 0(0\%) \\
\textbf{SimVascular}         & 2017-04-25 & 2017-09-30 & 92      & 71          & 8(9\%) \\
TAS                          & 2012-04-27 & 2017-09-30 & 141,768 & 1,561,216   & 0(0\%) \\
Tera3D                       & 2011-07-01 & 2015-10-27 & 41,986  & 5,729,699   & 55(0\%) \\
UCI Social Science Gateway   & 2013-11-04 & 2017-09-03 & 24,906  & 843         & 24,421(98\%) \\
Ultrascan                    & 2011-07-01 & 2016-07-08 & 2,306   & 916,757     & 0(0\%) \\
\textbf{Ultrascan3}          & 2011-07-20 & 2017-09-30 & 40,617  & 2,421,479   & 1,814(4\%) \\
vdjserver                    & 2016-10-12 & 2016-12-02 & 57      & 25          & 15(26\%) \\
\textbf{waterhub}            & 2014-04-29 & 2017-09-30 & 859     & 206,562     & 727(85\%) \\
\textbf{I-TASSER (Zhanglab)} & 2016-07-06 & 2017-09-30 & 723,577 & 2,973,032   & 29,949(4\%) \\
\bottomrule
\end{tabular}
\end{table}

\subsection{Job Performance Data}\label{appendix:taccstats}

Job performance data was obtained from \texttt{tacc\_stats} \cite{evans2014comprehensive} running on the
compute nodes of \ranger{}, \lonestar{}, \stampede{}, \stampedetwo{},
\texttt{LSU SUPERMIC}, \texttt{SDSC GORDON}
and \texttt{SDSC COMET}.  The performance data from \darter{} was obtained from the Cray
RUR software~\cite{Barry:2013}

The \texttt{tacc\_stats} software runs on every compute node and records a large
number of metrics. The data are written to an archive file and there is one file per compute node per day. The software is executed at
10 minute intervals using \texttt{cron} and is also called from the job scheduler prolog
and epilog scripts to record the metrics immediately before each HPC job begins and
just after each job ends.
These data are used to generate summary information for each HPC job using the
data collected on the compute nodes on which each job ran.

The summarization software uses the accounting data from the resource manager and the
compute node level archives and generates various metrics for each job.
There are two semantic metric types, instantaneous metrics and counter metrics. 
Examples of instantaneous metrics are memory usage or number of processes in the
O/S run queue. Examples of counter metrics are number of floating point operations
or number of clock ticks in CPU user. 
The summarization software treats the two semantic metric types differently. For counter
metrics, the total change in the counter value is typically used and is calculated as
the value recorded after the end of the job minus the value recorded immediately
before the start.  The instantaneous metrics are sub-sampled so any derived data from these
are estimates. The values of any instantaneous metric before job begin and after job end 
are not used when computing derived data.

In the following sections we describe the job level metrics in more detail.

\subsubsection{CPU}

The CPU usage data is obtained from the CPU statistics reported
in \texttt{/proc/stat}. The various CPU metrics
are computed as the ratio of time spent in the different modes (user, system, etc.) to the overall time
as reported in the proc file. The CPU metrics are counter metrics so the value for the HPC job
is the value of the counter at the end minus the value at the beginning.

\subsubsection{Runnable threads}

The number of runnable threads is obtained from the \texttt{procs\_running} field in \texttt{/proc/stat}.
This value is an instantaneous metric so the measurements made at 10 minute intervals during the
job are used.

\subsubsection{Memory}
The memory data is obtained from the \texttt{/sys/devices/system/node/node*/meminfo} files for
each numa node on the compute node. The value of the memory usage per core $m$ at time $i$ is given by
\begin{equation*}
    m_i = \frac{1}{n_{\text{cores}}} \sum_{\text{numa nodes}} \left( \text{MemTotal} - \text{MemFree} - \text{FilePages} - \text{Slab} \right)
\end{equation*}
where $n_{\text{cores}}$ is the total number of CPU cores per compute node, MemTotal is the total amount of memory, MemFree the amount of unused memory, FilePages the amount
of memory used by the kernel page cache and Slab the amount of memory used by the Linux kernel
data structures cache.

The memory metrics are instantaneous metrics so the measurements at the beginning and end of
the job are not used to compute the job summary values. The average memory usage is
the mean value of all of the measurements. The maximum memory usage is the value of the largest
measurement. All of the memory metrics are estimates of the actual usage since the memory
statistics are only sampled at 10 minute intervals during a job.

\subsubsection{I/O}

All of the resources that run \texttt{tacc\_stats} use the Lustre parallel filesystem. The metrics for Lustre are obtained from
the \texttt{/proc/sys/lnet/stats} file. The \texttt{tx\_bytes} and \texttt{rx\_bytes} values are used
for the data transmit and receive metrics respectively.

\subsubsection{Interconnect}

All of the resource use InfiniBand. The \texttt{ibmad} library is used to query the port statistics
on the InifiniBand switch. 

\subsubsection{Applications}
\label{appendix:appident}

The application associated with a job is derived from the executable path information. 
If Lariat~\cite{Lariat2013} or XALT~\cite{Agrawal:2014:UET:2691136.2691140} information is present then this is used as the source of the 
executable information\footnote{Lariat is available on \ranger{}, \lonestar{}
and \stampede{}. XALT data is available on \stampede{} and \stampedetwo{},
however Lariat is the source for the data presented in this report.}. If Lariat
or XALT data is absent then the process information from \texttt{tacc\_stats}
is used. If neither Lariat, XALT or \texttt{tacc\_stats} data is available then
the application is marked as ``NA''.

Lariat or XALT provide the path name of the main parallel process in the job. 
\texttt{tacc\_stats} does not have that information and instead
records the list of processes running on the compute nodes (ignoring common system
daemons). When this process data is used, there are potentially multiple different processes 
to choose from some of which may be software run for setup or tear down steps and not the main
job step. To attempt to identify the main step for the job, the classification algorithm 
records the number of unique process ids (PIDs) for each process name and checks
them from most to fewest until it finds a process name that is not in the ignore list. The
ignore list contains common unix processes such as \texttt{cp} and \texttt{bash}. This
algorithm therefore typically picks up the main parallel processing step in an HPC job rather than, say, the data copy in or copy out steps.

Once the main executable is identified then this string is checked against reference database
of known community applications. The reference database has 188 known applications and several
application categories, such as debugging tools (e.g.\ \texttt{ddd}, \texttt{gdb}) and 
interactive session commands (e.g.\ \texttt{xterm}). Applications are identified
by regular expressions. If no application is matched then the job is placed in the
``uncategorized'' category. A machine learning based study was done previously on the ``uncategorized'' jobs~\cite{Gallo2015}.  A model was developed that was able to classify jobs with better than 90\% accuracy.  It was found that 80-90\% of the ``uncategorized'' jobs were in fact custom user code and only 10-20\% of the ``uncategorized'' jobs were community applications that had been missed by our regular expression classification scheme.  

\newpage
\printbibliography

\end{document}